***INSTITUT NATIONAL POLYTECHNIQUE DE GRENOBLE***

***<u>T H E S E</u>***

pour obtenir le grade de

**DOCTEUR DE L'INP Grenoble**

***Spécialité :*** *Optique et Radiofréquences*

préparée au Département Optronique du
Laboratoire d'Electronique et des Technologies de l'Information (CEA-LETI)

dans le cadre de **l'Ecole Doctorale Electronique, Electrotechnique, Automatique, Traitement du Signal**

présentée et soutenue publiquement

par

**Pierre Labeye**

Le 06 Février 2008

# Composants optiques intégrés pour l'interférométrie astronomique

Directeur de thèse : **Pr. Jean-Emmanuel Broquin**

**JURY**

M. Pierre Saguet , Président
M. Guy Perrin , Rapporteur
M. François Ladouceur , Rapporteur
M. Jean-Emmanuel Broquin , Directeur de thèse
M. Jean-Philippe Berger , Examinateur
M. Serge Valette , Examinateur

## Composants intégrés optiques pour l'interférométrie astronomique

Il a été récemment démontré que l'optique intégrée pouvait considérablement améliorer les résultats obtenus en interférométrie astronomique en terme de précision, de stabilité, et de facilité de mise en œuvre.
Cette thèse porte sur l'étude d'un composant optique à base de guides d'onde monomodes destiné à recombiner de manière cohérente les faisceaux provenant de quatre télescopes. L'architecture proposée permet de fournir simultanément une mesure instantanée de la visibilité et de la phase des franges pour les six paires possibles de télescopes.
Le composant étudié est optimisé de manière à être achromatique sur la bande H de transparence de l'atmosphère et comporte un élément déphaseur achromatique original permettant d'obtenir quatre états de franges d'interférence en quadrature de phase.
Les résultats de caractérisation optique obtenus sur une série de dispositifs réalisés par dépôt et gravure de couches de silice sur substrat de silicium confirment les prédictions théoriques et permettent l'étude de composants plus complexes destinés aux projets d'instrumentation de seconde génération du VLTI (Very Large Telescope Interferometer).
Cette étude montre que l'architecture proposée est extensible aux bandes d'observation J et K, extensible à la recombinaison de faisceaux issus de six voire huit télescopes et peut être aussi avantageusement utilisée pour réaliser un suiveur de franges.

*Mots clés :* Optique intégrée, interférométrie, interférométrie astronomique

## Integrated optics components for stellar interferometry

It has been recently demonstrated that integrated optics could enhance accuracy, stability, and ease of use of stellar interferometry techniques.
The subject of this thesis is the study of an optical component based on singlemode waveguides for the coherent recombining of optical beams coming from four telescopes. Proposed architecture provides a simultaneous and instantaneous measurement of complex visibility of interferometric signals of the six possible pairs of telescopes.
The component is optimized for achromatic behaviour over the H spectral transparency band of atmosphere and integrates an original achromatic phase shifter in order to obtain four phase quadrature states of interferometric fringes.
Optical characterisation results obtained on devices realized by deposition and etching of silica layers on silicon substrate confirm theoretical predictions et enabled the study of more complex components at the heart of second generation instrumentation projects of the VLTI (Very Large Telescope Interferometer)
This study shows that proposed architecture can be extended to J and K spectral bands, can be applied for the recombination of beams coming from six and even eight telescopes and could also be applied to realise a fringe tracker.

*Keywords:* Integrated optics, interferometry, stellar interferometry

*A mes filles*

*à ma famille et à mes proches*

*« Le véritable voyage de découverte ne consiste pas à chercher de nouveaux paysages,*

*mais à avoir de nouveaux yeux. »*

*Marcel Proust*

# Remerciements

*Ce travail a été initié dans un cadre un peu particulier : étant à l'origine ingénieur en charge des projets optique intégrée pour l'astronomie au LETI, j'ai voulu approfondir le coté scientifique de l'un de ces projets en lui donnant cet aspect de thèse de doctorat.*

*Je tiens donc tout d'abord à remercier Claude Massit, chef du département intégration hétérogène sur silicium, pour avoir autorisé cette démarche qui n'est pas si habituelle au LETI. Je remercie ensuite Pascale Berruyer, qui, à l'origine, m'a confié cette activité au sein de son laboratoire. Enfin, je remercie particulièrement Xavier Hugon, en charge du département optronique du LETI au sein duquel j'ai pu terminer ce travail, et qui a continué à me soutenir lorsque j'ai rejoint son département.*

*Jean-Emmanuel Broquin a accepté d'encadrer scientifiquement ce travail. Qu'il trouve ici l'expression de ma plus profonde gratitude pour le soutien qu'il m'a apporté, son indulgence vis à vis de mes nombreuses bourdes administratives, son efficacité à les rattraper, ainsi que les nombreux et précieux conseils, scientifiques et humains, qu'il m'a donné aux moments les plus importants.*

*Je tiens aussi a exprimer ma gratitude à Monsieur Pierre Saguet, professeur à l'INPG qui m'a fait l'honneur d'accepter de présider le jury de cette thèse, ainsi qu'à Guy Perrin, directeur de recherche à l'observatoire de Paris Meudon, et à François Ladouceur, Professeur à l'université de Sydney, pour l'intérêt qu'ils ont porté à ce travail en acceptant d'en être les rapporteurs.*

*Je tiens aussi à remercier Serge Valette, d'une part pour m'avoir qui fait découvrir il y a déjà un certain temps le domaine de l'optique intégrée dont il est un des pionniers, d'autre part pour avoir accepté de participer à ce Jury.*

*Je tiens enfin à remercier tous mes collègues du LETI qui ont participé à cette étude :*

*Tout d'abord, Patrice Noël, technicien supérieur, dont la bonne humeur et la compétence n'ont d'égal que son astuce à utiliser au mieux les possibilités des salles blanches du LETI. Les résultats présentés témoignent de son œuvre. Un grand grand merci, Patrice !*

*Evelyne Desgranges, pour sa rigueur dans la réalisation des masques de photolithographie.*

*Guy Serra, qui a fait des miracles pour découper ces composants malgré un plan de découpe très particulier.*

*Tous les opérateurs de la plateforme silicium du bâtiment 41 pour la qualité de leur participation à la réalisation des composants.*

*Enfin, j'adresse ici un grand merci à mes autres collègues du LETI. Que nos chemins se soient croisés en DIHS ou DOPT, en LCEMO, en LCMS, en LMNO, ou encore en LPS..., c'est grâce à vous que ces dernières années ont été si riches pour moi.*

*Au cours de ce travail, j'ai eu l'honneur d'être invité à travailler au Laboratoire d'astrophysique de l'observatoire de Grenoble où je tiens à remercier un grand nombre de personnes qui m'ont beaucoup aidé à réaliser ce travail :*

*Tout d'abord, je remercie Thierry Montmerle, directeur du LAOG, pour m'avoir accueilli dans son laboratoire pendant ces six mois.*

*Ensuite je tiens à remercier Pierre Kern, directeur technique du LAOG, qui depuis les premiers jours, œuvre avec succès à la réussite de la collaboration entre le LAOG et le LETI, au gré de ses nombreuses et fréquentes réorientations et réorganisations. Merci Pierre pour ta gentillesse, ta patience, nos nombreuses discussions techniques ou autres, pour ta compréhension et ta faculté d'adaptation au fonctionnement du CEA-LETI !*

*Je remercie aussi Fabien Malbet et Karine Perraut, pour leur accueil, leur bonne humeur , et la patience dont ils ont fait preuve en répondant à toutes mes questions de néophyte sur l'instrumentation astronomique.*

*Je tiens à remercier tout particulièrement Jean-Philippe Berger, astronome, pour sa participation active et décisive à TOUS les aspects de ce travail. Merci, Jean-Philippe, de m'avoir fait découvrir l'instrumentation astronomique au sommet du Mont Hopkins ! Merci pour nos nombreuses discussions sur l'instrumentation en général aussi bien que sur la conception ou la caractérisation des puces en particulier! Merci pour ta relecture de ce manuscrit ! Merci de m'avoir fait l'honneur de participer au Jury !*

*Je tiens également à exprimer mes remerciements à Laurent Jocou pour son enthousiasme, sa jovialité, et pour l'exceptionnelle qualité de ses travaux, que ce soit sa participation à la conception des masques de photolithographie, le collage des puces, ou encore l'impressionnante qualité de son banc de caractérisation interférométrique.*

*Enfin, j'adresse un grand merci à toute l'équipe IONIC pour son accueil chaleureux, sa bonne ambiance, son enthousiasme, ainsi que pour l'extrême compétence dont ils font constamment preuve en mettant en œuvre ces composants intégrés et en les utilisant de la meilleure manière possible. Merci Myriam, Jean-Baptiste, Emilie, Alain, Lucas, Brahim, Etienne, ...*

*Pour finir, je ne saurais exprimer suffisamment l'immense gratitude que j'éprouve vis-à-vis de tous mes proches, dont le soutien sans failles, surtout aux heures les plus difficiles, ont permis à ce travail d'aboutir.*

*Merci à ma relectrice préférée...*
*et à tous ceux que j'ai oubliés...*

# Résumé

Il a été récemment démontré que l'optique intégrée pouvait considérablement améliorer les résultats obtenus en interférométrie astronomique en termes de précision, de stabilité, et de facilité de mise en œuvre.

Cette thèse porte sur l'étude d'un composant optique à base de guides d'onde monomodes destiné à recombiner de manière cohérente les faisceaux provenant de quatre télescopes à des fins d'imagerie par synthèse d'ouverture. L'architecture proposée permet de fournir simultanément une mesure instantanée de la visibilité et de la phase des franges pour les six paires possibles de télescopes.

Le composant étudié est optimisé de manière à être achromatique sur la bande H de transparence de l'atmosphère, comporte un tricoupleur pour séparer chaque faisceau incident en trois faisceaux, la recombinaison avec chacun des trois autres faisceaux se faisant à l'aide d'une cellule « ABCD » permettant de fournir quatre états de franges en quadrature de phase. Pour cela, la cellule « ABCD » comporte un élément déphaseur original optimisé pour fournir un déphasage de 90 degrés achromatique sur la bande H.

Les résultats de caractérisation optique obtenus sur une série de dispositifs réalisés par dépôt et gravure de couches de silice sur substrat de silicium confirment les prédictions théoriques avec une séparation en trois faisceaux parfaitement équilibrée et un déphasage moyen mesuré de 75 degrés avec une variation spectrale de 0,015 degrés par nanomètre.

Ces résultats validant les modèles de conception permettent le dimensionnement de composants plus complexes destinés aux projets d'instrumentation de seconde génération du VLTI (Very Large Telescope Interferometer). Ainsi, il est montré que l'architecture proposée peut être étendue aux bandes spectrales d'observation J et K, peut être étendue à la recombinaison de faisceaux issus de six voire huit télescopes pour laquelle une configuration alternative est aussi proposée et peut être aussi avantageusement utilisée pour réaliser un suiveur de franges optique intégré.

# Abstract

It has been recently demonstrated that integrated optics could enhance accuracy, stability, and ease of use of stellar interferometry techniques.

The subject of this thesis is the study of an optical component based on singlemode waveguides for the coherent recombining of optical beams coming from four telescopes. Proposed architecture provides a simultaneous an instantaneous measurement of complex visibility of interferometric signals of the six possible pairs of telescopes.

The component is optimized for achromatic behaviour over the H spectral transparency band of atmosphere, integrates a three waveguides evanescent coupler to split each incoming beam into three beams. The recombination of each beam is done within an "ABCD" scheme providing four quadrature states of interferometric fringes. For this purpose, an original phase shifter that provides an achromatic phase shift of 90 degrees over the H-band is introduced.

Optical characterisation results obtained on devices realized by deposition and etching of silica layers on silicon substrate confirm theoretical predictions with an average measured phase shift of 75° and a spectral variation of 0.015 degrees per nanometer.

These results validate the design method and enabled the study of more complex components at the heart of second generation instrumentation projects of the VLTI (Very Large Telescope Interferometer). This study shows that proposed architecture can be extended to J and K spectral bands, can be applied for the recombination of beams coming from six and even eight telescopes for which an alternate configuration is proposed and could also be applied to realise a fringe tracker.

# Table des matières

# Introduction

La recherche en instrumentation astronomique est gouvernée par deux demandes principales : plus de sensibilité pour détecter de nouveaux objets célestes moins brillants et plus de résolution pour obtenir des images plus complexes ou provenant d'objets plus éloignés. Ainsi de nombreuses thématiques scientifiques importantes comme l'étude de la formation des systèmes planétaires ou celle des trous noirs au cœur des galaxies actives ont un important besoin de résolution angulaire accrue. Si augmenter la taille de la surface collectrice d'un télescope permet d'améliorer en théorie ces deux points, la limitation principale de l'observation astronomique au sol provient en fait de l'instabilité de l'atmosphère qui dégrade la qualité des images obtenues. Même avec l'introduction ces dernières années de l'optique adaptative qui permet de s'affranchir partiellement en temps réel des turbulences atmosphériques, les plus gros télescopes actuels (10 m de diamètre) sont seulement capables de résoudre une petite dizaine d'étoiles parmi les plus grosses et les plus proches.

L'autre manière d'obtenir une meilleure résolution angulaire sur les sources lumineuses observées consiste à utiliser des méthodes de mesure interférométriques : en recombinant de manière cohérente la lumière provenant d'un ensemble de télescopes pointant la même source lumineuse, il est en effet possible d'obtenir des informations avec une résolution équivalente à celle d'un télescope virtuel qui engloberait les télescopes utilisés.

Si cette technique éminemment attractive fut envisagée dès le dix-neuvième siècle, elle connut des développements longtemps freinés par les difficultés instrumentales associées. En effet : pour produire des franges d'interférence, le réseau de télescopes doit contrôler les chemins optiques parcourus par les différents faisceaux à l'échelle d'une fraction de longueur d'onde, sachant que cette stabilité n'est déjà pas assurée par l'atmosphère avant même que la lumière ne parvienne à l'instrument.

Longtemps limitée à deux télescopes, cette technique connaît actuellement un nouvel essor avec la construction de plusieurs installations comprenant de deux à huit télescopes. Les progrès considérables en mécanique permettent maintenant d'assurer une bien meilleure stabilité. Les progrès non moins considérables en électronique permettent de réaliser les boucles de contre-réaction nécessaires à la correction de l'instabilité atmosphérique et des instabilités résiduelles de l'instrument. Enfin, la constante amélioration des détecteurs permet d'obtenir une meilleure sensibilité et un rapport signal sur bruit compatible avec le traitement de données nécessaire à la reconstruction d'images très haute résolution.

Ainsi, à ce jour, quatre installations ont été en mesure de recombiner plus de deux faisceaux optiques à des fins d'imagerie.

C'est dans ce contexte de constante amélioration de l'instrumentation qu'est apparue l'utilisation de l'optique guidée au cours de ces dernières années. La capacité d'un guide optique à filtrer spatialement les faisceaux optiques et à transformer les fluctuations de phase en fluctuations d'intensité mesurables permet d'accroître sensiblement la précision obtenue sur les mesures des deux observables de l'interférométrie que sont la visibilité et la phase des franges d'interférence obtenues.

En 1996, P. Kern & F. Malbet, du laboratoire d'astrophysique de Grenoble (LAOG), proposèrent d'utiliser des composants optiques intégrés pour la recombinaison elle-même, pressentant que les nombreuses fonctions développées pour les télécommunications optiques pourraient être judicieusement utilisées pour la recombinaison de plusieurs faisceaux en interférométrie astronomique. En effet, la possibilité de réaliser des schémas de guides complexes sur de petites surfaces permet d'envisager la recombinaison d'un plus grand nombre de faisceaux sans pour autant augmenter la complexité du banc complet de recombinaison. Cependant, si l'optique intégrée passive est depuis déjà quelques années une technologie mature et industrielle, un effort de recherche et développement fut nécessaire pour obtenir des composants utilisables en interférométrie astronomique où les contraintes sont différentes des télécommunications, notamment en terme de plage de longueur d'onde d'utilisation. Ainsi, à la suite de premiers travaux initiés par J.P Berger puis continués par P. Haguenauer, le principe de la recombinaison en optique intégrée fut validé aussi bien en laboratoire que sur le ciel pour deux puis trois télescopes.

C'est dans le contexte des excellents résultats obtenus avec ces dispositifs que s'inscrit le travail exposé dans ce document et qui a consisté à étendre ces concepts à la recombinaison de quatre ou plus de télescopes. Au début de cette étude, les résultats confirmaient le bon comportement de l'optique intégrée pour la recombinaison interférométrique, mais se limitaient à l'application directe de fonctions élémentaires optiques intégrées existantes : des jonctions Y utilisées pour la séparation et la recombinaison des faisceaux en technologie échange d'ions sur verre, et à l'utilisation de coupleurs asymétriques en technologie silice sur silicium. Mon travail a consisté, à partir des résultats obtenus sur ces dispositifs de première génération, à concevoir, réaliser, puis tester des dispositifs spécialement conçus pour la mesure interférométrique, d'une part en optimisant les fonctions optiques existantes et d'autre part en intégrant une nouvelle fonction de mesure instantanée de la phase des franges d'interférence. Ceci nous a permis de réaliser une puce recombinant simultanément quatre faisceaux en permettant la mesure instantanée de toutes les observables interférométriques (contraste et phase des franges sur chacune des six paires recombinées). La disponibilité de ce dispositif a contribué au développement d'un instrument de caractérisation interférométrique multi-faisceaux original au LAOG qui en retour a permis de caractériser finement ce dispositif. Enfin, les connaissances acquises à la suite de ces caractérisations m'ont permis d'étudier l'utilisation de ces dispositifs à la recombinaison à quatre ou six télescopes dans le cadre du projet VSI qui a pour but de définir l'instrument de recombinaison de prochaine génération du VLTi. Par ailleurs, les résultats obtenus sur le dispositif réalisé sont tels que son installation sur l'interféromètre CHARA est actuellement en cours d'étude afin d'obtenir des résultats sur le ciel.

La suite de ce document est organisée en quatre chapitres. Dans le premier chapitre, nous présentons l'interférométrie astronomique en exposant ses principes, ses avantages, et ses limitations puis nous expliquons les apports de l'optique guidée et intégrée dans cette discipline. Dans le deuxième chapitre, nous abordons le problème de la recombinaison de faisceaux en optique intégrée en détaillant les différentes architectures possibles pour la recombinaison de deux, trois, ou quatre faisceaux, puis nous exposons le principe de la recombinaison dite 'ABCD' permettant la mesure directe de la phase des franges d'interférence obtenues. Dans le troisième chapitre, nous exposons les réalisations effectuées au cours de cette étude et les résultats de caractérisation obtenus. Enfin, dans le quatrième chapitre, nous décrivons comment, à partir des résultats obtenus, nous avons étudié dans le cadre du projet VSI, les architectures possibles des prochains dispositifs qui pourraient être réalisés pour cet instrument.

# Chapitre I : Interférométrie astronomique et optique intégrée

## *I – A - Historique*

L'aventure débute avec Armand Hyppolyte Fizeau qui, le premier, en 1868 émit l'idée qu'il était possible d'obtenir des informations sur le diamètre angulaire des étoiles en réalisant des franges d'interférence au foyer des grands instruments d'observation astronomique de l'observatoire de Marseille. En 1873, Stephan mit l'idée de Fizeau en œuvre en réalisant des franges d'interférence au foyer du télescope de 80 cm de diamètre construit par Foucault 10 ans auparavant [*1*][*2*]. La séparation entre les faisceaux étant limitée à 65 cm, il ne put mesurer le diamètre d'une étoile, mais conclut que toutes les étoiles observées avaient un diamètre angulaire inférieur à 0,16 seconde d'arc. Outre Atlantique, Albert Michelson publia en 1890 une formalisation mathématique très précise de la mesure de diamètre angulaire de sources lumineuses [*3*], et publia les premières mesures de diamètre angulaire de corps célestes par interférométrie en 1891. Il utilisa pour cela un masque sur un télescope de 12 pouces (30,5 cm) et mesura la taille des quatre premiers satellites de Jupiter. En 1895, Karl Schwarzschild réalisa la première mesure sur une étoile binaire, mais il fallut ensuite attendre 1921 et la réalisation d'un interféromètre ayant une base de 20 pieds (6,1m) pour que Michelson et Pease publient la première mesure [*4*] de diamètre angulaire d'une étoile (Bételgeuse). Le projet suivant, un interféromètre ayant une base de 50 pieds se heurta aux difficultés inhérentes à l'interférométrie astronomique : une sensibilité accrue à la turbulence atmosphérique due à un plus grand diamètre des miroirs utilisés et une plus grande longueur de base, des vibrations mécaniques importantes, des problèmes de polarisation… C'est pourquoi cet instrument ne permit pas de fournir de résultats fiables.

Il fallut attendre 1974 pour que de tels problèmes instrumentaux soient résolus et que l'étape décisive suivante soit franchie : la réalisation de franges d'interférence en recombinant les faisceaux issus de deux télescopes séparés. Ainsi Johnson réalisa les premières franges à une longueur d'onde de 10µm [*5*] sur la planète Mercure et Labeyrie obtint les premières franges directes dans le domaine visible et proche infrarouge sur Vega [*6*]. Les progrès instrumentaux se succédèrent ensuite à un rythme plus rapide. En 1980 eurent lieu les premières mesures avec un système de suivi de franges [*7*] qui permet un temps d'intégration plus long sur le détecteur, donc une meilleure sensibilité. Au début des années 1980, l'idée d'utiliser des fibres optiques fut introduite par Froelhy [*8*], mise en oeuvre en laboratoire par Shaklan [*9*][*10*], et en 1996, Kern & Malbet introduisirent l'idée d'utiliser l'optique intégrée dans l'instrumentation pour recombiner plus efficacement les faisceaux [*11*].

Enfin, la première image par synthèse d'ouverture (de Capella) fut obtenue et publiée par Baldwin sur COAST (Cambridge Optical Aperture Synthesis Telescope) en 1996 [*12*] suivie de peu par Benson en 1997 sur NPOI (Navy Prototype Optical Interferometer) [*13*], et Monnier en 2004 a publié la première image obtenue à l'aide d'un instrument utilisant un recombineur optique intégré sur IOTA (Infrared Optical Telescope Array) [*14*]. Très récemment (2007), Monnier a publié une image de la surface d'Altair obtenue sur CHARA (Center for High Angular Resolution Astronomy) en utilisant quatre télescopes [*15*]. A ce jour, seuls ces quatre interféromètres ont fourni des images, mais les résultats scientifiques obtenus par interférométrie se multiplient ces dernières années : mesures d'étoiles naines ou géantes, évolution de la rotation des étoiles binaires, effet d'aplatissement des étoiles à rotation très rapide, présence d'enveloppes gazeuses autour de certaines étoiles, etc…

# I – B – Réseaux de télescopes existants

| Acronyme | Nom complet | Institution | Lieu | Date |
|---|---|---|---|---|
| CHARA | Center for High Angular Resolution Astronomy | Georgia State University | Mt Wilson, CA, USA | 2000 |
| COAST | Cambridge Optical Aperture Synthesis Telescope | Cambridge University | Cambridge, England | 1992 |
| GI2T | Grand Interféromètre à 2 Télescopes | Observatoire Côte d'Azur | Plateau de Calern, France | 1985 |
| IOTA | Infrared Optical Telescope Array | Smithsonian Astrophysical Observatory, University of Massachusetts (Amherst) | Mt Hopkins, AZ, USA | 1993 |
| ISI | Infrared Spatial Interferometer | University of California at Berkeley | Mt Wilson, CA, USA | 1988 |
| Keck-I | Keck Interferometer (Keck-I to Keck-II) | NASA-JPL | Mauna Kea, HI, USA | 2001 |
| MIRA-I | Mitake Infrared Array | National Astronomical Observatory, Japan | Mitaka Campus, Tokyo, Japan | 1998 |
| NPOI | Navy Prototype Optical Interferometer | Naval Research Laboratory, US Naval Observatory | Flagstaff, AZ, USA | 1994 |
| PTI | Palomar Testbed Interferometer | NASA-JPL | Mt Palomar, CA, USA | 1996 |
| SUSI | Sydney University Stellar Interferometer | Sydney University | Narrabri, Australia | 1992 |
| VLTI-UT (VIMA) | VLT Interferometer (Unit Telescopes) | European Southern Observatory | Paranal, Chile | 2001 |
| Keck* | Keck Auxiliary Telescope Array | NASA-JPL | Mauna Kea, HI, USA | ~2004? |
| LBTI* | Large Binocular Telescope Interferometer | LBT Consortium | Mt Graham, AZ, USA | ~2006 |
| MRO* | Magdalena Ridge Observatory | Consortium of New Mexico Institutions, Cambridge University | Magdalena Ridge, NM, USA | ~2007 |
| OHANA* | Optical Hawaiian Array for Nanoradian Astronomy | Consortium (mostly French Institutions, Mauna Kea Observatories, others) | Mauna Kea, HI, USA | ~2005 |
| VLTI-AT* (VISA) | VLT Interferometer (Auxiliary Telescopes) | European Southern Observatory | Paranal, Chile | ~2004 |

*Tableau I-1: les principaux interféromètres stellaires dans le monde.*

Le tableau I-1 donne la liste des interféromètres stellaires actuellement en fonctionnement avec leurs principales caractéristiques (les * indiquent les projets en cours).

Le développement de ces interféromètres est actuellement très actif [*16*][*17*][*18*][*19*][*20*][*21*][*22*][*23*][*24*], notamment avec l'introduction de l'optique adaptative, et de l'optique guidée, mais à ce jour, seuls quatre interféromètres ont fonctionné à plus de deux télescopes (COAST, NPOI, IOTA, CHARA). On est donc encore en retrait si l'on compare ces résultats au grand interféromètre radiométrique VLA (Very Large Array) qui recombine 27 télescopes.

Le tableau I-2 reprend ces différents interféromètres pour en donner les principales caractéristiques techniques.

| Acronyme | Nombre de télescopes | Taille des télescopes(m) | Séparation max des télescopes(m) | Couverture spectrale |
|---|---|---|---|---|
| CHARA | 6* | 1.0 | 330 | visible*, proche IR |
| COAST | 5 | 0.40 | 47(100*) | visible, proche IR |
| GI2T | 2 | 1.52 | 65 | visible, proche IR |
| IOTA | 3 | 0.45 | 38 | visible, proche IR |
| ISI | 3* | 1.65 | 85(>100*) | IR moyen |
| Keck-I | 2 | 10.0 | 85 | procheIR,IRmoyen* |
| MIRA-I | 2 | 0.25 | 30 | visible |
| NPOI | 6 | 0.12 | 64(>250*) | visible |
| PTI | 3 | 0.40 | 110 | Proche IR |
| SUSI | 2 | 0.14 | 64(640*) | visible |
| VLTI-UT | 4 | 8.0 | 130 | Proche IR, IR moyen |
| Keck* | 4* | 1.8 | 140* | Proche IR |
| LBTI* | 2* | 8.4 | 23* | Proche IR, IR moyen |
| MRO* | ~10 | ~1.5 | ~1000 | visible, proche IR |
| OHANA* | ~6 | 3.5-10 | 756 | proche IR |
| VLTI-AT* | 3* | 1.8 | 202 | procheIR, IR moyen |

*Tableau I-2 : les caractéristiques instrumentales des principaux interféromètres.*

# *I – C – Principe de l'interférométrie astronomique*

## I – C – 1 – Résolution d'un télescope

La limite de résolution angulaire d'un télescope est très simplement régie par les lois de la diffraction : elle est proportionnelle à la longueur d'onde d'observation $\lambda$ et inversement proportionnelle au diamètre $D$ du télescope : lorsque l'on éclaire une pupille circulaire d'un télescope par une onde plane provenant de l'infini, il se forme au foyer du télescope une tâche de diffraction d'Airy de rayon $r$ :

$$r = 1.22 \frac{\lambda f}{D} \qquad \text{(I-1)}$$

où $f$ est la distance focale du télescope. La résolution angulaire du télescope est alors simplement le diamètre de cette tâche d'Airy ramenée dans le plan objet :

$$\theta = 1.22 \frac{\lambda}{D} \qquad \text{(I-2)}$$

Il est, de nos jours, difficile technologiquement de dépasser des diamètres de miroirs de télescopes d'une dizaine de mètres. De plus, même s'ils sont atténués par l'utilisation de plus en plus répandue de systèmes d'optique adaptative, les phénomènes de turbulences atmosphériques viennent dégrader considérablement cette résolution théorique.

## I – C – 2 - Turbulence atmosphérique

L'atmosphère est en effet constituée de masses d'air ayant des températures et des densités différentes et se déplaçant les unes par rapport aux autres. L'indice de réfraction de l'air qui dépend de la pression et de la température varie, donc les faisceaux lumineux traversent des couches d'air successives d'indice légèrement différent et sont déviés au cours de leur traversée de l'atmosphère. Il en résulte une formation d'image dégradée et en perpétuel mouvement au foyer du télescope.

Un traitement statistique des effets de turbulences atmosphériques [25] permet de décrire de manière simple le comportement de la formation d'image au foyer d'un télescope à l'aide d'un nombre réduit de paramètres. Ainsi, on définit le paramètre de Fried $r_0$ qui correspond au rayon de corrélation de l'atmosphère : une surface de rayon $r_0$ définit une zone dans laquelle on peut considérer que les rayons lumineux sont suffisamment peu déviés pour ne pas induire de défauts dans la formation d'image. Ceci revient à dire qu'un télescope de diamètre inférieur ou égal à $r_0$ ne verra pas d'effet dû à la turbulence atmosphérique et que la résolution des images formées par ce télescope sera bien limitée par la diffraction. A contrario, un télescope de diamètre supérieur à $r_0$ sera influencé par les turbulences atmosphériques et formera des images ayant une résolution dégradée. Les valeurs typiques de $r_0$ sur de bons sites astronomiques sont de 10 cm dans le visible et d'environ 40 cm dans le proche infrarouge. La variation de ce paramètre $r_0$ en fonction de la longueur d'onde est donnée par [25] :

$$r_0 \propto \lambda^{\frac{6}{5}} \qquad \text{(I-3)}$$

Du point de vue de la meilleure résolution atteignable, il est donc plus intéressant d'observer dans l'infrarouge plutôt que dans le visible.

L'autre paramètre important permettant de décrire la turbulence atmosphérique est le temps de corrélation $\tau_0$ pendant lequel on peut considérer que la turbulence atmosphérique est fixe. Bien que ce paramètre varie de manière importante en fonction du site d'observation et des conditions atmosphériques, si l'on considère une vitesse moyenne du vent $<v>$ d'environ 20m/s, on peut définir $\tau_0$ comme étant simplement le rapport :

$$\tau_0 = \frac{r_0}{\langle v \rangle} \qquad \text{(I-4)}$$

Ce qui donne un temps de corrélation d'environ 5 ms dans le visible et de 20 ms dans le proche infrarouge. On parle aussi de temps de cohérence pour décrire ce paramètre qui définit en fait le temps pendant lequel la répartition de phase peut être considérée comme constante.

A cause de ces phénomènes, la résolution des télescopes est donc longtemps restée limitée à la seconde d'arc, c'est à dire la résolution théorique d'un télescope de diamètre $r_0$. Mais ces dernières années ont vu le développement de l'optique adaptative qui permet de compenser partiellement les effets de cette turbulence atmosphérique. Sur un télescope de diamètre supérieur à $r_0$, un système de mesure va prélever une partie du signal lumineux afin de mesurer en temps réel les variations du front d'onde. Ces mesures vont permettre de piloter la déformation d'une surface optique à l'intérieur du système qui va compenser ces variations et ainsi redresser le front d'onde. Ces systèmes sont actuellement en plein développement et permettent d'atteindre la résolution théorique des grands télescopes dans l'infrarouge proche et au delà comme par exemple ceux de 8,2 m de diamètre installés sur le mont Paranal au Chili, soit une résolution d'environ 50 mas (milliarcsecondes).

Si l'on désire toujours plus de résolution, il faut alors développer des télescopes de plus grand diamètre, et le projet ELT (Extremely Large Telescope) qui est à l'heure actuelle le projet le plus ambitieux, vise à construire un télescope de 42 mètres de diamètre. Cependant, un tel projet reste un véritable challenge technologique, le développement du système d'optique adaptative permettant de l'utiliser à son plein potentiel n'étant pas un des moindres.

## I – C – 3 – Interférométrie astronomique

L'autre manière d'obtenir de meilleures résolutions est de faire de l'interférométrie astronomique longue base qui consiste à recombiner de manière cohérente les faisceaux issus de plusieurs télescopes individuels éloignés les uns des autres. Ceci revient en fait à synthétiser un télescope de très grand diamètre égal à l'écartement maximal entre les télescopes. Pour cela, on effectue des franges d'interférence entre les faisceaux issus des télescopes. Les deux quantités observables sont la visibilité et la phase des franges d'interférence obtenues qui, pour simplifier, dépendent de la répartition d'intensité angulaire de la source lumineuse. Le schéma de principe d'un interféromètre stellaire est représenté sur la figure I-1 :

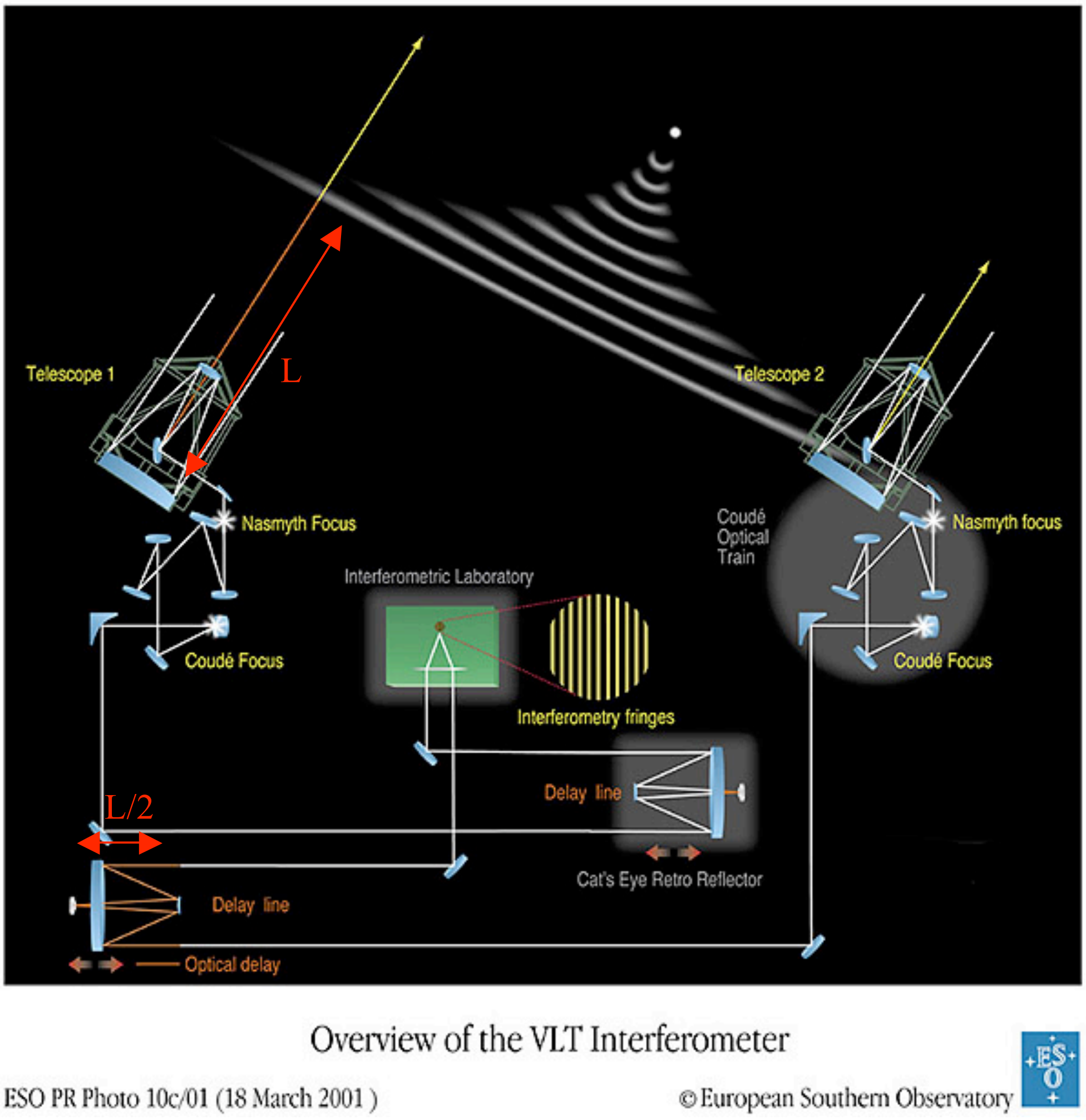

*Figure I-1 : schéma général d'un interféromètre stellaire (www.eso.org).*

Il s'agit du schéma de principe de l'interféromètre du VLTi décrivant les trois principaux sous ensembles d'un interféromètre : les télescopes, les lignes à retard, et le banc de recombinaison. Le front d'onde provenant de la source lumineuse observée est corrigé par des systèmes d'optique adaptative à l'entrée de chaque télescope, et est donc considéré comme plan. Il est récupéré par les deux télescopes puis transmis à un banc de recombinaison qui effectue les franges d'interférence. Lorsque l'objet observé n'est pas situé au zénith, l'onde parcourt une distance supplémentaire $L$ pour parvenir à l'un des deux télescopes et les voies comportent donc des lignes à retard pour compenser ce trajet optique différent sur les deux voies. L'ensemble du train optique est généralement symétrique pour éviter tout effet instrumental différentiel.

Il existe plusieurs types de recombinaisons possibles et nous y reviendrons plus en détail dans le chapitre suivant. Afin de décrire simplement le principe de la mesure, intéressons nous à un des cas les plus simples qu'est la recombinaison coaxiale (figure I-2).

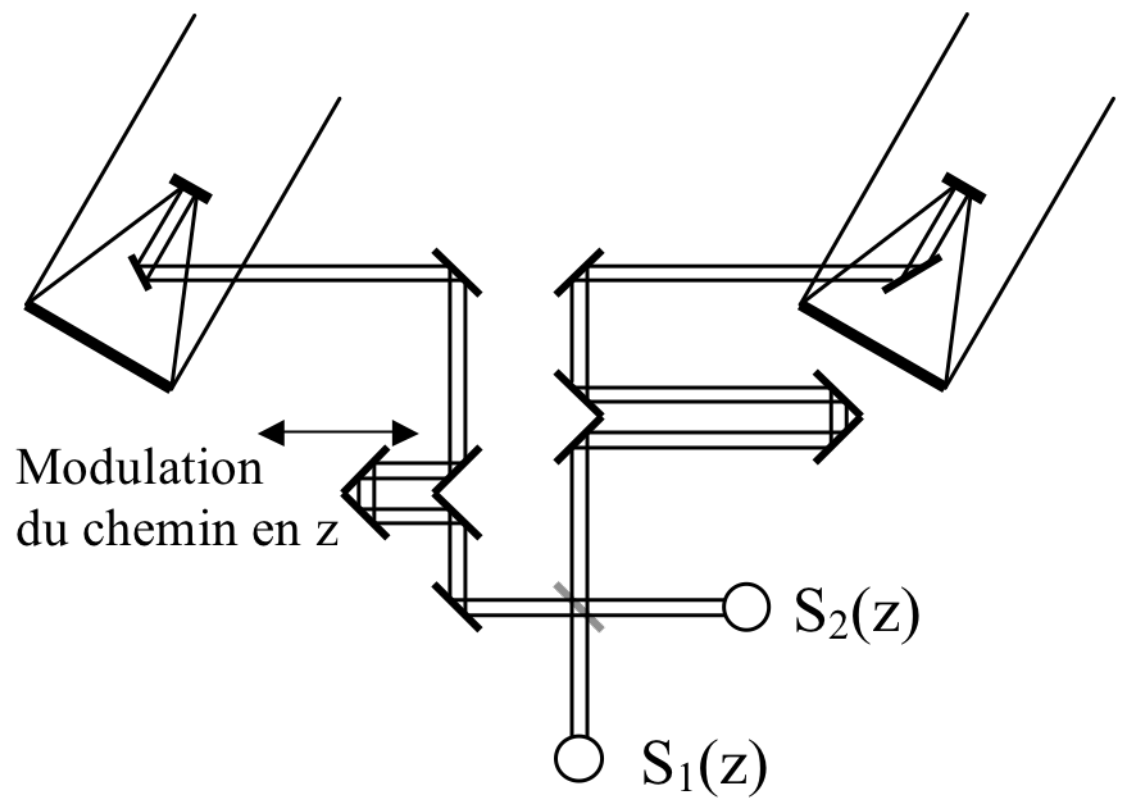

*Figure I-2 : schéma de recombinaison coaxiale à deux faisceaux.*

Les deux faisceaux issus des télescopes sont envoyés sur une lame semi-transparente puis récupérés par deux détecteurs. Afin d'effectuer la mesure, il est nécessaire d'introduire une modulation du chemin optique pour faire défiler les franges.

Si $I_1$ et $I_2$ sont les intensités des deux faisceaux, la visibilité $V$ et la phase $\varphi$ des franges sont reliées aux signaux $S_1(z)$ et $S_2(z)$ par la relation :

$$S_1(z) = I_1 + I_2 + 2\sqrt{I_1 I_2}V \sin\left(\frac{2\pi z}{\lambda} + \varphi\right) \quad \text{(I-5)}$$

$$S_2(z) = I_1 + I_2 - 2\sqrt{I_1 I_2}V \sin\left(\frac{2\pi z}{\lambda} + \varphi\right) \quad \text{(I-6)}$$

où $\lambda$ est la longueur d'onde et $z$ est la différence de chemin optique entre les deux faisceaux.

Les signaux sont complémentaires par conservation d'énergie, ils comportent essentiellement un terme continu proportionnel à la somme des intensités provenant des deux voies et un terme sinusoïdal décrivant les franges d'interférence en fonction de la différence de marche $z$. A deux voies, $\varphi$ est aléatoire à cause de la turbulence atmosphérique et l'on s'intéresse à la mesure du paramètre $V$ caractéristique de la source observée.

## I – C – 4 – Cohérence

Les équations (I-5, I-6) décrivent les signaux lorsque la source émet une seule longueur d'onde. Mais dans la pratique, on observe la source sur un domaine de longueur d'onde de largeur $\Delta\lambda$ sélectionné par un filtre spectral. Les franges d'interférence sont donc la somme des franges obtenues pour chaque longueur d'onde et ne sont visibles que lorsque la différence de marche z entre les faisceaux est inférieure à une certaine distance $L_c$ appelée longueur de cohérence et qui est égale à :

$$L_c = \frac{\lambda^2}{\Delta\lambda} \quad \text{(I-7)}$$

La figure I-3 montre les interférences obtenues pour différentes longueurs d'onde et l'interférogramme qui en résulte sur un certain domaine de longueur d'onde.

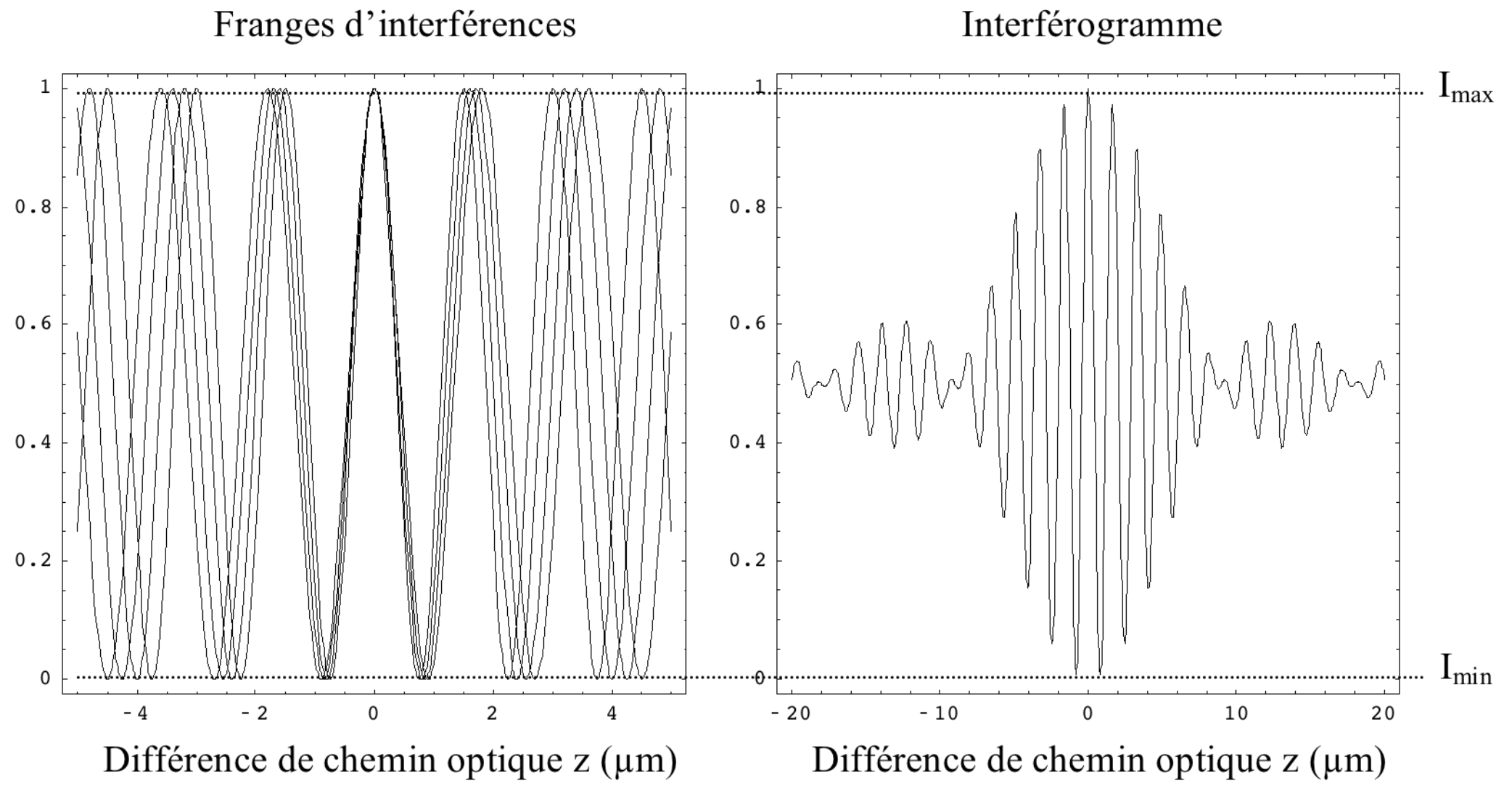

*Figure I-3 : franges d'interférence en lumière monochromatique et polychromatique. A gauche, franges pour des longueurs d'onde entre 1.5µm et 1.8µm. A droite, franges pour un spectre continu entre 1.5µm et 1.8µm.*

A partir de l'interférogramme, la visibilité des franges sera obtenue simplement à partir du minimum et du maximum de l'enveloppe de la courbe de franges :

$$V = \frac{I_{\max} - I_{\min}}{I_{\max} + I_{\min}} \qquad \text{(I-8)}$$

La visibilité est reliée à la répartition d'intensité spatiale de la source par une transformée de Fourier, c'est le théorème de Zernike Van-Cittert [*26*]. Ce théorème permet en effet d'écrire que le degré de cohérence spatiale *Γ(B/λ)* est relié à la répartition d'intensité angulaire *S(θ)* par la relation :

$$\Gamma\left(\frac{B}{\lambda}\right) = \int S(\theta) e^{-2i\pi\theta\frac{B}{\lambda}} d\theta = TF\left[S(\theta)\right] \qquad \text{(I-9)}$$

où $\lambda$ est la longueur d'onde, et $B$ la base de l'interféromètre. Le module de $\Gamma$ n'est autre que la visibilité des franges d'interférence $V$. Une mesure de visibilité pour une base $B$ donnée est donc une information sur la source $S(\theta)$ à la fréquence spatiale $B/\lambda$.

Pour déterminer certaines caractéristiques de la source, il faut faire des hypothèses sur sa géométrie, puis en utilisant l'équation ci-dessus, déterminer comment varie la visibilité en fonction de la base de l'interféromètre. Le cas le plus simple consiste à décrire la source par un disque circulaire d'intensité uniforme de rayon $\theta_0$. L'équation (I-9) montre alors que la visibilité n'est autre que le module de la transformée de Fourier du disque circulaire uniforme :

$$V\left(\frac{B}{\lambda}\right) = \left|\Gamma\left(\frac{B}{\lambda}\right)\right| = \left|\frac{2J_1\left(\pi\theta_0\frac{B}{\lambda}\right)}{\pi\theta_0\frac{B}{\lambda}}\right| \qquad \text{(I-10)}$$

où $J_1$ représente la fonction de Bessel de première espèce et du premier ordre. Cette fonction $V(B/\lambda)$ est représentée sur la figure I-4 pour deux rayons $\theta_0$ différents.

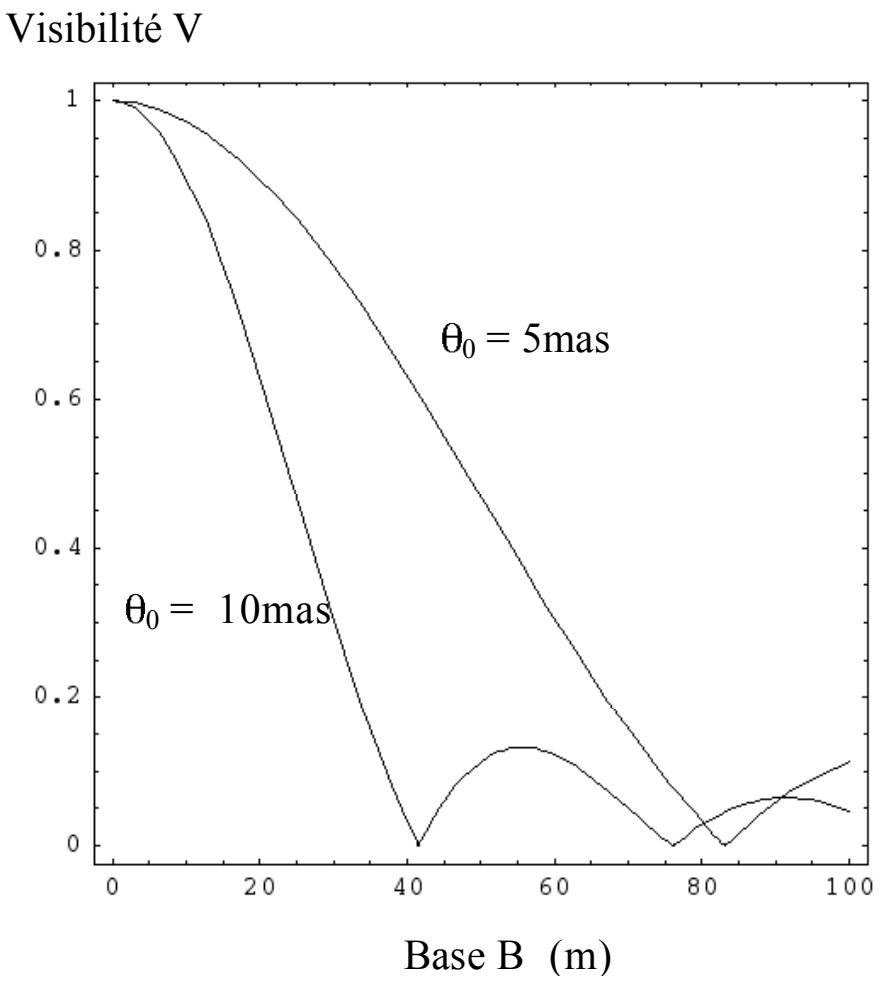

*Figure I-4 : visibilité en fonction de la base de l'interféromètre pour un disque uniforme à une longueur d'onde de 1,65µm.*

Dans la pratique, on mesure la visibilité des franges *V* pour différentes valeurs de base *B*, et le rayon de l'étoile $\theta_0$ sera celui permettant de superposer au mieux une courbe du type de celle la figure I-4 à ces points de mesure.

Enfin, puisqu'une mesure de visibilité $V(B/\lambda)$ est en fait une information sur la source à la fréquence spatiale $B/\lambda$, la résolution angulaire d'un interféromètre est simplement donnée par :

$$\theta = \frac{\lambda}{B} \qquad \text{(I-11)}$$

On obtient donc une information sur la source avec une résolution bien meilleure que celle d'un simple télescope.

## I – C – 5 - Influence de l'atmosphère et filtrage spatial

La visibilité mesurée est affectée par des effets instrumentaux et atmosphériques qui font que sa valeur exacte est perdue. Il faut donc la calibrer si l'on veut obtenir une information fiable. On peut en fait décomposer la visibilité observée suivant l'équation suivante :

$$V_{obs} = V_{inst} V_{atm} V_{source} \tag{I-12}$$

$V_{source}$ est la visibilité de la source que l'on cherche à mesurer. $V_{inst}$ est la visibilité de l'instrument qui peut être connue en effectuant une mesure sur une source de référence connue. L'important est que cette visibilité instrumentale $V_{inst}$ reste stable au cours du temps afin de ne pas avoir à effectuer la calibration trop souvent. Enfin $V_{atm}$ est la visibilité atmosphérique qui est beaucoup plus difficile à évaluer, mais concerne principalement la phase de l'onde. Comme nous l'avons vu, le front d'onde est déformé avant l'arrivée sur les télescopes. L'analyse de ce front d'onde se fait en décomposant cette déformation en polynômes de Zernike où l'on regroupe les termes de degré supérieur ou égal à deux. On obtient donc les trois termes suivants :

- *Le piston :* à l'ordre zéro (en moyenne), les fronts d'onde arrivant sur chacun des télescopes ont subi un déphasage constant sur toute leur pupille. Ce déphasage induit un décalage des franges observées. Afin de s'affranchir de ce décalage, on peut soit observer les franges sur une durée inférieure à la période de ce décalage qui est comme on l'a vu de l'ordre de quelques millisecondes dans le visible, soit intégrer au système un suiveur de franges qui permet de compenser en temps réel cette fluctuation de phase afin de pouvoir moyenner le signal et obtenir ainsi une meilleure sensibilité.
- *Le basculement (tip-tilt) :* à l'ordre 1, les fronts d'onde subissent un basculement angulaire sur chacun des télescopes que l'on compense généralement par l'introduction d'un miroir oscillant dans le système.

- *Les fluctuations de phase :* lorsque le diamètre des télescopes est supérieur à la largeur moyenne des turbulences atmosphériques, le front d'onde à l'entrée de chaque pupille de télescope ne peut plus être considéré comme plan. Il est alors nécessaire de faire appel à un système d'optique adaptative pour compenser ces déformations, ou de filtrer le faisceau optique.

En d'autres termes, lorsque le diamètre du télescope est inférieur au paramètre de Fried $r_0$ défini précédemment, seuls, le piston et le basculement ont besoin d'être corrigés pour visualiser des franges, tandis que lorsque le diamètre du télescope est supérieur à $r_0$, les fluctuations de phase doivent aussi être prises en compte. Dans les deux cas, la précision des mesures est affectée par les fluctuations de phase. Un moyen de calibrer ces fluctuations de front d'onde est de procéder à un filtrage spatial permettant de ne sélectionner que l'onde passant rigoureusement par le foyer du télescope. Un tel filtrage transforme les fluctuations de phase en fluctuations d'intensité qu'il est possible de mesurer en prélevant une partie des faisceaux incidents. Ainsi, la visibilité des franges $V$ n'est plus perturbée par la visibilité de l'atmosphère $V_{atm}$.

Comme nous le verrons en détail dans le paragraphe de ce chapitre qui lui est consacré, l'optique guidée est un moyen très efficace de filtrer spatialement un faisceau. Cependant, intéressons nous d'abord à la reconstruction d'images par interférométrie optique.

# *I – D – synthèse d'ouverture*

## I – D – 1 – Plan u,v

Le théorème de Zernike-Van Cittert (équation I-9) relie la répartition angulaire d'intensité de la source $I(\theta)$ à la visibilité complexe des franges $\Gamma(B/\lambda)$ par une relation de transformée de Fourier. Si l'on cherche à faire une image à deux dimensions, on définit deux angles apparents $\alpha$ et $\beta$ qui vont être les projections de l'angle $\theta$ dans deux directions orthogonales du plan x,y et on réécrit l'équation I-9 sous la forme :

$$I(\alpha,\beta) = TF\left[\Gamma(u,v)\right] \tag{I-13}$$

$u$ et $v$ sont les variables conjuguées de $\alpha$ et $\beta$ et représentent les fréquences spatiales de la source.

$$u = \frac{B_x}{\lambda} \tag{I-14}$$

$$v = \frac{B_y}{\lambda} \tag{I-15}$$

où $B_x$ et $B_y$ sont les projections de la base $B$ sur les axes x et y. Pour chaque base, une mesure de visibilité complexe est en fait une mesure de fréquence spatiale de la source. La synthèse d'ouverture consiste à effectuer des mesures de visibilités complexes pour une multitude de lignes de base afin de reconstituer le spectre spatial de la source. Chaque ligne de base représente un point d'échantillonage du plan u,v.

Pour obtenir ces différentes lignes de base u,v, on reconfigure la position des télescopes et on utilise la rotation de la terre, donc la modification de la base projetée sur le ciel. Bien évidemment, plus on dispose de télescopes, plus il est facile d'obtenir une bonne couverture du plan u,v. Le gain est à peu près quadratique en terme de nombre d'informations sur la source : passer de 3 à 8 télescopes permet de multiplier le nombre de mesures par 10.

A titre d'exemple, la figure I-5 montre différentes couvertures de plan u,v obtenues pour les différents réseaux de télescopes existants.

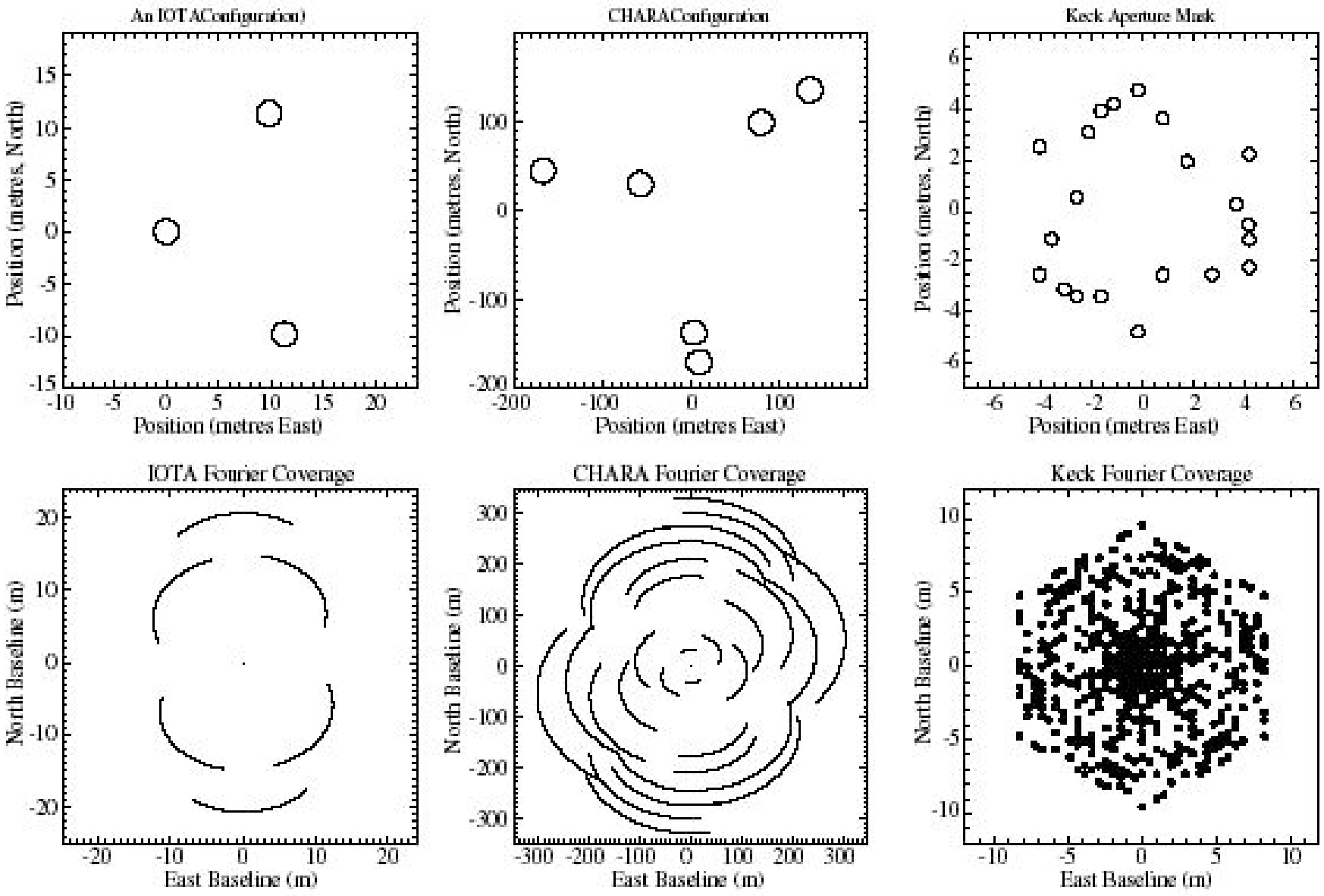

*Figure I-5 : Couvertures de plan u,v pour différents réseaux de télescopes existants. A gauche se trouve IOTA avec ses trois télescopes, au centre se trouve CHARA avec ses six télescopes, à droite se trouve le KECK où l'on simule plusieurs petits télescopes en mettant des caches sur un gros télescope de 10m.*

Les graphes du haut montrent la position des télescopes au sol vue du ciel. Les graphes du bas montrent les bases obtenues à partir de ces positions. Pour CHARA et IOTA, la rotation de la Terre multiplie le nombre de base obtenues.

## I – D – 2 – Clôture de phase

Un interféromètre à deux voies permet de mesurer la visibilité des franges d'interférence et ainsi d'évaluer le diamètre angulaire de la source observée, mais l'information comprise dans la phase des franges est noyée dans le bruit de phase introduit à la traversée de l'atmosphère. A partir de trois télescopes, une mesure partielle de la phase redevient possible par la technique de clôture de phase, décrite en détail dans [*27*], et dont nous ne donnons ici que les grandes lignes.

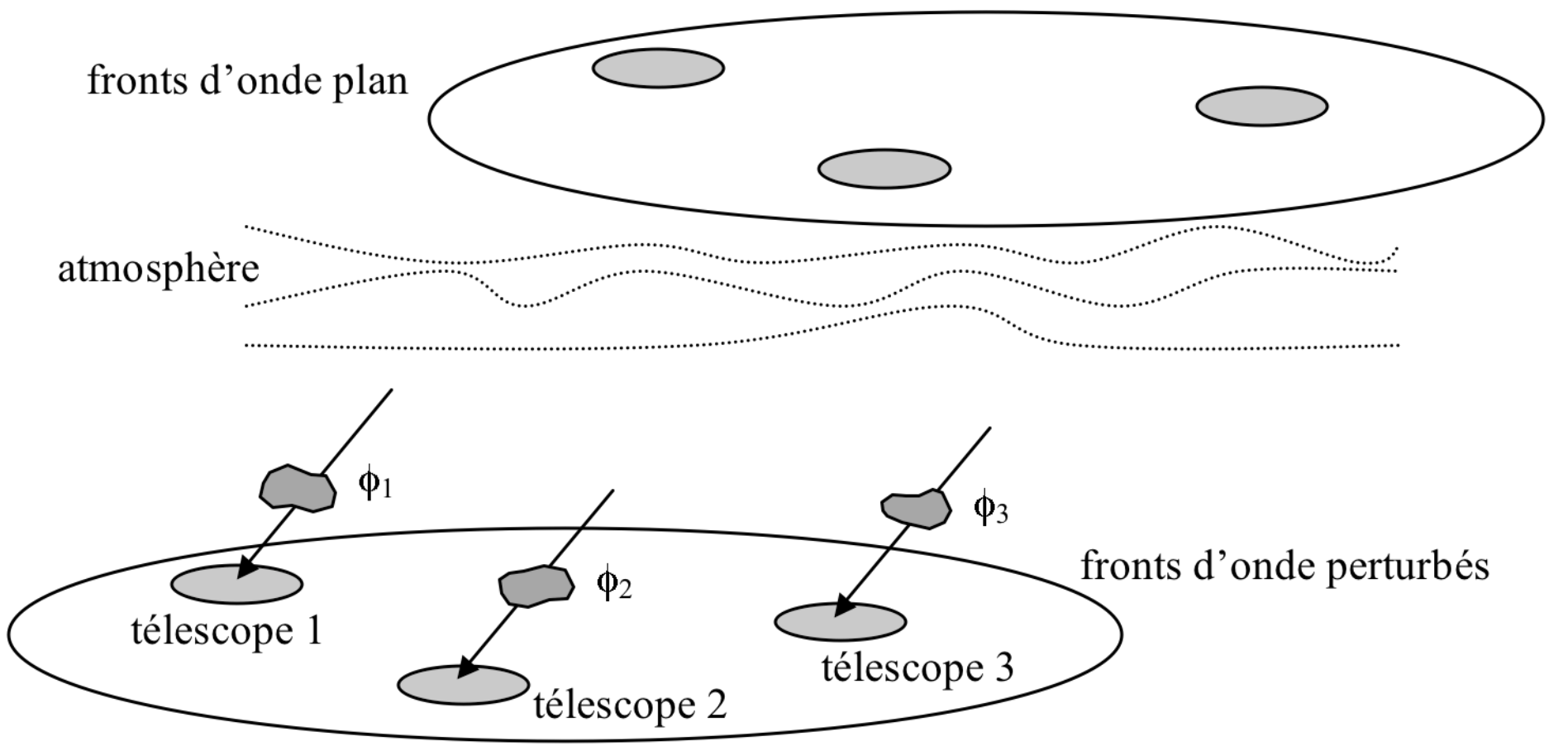

*Figure I-6 : influence de l'atmosphère sur les fronts d'onde de trois télescopes.*

A trois télescopes, si l'on recombine chaque paire de télescopes, on peut écrire que chaque phase de frange peut se mettre sous la forme :

$$\Phi_{12} = \varphi_{12} + \phi_2 - \phi_1 \tag{I-16}$$

$$\Phi_{23} = \varphi_{23} + \phi_3 - \phi_2 \tag{I-17}$$

$$\Phi_{31} = \varphi_{31} + \phi_1 - \phi_3 \tag{I-18}$$

où $\varphi_{ij}$ est la phase intrinsèque des franges entre les télescopes i et j contenant l'information sur la source et $\phi_i$ la phase aléatoire introduite par l'atmosphère sur chaque pupille de télescope i. La clôture de phase $\varphi_c$ est la somme de ces trois phases.

$$\varphi_c = \Phi_{12} + \Phi_{23} + \Phi_{31} = \varphi_{12} + \varphi_{23} + \varphi_{31} \tag{I-19}$$

On s'aperçoit que les termes dus à l'atmosphère s'annulent mutuellement, et que l'on retrouve une information sur les phases intrinsèques. On obtient donc une information partielle de la phase des franges (1/3 en l'occurrence pour trois télescopes).

Si l'on généralise la méthode à $N$ télescopes, le nombre de phases de Fourier possibles $N_{ph}$ est égal au nombre de paires possibles de télescopes parmi $N$, soit :

$$N_{ph} = \frac{N(N-1)}{2} \qquad \text{(I-20)}$$

Le nombre de triplets de télescopes possibles $N_{tr}$ est lui donné par :

$$N_{tr} = \frac{N(N-1)(N-2)}{6} \qquad \text{(I-21)}$$

Et le nombre de clôtures de phase $N_{cl}$ sera donné par le nombre de triplets de télescopes indépendants soit :

$$N_{cl} = \frac{(N-1)(N-2)}{2} \qquad \text{(I-22)}$$

Le tableau I-3 donne les valeurs pour un nombre croissant de télescopes.

| Nombre de télescopes | Nombre de phases de Fourier | Nombre de triangles | Nombre de clôtures | Pourcentage d'information |
|---|---|---|---|---|
| N | Nph | Ntr | Ncl | Ncl/Nph |
| 3 | 3 | 1 | 1 | 33% |
| 4 | 6 | 4 | 3 | 50% |
| 6 | 15 | 20 | 10 | 67% |
| 8 | 28 | 56 | 21 | 75% |
| 27 | 351 | 2925 | 325 | 93% |

*Tableau I-3 : information de phase récupérable en fonction du nombre de télescopes utilisés pour la synthèse d'ouverture*

Le VLTi utilisera bientôt ses 4 télescopes auxiliaires pour la recombinaison interférométrique, et envisage d'utiliser tous ses télescopes simultanément (4 grands télescopes + 4 télescopes auxiliaires). CHARA utilise 6 télescopes. A titre indicatif, le radio interféromètre VLA utilise 27 télescopes. On s'aperçoit que le pourcentage d'information de phase récupérable croît assez vite en fonction du nombre de télescopes utilisés.

### I – D – 3 – Reconstruction d'images

Puisque l'on n'a pas accès à la totalité de l'information sur la phase, il est nécessaire, pour reconstruire une image à partir des mesures de visibilités et clôtures de phase, de faire appel à des algorithmes complexes qui introduisent de l'a priori dans l'image observée (par exemple, il n'est pas réducteur d'imposer que la répartition d'intensité de l'image soit positive) et utilisent un mécanisme de convergence pour remonter à l'image qui correspond le mieux aux mesures effectuées. Bien évidemment, plus on dispose de télescopes, plus l'image sera précise, mais la qualité du résultat obtenu est aussi très dépendante de la précision des mesures de visibilités et de clôtures de phase. Ces mesures s'étalant dans le temps (à trois télescopes, il faut plusieurs nuits pour obtenir une couverture du plan u,v suffisante) la stabilité de l'instrument est cruciale.

## *I – E – Apport de l'optique guidée en interférométrie astronomique*

### I – E – 1 - Rappels

De manière générale, un guide optique diélectrique comporte un cœur, zone d'indice de réfraction $n_2$ entourée d'une gaine, zone diélectrique d'indice de réfraction $n_1$. Pour que le guidage soit possible, il est nécessaire que l'on ait $n_2$ supérieur à $n_1$.

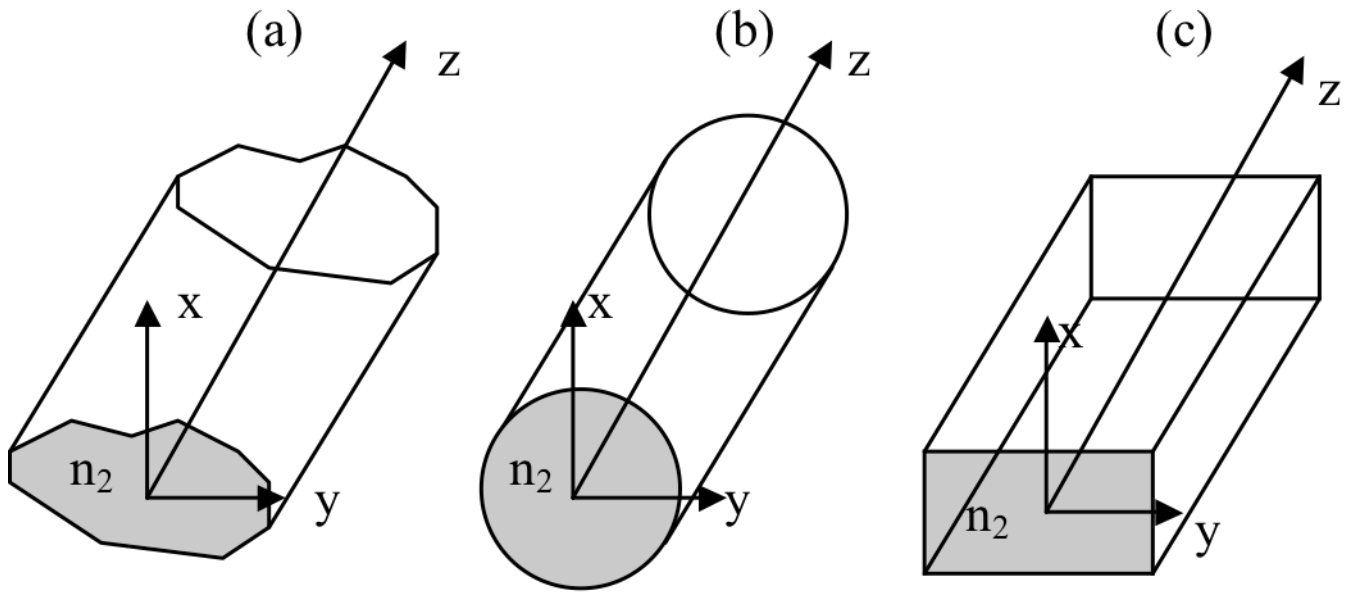

*Figure I-7 : géométrie d'un guide d'onde diélectrique. (a) cas général. (b) cas de la fibre optique.(c) cas des guides intégrés utilisés au cours de ce travail.*

Nous n'allons pas ici redétailler le calcul de la propagation de la lumière dans les guides d'onde, celui ci ayant été amplement traité par exemple dans [*28*][*29*]. On décrit la propagation dans un guide d'onde comme une propagation modale, c'est à dire qu'il existe un certain nombre d'ondes possédant un plan de phase et se propageant dans le guide avec une répartition d'amplitude transverse invariante selon l'axe de propagation. Si on considère une onde monochromatique de pulsation $\omega$, en conservant les notations de la figure I-7, un mode de propagation ***E,H*** peut se mettre sous la forme :

$$\vec{E} = \vec{E}(x,y)e^{i\beta z - \omega t} \qquad \text{(I-23)}$$

$$\vec{H} = \vec{H}(x,y)e^{i\beta z - \omega t} \qquad \text{(I-24)}$$

$\beta$ est la constante de propagation du mode considéré et est déterminée par la géométrie du guide. Un guide monomode est un guide pour lequel il n'existe qu'un seul mode de propagation possible.

## I – E – 2 – Filtrage en optique guidée

En première approximation, un guide d'onde monomode peut donc être considéré comme un filtre spatial parfait ne laissant passer que son mode fondamental. Dans la pratique, il faut tenir compte de la zone transitoire à l'entrée du guide dans laquelle la lumière filtrée qui n'a pas été couplée au mode fondamental fuit progressivement dans la gaine.

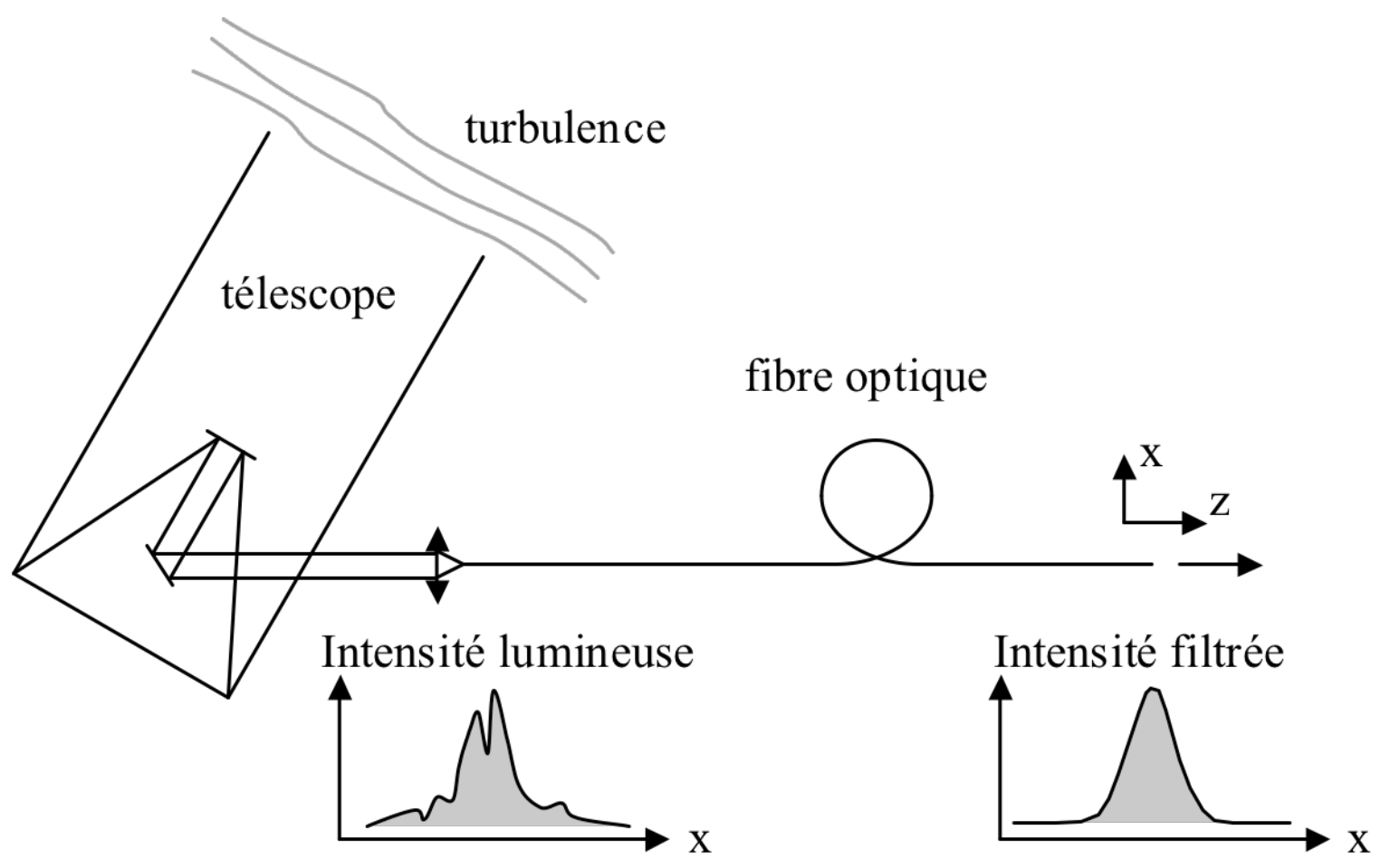

*Figure I-8 : couplage de la lumière dans un guide d'onde. La lumière focalisée par le télescope est injectée (« couplée ») dans la fibre optique qui la transporte à la fonction de recombinaison proprement dite.*

Si l'on considère le cas simplifié de la figure I-8, la tâche lumineuse focalisée par le télescope est une tache d'Airy (lorsque l'étoile est trop petite pour être résolue par le télescope) déformée par la turbulence atmosphérique. Une partie de cette tache va être couplée au mode fondamental de la fibre optique, tandis que le reste de la lumière va progressivement être dissipé (« rayonné ») dans la gaine de la fibre optique. Si l'on considère le cas optimisé où la taille de la tache d'Airy est adaptée à la taille du mode fondamental de la fibre optique, on peut coupler jusqu'à 80% de la lumière dans le mode fondamental [*30*]. Dans la pratique, la turbulence atmosphérique dégrade la tache de diffraction au foyer du télescope et le taux de couplage varie donc de manière aléatoire dans le temps. Les fluctuations de phase aléatoires introduites par l'atmosphère ont été transformées en fluctuations d'intensité aléatoires.

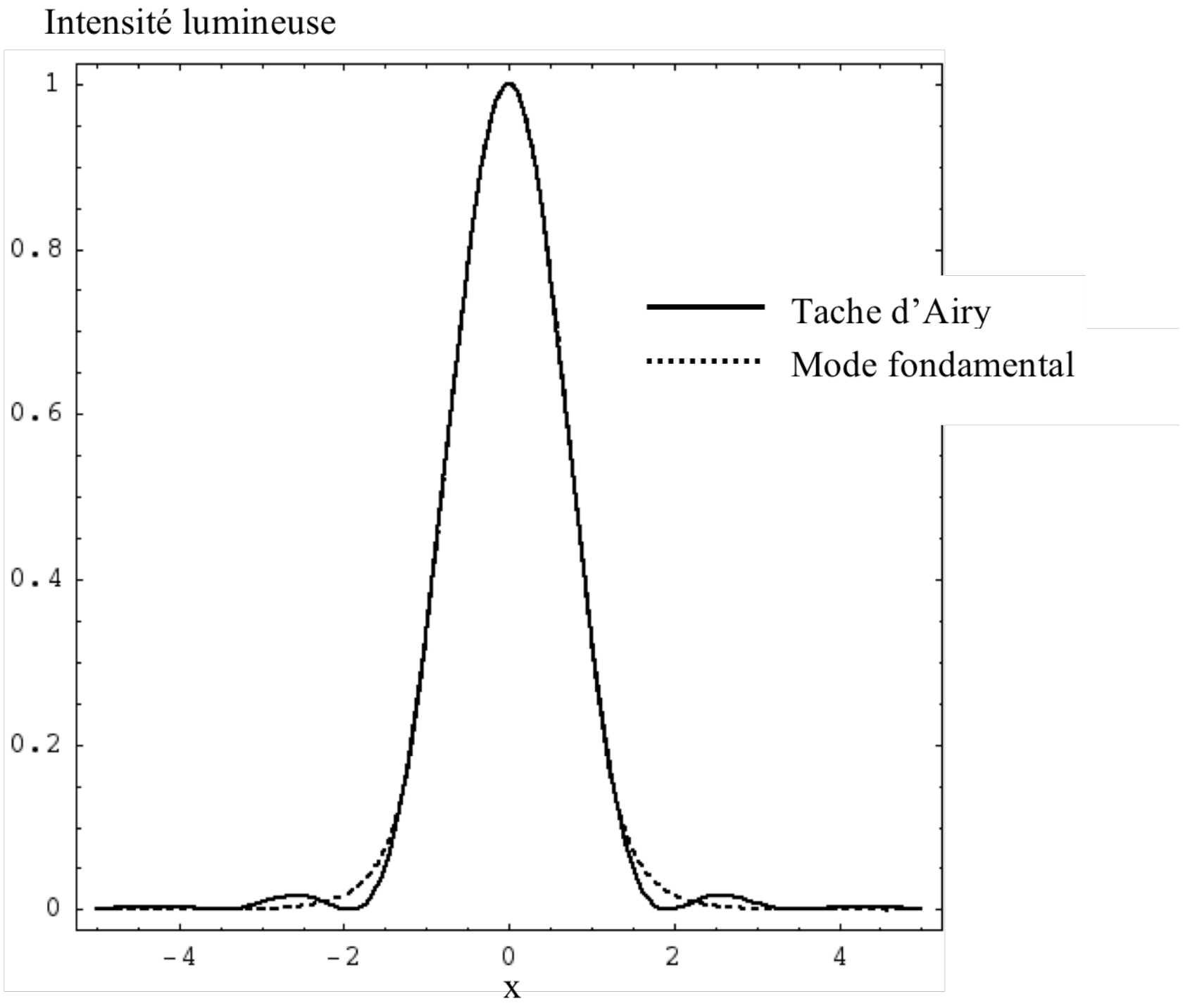

*Figure I-9 : tache d'Airy au foyer du télescope et mode guidé de la fibre monomode. Dans le meilleur des cas, le couplage est d'environ 80%.*

Une étude détaillée du filtrage par guide monomode peut être obtenue dans la littérature [*31*][*32*] et montrent que pour obtenir un filtrage spatial ayant un taux de réjection d'environ $10^{-6}$, il est nécessaire de se propager dans environ un mètre de fibre optique ou un centimètre de guide optique intégré.

## I – E – 3 – Recombinaison en optique guidée

### I – E – 3 – a - Le cas de deux télescopes

Outre l'intérêt du filtrage, l'optique guidée permet de prélever une partie des faisceaux couplés dans le guide d'entrée afin de suivre en temps réel la variation aléatoire du taux de couplage. Si l'on utilise des fibres optiques, comme c'est le cas dans l'instrument FLUOR [*33*][*34*] le schéma de la recombinaison est alors le suivant :

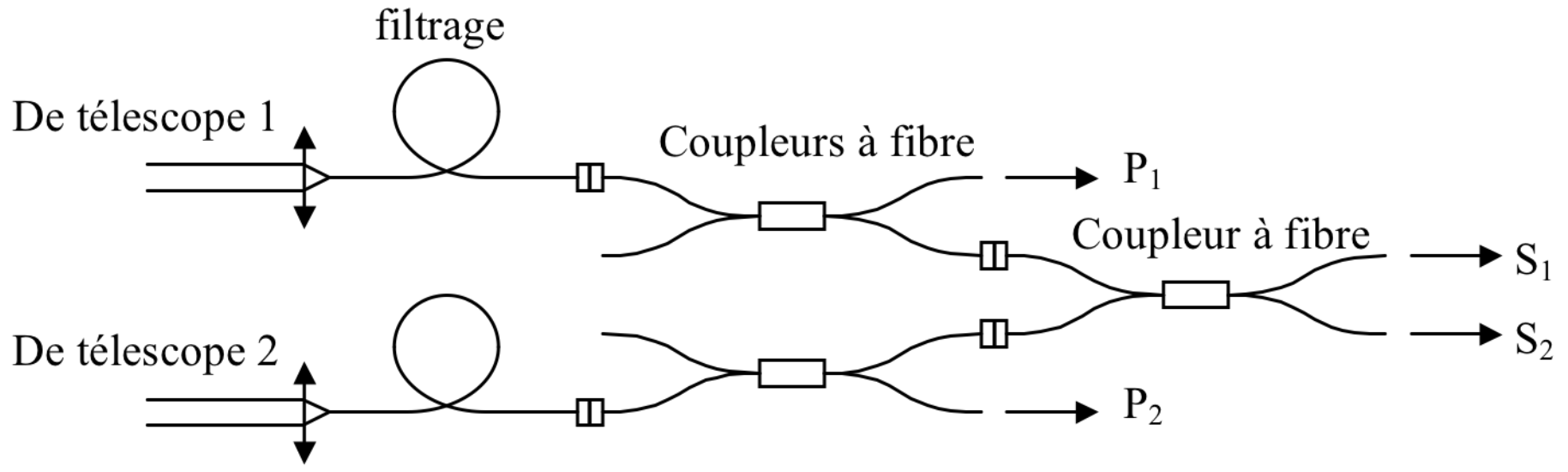

*Figure I-10 : principe de la recombinaison en utilisant des fibres optiques*

Les signaux issus des télescopes sont injectés dans deux fibres optiques. Sur chacune des voies, on prélève à l'aide de coupleurs à fibres optiques partie des signaux afin de calibrer chaque intensité $I_1$ et $I_2$. Les signaux photométriques $P_1$ et $P_2$ sont proportionnels aux intensités $I_1$ et $I_2$ entrant dans chaque guide :

$$P_i = \alpha_i I_i \qquad i = 1,2 \tag{I-25}$$

Afin de mesurer la visibilité V des équations (I-5, I-6), en utilisant les mêmes notations, on corrige le signal interférométrique des variations photométriques :

$$S_{cor}(z) = \frac{S_1(z) - \frac{1-\alpha_1}{\alpha_1}P_1 - \frac{1-\alpha_2}{\alpha_2}P_2}{2\sqrt{\frac{1-\alpha_1}{\alpha_1}P_1\frac{1-\alpha_2}{\alpha_2}P_2}} = V\sin\left(\frac{2\pi z}{\lambda} + \varphi\right) \tag{I-26}$$

On obtient ainsi une mesure de *V* indépendante des fluctuations introduites par l'atmosphère. Les expériences menées sur l'instrument FLUOR ont permis de montrer que la perte de flux introduite par le filtrage est largement compensée par l'augmentation de la précision de la mesure de visibilité [35][36].

A la place de coupleurs à fibres optiques, il est naturel d'envisager l'utilisation de guides optiques intégrés pour réaliser ce schéma de recombinaison. L'optique intégrée offre plus de choix sur les fonctions élémentaires utilisées qui peuvent être de plus optimisées suivant l'application.

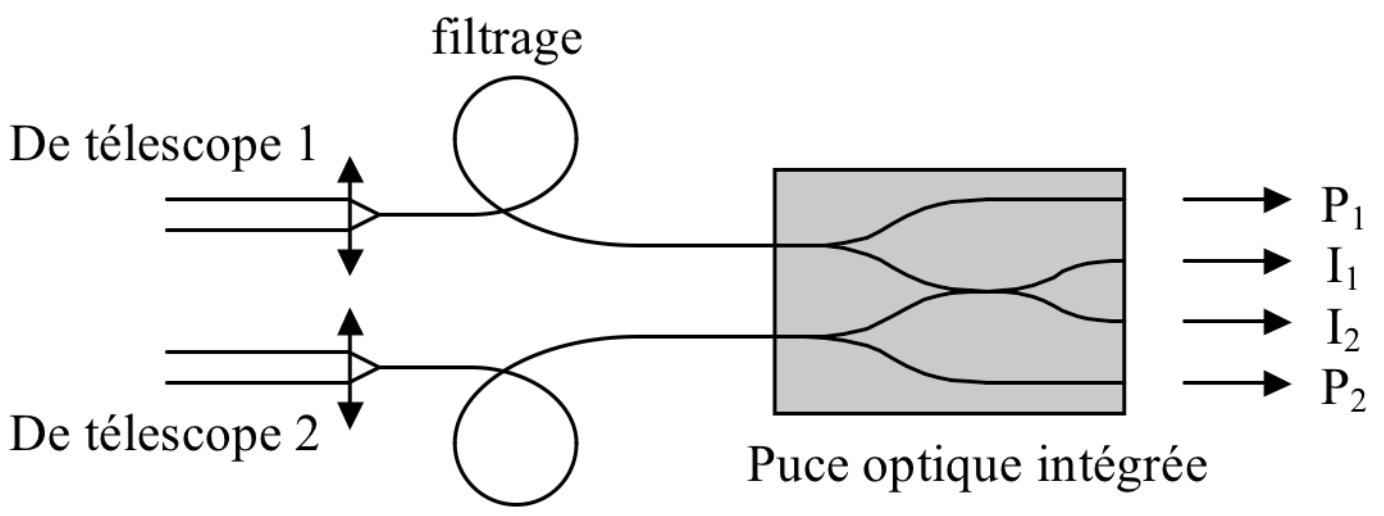

*Figure I-11 : recombinaison en optique intégrée.*

Le prélèvement du signal dans chaque voie photométrique peut être réalisé à partir de jonctions Y qui ont l'avantage d'être parfaitement achromatiques alors qu'un coupleur à fibre optique nécessitera une calibration spectrale. La recombinaison peut être effectuée par jonction Y, coupleur achromatisé, tricoupleur, etc… et sera discutée en détail dans le chapitre II. La figure I-12 montre des franges obtenues sur l'instrument IONIC alors qu'il utilisait un composant d'optique intégrée sur verre basé sur des jonctions Y [37].

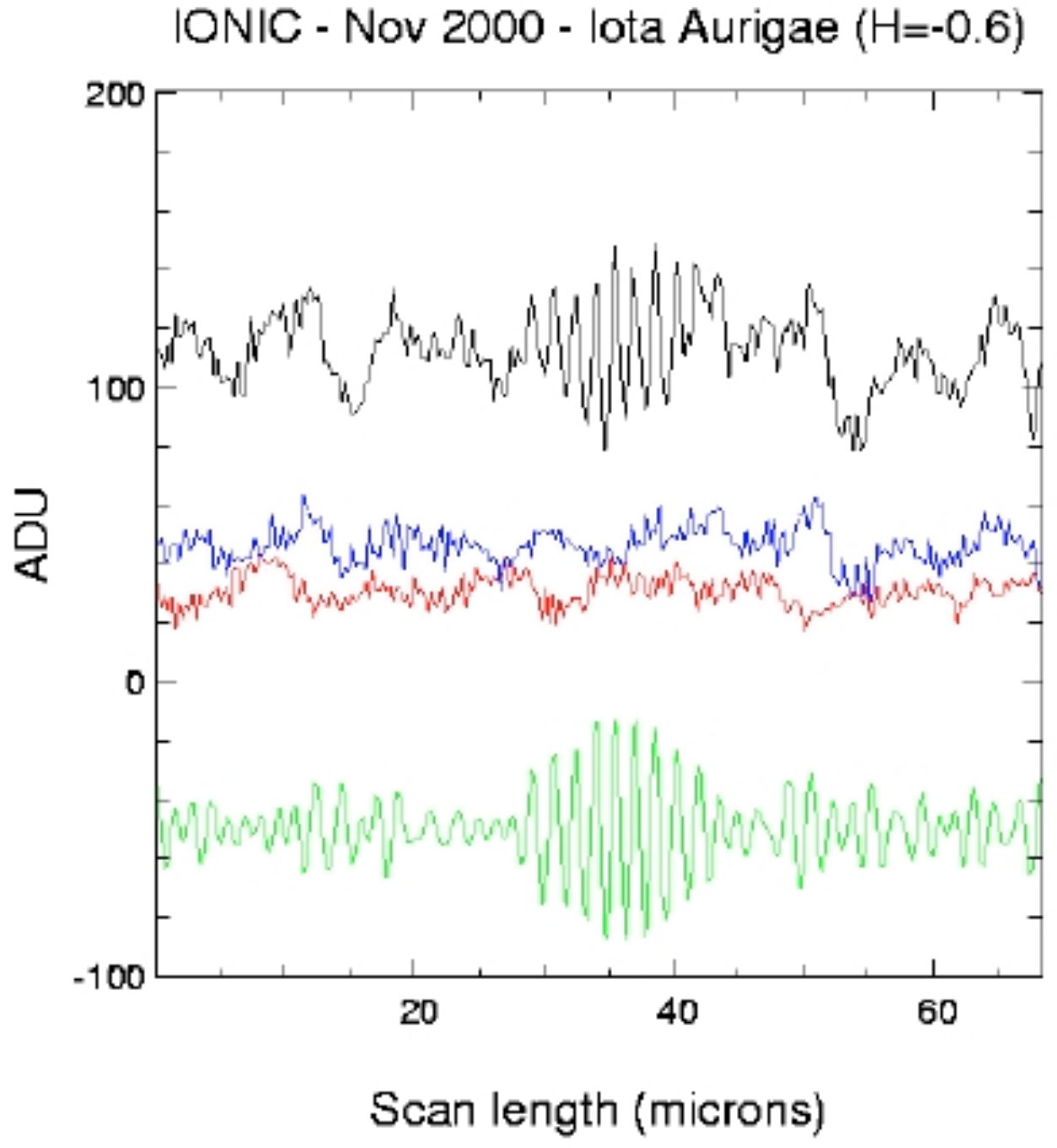

*Figure I-12 : Franges obtenues sur IOTA. On observe les signaux de chaque voie photométrique en rouge et bleu, le signal interférométrique brut en noir, et le signal interférométrique corrigé en vert. A partir de l'enveloppe des franges de ce signal interférométrique, on remonte au diamètre de la source observée. D'après [37]*

### I – E – 3 – b - Le cas de trois télescopes et plus

L'intérêt de l'optique intégrée croît avec le nombre de télescopes à recombiner. Dans le cas d'une recombinaison par paires, le schéma est alors le suivant :

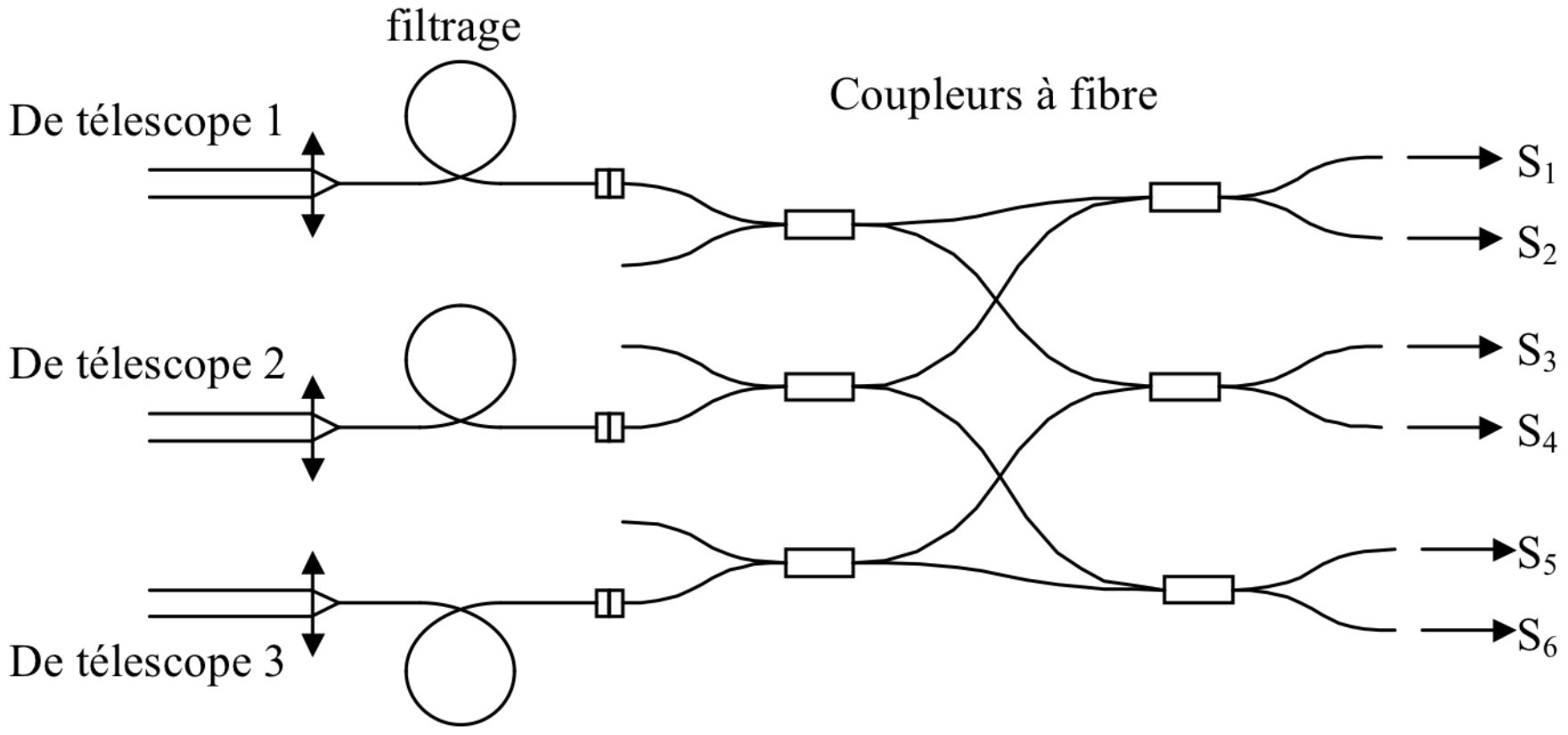

*Figure I-13 : recombinaison à trois télescopes. On sépare chaque voie en deux à l'aide d'un coupleur à fibre optique puis on recombine les voies deux à deux, à nouveau à l'aide de coupleurs.*

Bien que la figure I-13 montre qu'il soit encore possible à trois télescopes de réaliser un schéma à fibres optiques à l'aide de coupleurs fibrés, il va être extrêmement difficile d'extraire des signaux utilisables d'un tel dispositif. En effet, si la longueur des fibres optiques n'est pas rigoureusement égalisée après la première séparation de chaque faisceau en deux voies, la valeur de la clôture de phase sera biaisée. De plus on peut s'attendre à des problèmes de stabilité, la propagation de la phase dans une fibre optique étant particulièrement sensible aux conditions environnementales. En ce sens, il est plus judicieux de réaliser un tel schéma sur une puce optique intégrée où la longueur de chaque guide peut être rigoureusement égalisée et dont il est plus facile de stabiliser l'environnement.

On remarque que ce schéma ne comporte pas de voies photométriques. En effet, dans le cas de la recombinaison par paires, si l'on utilise des coupleurs pour réaliser la recombinaison, on obtient pour chaque paire recombinée, les deux signaux de franges en opposition de phase. Prenons le cas idéal de coupleurs parfaitement équilibrés pour les deux premières sorties :

$$S_1 = \frac{I_1}{2} + \frac{I_2}{2} + \sqrt{I_1 I_2} V_{12} \cos\left(\frac{2\pi z}{\lambda} + \varphi_{12}\right) \qquad \text{(I-27)}$$

$$S_2 = \frac{I_1}{2} + \frac{I_2}{2} - \sqrt{I_1 I_2} V_{12} \cos\left(\frac{2\pi z}{\lambda} + \varphi_{12}\right) \qquad \text{(I-28)}$$

Si l'on somme ces deux signaux, on obtient les sommes des signaux photométriques soit pour chaque paire recombinée :

$$P_1 = I_1 + I_2 \qquad \text{(I-29)}$$

$$P_2 = I_1 + I_3 \qquad \text{(I-30)}$$

$$P_3 = I_2 + I_3 \qquad \text{(I-31)}$$

Il est alors possible de remonter à la photométrie de chaque voie par combinaison linéaire de ces signaux :

$$I_1 = \frac{P_1 + P_2 - P_3}{2} \qquad \text{(I-32)}$$

$$I_2 = \frac{P_1 - P_2 + P_3}{2} \qquad \text{(I-33)}$$

$$I_3 = \frac{-P_1 + P_2 + P_3}{2} \qquad \text{(I-34)}$$

Dans la pratique, les valeurs exactes de chaque coupleur d'un tel dispositif sont plutôt mesurées afin d'obtenir une matrice de transmission précise de la puce et ainsi remonter aux bons coefficients photométriques.

Enfin dernière remarque : l'optique intégrée étant par nature planaire, il est nécessaire pour réaliser le schéma ci-dessus d'utiliser des croisements de guides optiques. Les guides optiques diélectriques possèdent la propriété de ne pas échanger d'énergie entre eux lorsqu'ils se croisent avec un angle suffisamment important. Nous verrons plus loin que ce détail est en grande partie responsable de la taille assez importante des puces que nous avons réalisées au cours de cette étude.

## I – F – Objectifs de la thèse

Lorsque nous avons débuté cette étude, l'expérience IONIC utilisait sa première puce intégrée pour recombiner les faisceaux issus des trois télescopes d'IOTA. Les résultats extrêmement encourageants obtenus incitaient à pousser plus loin le concept de l'utilisation de l'optique intégrée en interférométrie astronomique. Principalement deux voies de recherche devaient être explorées : comment recombiner plus de télescopes, et étendre l'utilisation de l'optique guidée à d'autres domaines de longueur d'onde.

Plusieurs projets de recherche ont alors été initiés en parallèle. Ainsi, le LETI a participé avec l'IMEP, le LAOG, et ALCATEL ALENIA au développement de guides monomodes pour la recombinaison interférométrique dans le domaine de l'infrarouge thermique [*38*]. Toujours avec le LAOG, le LETI a aussi participé à la réalisation d'un recombineur à deux télescopes dans la bande K (2,0-2,4µm) qui a été installé au VLTi [*39*].

L'objectif de ce travail de thèse s'est plutôt situé dans la première voie. Il était donc d'étudier puis de réaliser un recombineur à quatre télescopes possédant les fonctionnalités suivantes :

- Filtrage spatial des faisceaux par guidage monomode.
- Fonctionnement achromatique dans la bande de transparence H de l'atmosphère (de 1,47µm à 1,78µm).
- Transmission globale la plus élevée possible pour ne pas limiter les performances de l'interféromètre en terme de sensibilité.
- Isolation des voies à l'intérieur de la puce la plus élevée possible pour éviter des phénomènes d'interférences parasites.

- Conservation de la polarisation.
- Contrôle rigoureux de la phase à l'intérieur de la puce pour pouvoir mesurer une clôture de phase sans biais.
- Contraste instrumental le plus élevé possible
- Mesure instantanée de la phase des franges.

les résultats obtenus ayant été concluants, ce travail a aussi consisté à explorer l'extension de ces résultats à plus de télescopes et à d'autres bandes spectrales dans le cadre du projet VSI qui a pour but l'étude et la réalisation du prochain instrument recombineur de l'interféromètre du VLTi.

# *I – G – Conclusion*

Dans ce chapitre, nous avons présenté le contexte dans lequel s'inscrit cette étude. Après avoir présenté les principes généraux de l'interférométrie astronomique, nous avons montré comment, à partir des franges d'interférence mesurées, nous pouvions obtenir des informations sur la source observée avec une résolution bien meilleure que celle d'un simple télescope. Nous avons alors exposé comment la turbulence atmosphérique perturbait l'observation astronomique en général et l'interférométrie en particulier. Enfin nous avons montré l'intérêt de l'optique guidée et intégrée qui apporte un gain considérable en terme de précision de mesure et permet de s'affranchir de ces problèmes par le filtrage spatial.

Le chapitre II traite de l'étude de la recombinaison en optique intégrée, en présentant une classification des différentes méthodes, et de la conception du dispositif.

# Chapitre II : Recombinaisons en optique intégrée

## *II – A – Introduction*

Dans le premier chapitre, nous avons présenté le contexte et l'intérêt d'utiliser l'optique intégrée pour recombiner des faisceaux en interférométrie astronomique. Notre but est maintenant de concevoir un dispositif destiné à la recombinaison de quatre faisceaux et se rapprochant le plus possible des caractéristiques requises.

Dans un premier temps, nous passerons en revue les possibilités offertes par l'optique intégrée en termes de fonctions élémentaires afin de sélectionner les plus adaptées à un comportement achromatique sur une large bande spectrale. Dans un deuxième temps, nous présenterons les différentes architectures possibles de recombinaison à quatre télescopes, puis nous décrirons les différents recombineurs possibles à deux, trois, puis quatre télescopes. Enfin, nous présenterons l'étude détaillée d'un déphaseur achromatique permettant d'obtenir des franges d'interférence en quadrature de phase sur une large bande spectrale et son intégration dans le dispositif complet.

# *II – B – Fonctions optiques intégrées*

## II – B – 1 – Introduction

L'optique guidée et intégrée a connu un développement important grâce aux télécommunications optiques à longue distance. Les fibres optiques ont alors vu leur transmission améliorée jusqu'à atteindre quasiment la limite théorique de la transmission de la silice avec laquelle elles sont fabriquées [*40*]. Rapidement, l'optique intégrée a été pensée pour réaliser les composants d'extrémité permettant de traiter l'information issue de ces réseaux de fibres optiques [*41*]. Il est certes possible de réaliser des coupleurs à fibres optiques permettant de séparer ou recombiner les signaux lumineux, mais l'égalité des chemins optiques à l'intérieur du réseau ainsi formé reste un problème. L'optique intégrée offre une plus grande diversité de fonctions élémentaires et permet de réaliser des dispositifs plus complexes. Parmi les nombreuses fonctions possibles, voici celles qui apparaissent naturellement lorsque l'on envisage une recombinaison optique intégrée (figure II-1) :

- *Guides droits, guides courbes, croisements de guides :* ils sont bien évidemment présents sur la puce pour relier les différentes fonctions. La technologie optique intégrée en silice sur substrat de silicium étant par nature planaire, des croisements de guides sont nécessaires pour réaliser un routage complexe.
- *Jonctions « Y » :* un guide se séparant en deux guides est un excellent moyen pour prélever une information photométrique. Les jonctions « Y » peuvent aussi être utilisées à l'envers pour recombiner deux signaux et fournir un signal interférométrique. Les premières franges obtenues sur IOTA en optique intégrée utilisaient des jonctions Y [*17*].
- *Coupleurs à ondes évanescentes :* une jonction Y ne permet de fournir qu'une sortie interférométrique alors qu'un coupleur à ondes évanescentes permet de fournir les deux sorties complémentaires. En contrepartie, son comportement dépend de la longueur d'onde d'utilisation.
- *Tricoupleurs à ondes évanescentes :* ils permettent de séparer un signal en trois et peuvent permettre de contourner les défauts d'un coupleur à ondes évanescentes dans le cas d'une recombinaison à deux faisceaux de très haute performance [*42*][*43*].

- *Coupleurs interférentiels multimode (MMI) :* ils ont été envisagés pour la recombinaison à plusieurs faisceaux [*44*]. L'avantage principal est qu'ils sont applicables à une recombinaison de plus de deux faisceaux. L'inconvénient est qu'ils présentent une forte dépendance spectrale et nécessitent une bande d'observation restreinte pour le moment.
- *Recombineurs à zone planaire :* les deux voies à recombiner sont injectées dans une zone planaire afin de fournir des franges spatialement étalées. Ils sont plus facilement extensibles à une recombinaison multitélescopes [*45*].
- *Recombinaison ABCD :* il peut être très avantageux de fournir, pour une paire de faisceaux recombinés, non pas deux sorties en opposition de phase mais quatre sorties en quadrature de phase. Diverses solutions ont été proposées [*46*][*47*] pour réaliser cette fonction, mais dans le cadre de capteurs utilisant une lumière monochromatique. Nous verrons ici comment étendre cette fonction à une large bande spectrale.

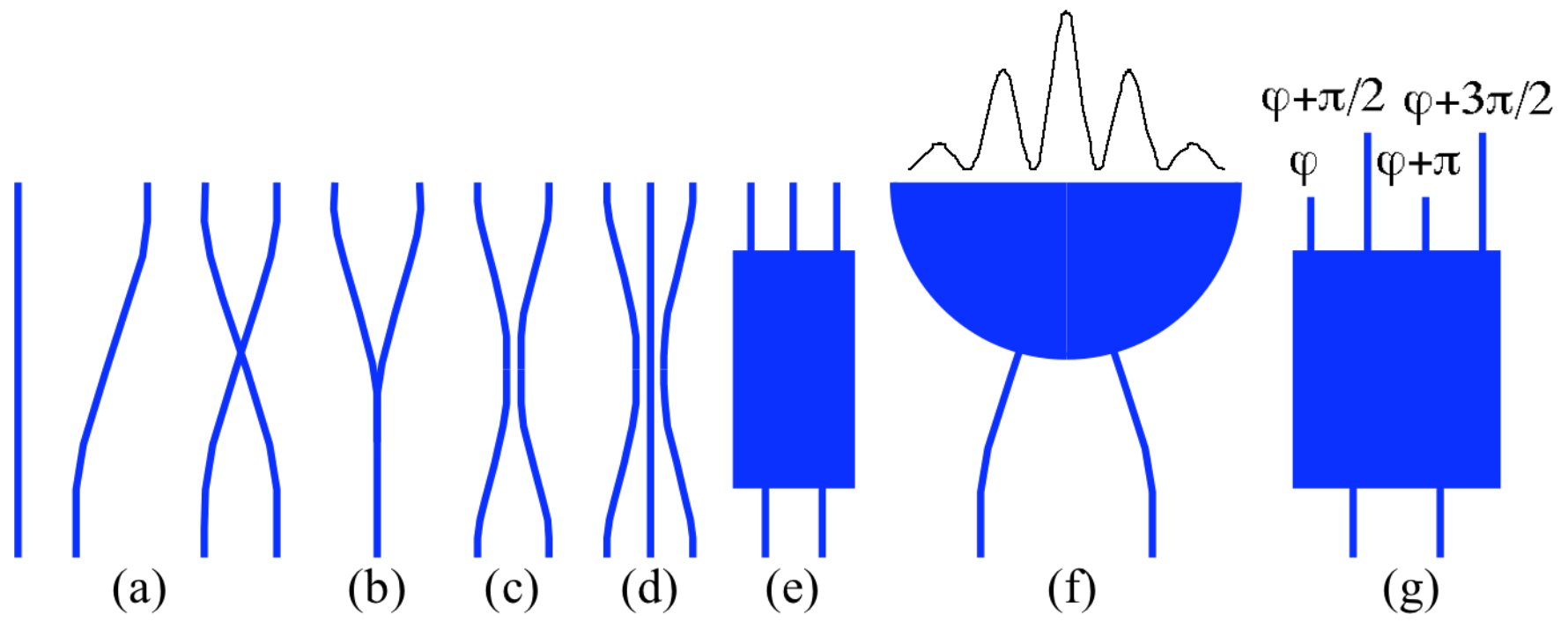

*Figure II-1 : fonctions élémentaires en optique intégrée. (a) guides droits, courbes, croisements. (b) jonction « Y ». (c) coupleur à ondes évanescentes. (d) tricoupleur. (e) coupleur multimode (MMI). (f) recombineur planaire. (g) recombineur « ABCD ».*

Ces fonctions ont été pour la plupart développées pour les télécommunications optiques dans un domaine de longueur d'onde assez restreint : autour de 1550nm et 1300nm qui sont les deux domaines de meilleure transparence de la silice. L'observation astronomique, quant à elle, utilise les bandes de transparence de l'atmosphère. Du visible à l'infrarouge moyen, le tableau II-1 donne les différentes bandes spectrales utilisées.

| **Bande spectrale** | $\lambda_{méd}$ µm | $\Delta\lambda$ µm | **Fo** $W/m^3$ | $\lambda_{min}$ µm | $\lambda_{max}$ µm | $\Delta\lambda/\lambda$ % |
|---|---|---|---|---|---|---|
| **U** | 0.367 | 0.066 | $3.981\ 10^{-2}$ | 0.334 | 0.400 | 17.98 |
| **B** | 0.436 | 0.094 | $6.310\ 10^{-2}$ | 0.389 | 0.483 | 21.56 |
| **V** | 0.545 | 0.088 | $3.631\ 10^{-2}$ | 0.501 | 0.589 | 16.15 |
| **R** | 0.638 | 0.138 | $2.239\ 10^{-2}$ | 0.569 | 0.707 | 21.63 |
| **I** | 0.797 | 0.149 | $1.148\ 10^{-2}$ | 0.7225 | 0.8715 | 18.70 |
| **J** | 1.220 | 0.213 | $3.162\ 10^{-3}$ | 1.1135 | 1.3265 | 17.46 |
| **H** | 1.630 | 0.307 | $1.148\ 10^{-3}$ | 1.4765 | 1.7835 | 18.83 |
| **K** | 2.190 | 0.390 | $3.981\ 10^{-4}$ | 1.995 | 2.385 | 17.81 |
| **L** | 3.450 | 0.472 | $7.079\ 10^{-5}$ | 3.214 | 3.686 | 13.68 |
| **M** | 4.750 | 0.460 | $2.042\ 10^{-5}$ | 4.520 | 4.980 | 09.68 |
| **N** | 10.200 | 4.000 | $1.230\ 10^{-6}$ | 8.200 | 12.200 | 39.22 |
| **Q** | 21.000 | 5.000 | $6.761\ 10^{-8}$ | 18.500 | 23.500 | 23.81 |

*Tableau II-1 : bandes spectrales d'observation astronomique au sol. Les longueurs d'onde médianes, min. ,et max. sont en µm.* Fo *est le flux correspondant à une magnitude* 0 *pour la bande considérée.*

Le travail décrit dans ce chapitre concerne la réalisation de composants utilisés dans la bande H. Une partie du chapitre IV est consacrée à l'extension de ces travaux aux autres bandes spectrales J, et K.

## II – B – 2 – Guidage unimodal ou « monomode »

La géométrie des guides d'onde intégrés avec lesquels nous travaillons est représentée sur la figure II-2.

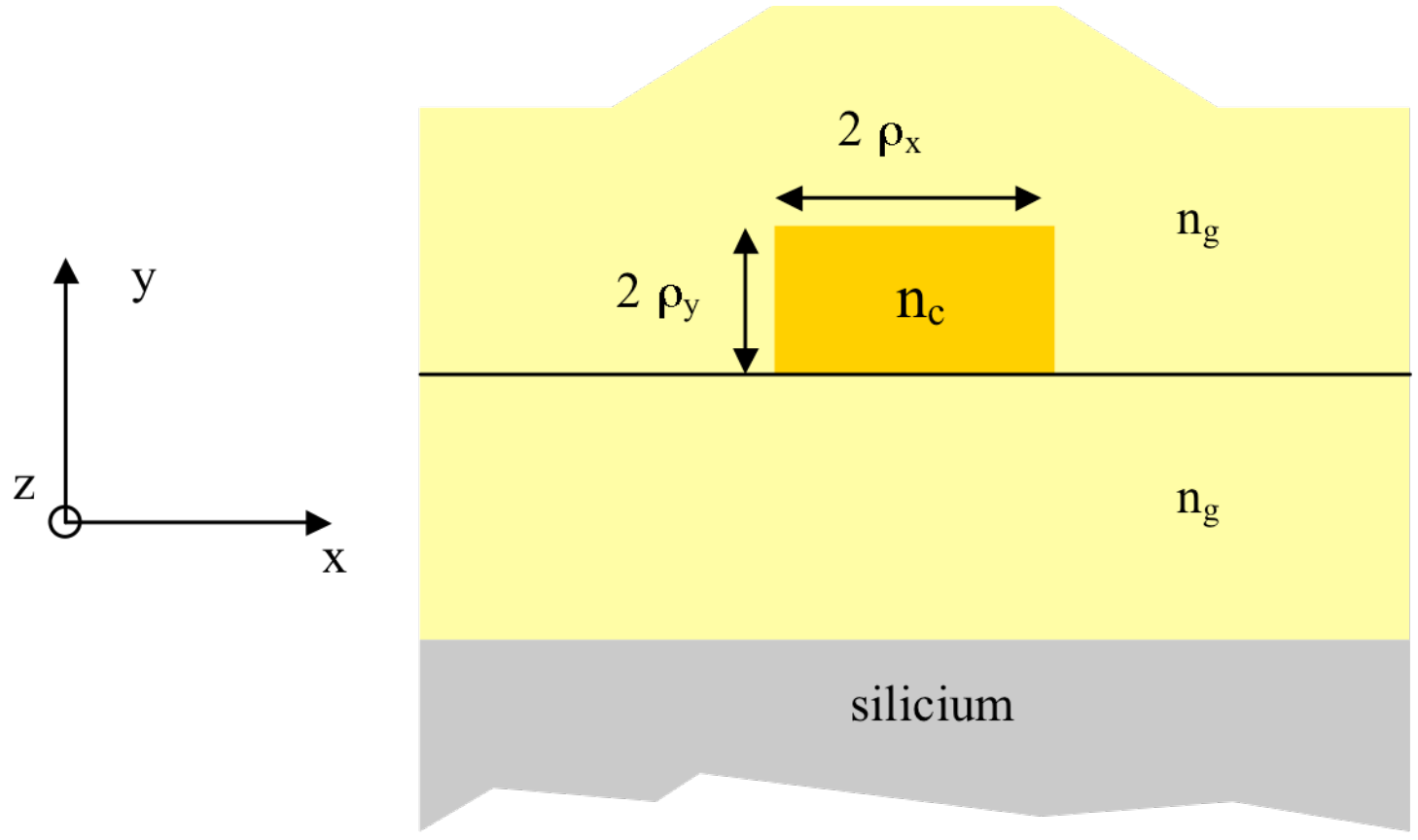

*Figure II-2 : guides optiques intégrés utilisés dans cette étude.*

Ils sont composés d'un cœur rectangulaire en silice dopée d'indice $n_c$ entouré d'une gaine en silice d'indice $n_g$. La largeur du guide d'onde peut être fixée en fonction des besoins. La hauteur maximale du cœur du guide d'onde est limitée par les contraintes technologiques. Le cône d'acceptance d'angle $\alpha$ de la lumière pouvant rentrer dans ce type de guide d'onde est donné par l'ouverture numérique *ON* :

$$ON = \sin(\alpha) = \sqrt{n_c^2 - n_g^2} \qquad \text{(II-1)}$$

Afin d'être compatible avec les ouvertures numériques des principaux télescopes et des fibres optiques de filtrage utilisées, on utilise typiquement un écart d'indice entre le cœur et la gaine du guide d'onde d'environ 0,01. On parle alors de « faible écart d'indice ».

Une étude très complète des guides d'onde rectangulaires de ce type peut être trouvée dans [*48*]. Nous reprendrons ici les principaux résultats de cette étude pour les appliquer à notre cas particulier. Afin de décrire le comportement de nos guides d'onde, nous allons utiliser le paramètre normalisé *V* du guide défini par :

$$V = \frac{2\pi}{\lambda}\sqrt{\rho_x \rho_y}\sqrt{n_c^2 - n_g^2} \qquad \text{(II-2)}$$

Ce paramètre permet de décrire complètement le comportement modal du guide. En première approximation, le nombre de modes guidés dépend de manière quadratique du paramètre *V*. Plus le guide est large, plus l'écart d'indice est élevé, ou plus la longueur d'onde d'utilisation est petite, plus le guide d'onde comporte de modes de propagation. Si l'on définit le paramètre ε comme étant :

$$\varepsilon = \frac{\rho_y}{\rho_x} \qquad \text{(II-3)}$$

alors la limite de monomodicité de ces guides peut être donnée par la relation[*48*] :

$$V_{co} = \frac{1.4464 + 1.03\varepsilon}{1 + 0.1594\varepsilon} \qquad \text{(II-4)}$$

Relation reportée sur le graphe de la figure II-3.

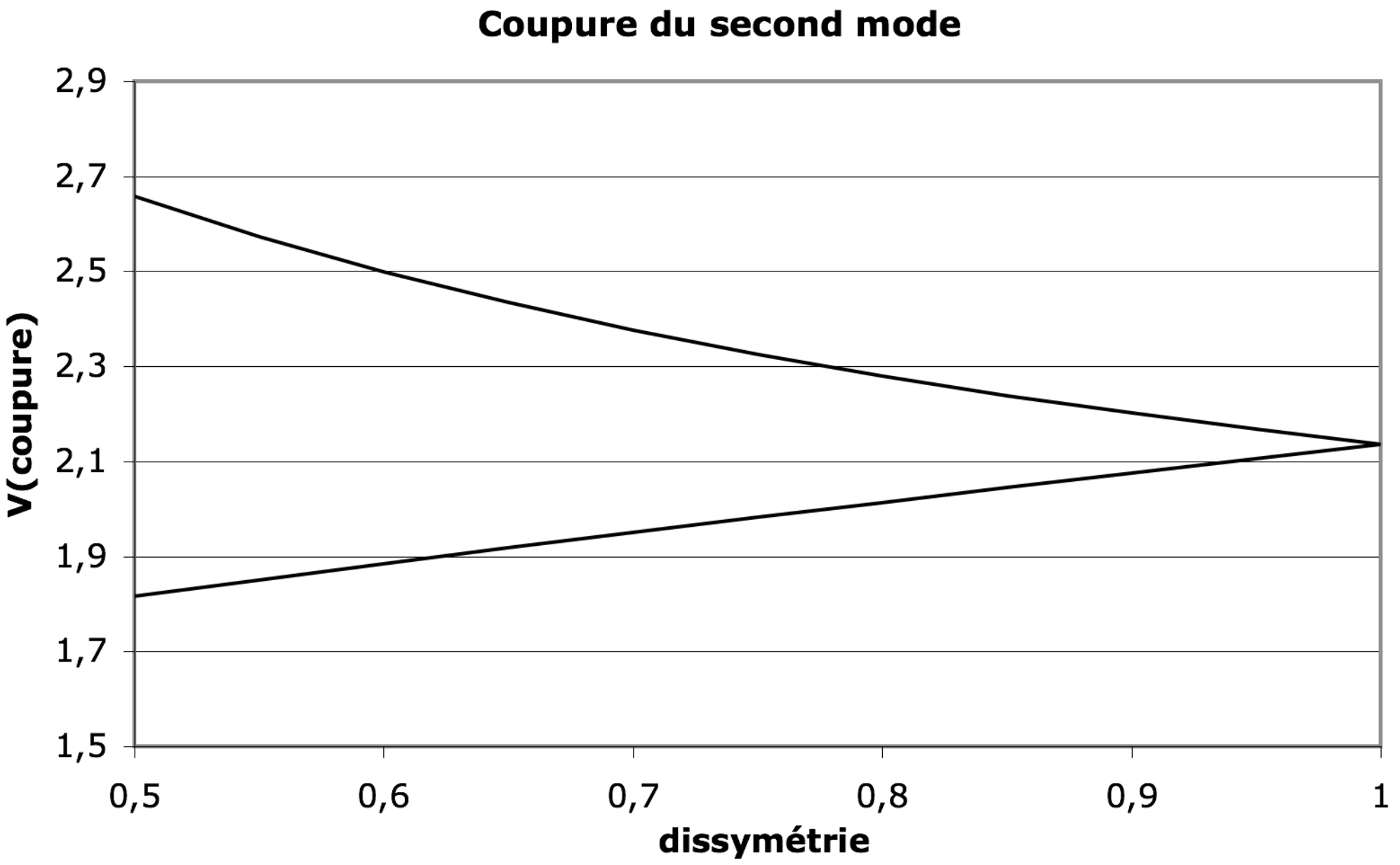

*Figure II-3 : limite de monomodicité pour des guides d'onde rectangulaires à faible écart d'indice. Les deux courbes sont liées par une rotation de guide de 90 degrés et donc par un changement de ε en 1/ε.*

On fixe ainsi la largeur de nos guides d'onde de manière à obtenir la coupure du second mode de propagation légèrement en dessous de la longueur d'onde minimum de la bande H.

## II – B - 3 – Modes guidés

La variation d'indice entre le cœur et la gaine du guide étant faible, le mode fondamental peut être considéré comme polarisé linéairement suivant une des directions privilégiées du guide. Il peut se mettre sous la forme :

$$\vec{E} = A_0 \psi(x,y) e^{j\beta z} \vec{u} \qquad \text{(II-5a)}$$

$$\vec{H} = B_0 \psi(x,y) e^{j\beta z} \vec{v} \tag{II-5b}$$

Ou *u et v* représentent deux directions transverses orthogonales x, ou y, du guide, et $A_0$ et $B_0$ définissent la quantité de puissance transportée par le mode fondamental. $\psi(x,y)$ est alors solution de l'équation d'onde scalaire :

$$\frac{\partial^2 \psi}{\partial x^2} + \frac{\partial^2 \psi}{\partial x^2} + \left(k^2 n(x,y)^2 - \beta^2\right)\psi = 0 \tag{II-6}$$

*k* est la norme du vecteur d'onde et *n(x,y)* est la répartition d'indice transverse à la propagation. Si l'on multiplie cette équation par $\psi^*$ et que l'on intègre sur une section infinie transverse à la direction de propagation, on obtient après calcul une autre forme de cette équation permettant de calculer la constante de propagation $\beta$ :

$$\beta^2 = \frac{\int_{A_\infty} \left[ k^2 n^2 |\psi|^2 - \left|\frac{\partial \psi}{\partial x}\right|^2 - \left|\frac{\partial \psi}{\partial y}\right|^2 \right] dA}{\int_{A_\infty} |\psi|^2 dA} \tag{II-7}$$

Dans le cas de guides rectangulaires, il n'existe pas de solution analytique à l'équation (II-6), celle-ci doit donc être résolue numériquement. Cependant, le résultat fait apparaître que la répartition transverse est très proche d'une gaussienne de la forme :

$$\psi(x,y) \approx \psi_0 \exp\left(-\frac{x^2}{\omega_x^2} - \frac{y^2}{\omega_y^2}\right) \tag{II-8}$$

Il est bien connu [*29*] que le mode fondamental $\psi(x,y)$ maximise la valeur de $\beta$ dans l'équation (II-7). La meilleure gaussienne (c'est-à-dire les meilleurs paramètres $\omega_x$, $\omega_y$) décrivant la propagation dans le guide sera celle maximisant aussi cette valeur de $\beta$. On obtient alors facilement ces paramètres $\omega_x$, $\omega_y$ en posant :

$$\frac{\partial \beta^2}{\partial \omega_x} = \frac{\partial \beta^2}{\partial \omega_y} = 0 \qquad \text{(II-9)}$$

Après calcul, on obtient les deux équations couplées suivantes qui permettent de relier $\omega_x$ et $\omega_y$ aux paramètres optogéométriques des guides d'onde :

$$\exp\left[-\frac{2\rho_x^2}{\omega_x^2}\right] erf\left[\frac{\sqrt{2}\rho_y}{\omega_y}\right] = \sqrt{\frac{\pi}{2}}\frac{\rho_x}{V_x^2\omega_x} \qquad \text{(II-10)}$$

$$\exp\left[-\frac{2\rho_y^2}{\omega_y^2}\right] erf\left[\frac{\sqrt{2}\rho_x}{\omega_x}\right] = \sqrt{\frac{\pi}{2}}\frac{\rho_y}{V_y^2\omega_y} \qquad \text{(II-11)}$$

où $V_x$ et $V_y$ sont définis par :

$$V_u = \frac{2\pi}{\lambda}\rho_u\sqrt{n_c^2 - n_g^2} \qquad \text{avec u = x, y} \qquad \text{(II-12)}$$

On obtient alors l'ouverture numérique pour chaque direction transverse du faisceau gaussien lumineux entrant ou sortant du guide par la relation :

$$ON_u = \sin(\theta_u) = \sin\left(\frac{\lambda}{\pi n \omega_u}\right) \qquad \text{avec u = x, y} \qquad \text{(II-13)}$$

où $n$ est l'indice du cœur ou de la gaine (puisque la différence d'indice entre les deux est petite). La connaissance de ces largeurs de mode ainsi que des ouvertures numériques permet alors de dimensionner correctement les optiques de transport de flux en amont et en aval du composant intégré.

## II – B – 4 – Courbures

Toujours dans [*48*], Ladouceur et Love ont montré que les pertes par courbure des guides rectangulaires à faible écart d'indice pouvaient s'approximer par la relation :

$$P(z) = P_0 \exp(-\alpha z) \tag{II-14}$$

où le coefficient de pertes $\alpha$ vaut :

$$\alpha = \frac{1}{8}\frac{V^4 I}{\sqrt{\pi \rho R W^3}} \exp\left(-\frac{4}{3}\frac{W^3 \Delta}{V^2}\frac{R}{\rho}\right) \tag{II-15}$$

où :

$$\rho = \sqrt{\rho_x \rho_y} \tag{II-16}$$

$$W = \rho\sqrt{\beta^2 - k^2 n_g^2} \tag{II-17}$$

$$\Delta = \frac{n_c^2 - n_g^2}{2n_c^2} \tag{II-18}$$

$$I = \frac{\left|\int_{coeur}\left(n_c^2 - n_g^2\right)\psi dA\right|^2}{\int_{A_\infty}|\psi|^2 dA} \tag{II-19}$$

Le coefficient de perte varie principalement de manière exponentielle en fonction du rayon de courbure $R$. Plus le rayon de courbure est grand, plus les pertes seront faibles, mais plus la puce sera longue et présentera donc des pertes de propagation intrinsèques importantes. On pourrait ainsi en théorie calculer un rayon de courbure optimal dépendant du dessin détaillé de chaque dispositif. Dans la pratique, il n'est pas nécessaire de calculer un rayon de courbure limite avec une telle précision. Nous avons fixé comme règle de dessin un rayon de courbure imposant des pertes maximales de un décibel par mètre.

Mais, même lorsque la courbure est suffisamment faible pour que les pertes soient négligeables, le mode de propagation subit un décalage latéral vers l'extérieur de la courbure, décalage proportionnel à cette courbure. Afin de minimiser ce phénomène, toutes les puces présentées dans ce document utilisent des chemins à courbure continûment variable de manière à annuler ce décalage de mode entre guides droits et guides courbes. Cette technique de dessin, décrite en détail dans [*49*], a été brevetée et validée expérimentalement par le LETI [*50*]. Elle est la méthode la plus précise que nous connaissons pour assurer un mode fondamental le moins déformé possible et toujours centré. L'intérêt majeur est alors de pouvoir concaténer des chemins de courbures différentes les uns à la suite des autres sans avoir à se soucier des transitions entre les différentes sections. Ce point nous permet en particulier d'égaliser précisément les chemins optiques à l'intérieur des puces pour pouvoir obtenir une mesure de phase précise.

Un soin particulier a été apporté pour que ces chemins géométriques soient générés automatiquement sur le logiciel de conception des masques de photolithographie, ceci avec une précision dépendant de la grille définitive du masque de photolithographie.

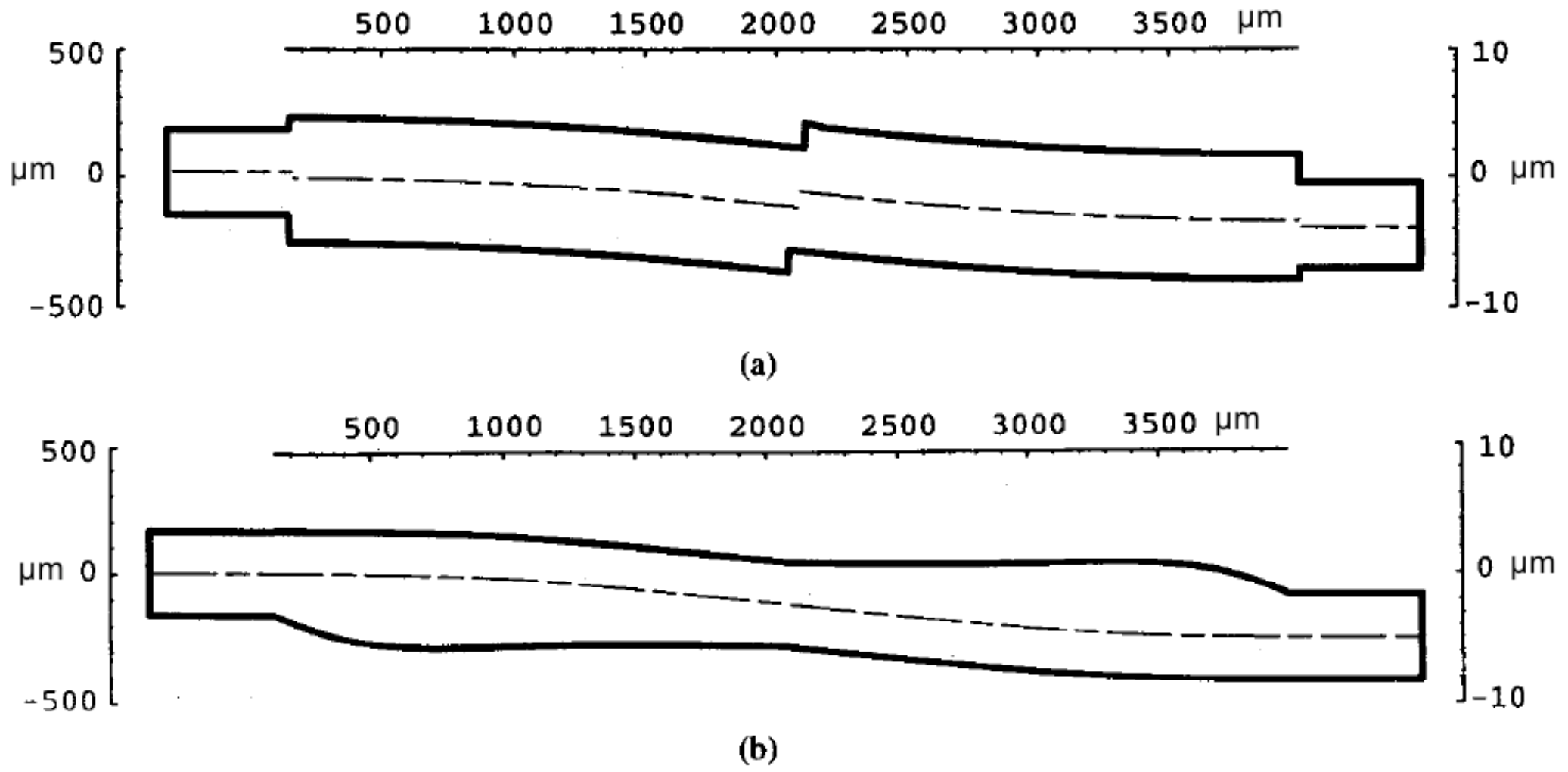

*Figure II-4 : exemple de guide en « S » optimisé de manière classique (a), et optimisé par courbures variant continûment (b). L'échelle verticale de gauche s'applique au chemin géométrique. L'échelle verticale de droite s'applique à la largeur du guide et a été dilatée pour plus de clarté.*

## II – B - 5 – Croisements de guides

Les guides d'onde optiques présentent la propriété de se croiser sans coupler d'énergie de l'un à l'autre si l'angle entre les guides est suffisamment grand. Une étude théorique des croisements de guides rectangulaires peut être trouvée dans [*51*].

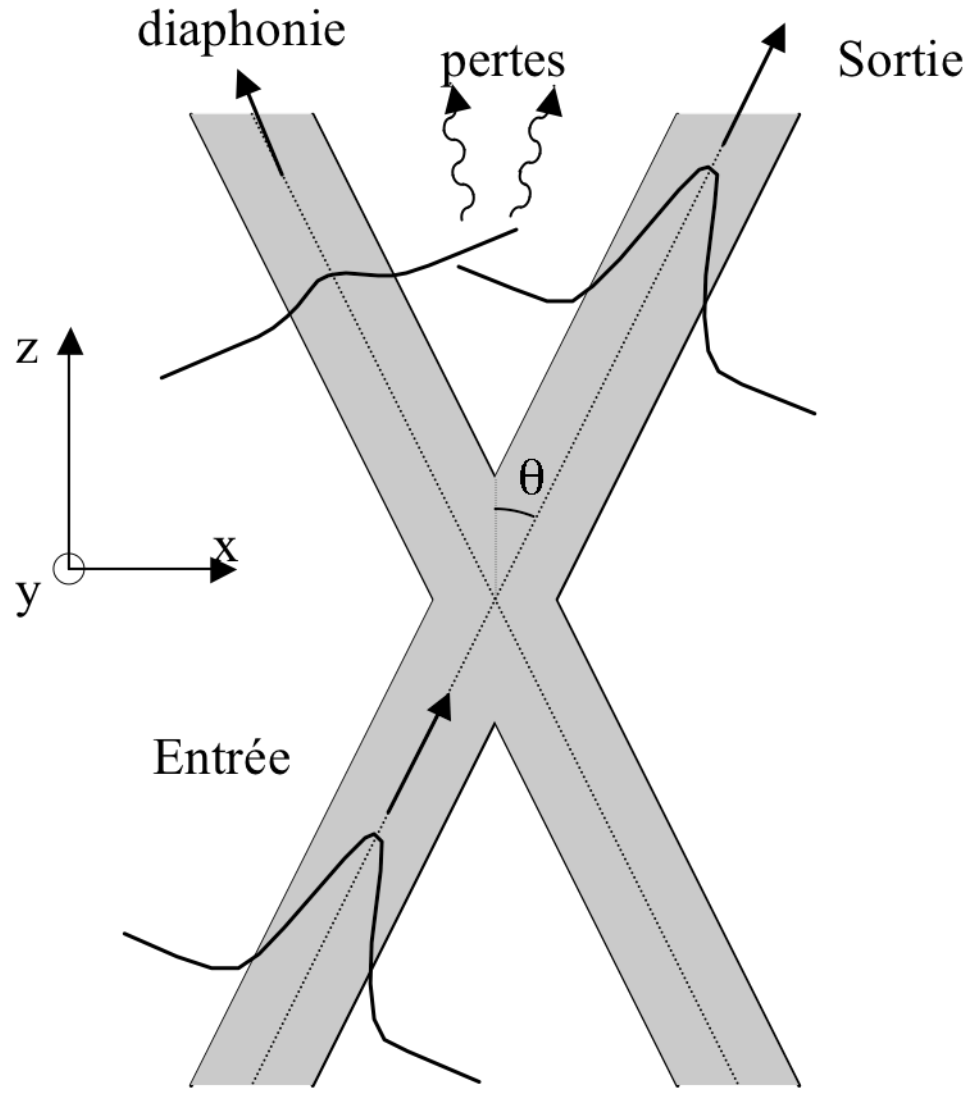

*Figure II-5 : croisement de guides : définition des termes.*

En première approximation, on peut supposer que si l'angle est supérieur à deux fois l'ouverture numérique de chaque mode de propagation, il n'y aura pas de recouvrement des cônes d'acceptance des modes guidés, et il ne pourra donc pas se coupler de lumière entre les modes. Si l'on regarde un peu plus précisément la géométrie d'un croisement (figure II-5), un mode se propageant verra une première interface après laquelle il divergera suivant son angle d'ouverture numérique, puis la majeure partie de l'énergie sera recouplée au mode du guide à la deuxième interface en subissant une légère diffraction sur les bords. On peut donc s'attendre à ce que la propagation subisse quelques pertes à chaque croisement. Afin d'évaluer ces pertes et vérifier qu'il n'y avait pas de couplage entre les guides, nous avons effectué des simulations numériques de type BPM (Beam Propagation Method) [*52*] pour des croisements de guides ayant des angles autour de la valeur déterminée par l'ouverture numérique du mode fondamental.

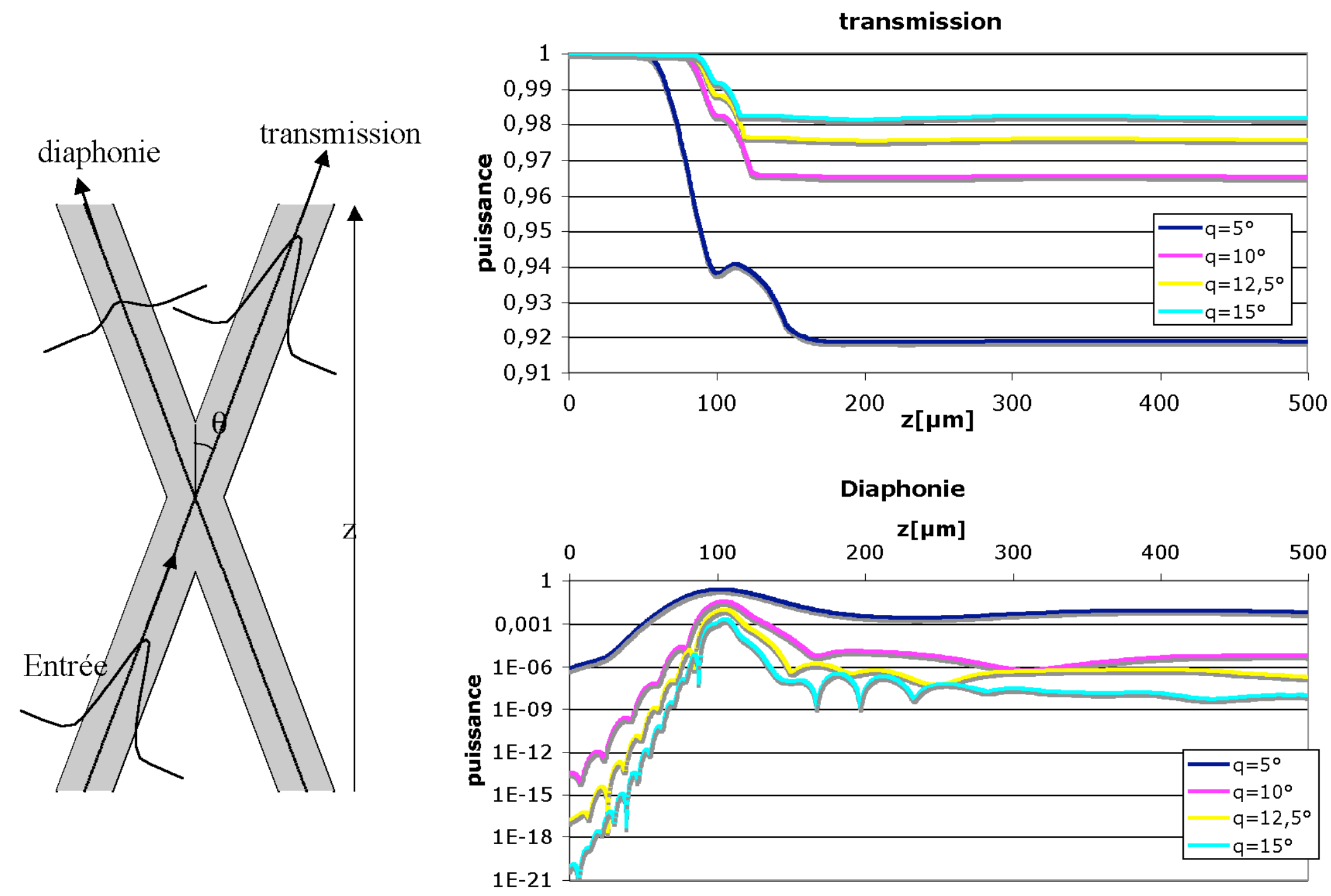

*Figure II-6 : résultats de simulations de transmission et de diaphonie pour les croisements de guides avec des demi angles entre guides compris entre 5° et 15°. Le croisement se situe à z = 100µm. Les graphes représentent la puissance transportée par le mode fondamental de chaque guide à une longueur d'onde de 1,65µm.*

Les résultats font apparaître que le couplage d'énergie entre les modes guidés décroît lorsque l'angle croît et qu'il faut au moins un angle de 30 degrés entre les guides pour que le couplage soit inférieur à $10^{-6}$. Nous avons conservé par la suite cet angle comme règle de dessin des puces. Cet angle relativement grand est en grande partie responsable de l'encombrement des puces dessinées.

Par ailleurs, les valeurs de pertes simulées sont faibles (de l'ordre de 0,08dB) mais sont à prendre avec précaution, la précision du calcul BPM étant du même ordre de grandeur. La caractérisation des puces réalisées semble donc être un meilleur moyen d'évaluer ces pertes.

## II – B - 6 – Jonctions « Y »

Une jonction « Y » peut être simplement décrite comme un guide se séparant progressivement en deux guides. Sa fonctionnalité réside principalement dans sa symétrie. Même lorsque des pertes se produisent à la séparation des guides, chaque guide de sortie recevra une puissance égale.

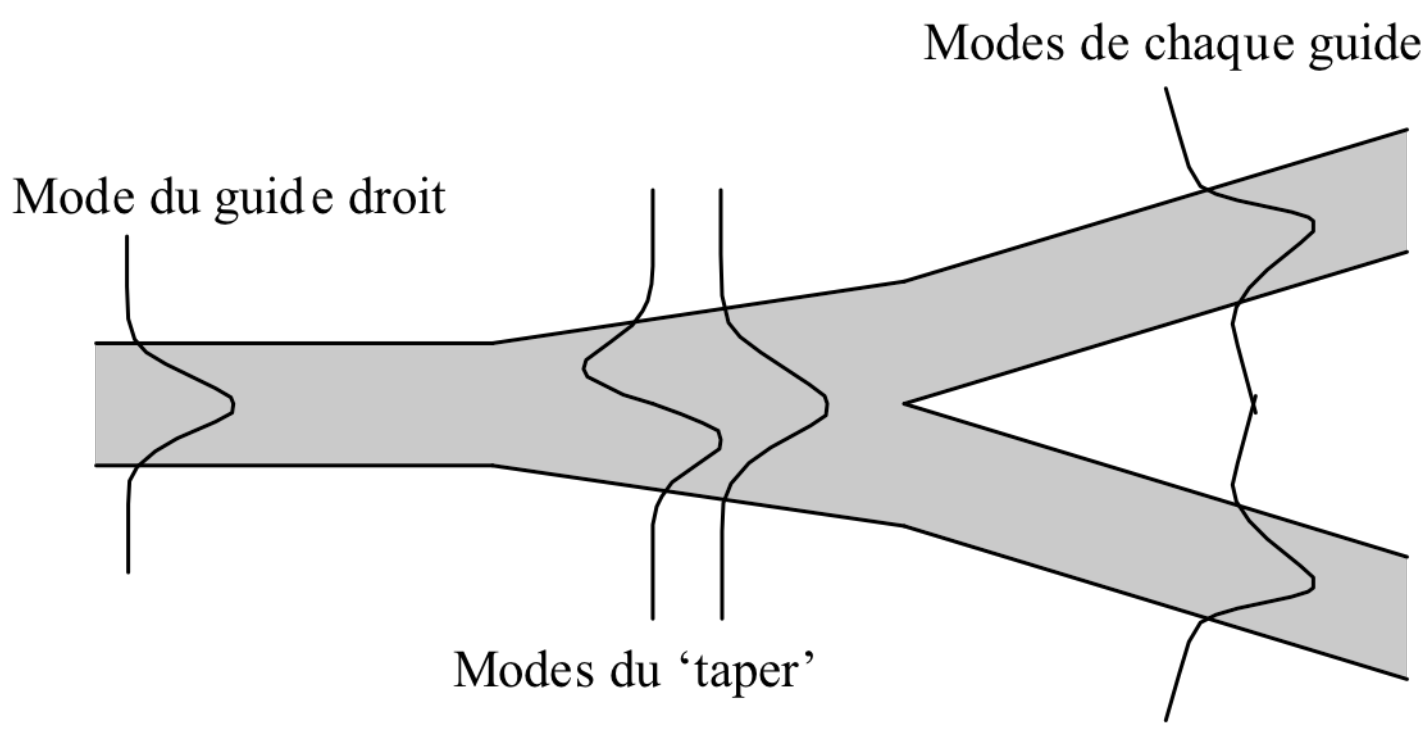

*Figure II-7 : jonction «Y ».*

Le comportement d'une jonction « Y » peut être bien expliqué en termes de propagation modale. A l'entrée, la structure est monomode, puis s'élargit. Elle devient alors bimode comportant un mode symétrique et un mode antisymétrique. Si cet élargissement est suffisamment lent, le champ électromagnétique conserve sa symétrie et se couplera dans le mode symétrique de la jonction. Dans la région située après la séparation des guides, la structure conserve sa structure bimode qui va en s'élargissant et la lumière sera donc couplée dans le mode symétrique, puis dans chacun des modes fondamentaux des guides de sortie lorsque ceux-ci seront suffisamment éloignés l'un de l'autre pour ne plus interagir.

Comme le montre la figure II-8, une jonction « Y » utilisée en sens inverse peut servir à recombiner deux faisceaux. Si ceux-ci sont en phase, ils se coupleront dans le mode symétrique de la jonction Y puis dans le mode fondamental du guide de sortie. S'ils sont en opposition de phase, ils se coupleront dans le mode antisymétrique de la structure qui ne sera plus guidé à la sortie de la jonction. La lumière sera rayonnée autour du guide.

Une jonction Y est donc un recombineur dont l'avantage est de fournir une sortie interférométrique à frange centrale brillante parfaitement achromatique. L'inconvénient est qu'elle ne fournit qu'une seule sortie : lorsque les signaux sont en opposition de phase, la lumière est rayonnée et n'est pas récupérable. En moyenne, on perd donc la moitié de la puissance du signal interférométrique.

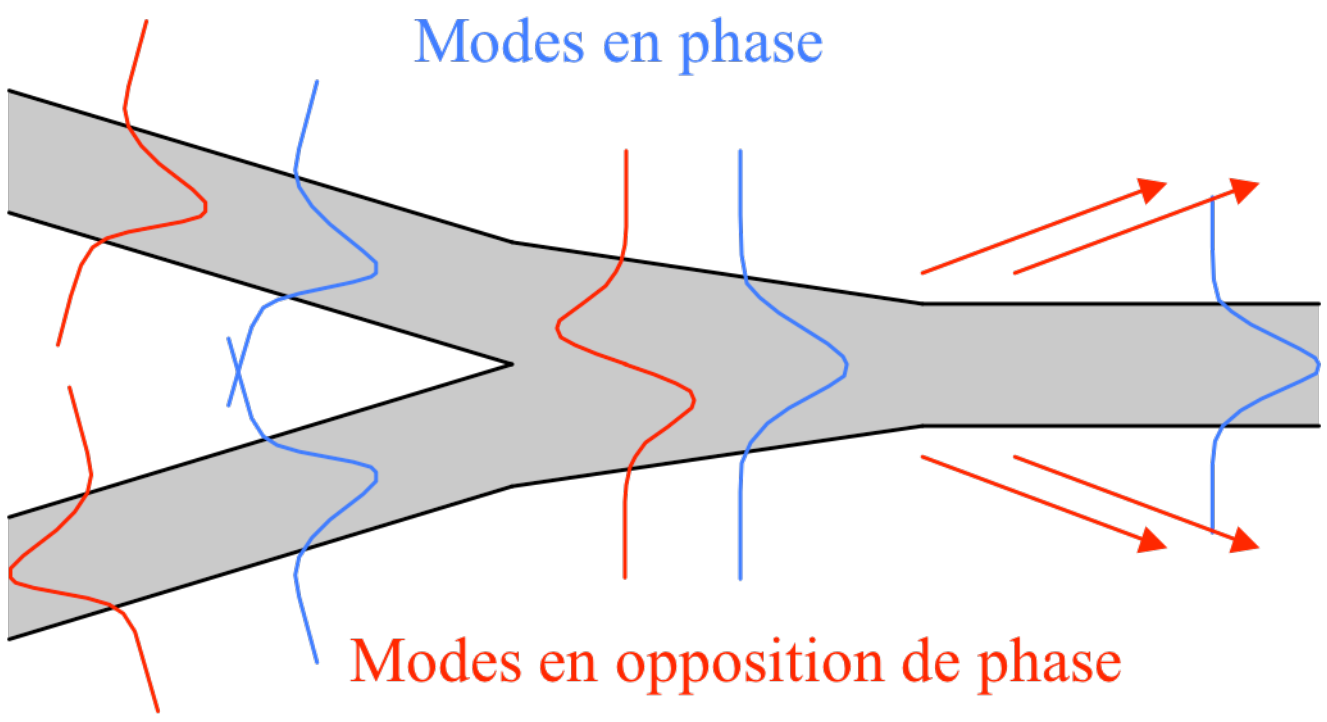

*Figure II-8 : recombinaison par jonction Y. Lorsque les modes sont en phase (bleu), la lumière se couple dans le mode fondamental du « taper » puis dans le mode du guide. Lorsque les modes sont en opposition de phase (rouge), la lumière se couple dans le second mode du taper, puis est ensuite rayonnée dans la gaine autour du guide.*

Au départ des deux guides se séparant, une jonction idéale comporte une pointe. Dans la pratique, la réalisation des guides inclut une étape de photolithographie dont la résolution est limitée. On ne peut donc pas réaliser une pointe infiniment fine. Pour contourner cette difficulté, on inclut dans le dessin une coche réalisable de manière plus fiable. Intuitivement cette coche étant par ailleurs une zone d'indice plus faible, la lumière se propage dans cette zone légèrement plus rapidement. Ainsi, le plan de phase se déforme pour s'adapter aux plans de phase des deux guides s'écartant l'un de l'autre à la sortie de la jonction.

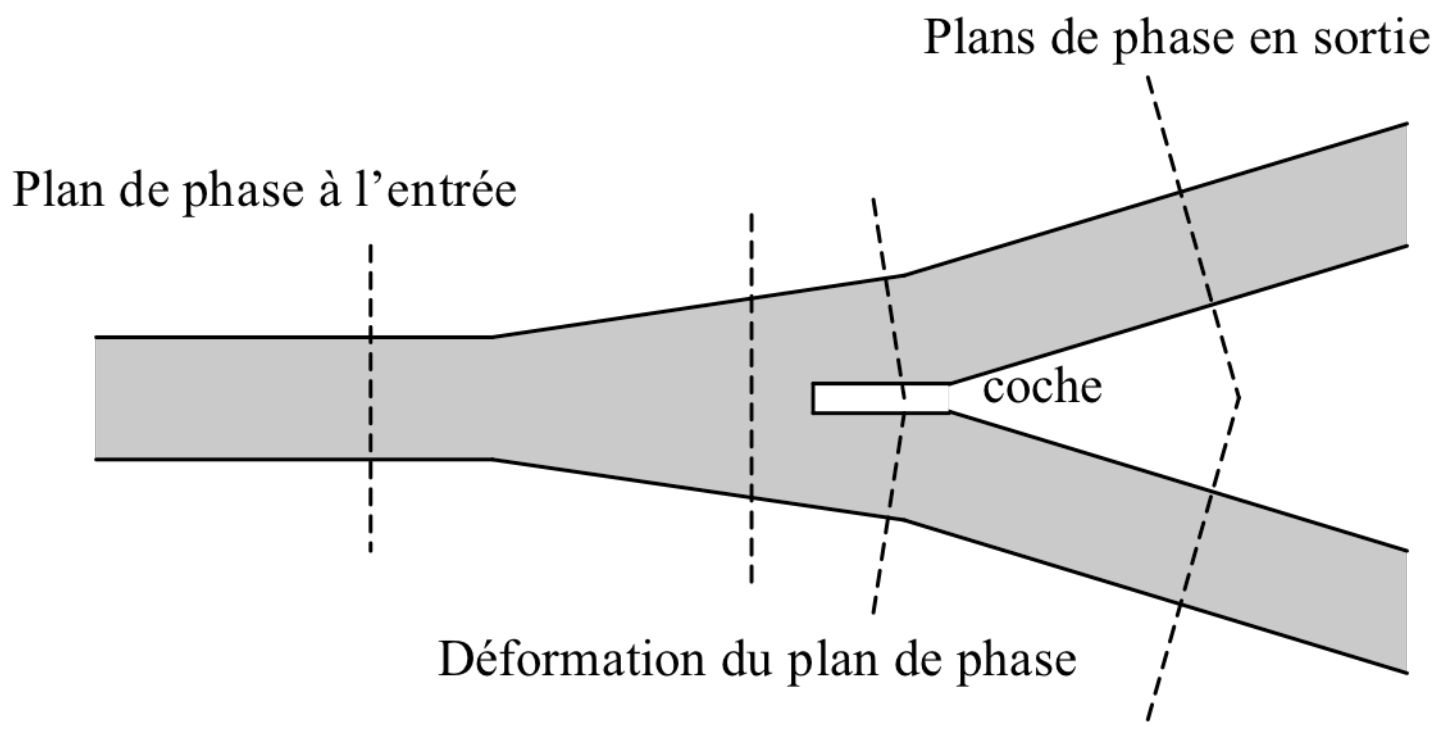

*Figure II-9 : jonction Y réalisée par photolithographie. La « coche » évite d'avoir à réaliser une pointe et permet de minimiser les pertes en adaptant le plan de phase au guides de sortie.*

L'optimisation d'une jonction Y se fait par simulations numériques. Cependant, on retient un critère pour faire cette optimisation. Dans la première zone, l'élargissement du guide doit être suffisamment faible pour ne pas introduire de pertes. Pour cela, on montre [*48*] que l'angle local du taper *θ(z)* doit être inférieur à une valeur limite donnée par :

$$\theta(z) < \frac{\rho(z)}{2\pi}\left(\beta - kn_g\right) \qquad \text{(II-20)}$$

Le calcul dans la seconde zone du Y (c'est-à-dire au niveau des guides de sortie) conduit à un angle similaire.

## II – B – 7 – Coupleurs asymétriques

Un coupleur est une structure comportant deux guides suffisamment proches pour pouvoir interagir et échanger de l'énergie entre leurs modes de propagation respectifs.

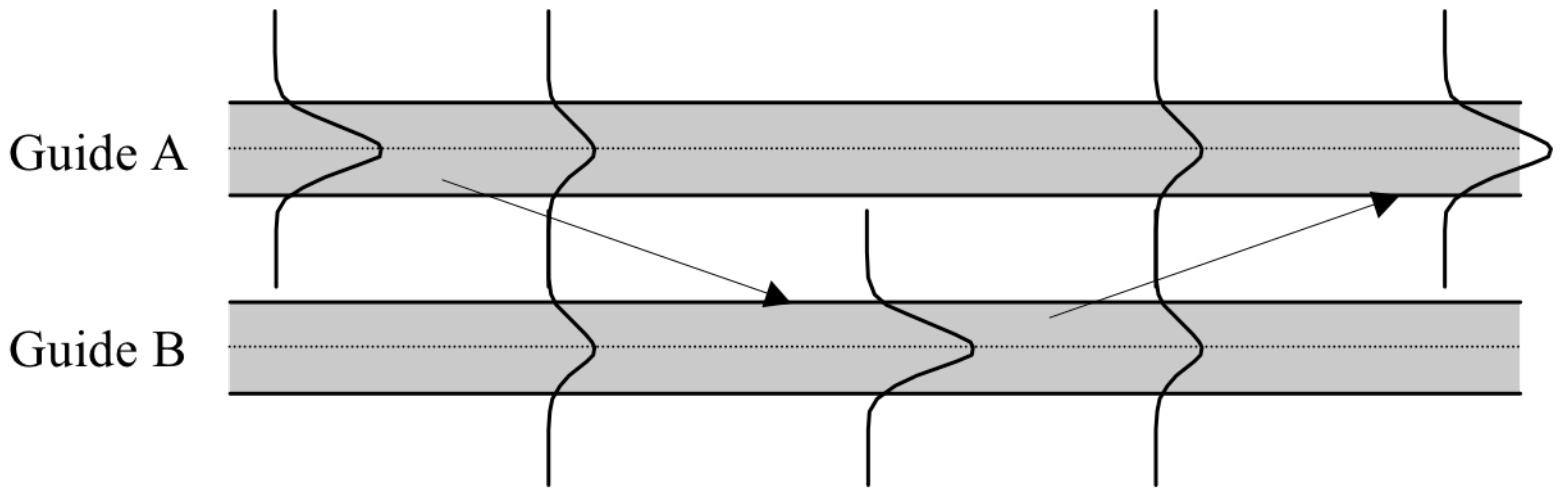

*Figure II-10 : schéma d'un coupleur. Lorsque les guides sont suffisamment proches, les modes fondamentaux de chaque guide peuvent échanger de l'énergie.*

Le calcul du comportement d'un coupleur par la méthode des perturbations peut être trouvé par exemple dans [29]. On considère que le champ total *E(z)* peut s'exprimer en fonction des modes $E_a$ et $E_b$ présents dans chaque guide :

$$E(z) = a(z)E_a + b(z)E_b \qquad \text{(II-21)}$$

Le problème est alors de trouver les coefficients *a(z)* et *b(z)* représentant la quantité de lumière couplée dans chaque mode au fur et à mesure de la propagation. On montre que ces coefficients sont régis par deux équations différentielles couplées :

$$\frac{da(z)}{dz} = j\delta a(z) + jKb(z) \qquad \text{(II-22)}$$

$$\frac{db(z)}{dz} = jKa(z) - j\delta b(z) \qquad \text{(II-23)}$$

où $\delta$ est l'écart entre les constantes de propagation de chaque mode :

$$\delta = \frac{\beta_a - \beta_b}{2} \qquad \text{(II-24)}$$

et *K* est le coefficient de couplage d'un mode sur l'autre :

$$K = \sqrt{\frac{\varepsilon_0}{\mu_0}}\frac{k}{4}\int_{A_\infty}\Delta\left(n^2\right)E_a E_b^* dA \qquad \text{(II-25)}$$

où $\Delta(n^2)$ est la perturbation d'indice qu'un guide représente sur l'autre. Après calcul, on obtient :

$$a(z) = a_0\cos\left(\sqrt{K^2+\delta^2}z\right) + a_1\sin\left(\sqrt{K^2+\delta^2}z\right) \qquad \text{(II-26)}$$

$$b(z) = b_0\cos\left(\sqrt{K^2+\delta^2}z\right) + b_1\sin\left(\sqrt{K^2+\delta^2}z\right) \qquad \text{(II-27)}$$

Les coefficients $a_0$, $a_1$, $b_0$, et $b_1$ s'obtiennent en fonction des conditions initiales, c'est à dire de l'injection. Les puissances transportées par chaque guide sont données par :

$$P_a(z) = |a(z)|^2 \qquad \text{(II-28)}$$

$$P_b(z) = |b(z)|^2 \qquad \text{(II-29)}$$

si l'on rentre une puissance unitaire dans le guide A, et rien dans le guide B, on obtient :

$$P_a(z) = 1 - F\sin^2(\mu z) \qquad \text{(II-30)}$$

$$P_b(z) = F\sin^2(\mu z) \qquad \text{(II-31)}$$

où l'on a :

$$F = \frac{K^2}{K^2+\delta^2} \qquad \text{(II-32)}$$

$$\mu = \sqrt{K^2+\delta^2} \qquad \text{(II-33)}$$

Dans le cas général, $F$ représente la puissance maximale qui peut être transférée d'un guide à l'autre, $\mu$ est inversement proportionnel à la distance pour laquelle a lieu ce transfert maximal de puissance. L'évolution de la puissance présente dans chaque guide au cours de la propagation est représentée sur la figure II-11 :

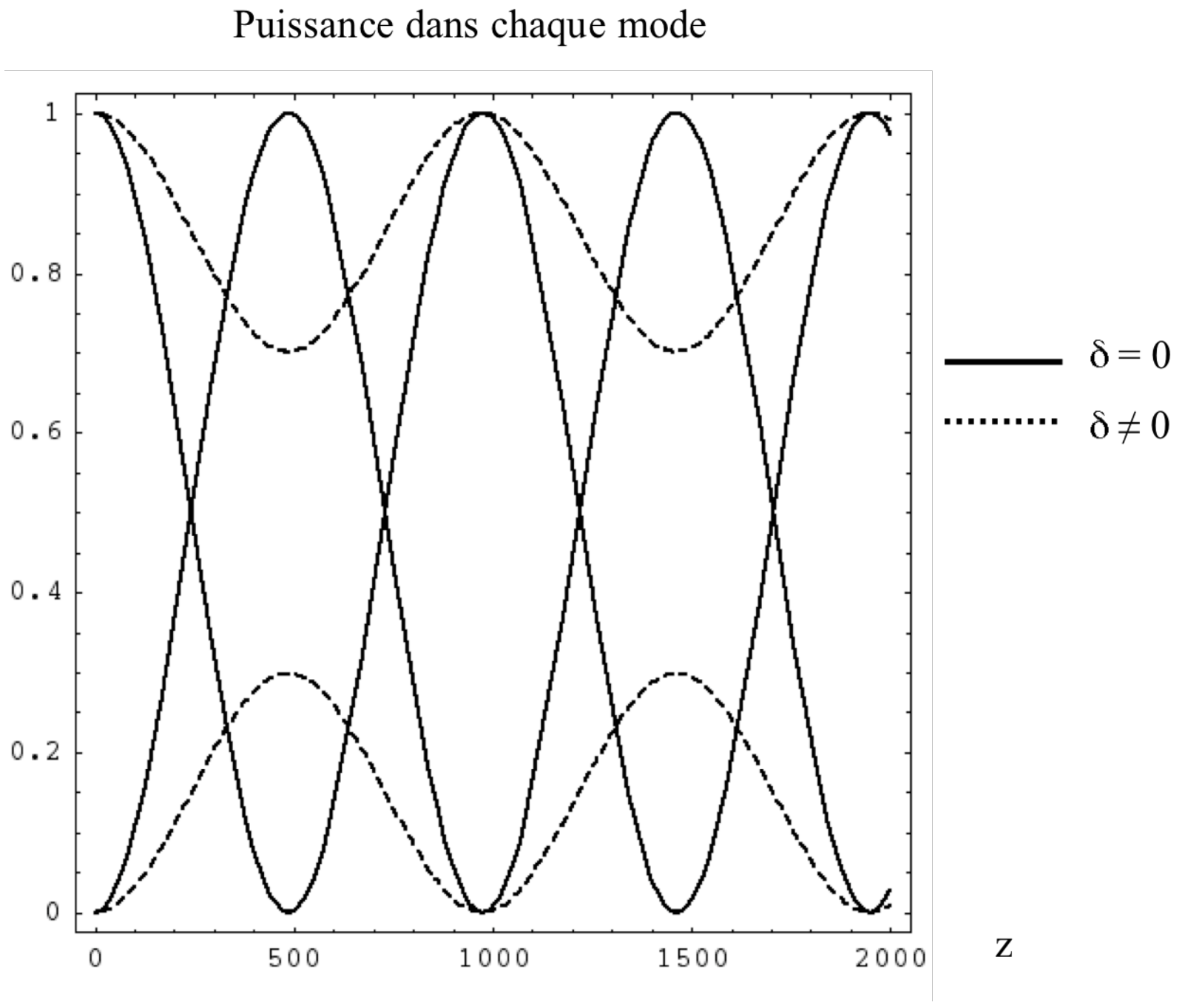

*Figure II-11 : puissance dans chaque guide du coupleur. Lorsque les guides sont identiques, toute la lumière peut être couplée d'un guide sur l'autre.*

Si les deux guides sont identiques, comme c'est le cas pour les coupleurs à fibres optiques, toute l'énergie peut être transférée d'un guide à l'autre. Les coupleurs à fibres 50/50 étant constitués de deux guides identiques, on choisit donc une longueur égale à un nombre impair de demi-longueur de couplage. Ce paramètre dépendant de la longueur d'onde, il en résulte qu'un coupleur symétrique est chromatique.

Lorsque l'on travaille avec des coupleurs asymétriques, les deux paramètres $F$ et $\mu$ varient en fonction de la longueur d'onde. Il est alors possible de trouver une combinaison de paramètres permettant de compenser le comportement chromatique du coupleur.

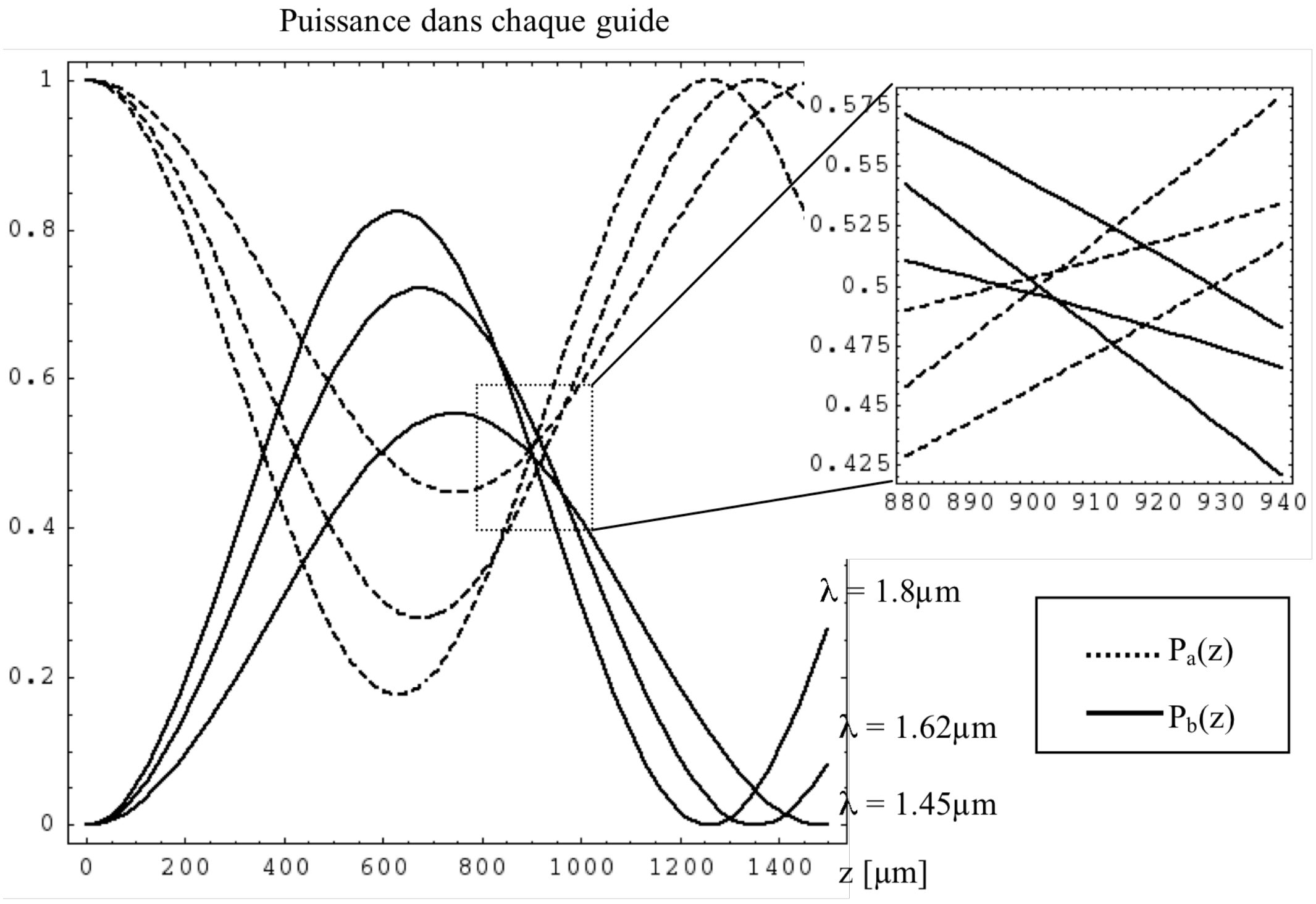

*Figure II-12 : puissance dans chaque guide pour un coupleur compensé. Les guides font 5µm et 4µm de large, et sont écartés de 3µm.* F *varie de 0,55 à 0,82 en fonction de la longueur d'onde et* µ *varie de 0,002µm*$^{-1}$ *à 0,0025µm*$^{-1}$*. On voit que pour une longueur de coupleur de 915µm, le couplage est compris entre 47% et 53% pour toute la bande H.*

On voit sur le graphique de la figure II-12 que les courbes de puissance dans chaque guide d'onde pour différentes longueurs d'onde se croisent quasiment toutes au même endroit. Pour cette longueur particulière d'interaction, le taux de couplage sera quasiment identique pour toutes les longueurs d'onde. On obtient une transmission spectrale du coupleur représentée sur la figure II-13 :

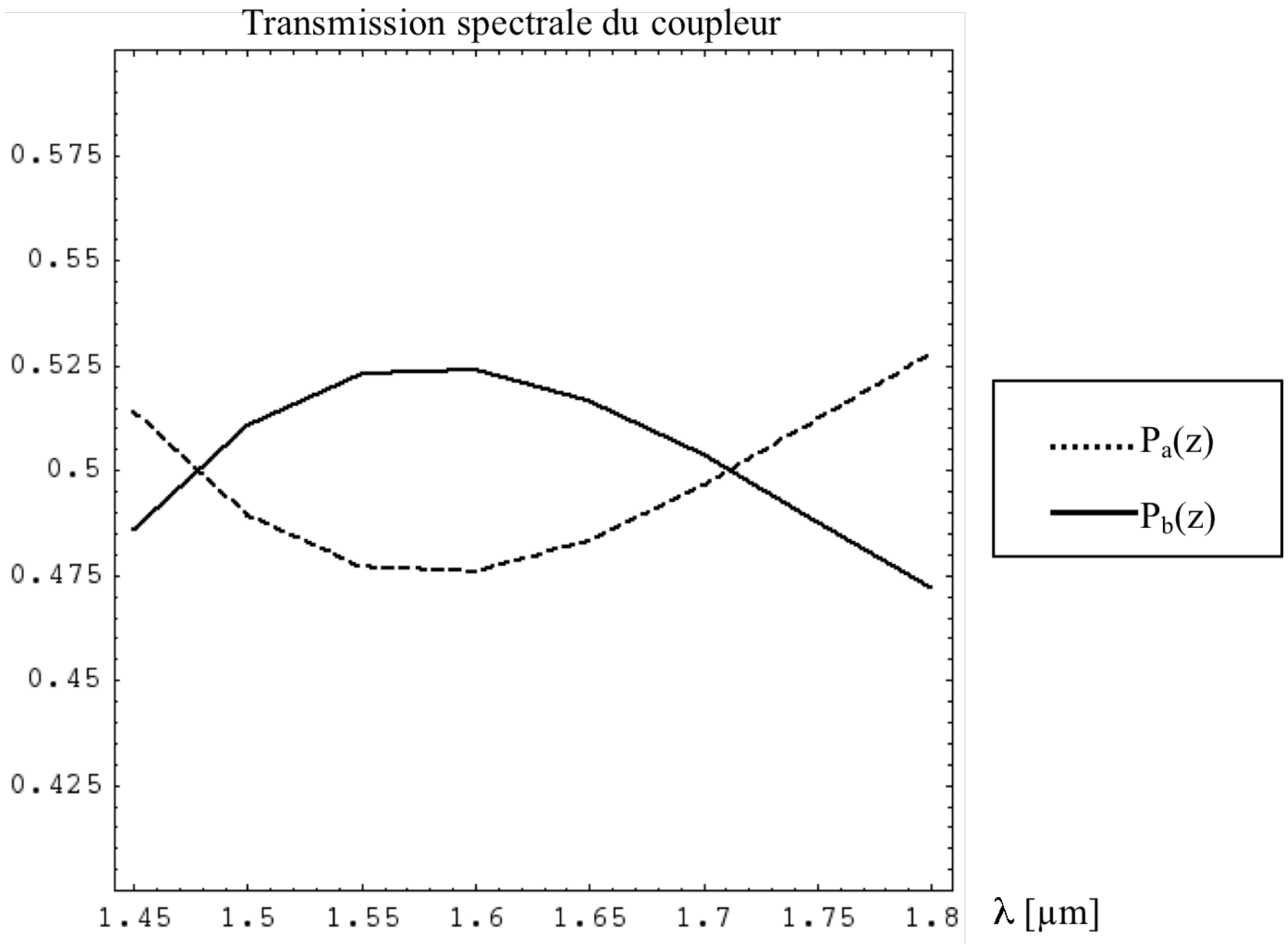

*Figure II-13 : transmission spectrale sur la bande H du coupleur pour une longueur d'interaction de 915µm.*

La théorie des modes couplés décrite ci-dessus permet d'obtenir des résultats analytiques très rapidement si l'on travaille sur un problème en deux dimensions. Lorsque l'on travaille sur un cas en trois dimensions, la théorie s'applique toujours mais la difficulté est d'évaluer les coefficients $K$ et $\delta$. Pour cela nous allons considérer le coupleur autrement : le coupleur à deux guides équidistants est une structure comportant deux modes propres de propagation $E_+$ et $E_-$. Les formes de ces modes propres sont schématisées sur la figure II-14 :

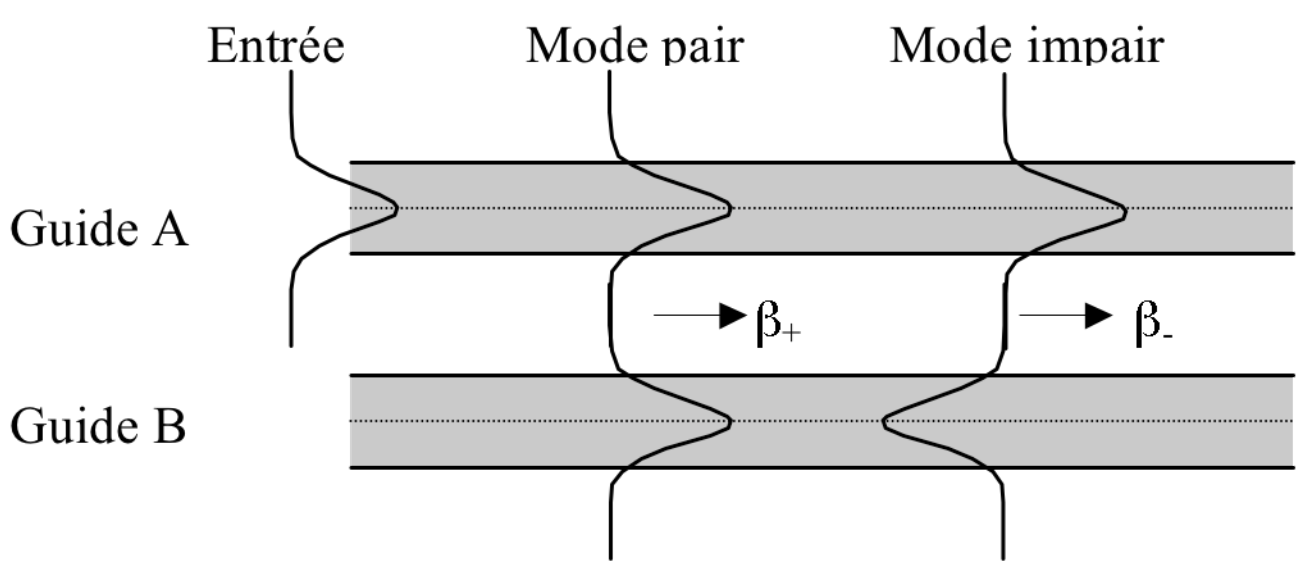

*Figure II-14 : modes propres du coupleur.*

A l'entrée du coupleur, le mode du guide A se décompose sur ces deux modes propres $E_+$ et $E_-$ par la relation :

$$E(z=0) = E_a = \frac{1}{\sqrt{2}}\left(E_+ + E_-\right) \tag{II-34}$$

La longueur de battement, c'est-à-dire la longueur à laquelle l'énergie sera au maximum transférée dans l'autre guide correspond à la longueur $L_b$ à laquelle on peut écrire :

$$E(L_b) = e^{j\varphi}E_b = \frac{1}{\sqrt{2}}\left(E_+e^{j\beta_+L_b} + E_-e^{j\beta_-L_b}\right) = \frac{e^{j\beta_+L_b}}{\sqrt{2}}\left(E_+ + E_-e^{-j(\beta_+-\beta_-)L_b}\right) = \frac{e^{j\varphi}}{\sqrt{2}}\left(E_+ - E_-\right) \tag{II-35}$$

On trouve alors la longueur de battement :

$$L_b = \frac{\pi}{\beta_+ - \beta_-} \tag{II-36}$$

Par substitution dans les équations des puissances trouvées dans le cas des modes couplés, on trouve après calcul :

$$\mu = \sqrt{\delta^2 + K^2} = \frac{\beta_+ - \beta_-}{2} \tag{II-37}$$

$$F = \frac{K^2}{\delta^2 + K^2} = 1 - \left(\frac{\beta_a - \beta_b}{\beta_+ - \beta_-}\right)^2 \tag{II-38}$$

Ces équations très pratiques permettent d'obtenir un calcul très précis du coupleur uniquement à partir du calcul de quatre constantes de propagation. Il n'est pas nécessaire de calculer la répartition de champ dans la structure 3D.

Dans la pratique, un coupleur est constitué de trois zones : une première zone dans laquelle les guides éloignés se rapprochent l'un de l'autre, une deuxième zone dans laquelle les guides sont parallèles et équidistants, et une zone (symétrique de la première) dans laquelle les guides s'éloignent l'un de l'autre. Si le calcul de la deuxième zone peut être facilement calculé par les équations données ci-dessus, les zones 1 et 3 contribuent en général au comportement du coupleur de manière non négligeable.

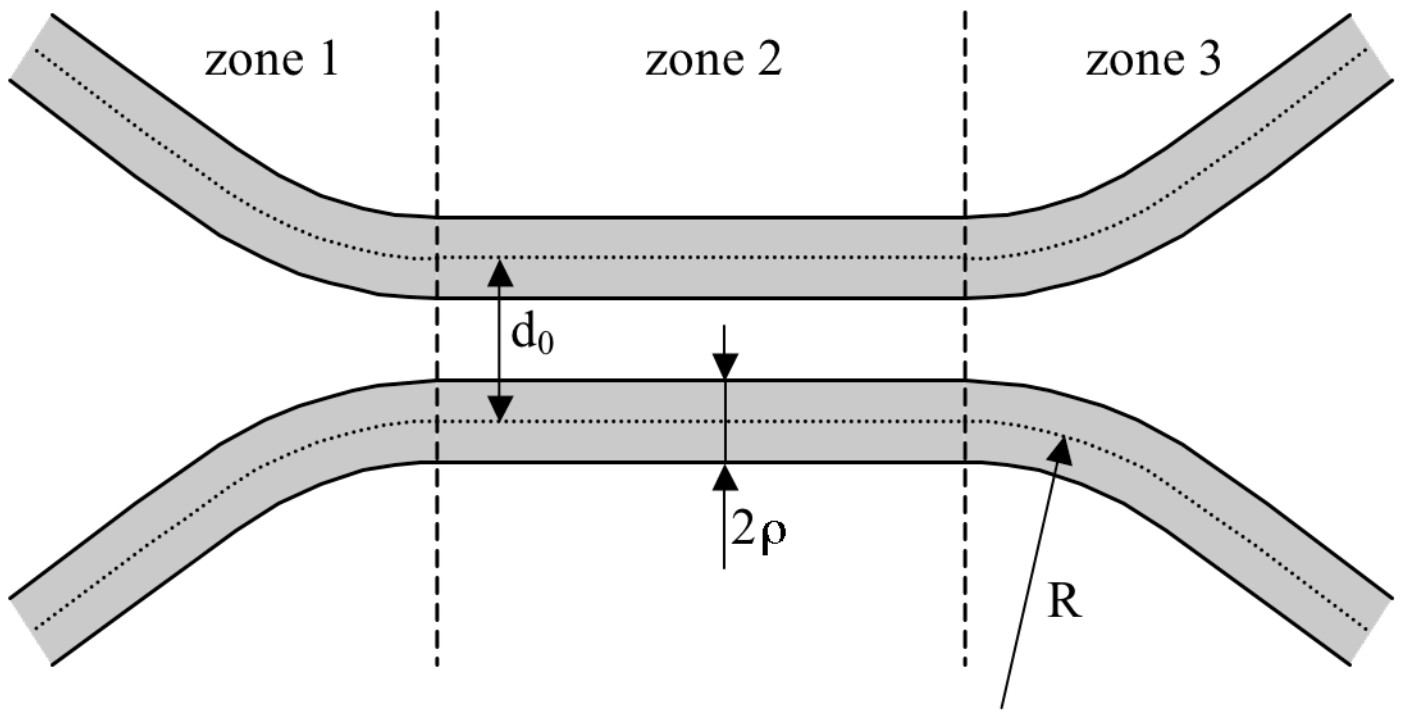

*Figure II-15 : coupleur avec zones d'entrée et de sortie comprenant des parties courbes.*

Dans ce cas, sous réserve que la géométrie du coupleur varie lentement suivant la direction de propagation, il est possible de montrer pour les coupleurs symétriques [*29*] qu'en première approximation, on peut mettre les puissances transportées dans chaque guide sous la forme :

$$P_a \approx 1 - \sin^2\left( \int_L K(z)dz \right) \qquad \text{(II-39)}$$

$$P_b \approx \sin^2\left( \int_L K(z)dz \right) \qquad \text{(II-40)}$$

Par ailleurs dans le cas de guides planaires, on peut aussi montrer que le coefficient de couplage $K$ d'un coupleur à deux guides peut se mettre sous la forme :

$$K = K_0 \exp(-K_1 d) \qquad \text{(II-41)}$$

où $d$ représente la distance entre les deux guides. Appliqué au cas de deux guides se rapprochant à une distance $d_0$, ou s'éloignant progressivement d'une distance $d_0$, on peut alors calculer l'influence des parties courbes du coupleur. Pour cela, on écrit :

$$K(z) = K_0 \exp(-K_1 d) = K_0 \exp\left(-K_1\left(d_0 + \Delta(z)\right)\right) = K(d_0)\exp\left(-K_1\Delta(z)\right) \quad \text{(II-42)}$$

où $\Delta(z)$ représente l'écart à la distance $d_0$ en un point $z$. On peut donc mettre l'expression de la puissance sous la forme :

$$P_b(z) \approx \sin^2\left(\int_L K(z)dz\right) = \sin^2\left(K(d_0)\int_L \exp\left(-K_1\Delta(z)\right)dz\right) = \sin^2\left(K(d_0)L_{eq}\right) \quad \text{(II-43)}$$

$$L_{eq} = \int_L \exp\left(-K_1\Delta(z)\right)dz \quad \text{(II-44)}$$

c'est à dire que l'on peut considérer que les parties courbes du coupleur se comportent comme un coupleur à guides équidistants séparés par une distance $d_0$ ayant une longueur équivalente $L_{eq}$ . Si par exemple, les parties courbes sont constituées de guides courbes suivant un arc de cercle de rayon $R$, on trouve rapidement :

$$L_{eq} = \sqrt{\frac{\pi R}{K_1}} \approx \sqrt{\frac{\pi \rho R}{W}} \quad \text{(II-45)}$$

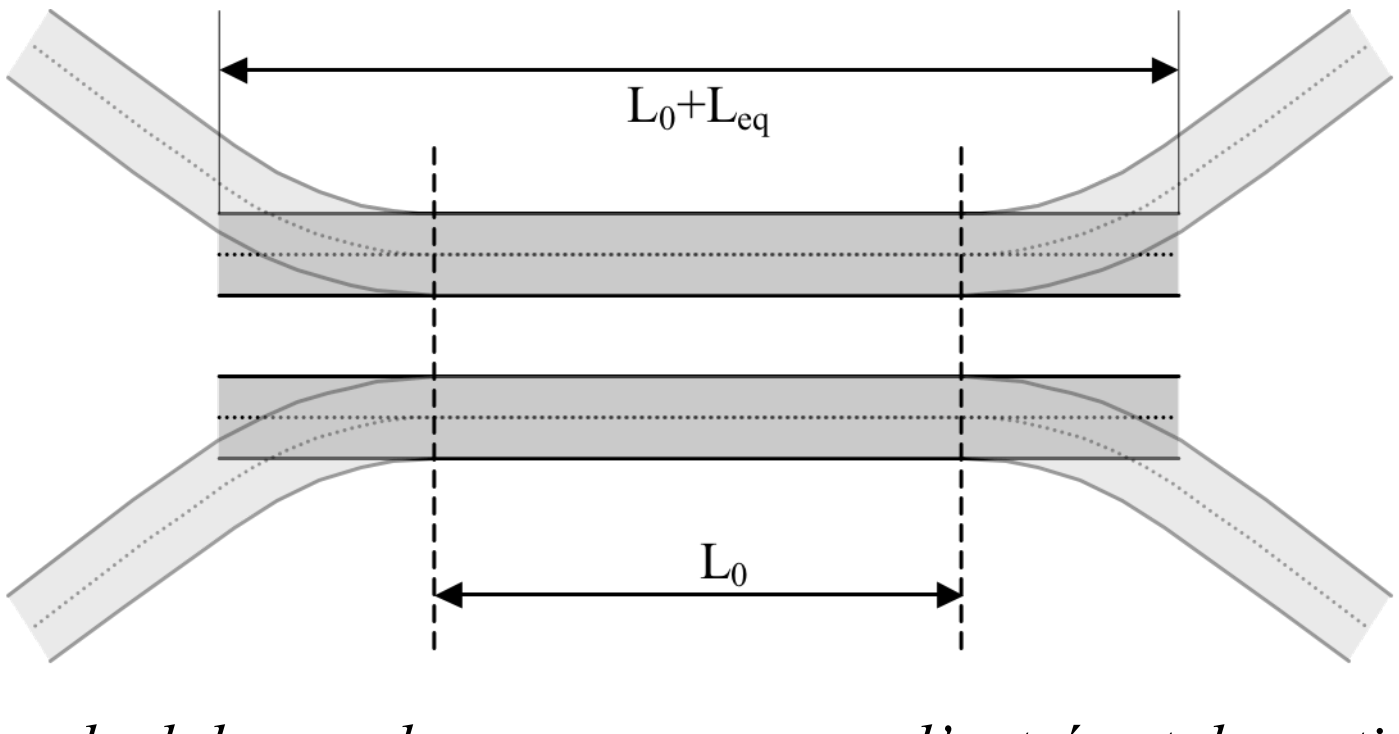

*Figure II-16 : calcul du coupleur avec ses zones d'entrée et de sortie. En première approximation, tout se passe comme si l'on avait un coupleur droit de longueur légèrement supérieure.*

Les courbes que nous utilisons sont plus complexes, et se mettent sous forme paramétrique [*x(t),z(t)*]. Dans ce cas, on trouve numériquement $L_{eq}$ par :

$$L_{eq} = \int_t \exp(-K_1 x(t)) z'(t) dt \qquad \text{(II-46)}$$

On trouve ainsi une valeur approchée de la longueur du coupleur à réaliser. Une étape d'optimisation par méthode numérique de type BPM est cependant nécessaire pour trouver la longueur optimale du coupleur. L'erreur faite sur le calcul approché de $L_{eq}$ par rapport à la BPM est d'environ 5%, ce qui est remarquable compte tenu des approximations effectuées.

La figure II-17 donne un exemple de transmission spectrale de coupleur obtenue après optimisation par BPM.

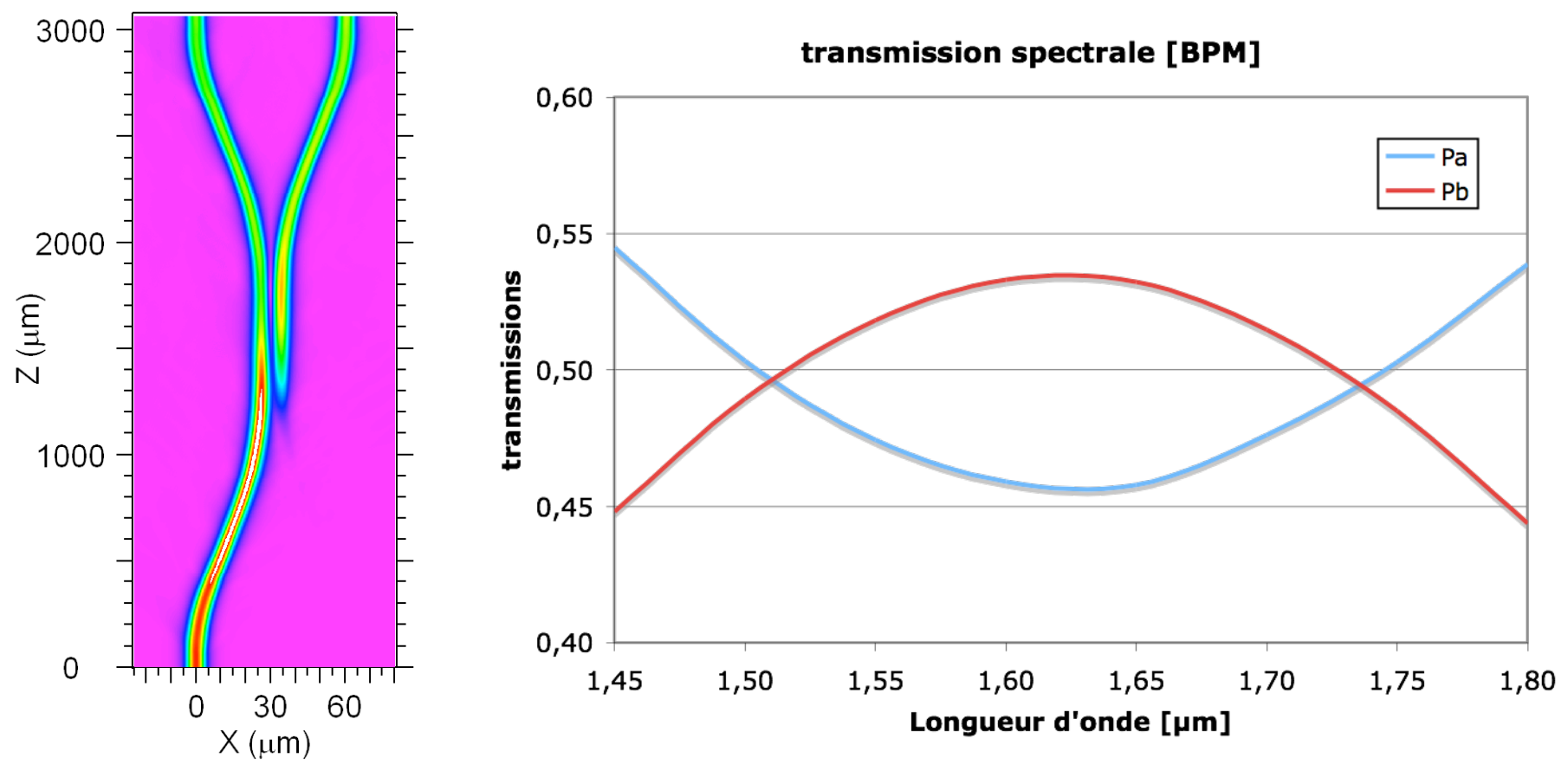

*Figure II-17 : exemple de simulation BPM scalaire du coupleur asymétrique (λ=1.625µm) et transmission spectrale obtenue. Les pertes globales sont de l'ordre de 0,05dB. La partie droite du coupleur fait 370µm de long, alors que le calcul par modes couplés 3D avec parties courbes prévoyait une partie droite de 390µm.*

La BPM présente l'avantage de fournir un résultat prenant en compte les pertes en excès de la fonction (de l'ordre de 0,05dB sur l'exemple, soit dans la précision de calcul de ce type de simulations), c'est pourquoi les courbes se croisent légèrement en dessous de 0,5. Grâce au dimensionnement effectué préalablement avec la théorie des modes couplés, quelques simulations suffisent pour arriver au coupleur optimisé.

Finalement, nous avons vérifié l'influence de la polarisation sur ce type de structure en effectuant un calcul numérique prenant en compte la nature vectorielle du champ électromagnétique. Les résultats sont reportés sur le graphe de la figure II-18 et font apparaître un bon comportement du coupleur asymétrique vis-à-vis de la polarisation. La différence entre les transmissions TE et TM croît avec la longueur d'onde. En effet, pour les plus grandes longueurs d'onde, les modes guidés sont moins confinés dans le cœur du guide et voient donc plus les interfaces cœur gaine où se manifestent les effets de polarisation. Néanmoins, cette différence reste toujours inférieure au pourcent.

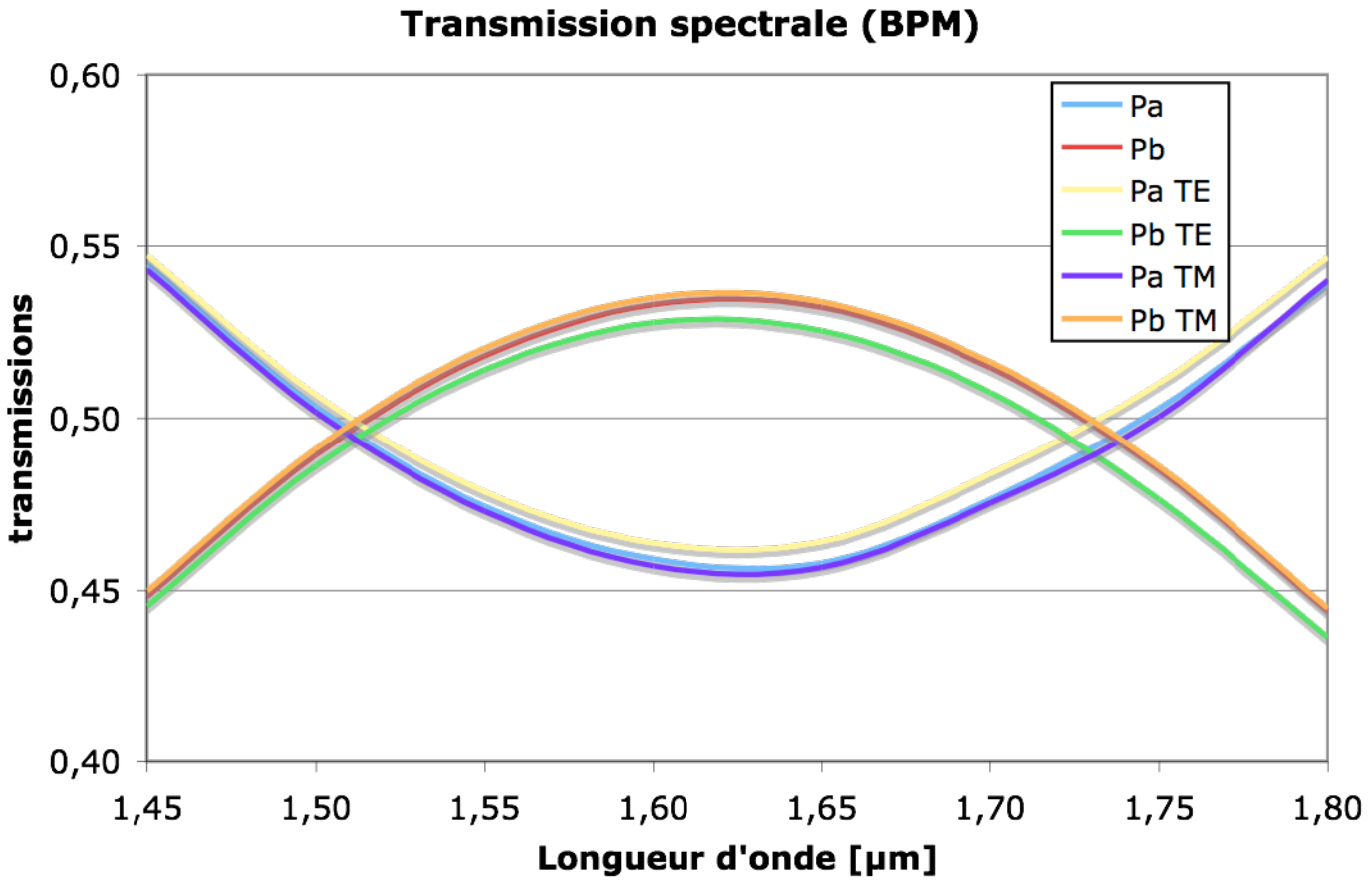

*Figure II-18 : influence de la polarisation. La simulation prend en compte la biréfringence introduite par la technologie de nos guides.*

D'un point de vue mathématique, si l'on développe la transmission spectrale des coupleurs présentés ici en série de Taylor autour de la longueur d'onde centrale de la bande spectrale considérée, l'optimisation a permis d'annuler le terme du premier ordre de la série. La dépendance spectrale est donc approximativement parabolique. Steblina & al. ont montré dans [*53*] qu'il était possible d'aller théoriquement plus loin et de trouver des configurations de coupleurs asymétriques compensés au second ordre (il est possible d'annuler le deuxième terme de la série de Taylor). Ces coupleurs théoriquement très performants sont malheureusement très difficiles à réaliser car ils correspondent à une configuration où les cœurs des deux guides sont très rapprochés l'un de l'autre. La difficulté est en quelque sorte la même que pour réaliser une pointe infiniment fine de jonction Y. La technologie actuelle que nous utilisons ne permettant pas de réaliser de tels coupleurs, nous n'avons pas jugé nécessaire de les présenter ici.

Du point de vue de la recombinaison interférométrique, un coupleur ayant deux entrées et deux sorties agit globalement de la même manière qu'une lame semi réfléchissante en optique classique. Son comportement est identique. Il comporte un ratio de transmission d'environ 50% variant faiblement en fonction de la longueur d'onde et de la polarisation.

## II – B - 8 – Tricoupleurs

Un tricoupleur fonctionne sur le même principe que le coupleur mais est constitué de trois guides d'onde placés côte à côte donc pouvant échanger de l'énergie lumineuse.

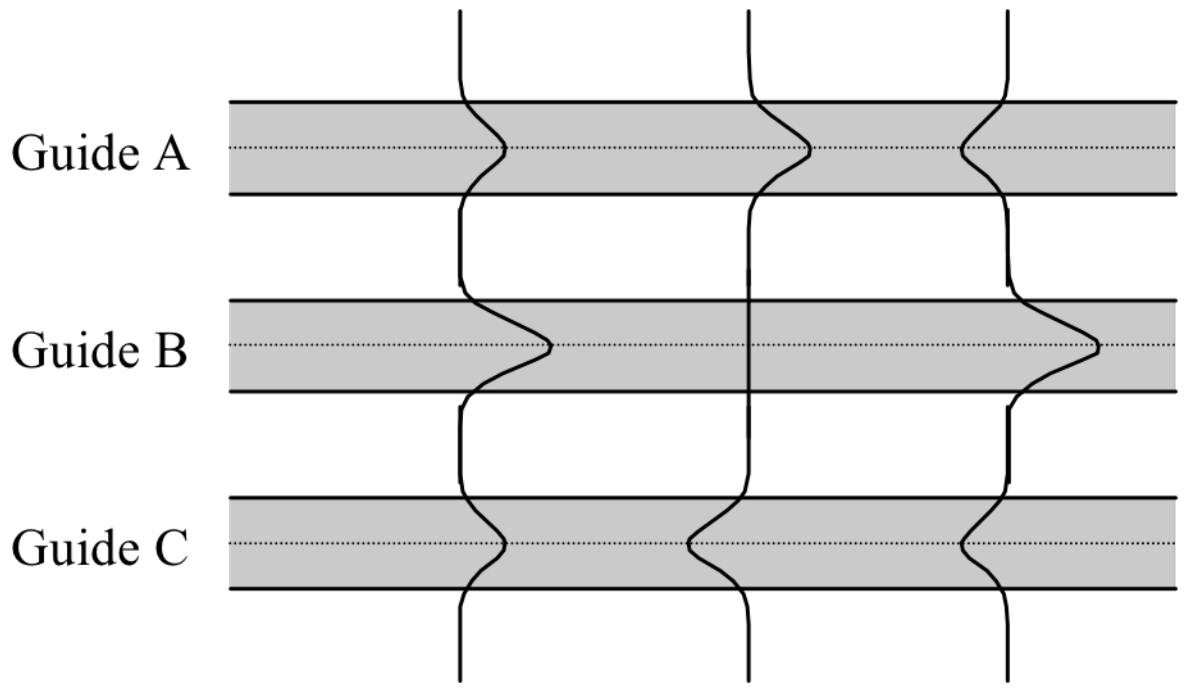

*Figure II-19 : tricoupleur, ou coupleur à trois guides avec ses modes propres.*

Le calcul du comportement du tricoupleur s'effectue de la même manière que le calcul du coupleur. En considérant un tricoupleur où les deux guides latéraux sont de largeur identique, et un guide central de largeur différente, les équations régissant le comportement du tricoupleur s'écrivent suivant le même déroulement. Le champ se met cette fois sous la forme :

$$E(z) = a(z)E_a + b(z)E_b + c(z)E_c \qquad \text{(II-47)}$$

Et les coefficients *a(z)*, *b(z)*, et *c(z)* s'obtiennent grâce aux équations différentielles suivantes :

$$\frac{da(z)}{dz} = -j\delta a(z) + jK_{cl}b(z) + jK_{ll}c(z) \tag{II-48}$$

$$\frac{db(z)}{dz} = jK_{cl}a(z) + j\delta b(z) + jK_{cl}c(z) \tag{II-49}$$

$$\frac{dc(z)}{dz} = jK_{ll}a(z) + jK_{cl}b(z) - j\delta c(z) \tag{II-50}$$

qui peut se mettre avantageusement sous forme matricielle :

$$\begin{bmatrix} \frac{da(z)}{dz} \\ \frac{db(z)}{dz} \\ \frac{dc(z)}{dz} \end{bmatrix} = j \begin{bmatrix} -\delta & K_{cl} & K_{ll} \\ K_{cl} & \delta & K_{cl} \\ K_{ll} & K_{cl} & -\delta \end{bmatrix} \bullet \begin{bmatrix} a(z) \\ b(z) \\ c(z) \end{bmatrix} \tag{II-51}$$

$\delta$ est la demi différence des constantes de propagation des modes du guide central et du guide latéral. $K_{cl}$ est le coefficient de couplage entre les guides centraux et latéraux. $K_{ll}$ est le coefficient de couplage entre les deux guides latéraux. Les guides latéraux étant beaucoup plus éloignés l'un de l'autre que du guide central, on a :

$$K_{ll} << K_{cl} \tag{II-52}$$

Après calcul, en négligeant $K_{ll}$ (l'expression complète est trop lourde pour être présentée à d'autres qu'un logiciel !) on trouve :

$$\begin{bmatrix} a(z) \\ b(z) \\ c(z) \end{bmatrix} = \begin{bmatrix} -1 & 0 & 1 \\ 1 & p & 1 \\ 1 & q & 1 \end{bmatrix}^{-1} \bullet \begin{bmatrix} \exp(irz) & 0 & 0 \\ 0 & \exp(irz) & 0 \\ 0 & 0 & \exp(-irz) \end{bmatrix} \bullet \begin{bmatrix} -1 & 0 & 1 \\ 1 & p & 1 \\ 1 & q & 1 \end{bmatrix} \bullet \begin{bmatrix} a_0 \\ b_0 \\ c_0 \end{bmatrix} \quad \text{(II-53)}$$

avec :

$$p = \frac{\delta - \sqrt{\delta^2 + 2K_{cl}^2}}{K_{cl}} \quad \text{(II-54)}$$

$$q = \frac{\delta + \sqrt{\delta^2 + 2K_{cl}^2}}{K_{cl}} \quad \text{(II-55)}$$

$$r = \sqrt{\delta^2 + 2K_{cl}^2} \quad \text{(II-56)}$$

De la même manière que l'on minimise le chromatisme d'un coupleur en utilisant des guides asymétriques, on va minimiser le chromatisme d'un tricoupleur en utilisant un guide central légèrement différent des guides latéraux (ceci permet de conserver la symétrie globale du tricoupleur). Le graphe de la figure II-20 montre l'évolution de la puissance dans le guide central et l'un des guides latéraux du tricoupleur lorsque l'on injecte une puissance unitaire sur le mode fondamental du guide central. Les guides font 6µm de large pour le guide central et 4,6µm de large, et sont écartés de 2,8µm. On voit que pour une longueur de coupleur de 620µm environ, le couplage est compris entre 30% et 40% pour toute la bande H. La sortie est symétrique dans chaque guide latéral.

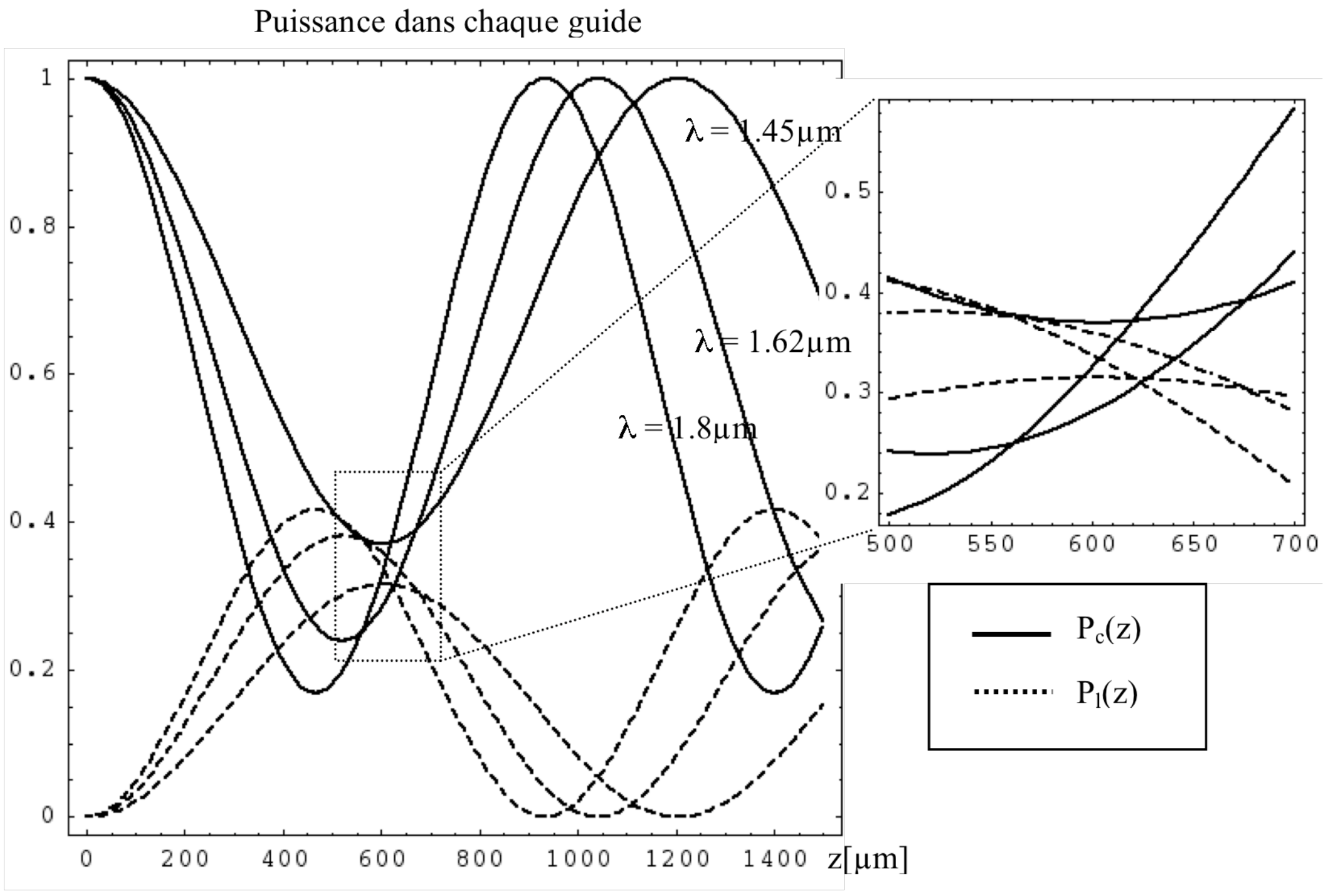

*Figure II-20 : Evolution de la puissance dans les bras central $P_c(z)$ et latéral $P_i(z)$ d'un tricoupleur en fonction de sa longueur pour différentes longueurs d'onde.*

Comme pour le coupleur à deux guides, on choisit les paramètres opto-géométriques du tricoupleur de manière à ce que les courbes se croisent sur une faible largeur spectrale afin d'obtenir un taux de couplage le plus indépendant possible de la longueur d'onde. On obtient alors une transmission du tricoupleur en fonction de la longueur d'onde représentée sur la figure II-21.

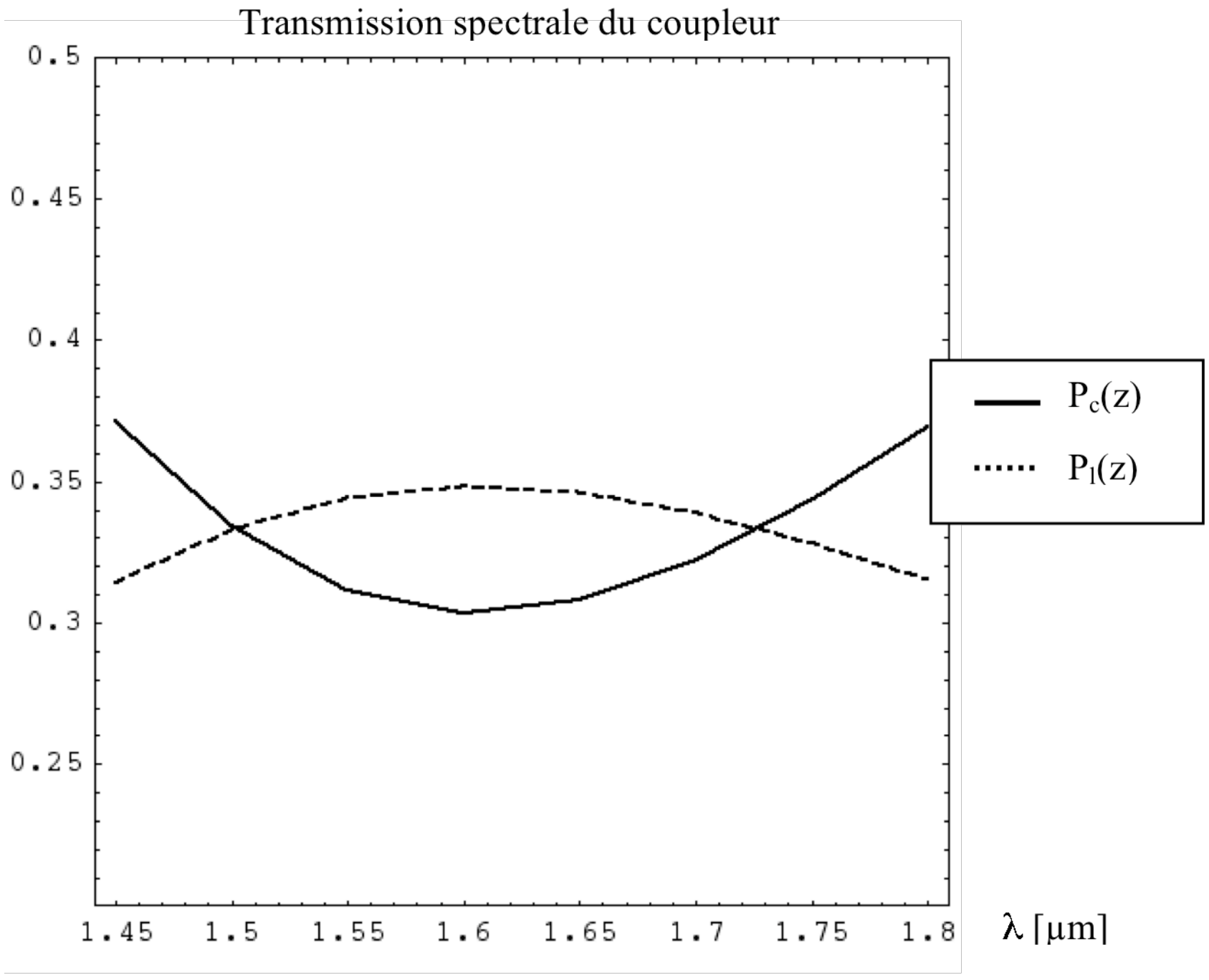

*Figure II-21 : transmission spectrale du tricoupleur lorsque l'on injecte une puissance unitaire sur le mode fondamental du guide central. Les deux sorties latérales sont identiques. La longueur d'interaction vaut 615µm.*

Après intégration des parties courbes et vérification à la BPM, on obtient le résultat légèrement plus chromatique montré sur la figure II-22 :

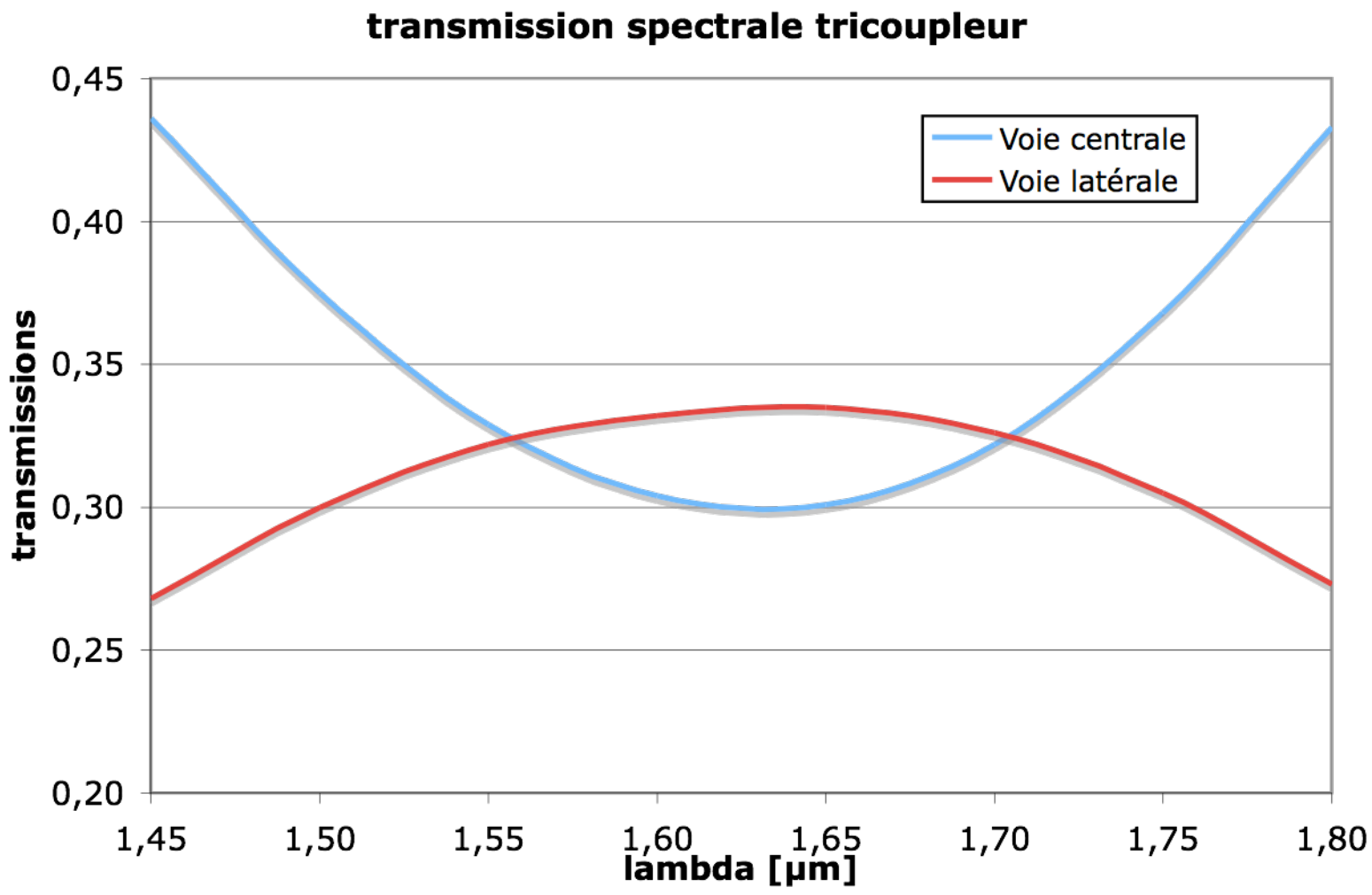

*Figure II-22 : simulation BPM du tricoupleur. Les parties courbes introduisent un léger chromatisme supplémentaire. Les pertes totales d'environ 0,1dB sont introduites dans les parties courbes des voies latérales plus « courbées » que dans le cas du coupleur.*

## II – B – 9 – Recombinaison planaire

Une recombinaison de type planaire est représentée sur la figure II-23. Elle est composée de deux guides entrant dans une zone de guidage planaire suffisamment large pour que l'intensité lumineuse n'interagisse pas avec ses bords et telle que le point d'intersection de leur axe de propagation se trouve à la sortie de cette zone planaire.

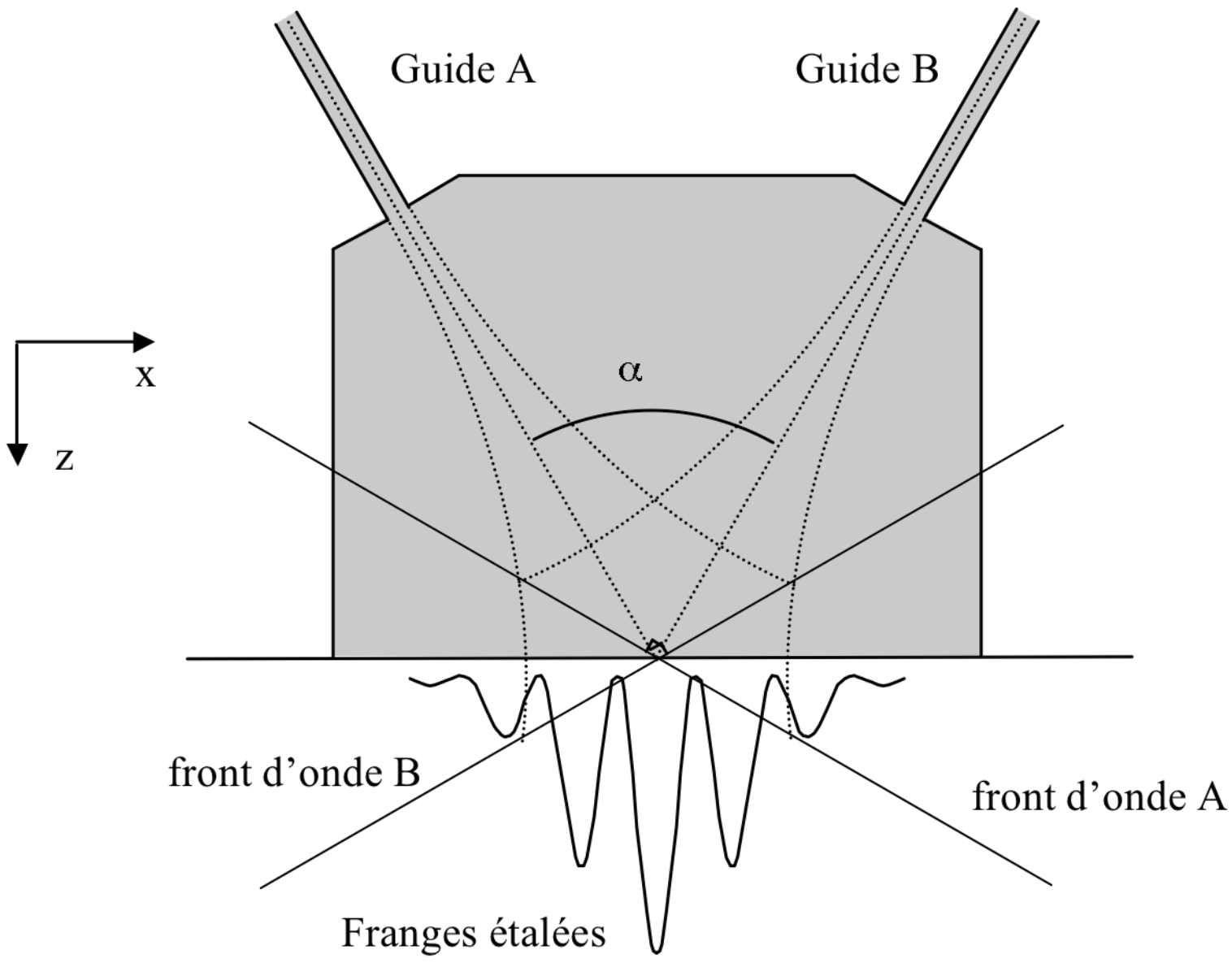

*Figure II-23 : recombinaison de faisceaux en zone planaire.*

Ce type de recombinaison présente l'avantage de réaliser un codage spatial des franges d'interférence dans la zone de guidage planaire, rendant ainsi inutile la modulation du chemin optique sur l'un des bras de l'interféromètre. En effet, selon les notations de la figure II-23, si l'on considère des ondes planes, il est aisé de vérifier que la répartition d'intensité *I(x)* suit la relation :

$$I(x) = I_e(x)\left(\frac{1+\cos(\phi(x))}{2}\right) \qquad \text{(II-57)}$$

$$\phi(x) = \frac{4\pi}{\lambda} n_{eff}\, x \sin(\alpha) \qquad \text{(II-58)}$$

où $I_e(x)$ représente l'enveloppe des franges et peut s'apparenter à une gaussienne. $n_{eff}$ est l'indice effectif du mode guidé dans la zone de recombinaison planaire.

En optique intégrée, si l'on ne prend aucune précaution, à la sortie des guides (ou à l'entrée de la zone planaire) la répartition d'amplitude lumineuse du mode fondamental va être diffractée dans la zone planaire et la relation de phase va être légèrement plus complexe. En faible guidage, nous avons vu que nous pouvions approximer la répartition transverse du mode par une gaussienne :

$$E(x) = E_0 \exp\left(-\frac{x^2}{w_0^2}\right) \qquad \text{(II-59)}$$

Si l'on considère la théorie de la propagation de faisceaux gaussiens [*28*], on obtient alors dans la zone planaire une répartition transverse de la forme :

$$E(x,z) = E_0 \sqrt{\frac{w_0}{w(z)}} \exp\big(jkz + \phi(z)\big) \exp\left(-\frac{x^2}{w^2(z)}\right) \exp\left(\frac{jkx^2}{2R(z)}\right) \qquad \text{(II-60)}$$

$$w(z) = w_0 \sqrt{1 + \left(\frac{z}{z_0}\right)^2} \qquad \text{(II-61)}$$

$$R(z) = z + \frac{z_0^2}{z} \qquad \text{(II-62)}$$

$$z_0 = \frac{kw_0^2}{2} \qquad \text{(II-63)}$$

$k$ représente le vecteur d'onde dans le guide planaire, c'est-à-dire la constante de propagation du mode de propagation du guide.

On voit que les franges ont bien une enveloppe gaussienne, mais qu'elles sont générées à partir de fronts d'onde circulaires et non plans. Si l'on considère deux fronts d'onde circulaires dont les axes principaux de propagation convergent à la sortie de la zone planaire selon la figure II-24 :

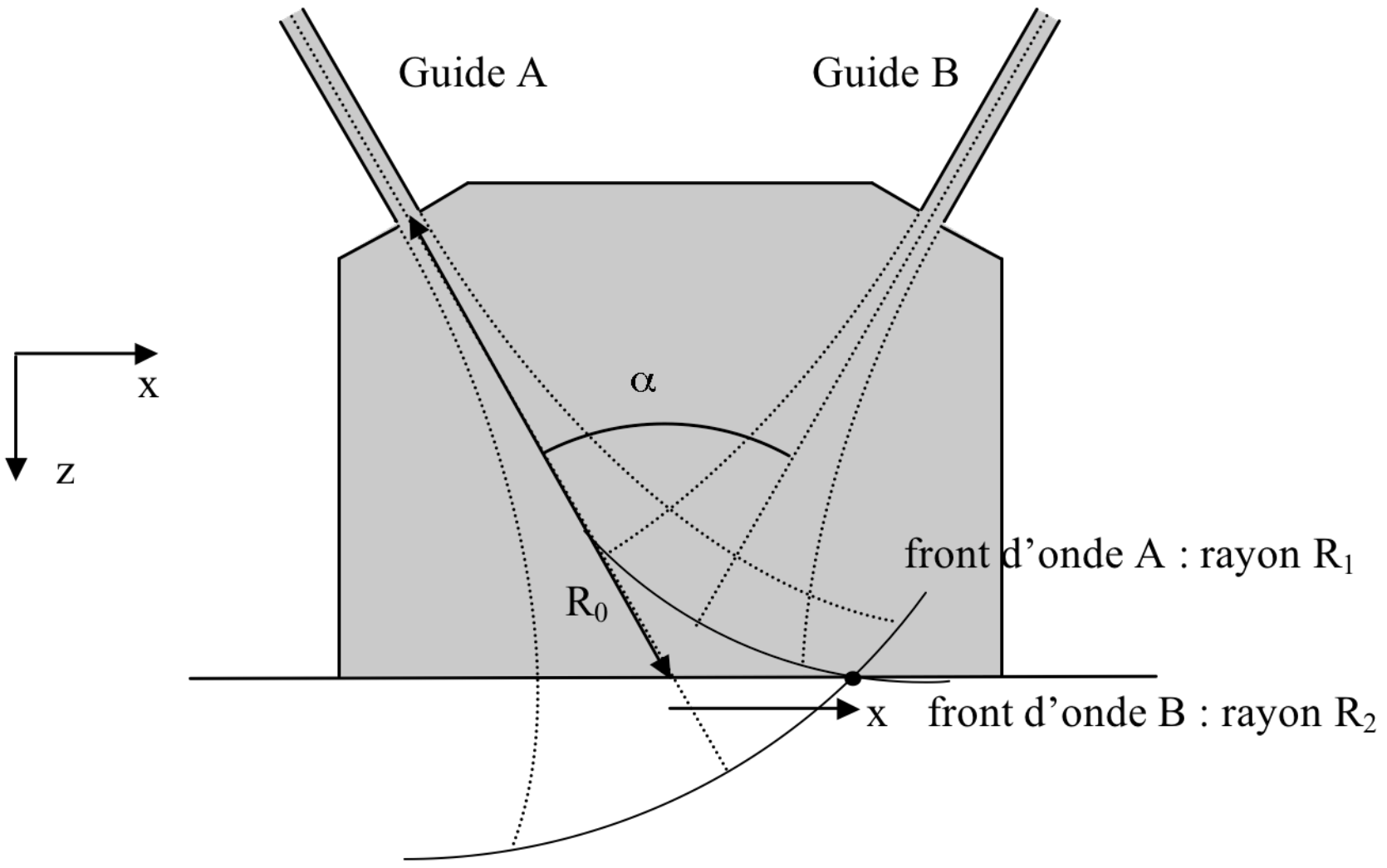

*Figure II-24 : prise en compte des fronts d'onde circulaires.*

On peut montrer que le déphasage entre les deux ondes s'écrit :

$$\phi(x) = \frac{4\pi n_{eff} x \sin(\alpha)}{\lambda} \frac{2}{\sqrt{1+\frac{2\sin(\alpha)x}{R}+\frac{x^2}{R^2}}+\sqrt{1-\frac{2\sin(\alpha)x}{R}+\frac{x^2}{R^2}}} \approx \frac{4\pi n_{eff} x \sin(\alpha)}{\lambda}\left(1-\frac{\cos^2(\alpha)x^2}{R^2}\right) \quad \text{(II-64)}$$

L'erreur sur la phase par rapport à des ondes planes est au second ordre en *x/R*. Cette erreur est très faible étant données les dimensions utilisées (comme nous le verrons au chapitre IV). Cependant, si l'on désire corriger cette erreur, on peut générer des ondes planes dans la zone planaire en réalisant un épanouisseur qui va élargir le mode et créer une gaussienne suffisamment large pour que sa divergence soit faible et que son front d'onde soit considéré comme plan. L'angle du taper doit être suffisamment petit pour que l'énergie reste bien dans le mode fondamental et ne se couple pas dans les modes d'ordre supérieur de la structure devenue multimode. Pour cela, on peut montrer [*29*] que l'angle du taper doit respecter la même condition que pour la jonction Y. L'emploi d'un taper adiabatique permet d'obtenir un codage spatial des franges plus précis, mais au prix d'un encombrement nettement plus important, la longueur du taper étant de l'ordre du centimètre.

## II – B - 10 – Recombinaison ABCD

Nous avons vu que la grandeur que l'on cherche à mesure en interférométrie astronomique est la visibilité complexe des franges d'interférence entre deux faisceaux, c'est à dire le contraste et la phase des franges. Si une sortie ou deux sorties en opposition de phase permettent de mesurer le contraste des franges en détectant le maximum et le minimum des franges, il reste une indétermination sur la mesure de la phase. Si, à la place de deux sorties en opposition de phase, on est en mesure de fournir quatre sorties en quadrature de phase, il devient alors possible d'extraire instantanément la phase des franges à partir des quatre signaux interférométriques. Ce principe fut proposé en astronomie par Shao dans [*54*].

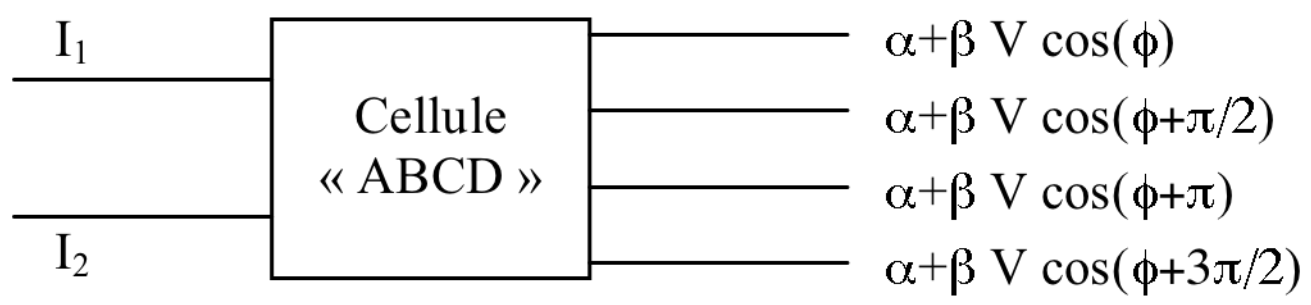

*Figure II-25 : principe de l'ABCD. α et β dépendent des caractéristiques de la cellule « ABCD ».*

En effet, les quatre signaux peuvent s'écrire en sortie (dans le cas idéal) :

$$A = \frac{1}{4}\left(I_1 + I_2 + 2\sqrt{I_1 I_2}V_{12}\cos(\varphi)\right) \qquad \text{(II-65)}$$

$$B = \frac{1}{4}\left(I_1 + I_2 + 2\sqrt{I_1 I_2}V_{12}\sin(\varphi)\right) \qquad \text{(II-66)}$$

$$C = \frac{1}{4}\left(I_1 + I_2 - 2\sqrt{I_1 I_2}V_{12}\cos(\varphi)\right) \qquad \text{(II-67)}$$

$$D = \frac{1}{4}\left(I_1 + I_2 - 2\sqrt{I_1 I_2}V_{12}\sin(\varphi)\right) \qquad \text{(II-68)}$$

On obtient alors directement la visibilité et la phase par :

$$V^2 = \frac{(A-C)^2 + (B-D)^2}{I_1 I_2} \quad \text{(II-69)}$$

$$\varphi = Arctg\left(\frac{B-D}{A-C}\right) \quad \text{(II-70)}$$

Il est nécessaire d'avoir les signaux photométriques pour obtenir la visibilité. La phase est quant à elle, obtenue à 2π près grâce aux signes de *B-D* et *A-C*.

En utilisant des guides monomodes, il est possible de réaliser cette fonction en utilisant le schéma de la figure II-26 :

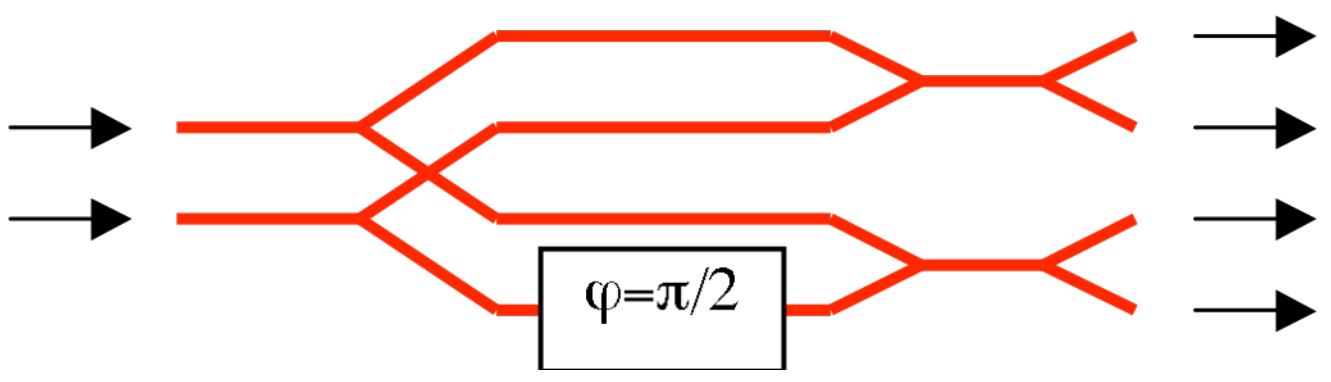

*Figure II-26 : sorties en quadrature de phase en utilisant des guides monomodes. A un coupleur fournissant deux signaux en opposition de phase, il suffit d'intercaler un déphaseur de π/2 sur une voie.*

Ce schéma fut proposé initialement par Severi [*47*] qui l'utilisa pour réaliser un capteur interférométrique de déplacement utilisant une lumière monochromatique. La difficulté de réalisation de cette fonction réside principalement dans le « déphaseur » qui doit être contrôlé précisément.

Il existe d'autres possibilités en optique intégrée pour réaliser cette fonction. Dans [*46*], Poupinet utilise une recombinaison planaire proposée à l'origine par Lang et al. [*55*] pour étaler les franges puis les échantillonner aux quatre positions permettant d'obtenir ces quatre valeurs de déphasage.

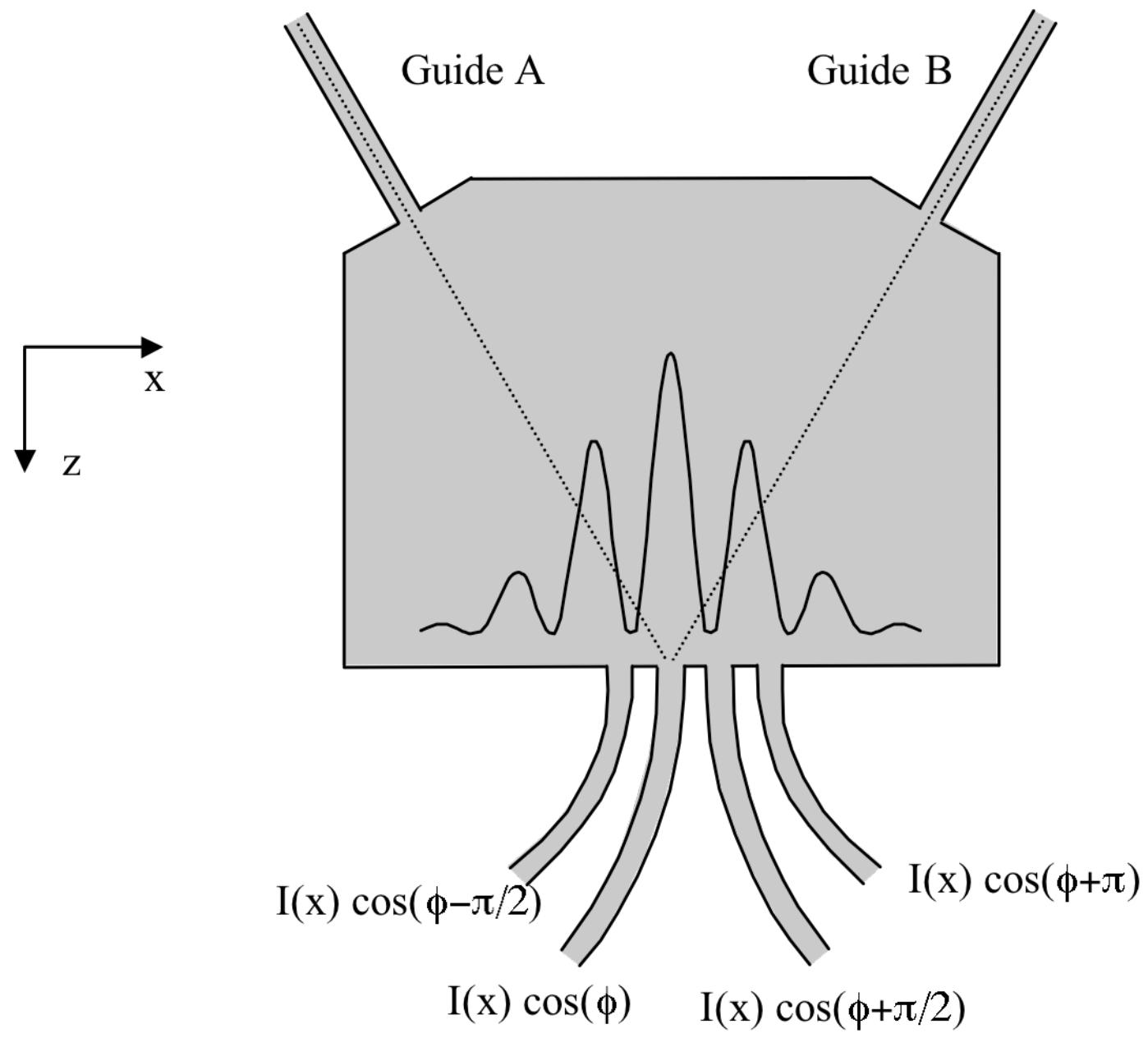

*Figure II-27 : schéma ABCD en utilisant un recombineur à zone planaire. Les guides de sortie sont placés de manière à échantillonner le signal des franges aux bons endroits.*

Les franges étant codées spatialement dans la zone planaire, ce principe marche pour une longueur d'onde monochromatique mais ne fournit pas des signaux tout à fait identiques au schéma précédent en spectre large. Une des difficultés de ce montage est que les guides de sortie doivent être suffisamment éloignés pour ne pas échanger d'énergie entre eux. Afin de résoudre ce problème, le capteur réalisé selon ce principe échantillonne le signal de sortie aux points 3π/2,  2x3π/2, 3x3π/2, 4x3π/2. Par ailleurs, la transmission globale d'un tel composant est forcément réduite, tout le signal des franges n'étant pas récupéré par les quatre guides de sortie. De plus cette transmission dépend de la phase des franges, donc de la phase entre les deux signaux d'entrée. Haguenauer a évalué un tel composant pour l'astronomie dans [*56*]. Il a notamment mesuré une transmission globale entre 20% et 46% suivant le déphasage entre les voies.

Nous avons passé en revue les différentes fonctions élémentaires utilisées en optique intégrée pour l'astronomie. Avant de décrire plus amplement leur implémentation dans des dispositifs, intéressons nous aux différents modes de recombinaison possibles pour l'interférométrie astronomique.

# *II – C – Recombinaison interférométrique*

## II – C – 1 - Modes de recombinaison en optique classique

Il existe une classification des différents modes de recombinaison interférométrique en optique classique [*57*]. Cette classification fait apparaître 3 critères :

- *La recombinaison par paires ou tout-en un :* dans la recombinaison par paires, chacune des N voies est séparée en N-1 canaux, chacun de ces canaux étant recombiné avec un canal de chacune des N-1 autres voies. Dans la recombinaison tout-en-un, les voies sont mélangées toutes ensemble puis le signal interférométrique de chaque paire recombinée est extrait par traitement du signal.
- *La recombinaison co-axiale ou multiaxiale :* dans la recombinaison co-axiale, les faisceaux sont recombinés suivant le même axe optique et si l'on considère des ondes planes, l'état de frange d'interférence est identique en tout point du faisceau. Afin d'obtenir différents états de frange, il est alors nécessaire d'effectuer une modulation temporelle du chemin optique. On parle de codage temporel des franges d'interférence. Dans la recombinaison multiaxiale, on introduit un angle entre les faisceaux à recombiner, on obtient alors des franges étalées spatialement à l'intérieur du faisceau. On parle de codage spatial des franges d'interférence. Il n'est plus nécessaire dans ce cas de moduler le chemin optique.
- *La recombinaison dans le plan pupille ou dans le plan image :* ce dernier critère très important dans le cas où on réalise des franges en champ large (multimode) perd complètement son intérêt lorsque l'on filtre les faisceaux avec de l'optique guidée monomode. Dans ce cas, en effet, on ne parle plus de plan image ou plan pupille, mais plutôt de plan modal dans lequel les deux plans sont confondus. P. Mège [*58*] a montré que dans ce cas, un instrument utilisant de l'optique guidée monomode mesure en fait une visibilité modale indépendante du plan de recombinaison.

## II – C – 2 – Modes de recombinaison en optique intégrée

La notion de plan pupille et plan image est difficilement applicable en optique intégrée où l'on peut considérer qu'ils sont tous deux confondus en un « plan modal ». Nous ne considérerons donc par la suite que les deux premiers critères. Par contre, l'optique intégrée apporte son lot de nouvelles fonctions qu'il est difficile de ranger dans cette classification. Lebouquin dans [*59*] propose une troisième dénomination pour contourner ce problème : la recombinaison « matricielle », s'appliquant à la recombinaison coaxiale, mais proposant aussi un codage de franges permettant de s'affranchir de la modulation temporelle. On peut alors classer les modes de recombinaison selon un tableau à deux colonnes et trois lignes, comme montré sur la figure II-28 dans le cas de la recombinaison à quatre télescopes.

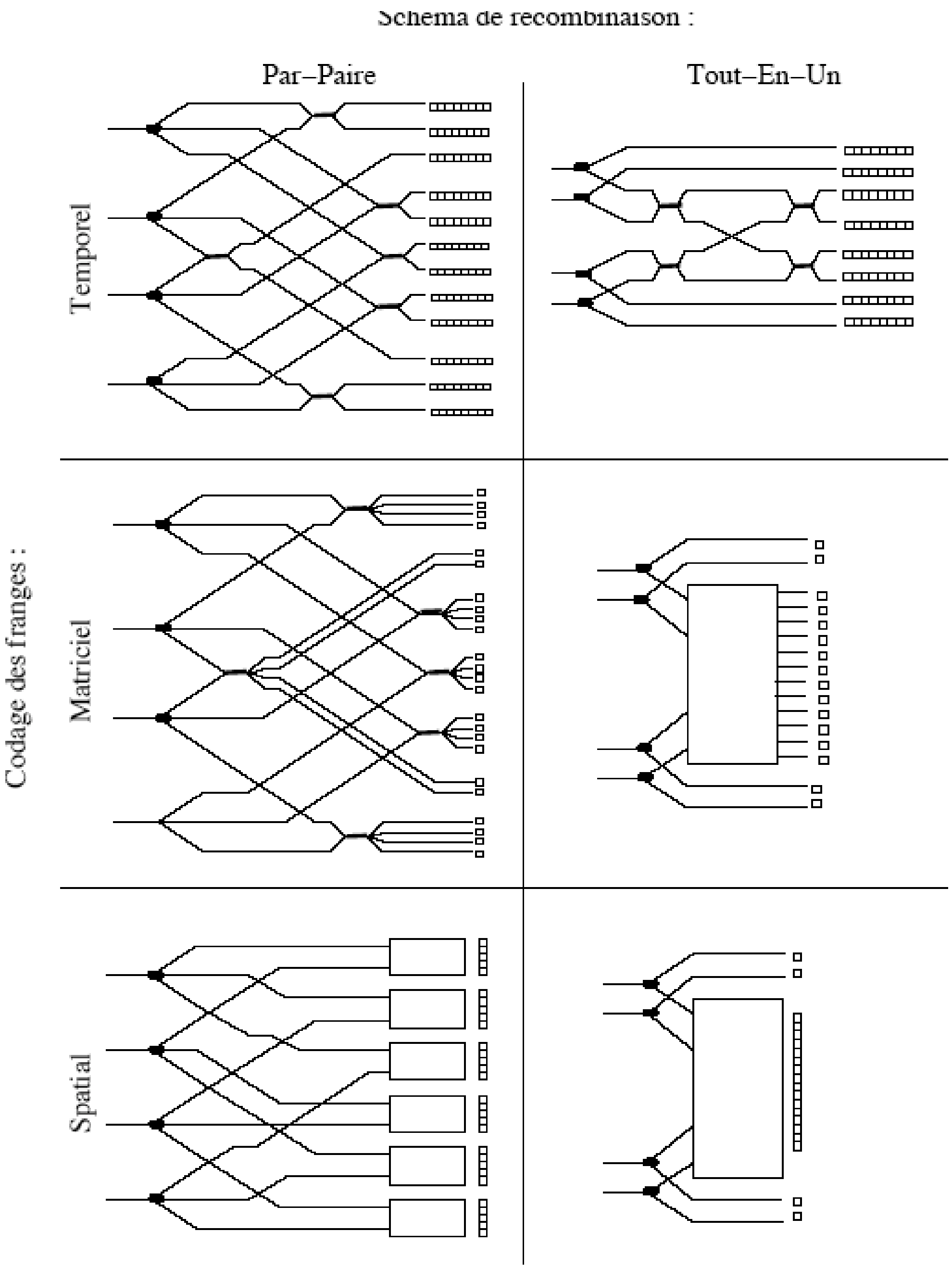

*Figure II-28 : classement des différents principes de recombineurs intégrés. D'après [59].*

La colonne de gauche regroupe les modes de recombinaison par paires tandis que la colonne de droite regroupe les modes de recombinaison tout-en-un. Chaque ligne correspond à un codage de franges différent qu'il est possible d'implémenter pour chaque mode.

En fait, puisque qu'une recombinaison matricielle fournit un codage de franges permettant de s'affranchir d'une modulation temporelle du chemin optique, on pourrait aussi considérer une recombinaison matricielle comme une recombinaison multiaxiale généralisée. Ceci reviendrait à considérer qu'une recombinaison multiaxiale n'est qu'un cas particulier de recombinaison matricielle comportant un codage spatial particulier. Mais au final, il est préférable de considérer les choses de manière fonctionnelle : un codage matriciel fournit en sortie un codage des franges qui est déjà discrétisé : chaque sortie peut être redirigée vers un détecteur mono pixel. Un codage multiaxial fournit un faisceau continu (unidimensionnel dans le cas de l'optique intégrée) qui est imagé sur une barrette de détecteurs ou une caméra qui effectue cette discrétisation. En ce sens, la classification n'est pas forcément pertinente d'un point de vue purement « optique intégrée », mais elle l'est du point de vue du traitement du signal nécessaire à l'obtention des mesures que sont les visibilités complexes de chaque base.

J-B. Lebouquin a détaillé dans [*59*] les performances de chaque type de recombineur pour 4, 6, et 8 télescopes. Il n'est pas aisé au final de classer ces différents concepts de recombinaison, car chacun présente des avantages et des inconvénients suivant le mode d'utilisation. D'une manière générale, le codage temporel nécessite deux fois plus de mesures que le codage spatial, qui nécessite lui-même environ six fois plus de mesures que le codage matriciel. Du point de vue du rapport signal sur bruit, pour un nombre réduit de télescopes (2 ou 3) tous les concepts sont sensiblement équivalents. A partir de quatre télescopes, le codage temporel a tendance à devenir moins performant que les codages spatiaux ou matriciels. Les deux concepts les plus intéressants sont alors : le concept par paires matriciel que nous présentons pour quatre télescopes dans ce chapitre et le composant tout-en-un spatial que nous détaillerons pour plus de télescopes dans le chapitre IV. Ces deux composants ont chacun leurs avantages et inconvénients dépendant des conditions d'observation et des fonctionnalités de l'interféromètre complet (analyse spectrale, suiveur de franges, etc…).

# *II – D – Recombineurs optiques intégrés*

## II – D – 1 – Recombineurs à deux télescopes

La notion de recombinaison par paires ou de recombinaison tout-en-un n'a évidemment pas de sens dans ce cas. Dans tous les schémas possibles, il est nécessaire de prélever une partie du flux pour obtenir les signaux photométriques de chaque voie, et le meilleur choix pour cela est d'utiliser une jonction « Y » qui présente l'avantage d'être achromatique. Le schéma général d'un recombineur intégré à deux télescopes est représenté sur la figure II-29.

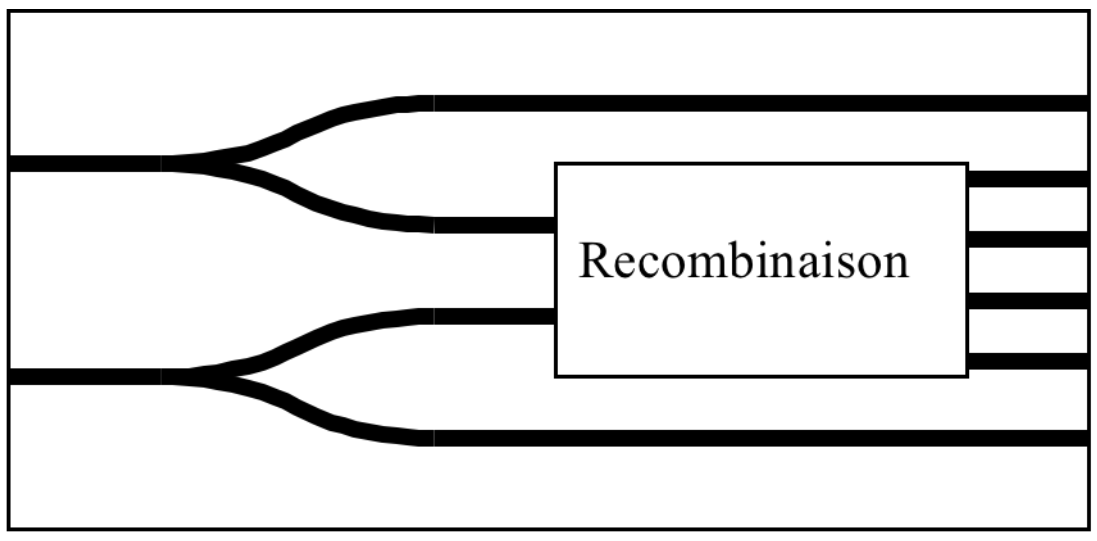

*Figure II-29 : schéma d'un recombineur à deux télescopes*

Pour la recombinaison elle même, on peut considérer une jonction « Y » utilisée en sens inverse. C'est le cas du premier recombineur historiquement validé en interférométrie astronomique [*37*] et réalisé en technologie échange d'ions sur verre[*60*]. La recombinaison est certes achromatique et les signaux de franges obtenus sont excellents, mais elle présente l'inconvénient de ne fournir qu'une seule voie interférométrique, perdant la moitié des photons. Il est donc préférable d'utiliser un coupleur asymétrique qui assure une meilleure transmission. Un tel composant réalisé par le LETI fut testé sur le ciel en même temps que le composant à jonction « Y » [*56*]. Il fournit des résultats comparables à ceux obtenus par jonction « Y » en terme de contraste et précision de mesure, mais fournit deux voies interférométriques complémentaires au lieu d'une. Il a ainsi permis de valider l'utilisation de coupleurs asymétriques pour la recombinaison en interférométrie astronomique.

Si l'on remplace le coupleur asymétrique par un dispositif de mesure ABCD, on obtient le schéma de la figure II-30.

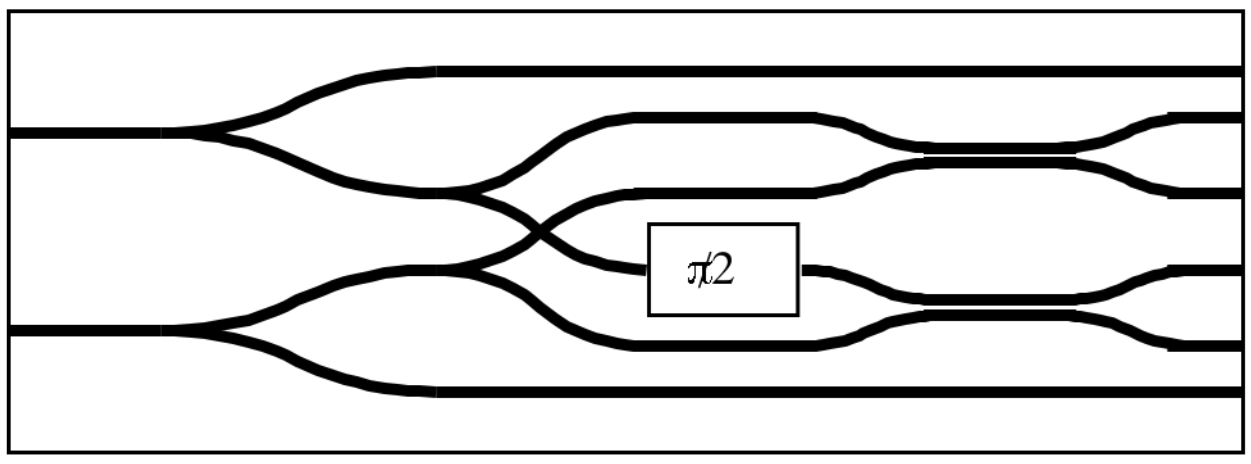

*Figure II-30 : recombineur à deux voies avec mesure ABCD.*

Nous présenterons dans le chapitre III les résultats que nous avons obtenus sur ce type de composant.

## II – D – 2 – Recombineurs à trois télescopes

Au début de ce travail, divers composants à trois télescopes avaient déjà été testés avec succès en laboratoire. Les différents concepts « par paires », « tout-en-un », et « matriciel » avaient été explorés par Haguenauer dans le cas des deux premiers [*56*], et par Rooms dans le cas du dernier[*61*].

L'étude de ces composants a été jusqu'à l'installation du meilleur composant sur un interféromètre à trois télescopes. Il s'agit de l'instrument IONIC installé à IOTA qui a fait de ce dernier, l'un des interféromètres les plus sensibles et les plus précis au monde, prouvant par là même l'intérêt de l'utilisation de l'optique intégrée en interférométrie astronomique.

Le composant intégré installé au cœur de l'instrument IONIC [*62*] est représenté sur la figure II-31.

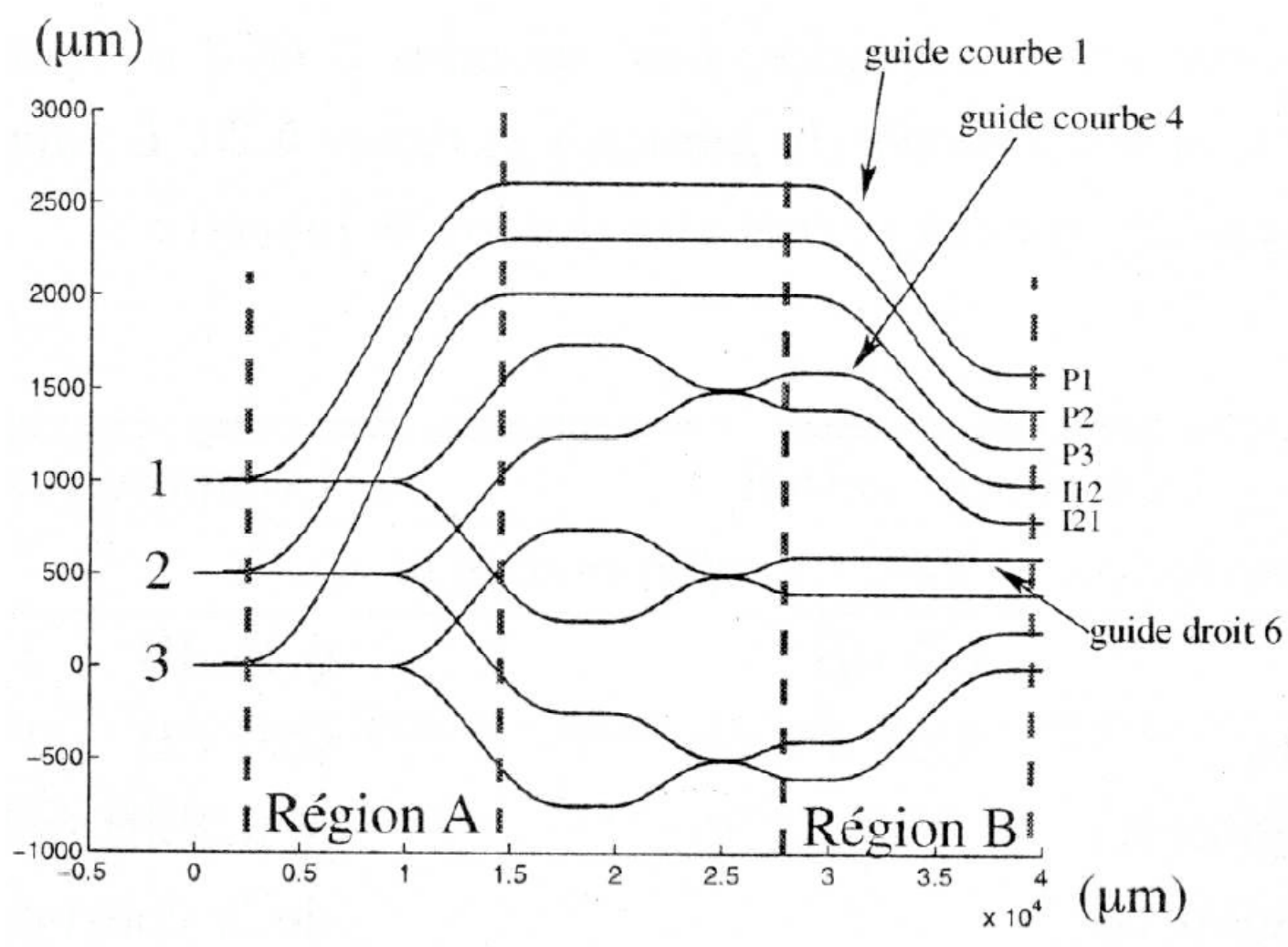

*Figure II-31 : composant 3T installé sur IONIC à IOTA*

On peut noter que le composant comporte des voies photométriques qui se sont avérées inutiles par la suite et qui ont été retirées dans une deuxième réalisation du dispositif.

Devant la quantité des dispositifs étudiés et la qualité des résultats obtenus, il ne nous a pas paru nécessaire d'étudier de nouvelle configuration à trois télescopes. L'intégration d'une fonction ABCD dans un schéma 3T par paires ferait sans doute un excellent composant, mais la demande astronomique se situe maintenant plutôt pour des recombineurs à quatre télescopes et plus.

## II – D – 3 – Recombineurs à quatre télescopes

Rooms [*61*] a montré que son concept de recombineur tout-en-un pouvait éventuellement être appliqué à la recombinaison de quatre télescopes. Cependant, le nombre de modes dans la zone multimode augmentant, la dépendance chromatique de ce type de recombineur risque d'être assez importante ou, pour le moins, délicate à minimiser. Nous nous sommes donc plutôt orientés vers la recombinaison par paires.

Dans ce cas, le schéma de recombinaison dépend du schéma de séparation en trois voies que l'on peut envisager de deux manières selon la figure II-32.

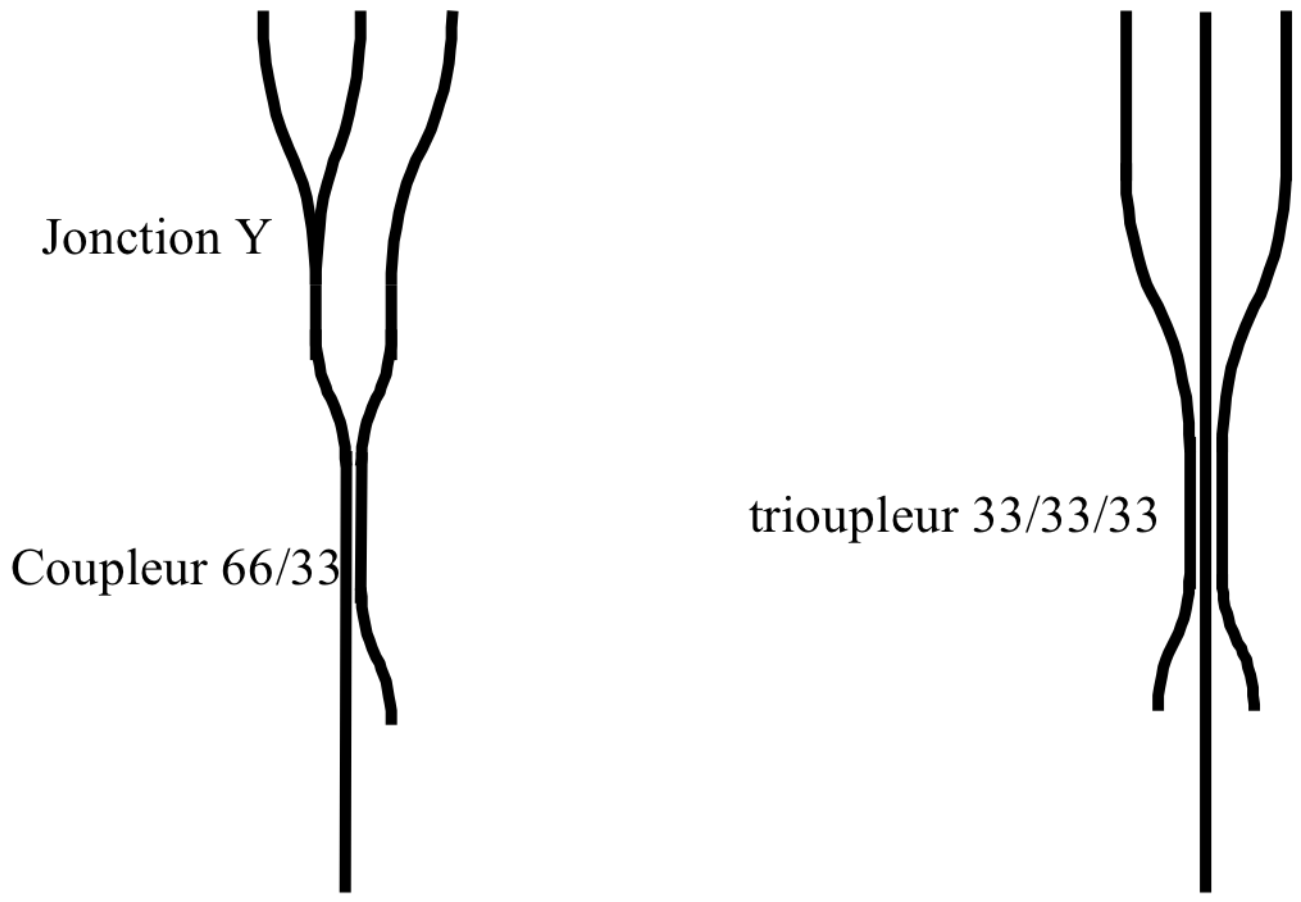

*Figure II-32 : séparation en trois voies : à gauche, séparation cascadée. A droite, séparation directe*

En effet, après séparation des voies, il est important que, pour chaque paire recombinée, chaque voie ait vu le même chemin optique, afin de garantir une bonne mesure de la phase des franges. Ainsi, une voie ayant traversé le coupleur 66/33 par la sortie 66 puis une jonction Y doit se recombiner avec une autre voie ayant vu un chemin identique. Dans le cas d'une séparation directe en trois voies, chaque sortie est soit centrale, soit latérale. De la même manière, les voies centrales vont se recombiner entre elles, et les voies latérales vont se recombiner avec les voies latérales.

En prenant en compte ces considérations, et en ne s'intéressant qu'aux configurations symétriques, dans le cas d'une séparation cascadée, on trouve deux schémas, l'un étant légèrement plus avantageux que l'autre en termes de croisements de guides. Dans le cas d'une séparation directe en trois voies, on trouve un schéma symétrique. Ces trois schémas sont reportés sur la figure II-33.

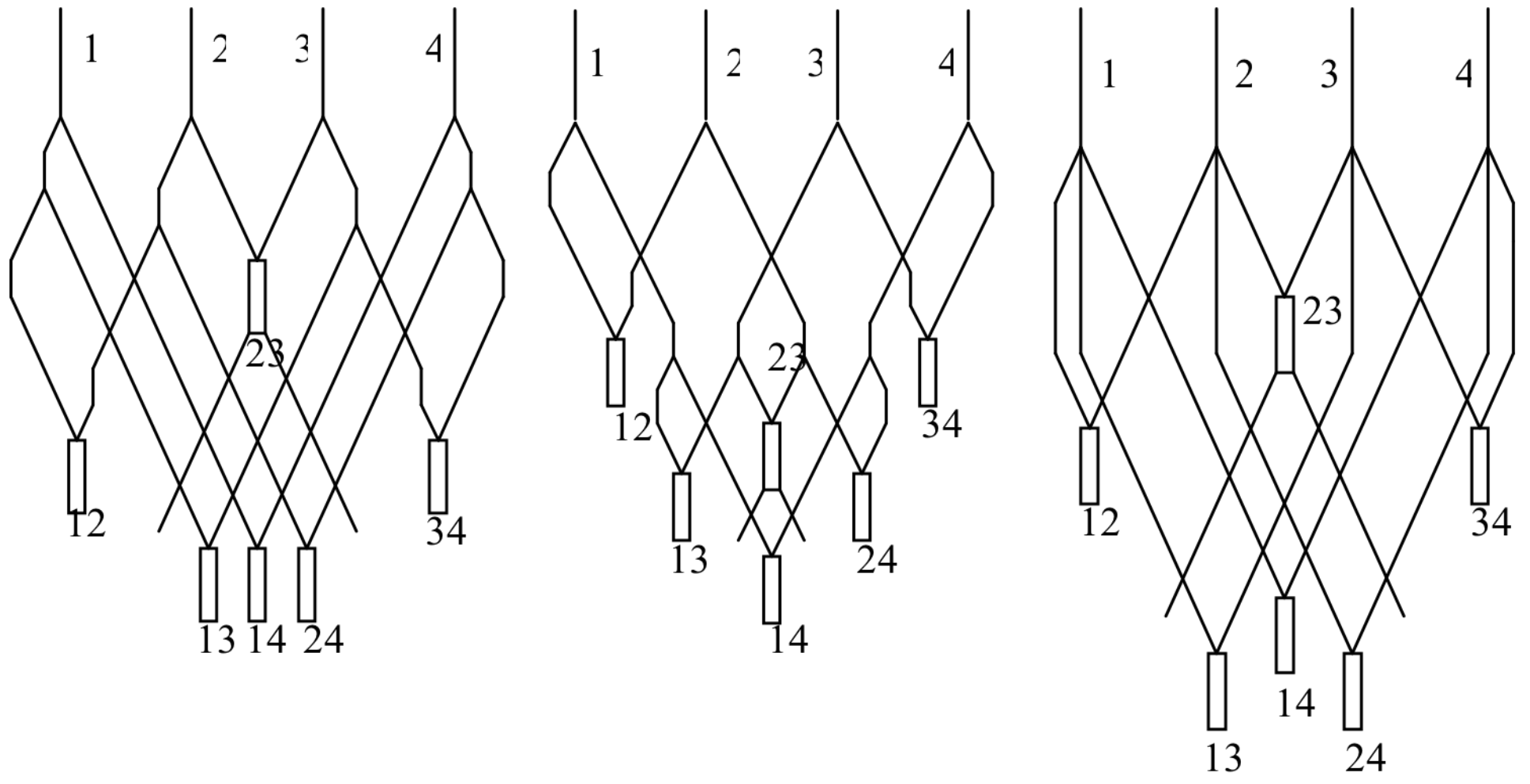

*Figure II-33 : recombineurs par paires à quatre télescopes. Ceux ci peuvent être temporels ou matriciels.*

Du point de vue des pertes, les schémas sont quasiment équivalents. Le schéma utilisant un tricoupleur comporte moins de pertes en excès, mais le schéma global nécessite un encombrement plus important, donc des pertes de propagation plus élevées. Nous avons finalement opté pour le schéma utilisant un tricoupleur car c'est celui qui présente le moins de lumière parasite dans la puce susceptible de perturber les mesures. Dans le cas de l'utilisation de jonctions Y, les pertes en excès de la jonction Y sont générées en un endroit localisé qui agit comme une source de lumière parasite cohérente avec les signaux à mesurer. Dans le cas de pertes de propagation, distribuées tout le long de la puce, l'éventuelle lumière parasite agira plutôt comme une source incohérente, donc générera moins de perturbations.

Avec un dessin détaillé de chaque fonction, on obtient finalement le schéma du recombineur de la figure II-34.

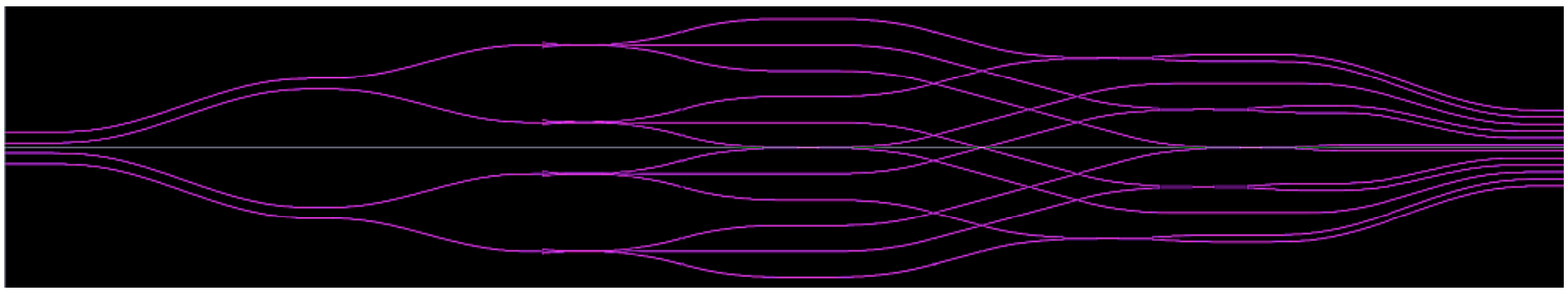

*Figure II-34 : recombineur à quatre télescopes utilisant un tricoupleur et des recombinaisons par paires temporelles. La puce fait environ 36mm de long pour 6mm de large et utilise un rayon de courbure maximal de 8mm pour les parties courbes.*

Le dispositif complet comporte en entrée un épanouisseur permettant d'écarter les guides les uns des autres tout en conservant l'égalité des chemins optiques. Les voies sont ensuite divisées en 3 par le tricoupleur présenté précédemment, puis recombinées par un coupleur asymétrique. Tous les chemins sont rigoureusement symétriques. Ainsi, des croisements de guide factices ont été rajoutés sur les voies extérieures afin de symétriser complètement chaque chemin. La puce fait 36mm de long pour 6mm de large. L'écartement des guides en entrée est de 250µm pour permettre un couplage avec une nappe de fibres standard. En sortie, les guides sont espacés de 160µm pour pouvoir être imagés avec un grandissement unitaire sur une caméra ayant des pixels de 40µm.

## II – E – Conception du déphaseur

Nous avons vu qu'un schéma de type ABCD permettait d'obtenir instantanément la phase des franges d'interférence sans avoir à moduler le chemin optique. Pour cela, il nous faut un élément déphaseur sur un des chemins optiques présents dans la puce. L'idée la plus simple qui vient à l'esprit pour réaliser un déphasage est simplement de considérer un chemin plus long sur l'une des voies. Le problème est que le déphasage introduit dépend alors de la longueur d'onde. Par ailleurs, le déphasage introduit dépendra de la valeur de l'indice effectif du mode fondamental, et sera donc très dépendant des incertitudes technologiques. Afin d'éviter ces problèmes, nous proposons d'utiliser un déphaseur basé sur un principe différent.

## II – E – 1 – Déphaseur optique intégré

Le principe utilise la variation de l'indice effectif (donc de la vitesse de phase) du mode fondamental en fonction de la largeur du guide. Le concept a été proposé en 98 [47] et permet d'obtenir un déphasage beaucoup moins sensible aux fluctuations technologiques car il est basé sur un effet différentiel.

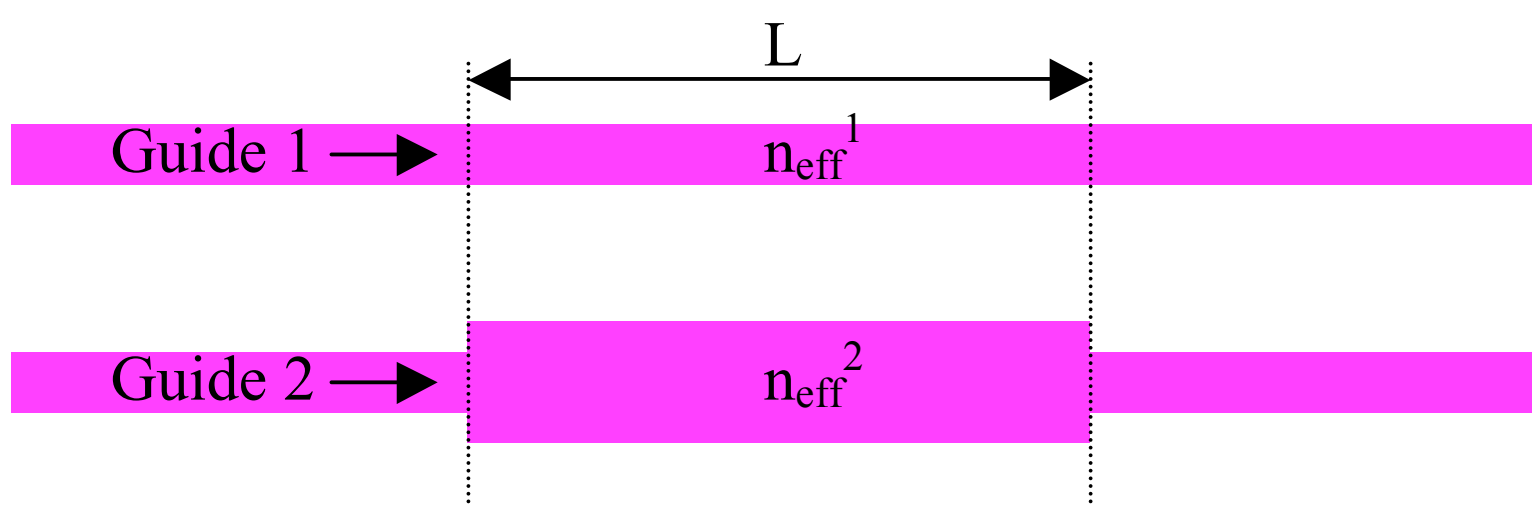

*Figure II-35 : principe du déphaseur optique intégré*

Alors que les chemins sont parallèles et de même longueur, le chemin optique est différent grâce à l'élargissement du guide sur une longueur donnée. Afin de calculer le déphasage $\Delta\varphi$, il suffit de calculer les indices effectifs des modes fondamentaux dans les guides de largeur différente. On obtient :

$$\Delta\varphi = \varphi_1 - \varphi_2 = \frac{2\pi}{\lambda}\left(n_{eff}^{2} - n_{eff}^{1}\right)L \qquad \text{(II-71)}$$

Si les largeurs des guides sont proches, la dépendance de leur indice effectif en fonction de la longueur d'onde est quasiment identique et on a donc un déphasage approximativement inversement proportionnel à la longueur d'onde. La figure II-36 montre les caractéristiques d'un tel déphaseur.

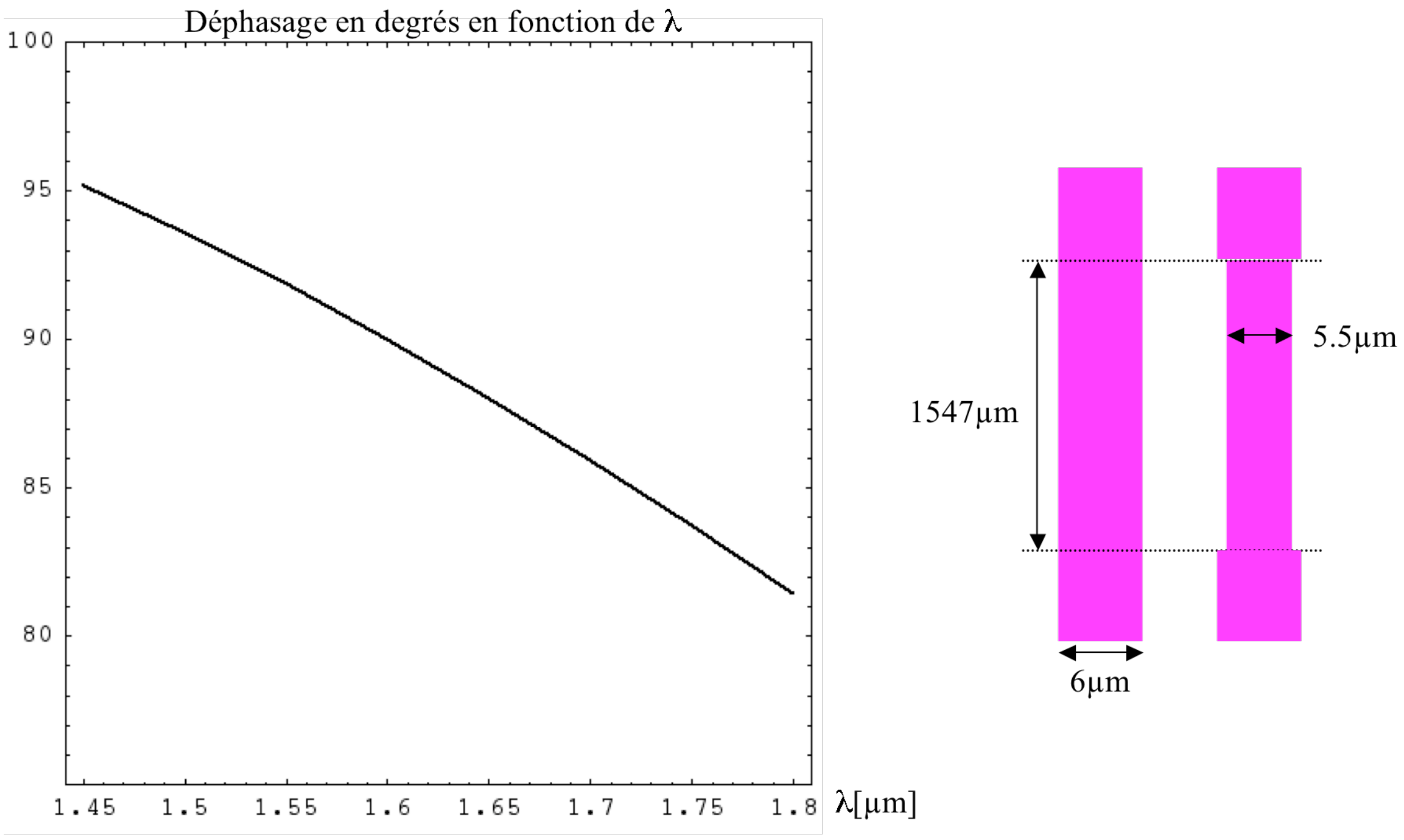

*Figure II-36 : déphaseur simple. Le déphasage décroît avec la longueur d'onde.*

Il est possible de « compenser » cette dépendance en longueur d'onde en concaténant plusieurs segments de guides, chacun de largeur différente :

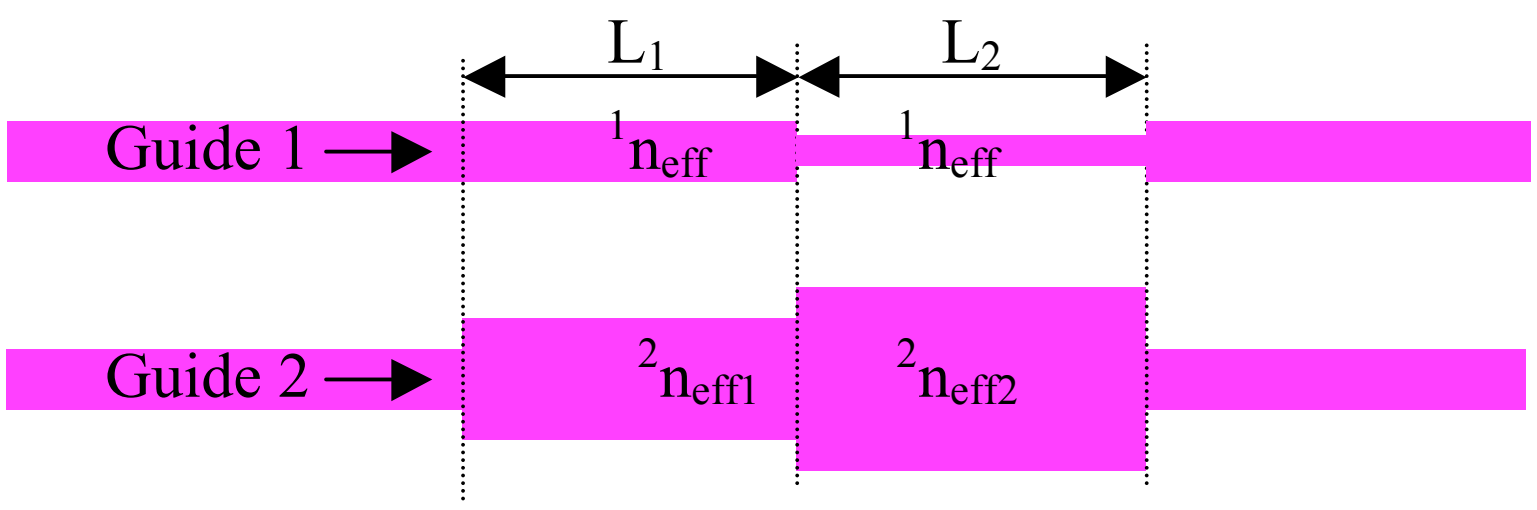

*Figure II-37 : déphaseur compensé en longueur d'onde*

En notant $\Delta n_i$ la différence d'indice effectif sur le segment « i » due à la différence de largeur des guides, on obtient un déphasage $\varphi$ :

$$\varphi = \frac{2\pi}{\lambda}\sum_{i=1}^{n}\Delta n_i L_i \qquad \text{(II-72)}$$

La dépendance en longueur d'onde d'une telle géométrie sera donnée par :

$$\frac{\partial\varphi}{\partial\lambda} = -\frac{2\pi}{\lambda}\sum_{i=1}^{n}\left(\frac{\Delta n_i}{\lambda} - \frac{\partial\Delta n_i}{\partial\lambda}\right)L_i \qquad \text{(II-73)}$$

Afin de calculer numériquement cette dépendance en longueur d'onde, plusieurs choix se présentent. Tout d'abord, on peut remarquer que dans le cas de calculs 2D (guides planaires), le calcul peut être quasiment analytique. En effet, si l'on utilise les notations de [29] :

$$V = \frac{2\pi\rho}{\lambda}\sqrt{n_c^2 - n_g^2} \qquad \text{(II-74)}$$

$$U = \frac{2\pi\rho}{\lambda}\sqrt{n_c^2 - n_{eff}^2} \qquad \text{(II-75)}$$

$$W = \frac{2\pi\rho}{\lambda}\sqrt{n_{eff}^2 - n_g^2} \qquad \text{(II-76)}$$

où $\rho$ est la demi largeur du guide, $\lambda$ la longueur d'onde, $n_c$ est l'indice du cœur et $n_g$ est l'indice de la gaine, on peut alors mettre la relation de dispersion des guides plans symétriques pour les modes pairs (donc en particulier pour le mode fondamental) sous la forme :

$$W = U \cdot Tan(U) \qquad \text{(II-77)}$$

Il est alors facile de démontrer que l'on a :

$$\frac{dU}{dV} = \frac{U}{V}\frac{1}{1+W} \qquad \text{(II-78)}$$

on peut alors obtenir facilement la dérivée en fonction de $\lambda$ :

$$\frac{dU}{d\lambda} = \frac{dU}{dV}\frac{dV}{d\lambda} = -\frac{U}{\lambda}\frac{1}{1+W} \qquad \text{(II-79)}$$

et de là la dérivée de $n_{eff}$ en fonction de $\lambda$. Dans le cas 2D, on peut donc calculer les indices effectifs ou les paramètres $U$, $V$ et $W$ des guides à une longueur d'onde donnée et ensuite calculer analytiquement les dérivées en fonction de la longueur d'onde.

Dans le cas général 3D, on peut se ramener au cas 2D par la méthode de l'indice effectif et ainsi à nouveau bénéficier du cas analytique, mais cela n'est pas très bien adapté au cas de nos guides rectangulaires en silice. Nous avons donc calculé numériquement un tableau d'indice effectif en fonction de la largeur du guide et pour quelques longueurs d'onde réparties dans la bande H. Pour chaque largeur de guide, nous avons ensuite fait un lissage numérique (ou « fit ») sur les indices effectifs obtenus. Etant donnée l'allure de la variation d'indice en fonction de la longueur d'onde, un fit polynomial d'ordre 2 ajusté par la méthode des moindres carrés s'est avéré suffisant. Considérons n guides de largeur $e_j$. En calculant numériquement leur indice effectif pour quelques longueurs d'onde réparties dans la bande H, on obtient après fit polynomial :

$$n_j = A_j + B_j\lambda + C_j\lambda^2 \qquad \text{(II-80)}$$

Si on considère un segment avec deux guides de largeur $e_j$ et $e_k$, on obtient alors l'écart d'indice effectif :

$$\Delta n_i = \left(A_j - A_k\right) + \left(B_j - B_k\right)\lambda + \left(C_j - C_k\right)\lambda^2 \qquad \text{(II-81)}$$

que l'on peut mettre sous la forme :

$$\Delta n_i = \overline{A_i} + \overline{B_i}\lambda + \overline{C_i}\lambda^2 \qquad \text{(II-82)}$$

La figure II-38 donne un exemple de lissage polynomial effectué sur des valeurs d'indice effectif calculés.

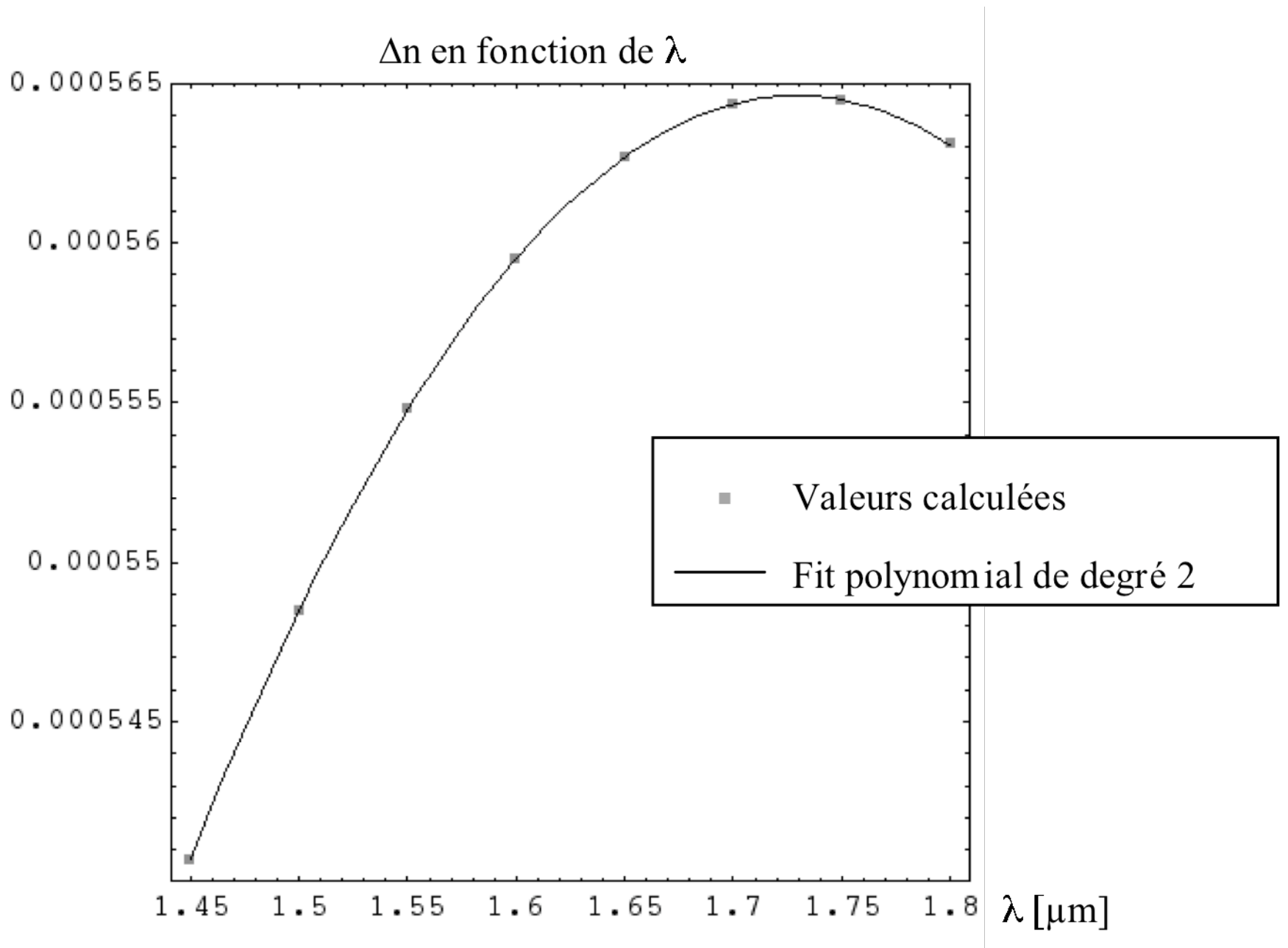

*Figure II-38 : différence d'indice effectif entre un guide de 6µm de large et un guide de 5µm de large en fonction de la longueur d'onde.*

Si on reporte cette fonction d'indice dans l'expression du déphasage, on obtient :

$$\varphi = \frac{2\pi}{\lambda}\sum_{i=1}^{n}\left(\overline{A_i} + \overline{B_i}\lambda + \overline{C_i}\lambda^2\right)L_i \qquad \text{(II-83)}$$

$$\frac{\partial\varphi}{\partial\lambda} = -\frac{2\pi}{\lambda}\sum_{i=1}^{n}\left(\frac{\overline{A_i}}{\lambda} - \overline{C_i}\lambda\right)L_i \qquad \text{(II-84)}$$

Si on veut réaliser un déphaseur achromatique de valeur $\varphi_0$, on va écrire :

$$\forall\lambda,\begin{cases}\varphi=\varphi_0\\ \dfrac{\partial\varphi}{\partial\lambda}=0\end{cases}\Leftrightarrow\begin{cases}\sum_{i=1}^{n}\overline{A_i}L_i=0\\ \sum_{i=1}^{n}\overline{B_i}L_i=\dfrac{\varphi_0}{2\pi}\\ \sum_{i=1}^{n}\overline{C_i}L_i=0\end{cases} \qquad \text{(II-85)}$$

Ce problème devient facilement inversible si l'on considère un déphaseur composé de trois segments de guides de longueurs respectives $L_1$, $L_2$, et $L_3$. Dans ce cas, on obtient en effet trois équations reliant les coefficients des polynômes et les trois longueurs de segment, ce qui peut se mettre sous forme matricielle :

$$\begin{bmatrix}\overline{A_1} & \overline{A_2} & \overline{A_3}\\ \overline{B_1} & \overline{B_2} & \overline{B_3}\\ \overline{C_1} & \overline{C_2} & \overline{C_3}\end{bmatrix}\bullet\begin{bmatrix}L_1\\ L_2\\ L_{31}\end{bmatrix}=\begin{bmatrix}0\\ \dfrac{\varphi_0}{2\pi}\\ 0\end{bmatrix} \qquad \text{(II-86)}$$

En notant :

$$M=\begin{bmatrix}\overline{A_1} & \overline{A_2} & \overline{A_3}\\ \overline{B_1} & \overline{B_2} & \overline{B_3}\\ \overline{C_1} & \overline{C_2} & \overline{C_3}\end{bmatrix} \qquad \text{(II-87)}$$

et :

$$[L]=\begin{bmatrix}L_1\\ L_2\\ L_3\end{bmatrix} \qquad \text{(II-88)}$$

on obtient alors simplement les longueurs permettant d'obtenir le déphaseur voulu par :

$$[L]=[M]^{-1}\cdot\begin{bmatrix}0\\ \dfrac{\varphi_0}{2\pi}\\ 0\end{bmatrix} \qquad \text{(II-89)}$$

Afin d'optimiser la longueur totale du déphaseur, nous avons calculé les indices effectifs des modes fondamentaux pour des guides allant de 3 à 9µm de large par pas de 0,5µm pour des longueurs d'onde comprises entre 1,45µm et 1,8µm par pas de 0,05µm. Nous avons calculé chacun des déphaseurs possibles puis sélectionné un des plus courts. On obtient finalement le déphaseur de la figure II-39. La longueur totale est de 5,5mm.

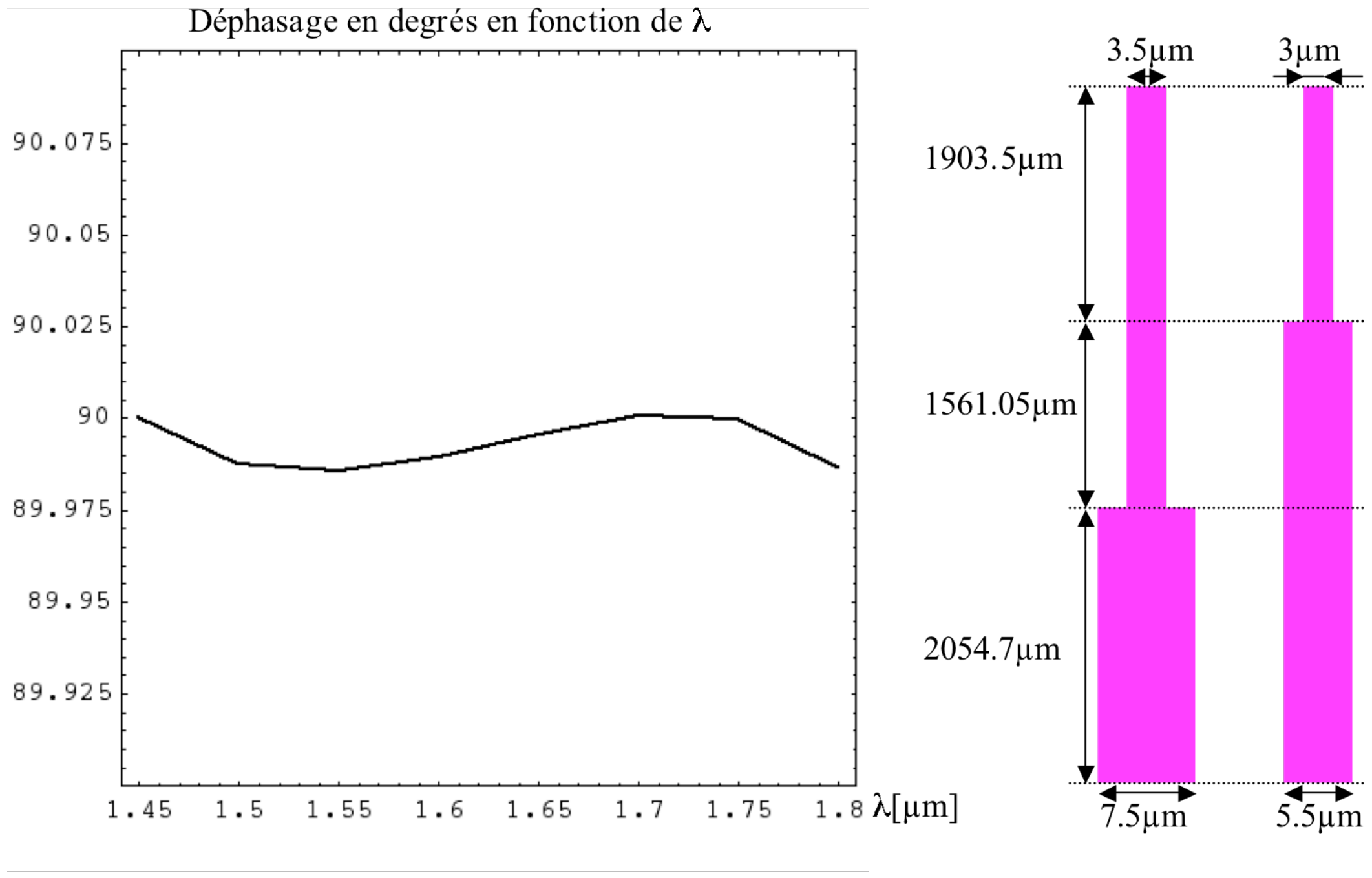

*Figure II-39 : déphaseur achromatique final obtenu. Les longueurs de guide ont été arrondies à la grille du masque de photolithographie.*

L'un des inconvénients liés à ce type de déphaseur est que l'on introduit des pertes à chaque discontinuité entre les différentes portions de guide (couplage aux modes rayonnés + réflexion parasite). Afin d'éviter ces pertes, il est possible d'introduire des épanouisseurs entre chaque segment comme montré sur la figure II-40.

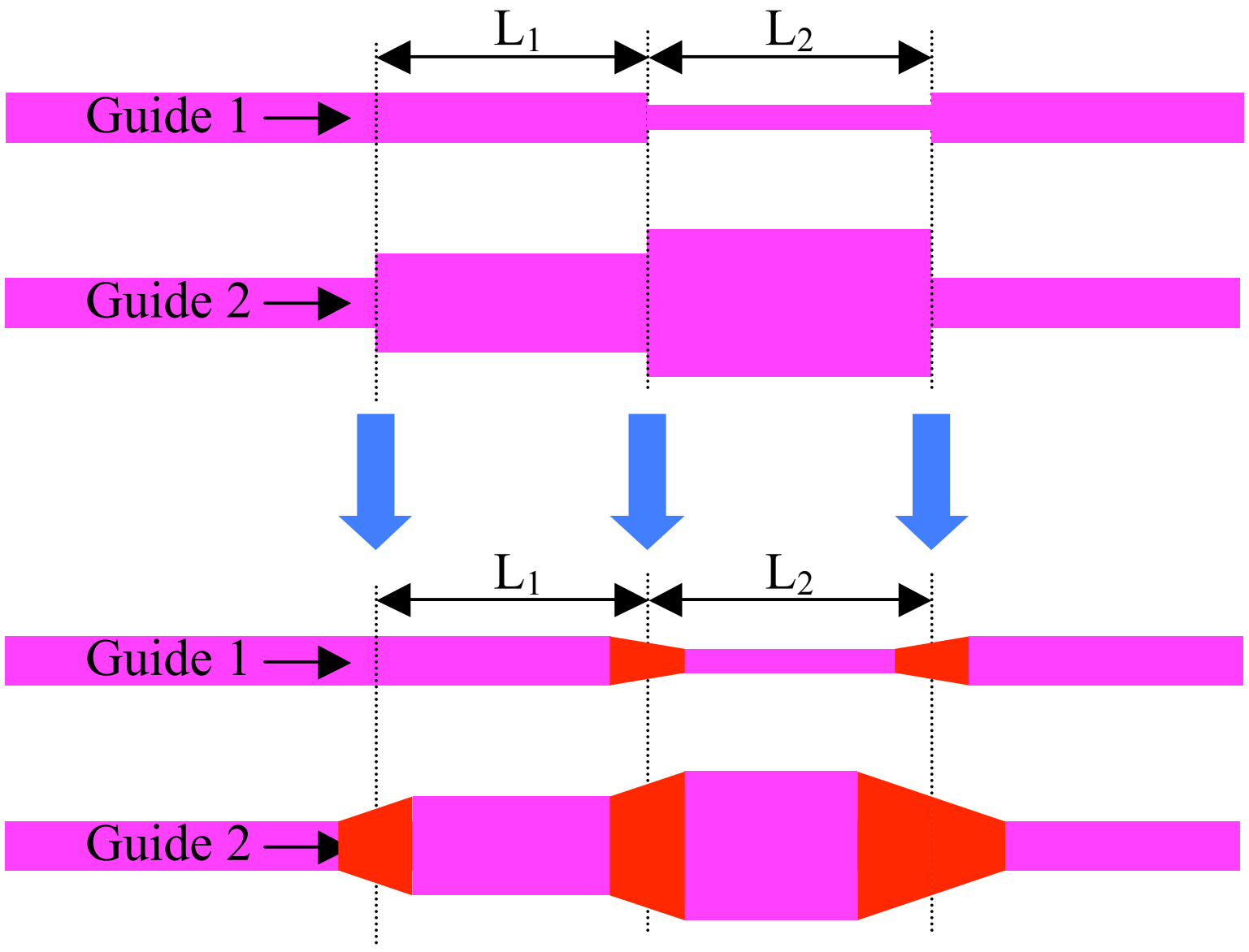

*Figure II-40 : intégration d'épanouisseurs (« tapers ») pour minimiser les pertes aux discontinuités.*

En première approximation, la partie où le « taper » est plus large que le guide qu'il remplace est compensée par la partie où le « taper » est plus étroit que le guide qu'il remplace. Par ailleurs, on a des pertes négligeables lorsque l'angle des « tapers » est inférieur au critère d'adiabaticité de variation des guides déjà cité précédemment :

$$\theta(z) < \frac{\rho(z)}{2\pi}\left(\beta - kn_g\right) \qquad \text{(II-90)}$$

Lorsque l'on ne veut pas faire d'approximation et que l'on a suffisamment de place sur la puce, on peut compenser l'erreur introduite par les « tapers » en symétrisant leur présence sur les deux guides :

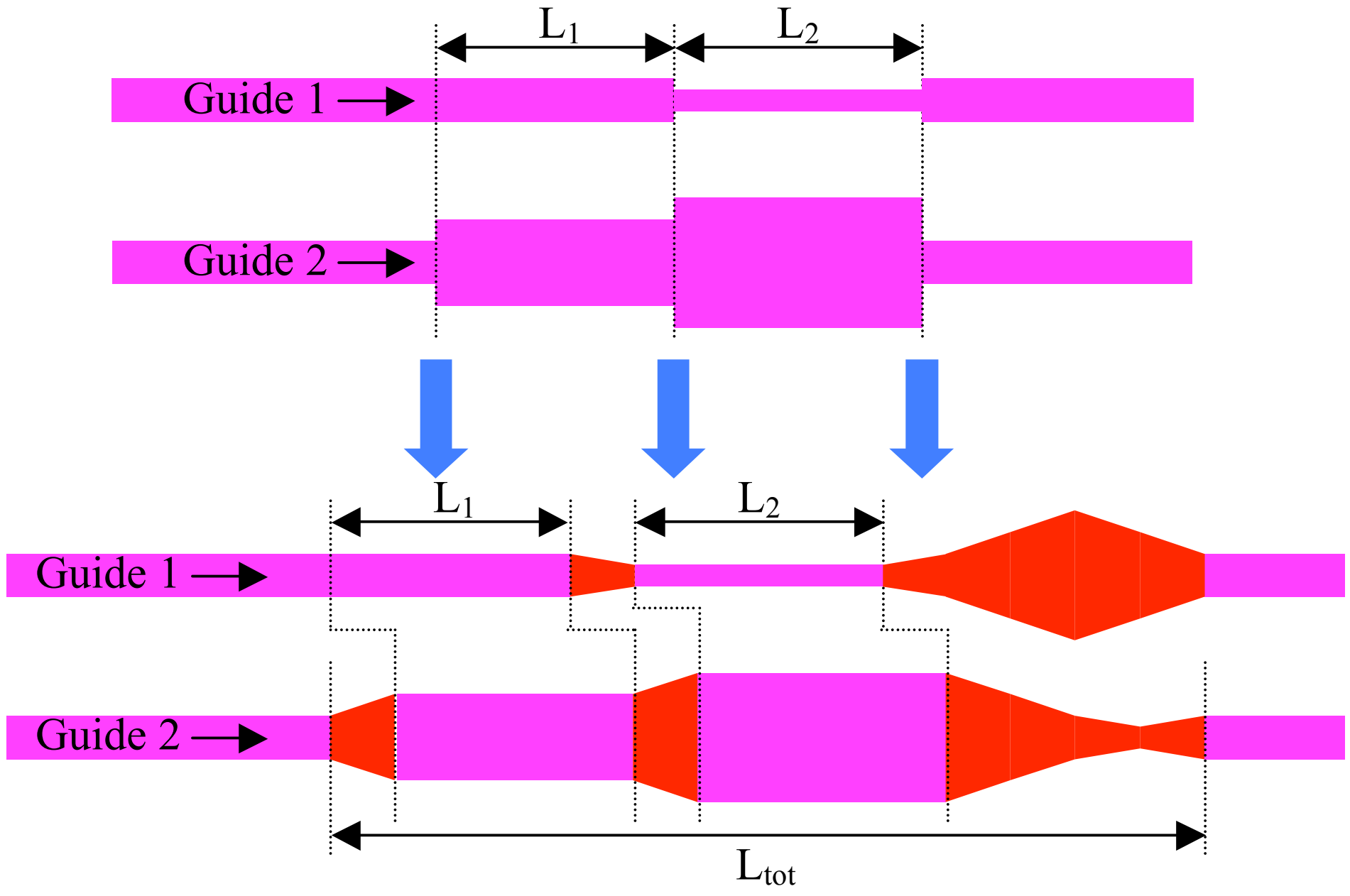

*Figure II-41 : intégration de « tapers » aux discontinuités sans modifier le déphasage calculé. La longueur totale $L_{tot}$ du déphaseur est alors plus grande. Dans le cas particulier du déphaseur de la figure II-39, la longueur totale du déphaseur passe de 5519.25µm à 7769.25µm.*

Ces derniers types de déphaseurs ont été vérifiés en BPM-3D scalaire et sont sans pertes et en très bon accord avec les résultats obtenus par le calcul ci-dessus. Nous allons voir au chapitre III suivant les résultats obtenus sur les deux types de déphaseurs présentés ici : les déphaseurs simples à un seul segment, et les déphaseurs compensés à plusieurs segments. Enfin, nous reviendrons sur ces déphaseurs au chapitre IV qui peuvent être conçus de manière légèrement différente.

## II – E – 2 – Intégration dans un recombineur à quatre télescopes

Tout en conservant le principe du recombineur 4T par paires, nous pouvons inclure pour chaque recombinaison une cellule ABCD de mesure instantanée de la phase. Chaque cellule ayant quatre sorties au lieu de deux, les chemins comportent un croisement de guide supplémentaire. Par ailleurs, dans chaque cellule ABCD, il y a aussi un croisement de guide, on obtient donc au final cinq croisements de guide par chemin. Par ailleurs, la puce est beaucoup plus longue, l'encombrement d'une cellule ABCD étant nettement plus important qu'un simple coupleur.

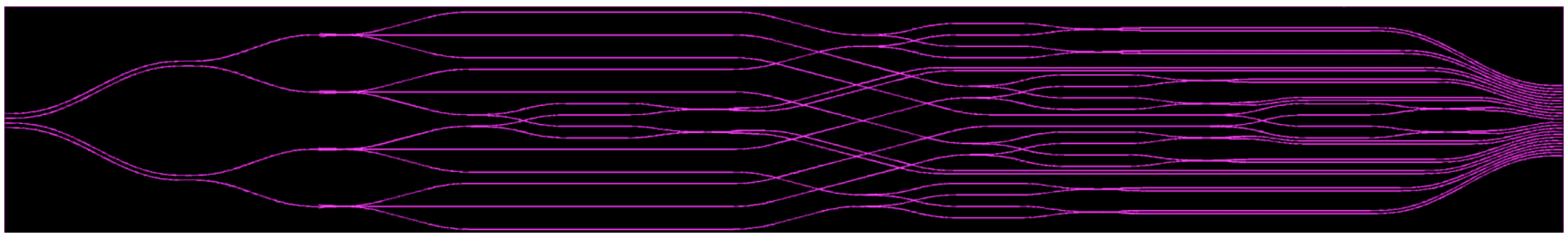

*FigureII-42 : recombineur à quatre télescopes utilisant un tricoupleur et des recombinaisons ABCD pour chaque paire. La puce fait 80mm par 12mm. Le rayon de courbure utilisé est de 8mm.*

En conséquence, la puce fait cette fois environ 80mm de long par 12mm de large. Il s'agit de la puce optique intégrée la plus longue jamais conçue et réalisée au LETI.

## II – F - Conclusion

Après avoir présenté les fonctions optiques intégrées élémentaires à notre disposition, nous avons étudié les différentes possibilités de recombinaisons jusqu'à quatre télescopes. L'étude s'est principalement portée sur l'étude de recombineurs à quatre télescopes et a permis de concevoir un recombineur par paires utilisant un tricoupleur pour séparer chaque faisceau entrant en trois voies qui se recombinent avec chacune des trois autres voies. Nous avons appliqué cette géométrie à deux types de recombineurs : les recombineurs de type temporel utilisant une recombinaison de chaque paire basée sur un simple coupleur achromatique et les recombineurs de type matriciel utilisant une cellule ABCD afin d'obtenir instantanément une mesure de phase des franges d'interférence sans avoir à moduler le chemin optique en amont du dispositif de recombinaison. Afin de réaliser des cellules ABCD performantes, nous avons conçu un nouveau type de déphaseurs optiques intégrés permettant d'introduire un déphasage achromatique de $\pi/2$ entre deux chemins optiques parallèles de même longueur. Ces déphaseurs sont certes encombrants mais permettent d'obtenir un comportement très plat sur toute la bande H. Ces études ont permis de concevoir des puces de recombinaison dont nous présentons la réalisation et la caractérisation dans le chapitre III.

# Chapitre III : Réalisations et caractérisations

## III – A – Introduction

Dans le chapitre II, nous avons passé en revue les différentes possibilités que nous offrait l'optique intégrée. Nous les avons comparées, puis avons sélectionné celles qui nous semblaient les plus intéressantes pour la recombinaison interférométrique. Cela nous a alors permis de concevoir complètement des dispositifs intégrés pour la recombinaison à quatre télescopes.

Dans ce chapitre, nous allons d'abord décrire la technologie de réalisation de ces dispositifs, puis présenter les résultats optiques que nous avons obtenus. Nous verrons dans un premier temps les transmissions spectrales des dispositifs et des différentes fonctions élémentaires qui les constituent. Puis nous détaillerons les résultats interférométriques que nous avons réalisés et nous nous intéresserons particulièrement aux performances des déphaseurs optiques intégrés que nous avons introduits à la fin du chapitre II.

# *III – B – Optique intégrée silice sur silicium*

## III – B – 1 – Introduction

Le LETI utilise un procédé de réalisation de guides optiques en silice sur substrat silicium basé sur des étapes technologiques standardisées de la microélectronique (communément appelées les « briques de base technologiques »), c'est-à-dire des dépôts, recuits et gravures de couches minces sur substrat silicium.

L'intérêt majeur de l'optique intégrée sur substrat de silicium est de pouvoir bénéficier de l'effort de recherche considérable effectué en microélectronique tiré par des marchés en forte croissance depuis de nombreuses années. Ainsi, on dispose de « briques élémentaires» stabilisées de très grande qualité, sur des équipements développés et optimisés spécifiquement pour cette technologie et garantissant une excellente répétabilité des procédés. En outre, le silicium est un substrat mécanique de très bonne qualité, cristallin (donc clivable), possédant une excellente qualité de surface, ainsi qu'une excellente conductivité thermique. L'inconvénient est que si l'on compare à d'autres technologies de réalisation de guides optiques intégrés dont le coût de fabrication peut être comparable, le coût de l'investissement initial en équipements de production est très élevé et ne peut donc être justifié que par un gros volume de marché permettant de rentabiliser ces derniers.

## III – B – 2 – Empilement technologique

L'enchaînement des étapes technologiques du procédé de réalisation des guides est représenté sur la figure III-1. Il est découpé de manière à montrer l'évolution de la structure au cours de la réalisation. Il est ensuite décrit plus en détail dans les paragraphes suivants.

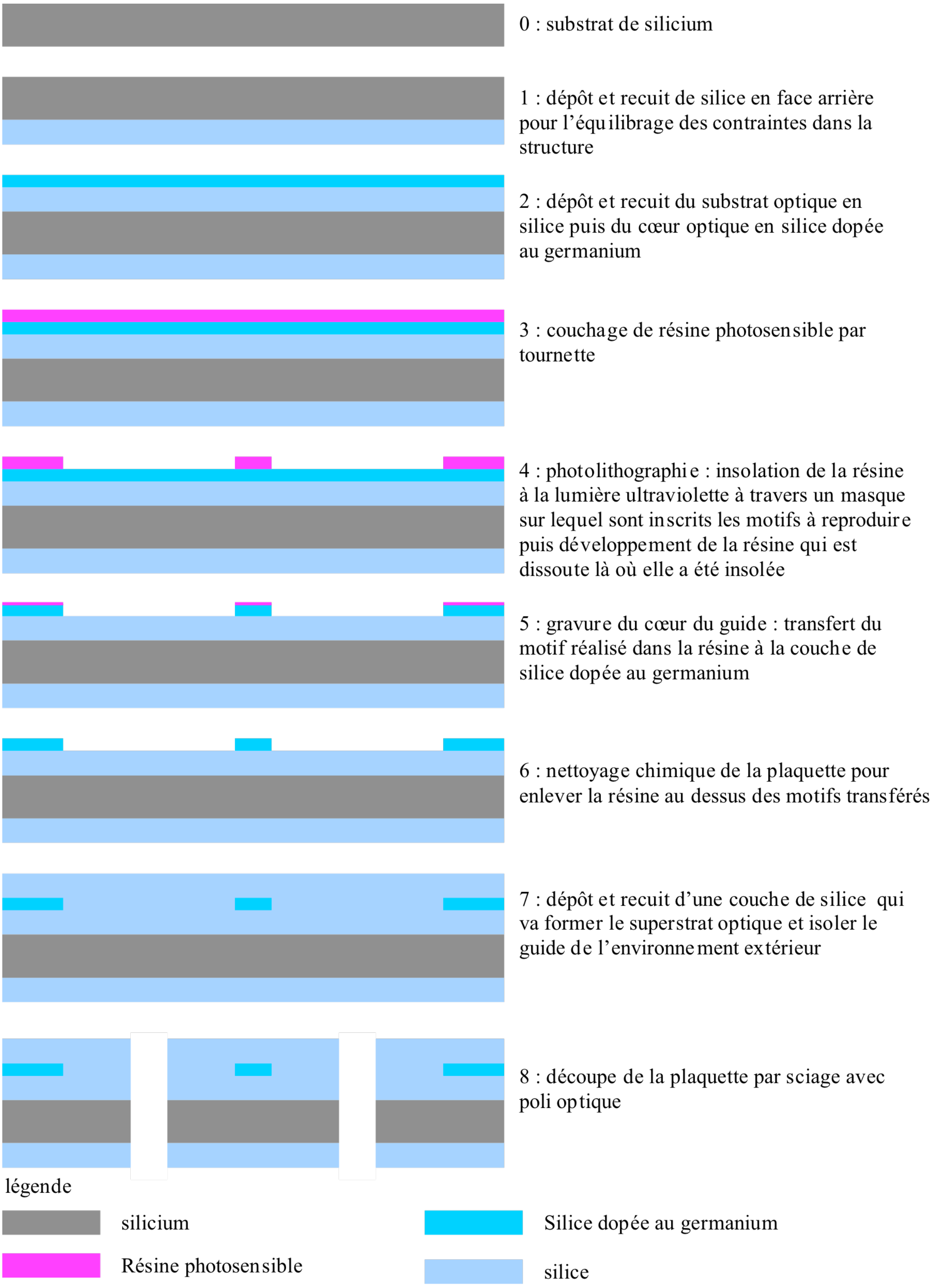

*Figure III-1 : empilement technologique de réalisation des guides d'onde optique en silice sur substrat de silicium.*

Il y a au total une vingtaine d'étapes élémentaires ce qui peut paraître peu élevé lorsqu'on compare à un procédé de réalisation de puce microélectronique comportant généralement plus d'une centaine d'étapes. Cet empilement présente cependant des difficultés spécifiques liées principalement à l'épaisseur totale de la structure (environ 30µm) qui est très supérieure à ce qui est communément fait en microélectronique. En effet, les couches de silice étant déposées à 400°C, la structure comporte des contraintes mécaniques dues à la différence entre les coefficients de dilatation thermique de la silice et du silicium (figure III-2).

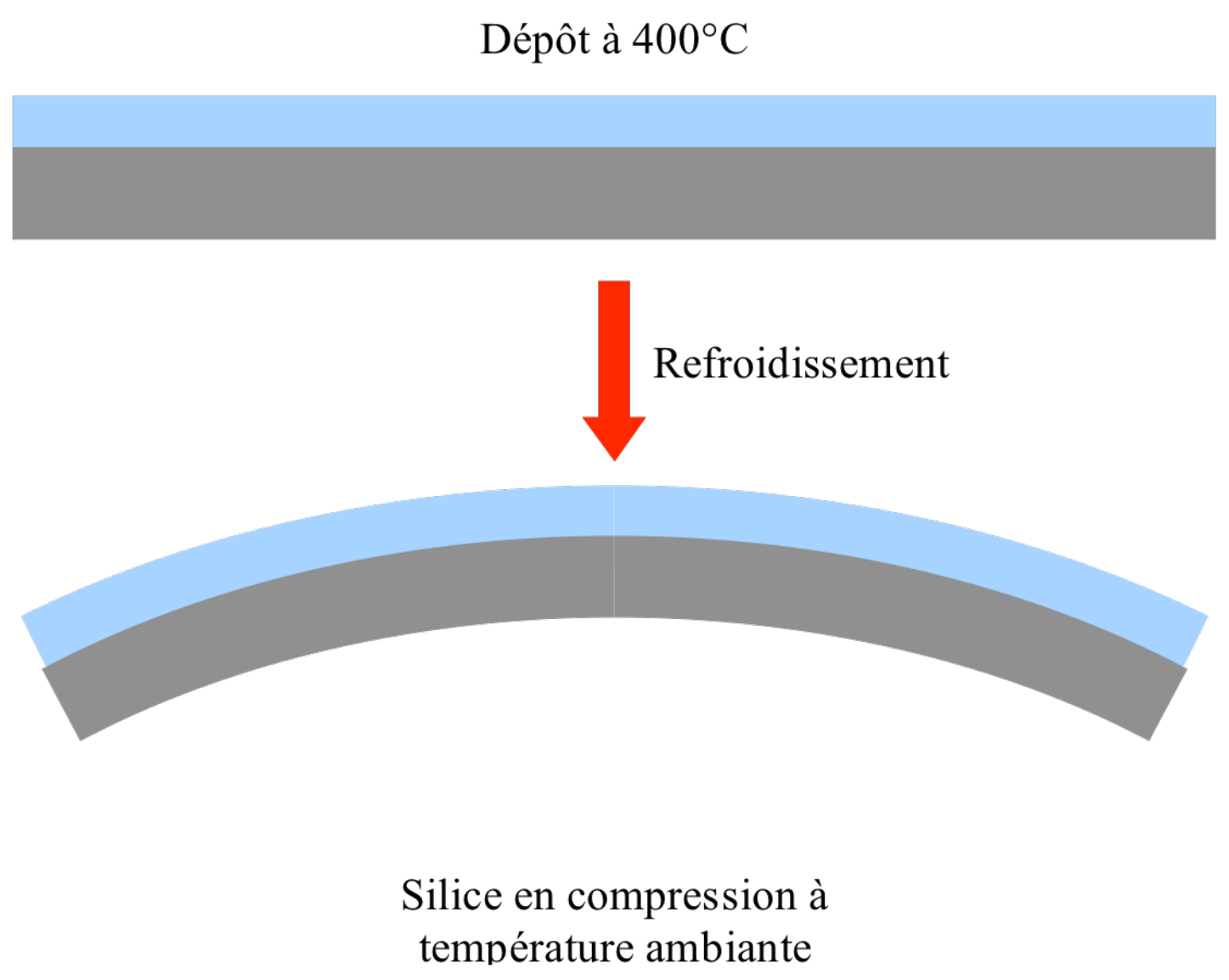

*Figure III-2 : contraintes générées dans la structure. La convention des couleurs est celle de la figure III-1. La couche de silice est déposée à 400°C. Lorsque la structure refroidit à température ambiante, le coefficient de dilatation du silicium étant supérieur à celui de la silice, la couche déposée est en compression et génère une déformation de la plaquette.*

Ces contraintes mécaniques engendrent une déformation de la plaquette proportionnelle à l'épaisseur de la couche déposée. C'est pourquoi l'empilement débute par un dépôt de silice face arrière afin d'équilibrer les contraintes et de minimiser la déformation de la plaquette.

## III – B – 3 – Procédé technologique des étapes de réalisation

Les couches de silice sont réalisées par dépôt chimique en phase vapeur assisté par plasma (PECVD pour Plasma Enhanced Chemical Vapor Deposition). Les précurseurs utilisés sont le silane ($SiH_4$) et l'oxygène ($O_2$). Le principe de la PECVD consiste à créer un plasma par l'application d'un champ radiofréquence. Les précurseurs ionisées condensent alors sur le substrat créant la couche de silice. La silice obtenue ainsi présente très peu d'impuretés et un indice de réfraction très proche de celui de la silice obtenue par fusion (n = 1,444 à une longueur d'onde de 1,55µm). Une couche de 12µm de cette silice constitue donc le substrat optique des guides d'onde. Le cœur du guide est réalisé par dépôt d'une couche de silice dopée au germanium grâce à l'ajout de germane ($GeH_4$) comme précurseur dans la chambre au moment du dépôt. L'écart d'indice entre le cœur de silice dopée et le substrat de silice pure utilisé pour nos guides d'onde est de $\Delta n = 0{,}01 \pm 0{,}0005$.

L'étape suivante est la photolithographie qui a pour but de transférer les motifs dessinés sur un masque dur sur la plaquette. Ce masque dur est généralement réalisé avec du chrome déposé sur un substrat de quartz puis gravé par faisceau laser ou faisceau d'électrons. Dans notre cas, le masque dur est réalisé avec une résolution de 50nm et une précision des motifs inférieure à 100nm. La photolithographie débute par le dépôt à la tournette d'une couche de résine photosensible. Afin d'améliorer la tenue de cette résine sur la silice, on utilise un promoteur d'adhérence avant dépôt : l'héxaméthyldisilasane (HMDS). La résine est ensuite stabilisée grâce à un recuit (par convection) notamment pour évacuer les solvants. L'insolation aux UV à une longueur d'onde de 365nm est ensuite effectuée à travers le masque dur. La dose d'énergie typique est de l'ordre de 100mJ/cm$^2$. Après insolation, la résine est « développée », c'est à dire immergée dans un milieu basique. On distingue les résines « positives », où ce sont les zones insolées qui sont dissoutes au développement, des résines « négatives » où ce sont les zones non insolées qui sont dissoutes. Le choix du type de résine dépend en général du matériau à graver et du taux d'ouverture du masque. Pour notre application, nous avons utilisé une résine photosensible positive dont l'épaisseur limite la résolution à environ 1µm. Finalement, la résine est à nouveau recuite (« post-bake ») principalement afin d'éliminer l'eau. La photolithographie est l'étape du procédé de réalisation la plus sensible aux conditions extérieures (poussières, température, humidité, etc…). C'est pourquoi elle nécessite un environnement le plus propre possible (salle blanche classe 100 ou inférieure) sous lumière jaune car la résine est sensible aux longueurs d'ondes inférieures à 450nm.

Le transfert des motifs de résine dans la couche de silice s'effectue par gravure ionique réactive (RIE pour Reactive Ion Etching) dont le principe consiste là encore à réaliser un plasma entre deux electrodes polarisées (la plaquette étant à la cathode) par l'application d'un champ radiofréquence. Les espèces utilisées pour graver la silice sont à base de composants fluorés ($CF_4 + O_2$ ou $CHF_3 + O_2$). En ajustant les paramètres du procédé (tension radiofréquence, température, et pression), on rend la gravure plutôt anisotrope (flancs verticaux) ou isotrope (flancs arrondis) [*63*].

Les étapes de photolithographie et gravure permettent de transférer les motifs dans la couche de silice dopée avec une résolution de l'ordre du micromètre. L'épaisseur du coeur des guides d'onde étant de 4,85µm ± 0,15µm, il est nécessaire de graver une profondeur de 5µm ce qui est une épaisseur relativement élevée pour un procédé microélectronique. En conséquence, l'épaisseur de résine utilisée est aussi assez élevée (3,35µm) et la perte de résolution provient du fait que l'insolation ne peut être parfaitement focalisée sur toute l'épaisseur de la résine. De plus, même lorsque la gravure est très anisotrope afin d'obtenir des flancs de guides les plus verticaux possible, on garde une légère gravure horizontale qui conduit à une perte de cote latérale des guides. Néanmoins, si cette perte de cote est répétable, elle peut être anticipée lors du dessin du masque.

Après nettoyage, la dernière étape consiste à déposer la couche supérieure de silice (le superstrat). On utilise là encore un dépôt PECVD mais en dopant la couche au Bore et au Phosphore. Ces deux dopants rendent le dépôt plus conforme, c'est à dire recouvrant le mieux possible les reliefs créés par la gravure du cœur du guide optique, l'augmentation de l'indice de réfraction par le phosphore étant compensée par la diminution de l'indice de réfraction par le bore. Cependant, dans le cas de motifs très petits (comme la pointe d'une jonction Y ou l'interstice entre deux guides composant un coupleur), le recouvrement peut s'effectuer de manière imparfaite et conduire à une silice localement moins dense, voire comportant des bulles d'air (« voids »). La conception exposée au chapitre II prend en compte ces limitations technologiques.

## III – B - 4 - Migration sur substrat de silicium de huit pouces

Au début de ce travail de thèse, nous avons fait migrer ce procédé entièrement développé pour des substrats de quatre pouces vers des substrats de huit pouces en utilisant des équipements de dernière génération et un masque de test comportant des guides droits. La principale difficulté que nous avons due alors résoudre était la gestion de la déformation plus grande de la plaquette. Les résultats en termes de pertes de propagation sur des guides droits sont reportés sur la figure III-3.

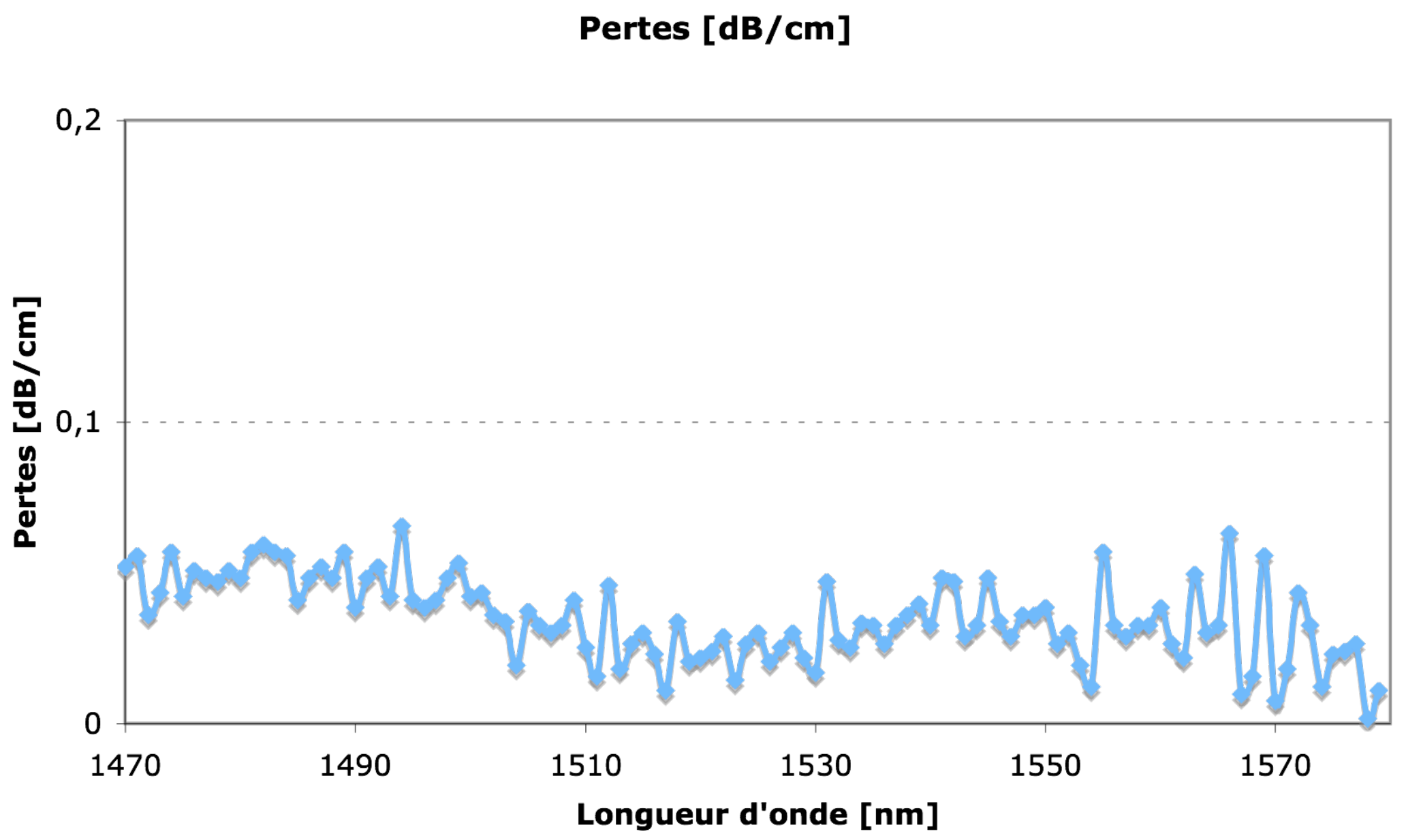

*Figure III-3 : pertes des guides silice sur silicium en technologie 8 pouces.*

Ces mesures s'obtiennent en injectant de la lumière par une fibre optique monomode et en récupérant la lumière par une fibre monomode connectée à un spectromètre. Le coefficient de pertes s'obtient en effectuant des mesures pour différentes longueurs de guide afin de s'affranchir des pertes de couplage en entrée et en sortie de guide. Si $T_e$ représente la transmission du couplage en entrée, $T_s$, la transmission du couplage en sortie, $L_1$ et $L_2$ les deux longueurs de guide mesurés, $T_1$, et $T_2$ les deux mesures de transmission effectuées, le coefficient de perte par propagation $\alpha$ est tel que :

$$T_1 = T_e e^{-\alpha L_1} T_s \quad \text{(III-1)}$$

$$T_2 = T_e e^{-\alpha L_2} T_s \quad \text{(III-2)}$$

En exprimant les longueurs en cm, on obtient facilement les pertes en dB/cm :

$$\alpha_{dB/cm} = 10 \frac{Log\left(\frac{T_1}{T_2}\right)}{\left(L_2 - L_1\right)_{cm}} \quad \text{(III-3)}$$

Dans la pratique, on effectue la mesure pour plus de deux longueurs afin de minimiser l'erreur sur le coefficient $\alpha$.

Les pertes sont de 5dB/m ± 1dB/m, c'est à dire équivalentes, voire légèrement plus faibles que les pertes obtenues en technologie quatre pouces sur la bande spectrale 1470–1580nm, ce qui valide la technologie silice sur silicium à base de substrats huit pouces.

Nous allons maintenant décrire les dispositifs réalisés avant de passer aux résultats de caractérisation des dispositifs.

## III – C - Puces réalisées

Quatre types de puces ont été réalisées :

- des recombineurs à quatre télescopes par paires que nous appellerons par la suite « puces 4T »
- des recombineurs à deux télescopes utilisant une cellule ABCD que nous appellerons par la suite « puces 2T-ABCD »

- des recombineurs à quatre télescopes utilisant une cellule ABCD que nous appellerons par la suite « puces 4T-ABCD »
- des puces de test

Pour chaque type de puce, afin de pouvoir gérer la perte de côte latérale due à la surgravure RIE du cœur, nous avons réalisé trois variantes avec des largeurs de guide d'onde sur le masque légèrement différentes. La largeur visée du cœur du guide d'onde étant de 6µm, chaque type de puce a été réalisé avec des motifs de masque de 6µm, 6,5µm et 7µm de large. Par ailleurs, pour chaque puce, l'écartement entre les guides d'entrée est de 250µm afin de pouvoir exciter simultanément toutes les entrées à l'aide d'une nappe de fibres au pas de 250µm. En sortie, l'écartement entre les guides est de 160µm afin de pouvoir les imager avec un grandissement unitaire sur une caméra infrarouge comportant des pixels de 40µm.

## III – C – 1 – Recombineurs à quatre télescopes par paires

Les « puces 4T » sont représentées sur la figure III-4. Chaque dispositif présente quatre entrées. Chaque entrée est divisée en trois voies à l'aide d'un tricoupleur 33% et se recombine avec l'une des autre voies grâce à un coupleur asymétrique 50/50. Les chemins optiques comportent deux ou trois croisements de guides suivant les cas. Les dispositifs font 36 mm de long et comportent des guides droits afin d'évaluer la perte globale des fonctions du dispositif.

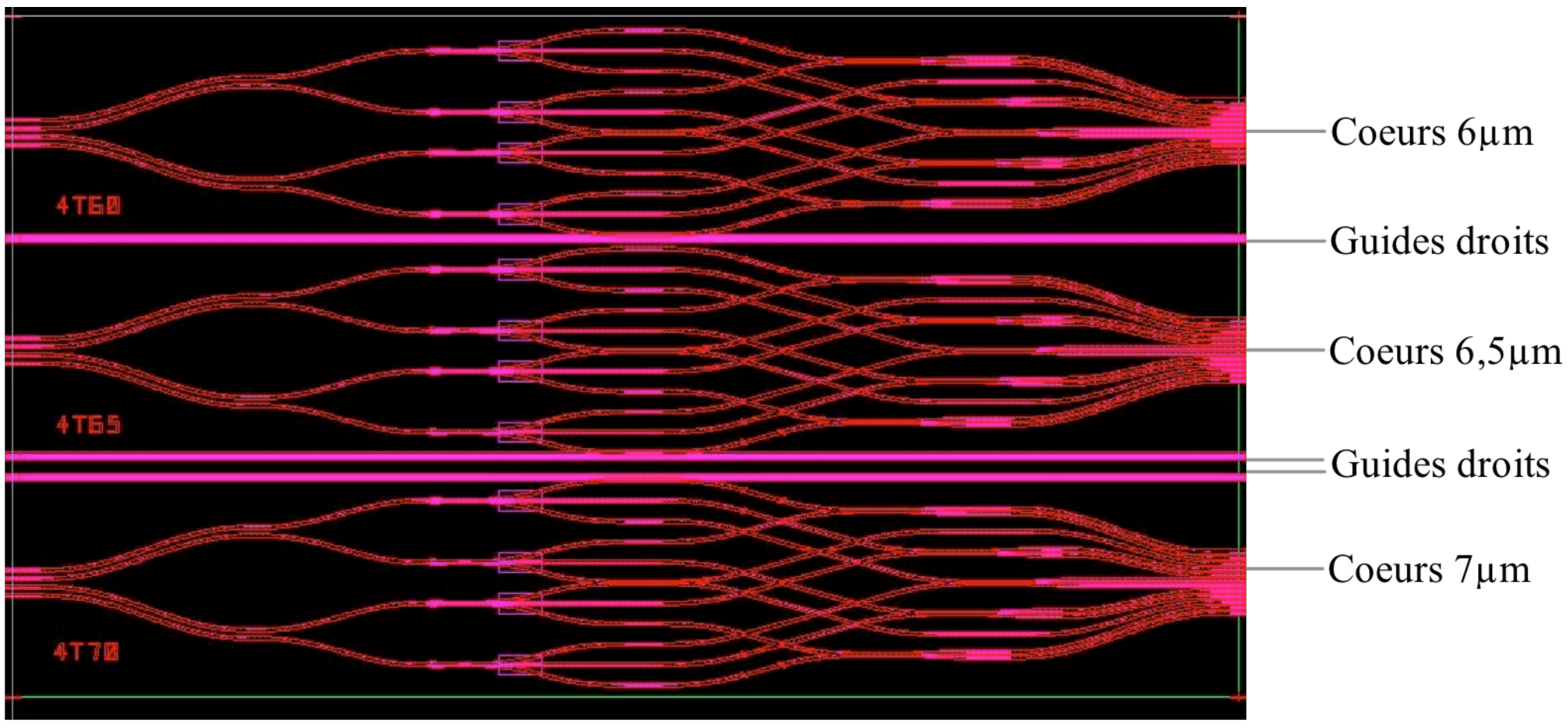

*Figure III-4 : puce 4T simple.*

## III – C – 2 – Recombineurs à deux télescopes ABCD

Les « puces 2T-ABCD » sont représentées sur la figure III-5. Chaque dispositif comporte deux entrées. Chaque entrée comporte une première jonction Y, l'une des voies servant à la mesure de la quantité de lumière injectée dans le guide, l'autre voie servant à la recombinaison interférométrique. Sur les voies interférométriques, les chemins comportent donc deux jonctions Y et un croisement de guide avant le coupleur asymétrique de recombinaison. Les dispositifs ont été légèrement allongés à une longueur de 36 mm de long pour obtenir des puces de la même taille que les « puces 4T » et ainsi faciliter la découpe.

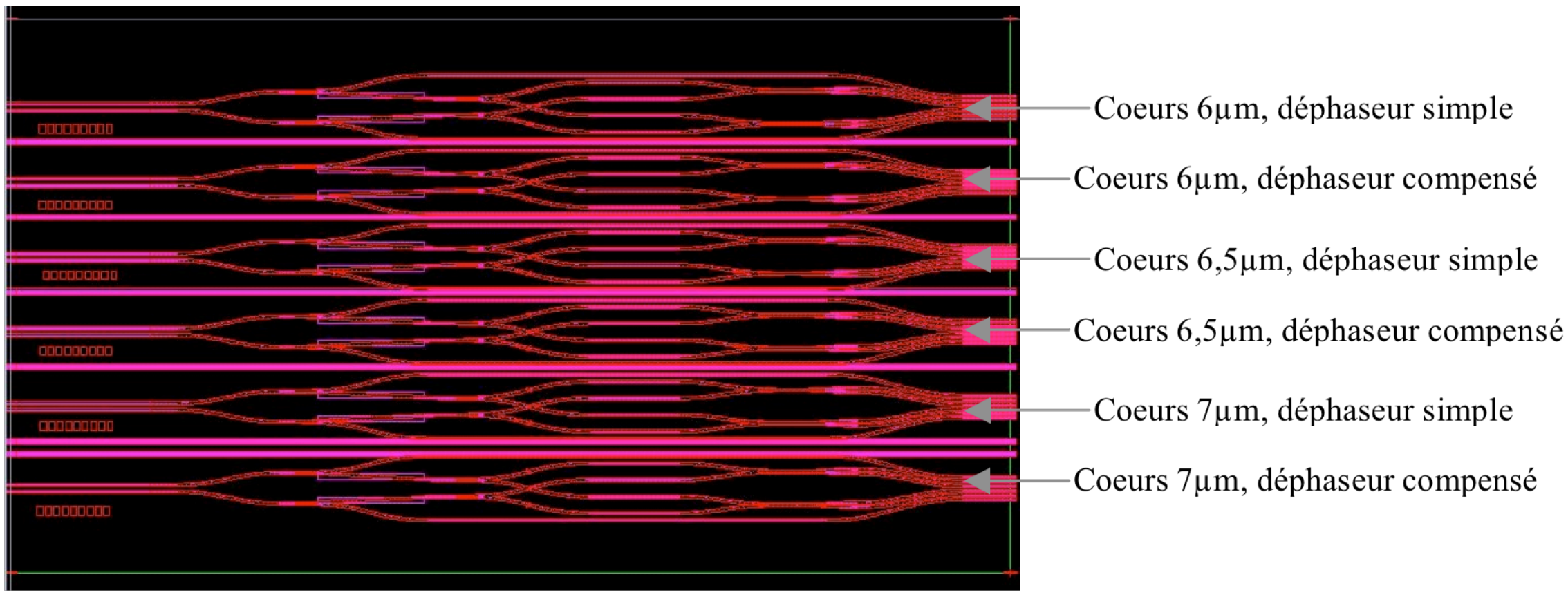

*Figure III – 5 : puce 2T-ABCD.*

Pour chaque dispositif, deux types de déphaseurs ont été réalisés : un déphaseur simple comportant une seule section de guide élargi et un déphaseur compensé comportant deux sections de guides élargis afin d'être achromatique.

## III – C – 3 – Recombineurs à quatre télescopes et cellule ABCD

Enfin, la « puce 4T-ABCD » est représentée sur la figure III-6.

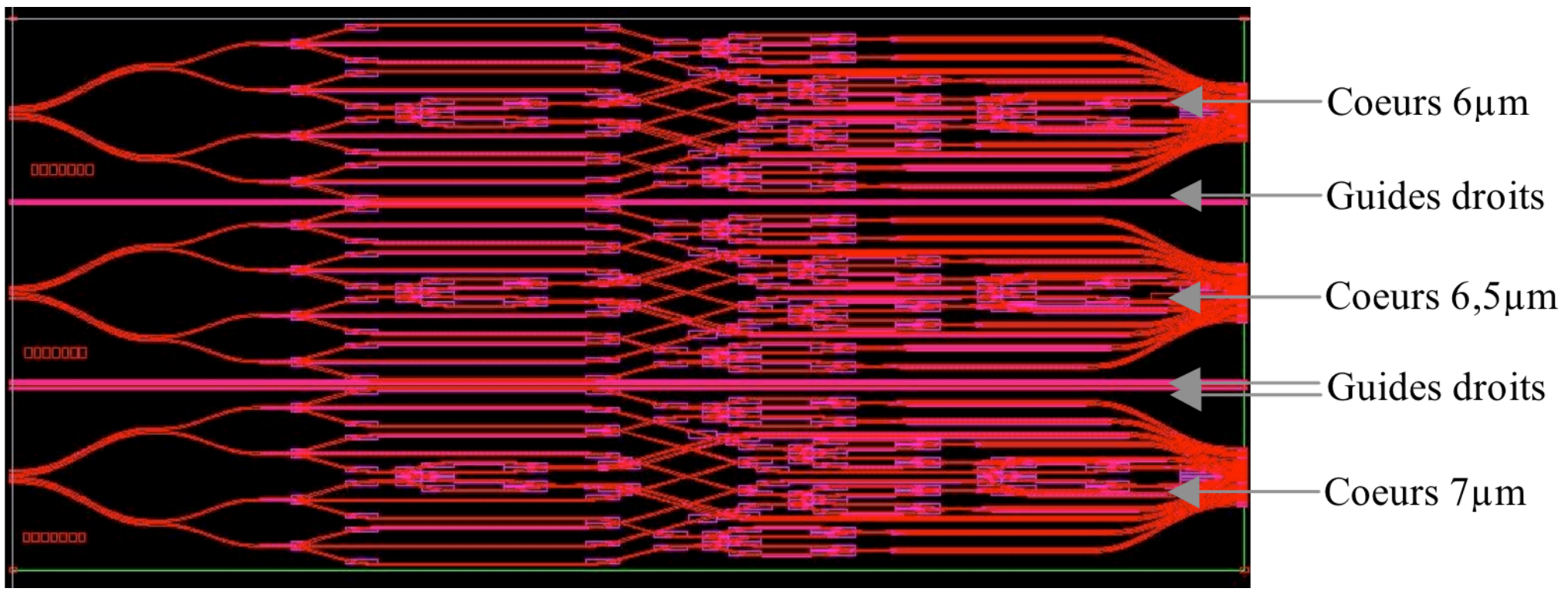

*Figure III-6 : puce 4T-ABCD.*

Le principe est le même que pour la « puce 4T » mais chaque recombinaison par paire a simplement été remplacée par une cellule ABCD. La puce est réalisée en trois largeurs de guides et comporte des guides droits afin d'évaluer la perte globale des fonctions optiques. Chaque chemin optique comporte un tricoupleur, quatre ou cinq croisements de guides suivant les cas, une jonction Y et un coupleur asymétrique. La longueur totale de la puce est de 80mm.

# III – D - Mesures photométriques fibrées

## III – D – 1 – Dispositif expérimental utilisé

Dans un premier temps, nous avons effectué des mesures photométriques fibre à fibre soit en utilisant une source monochromatique à 1,55µm, soit en utilisant une source blanche et un spectromètre fibré en sortie. Le schéma du montage utilisé est représenté sur la figure III-7.

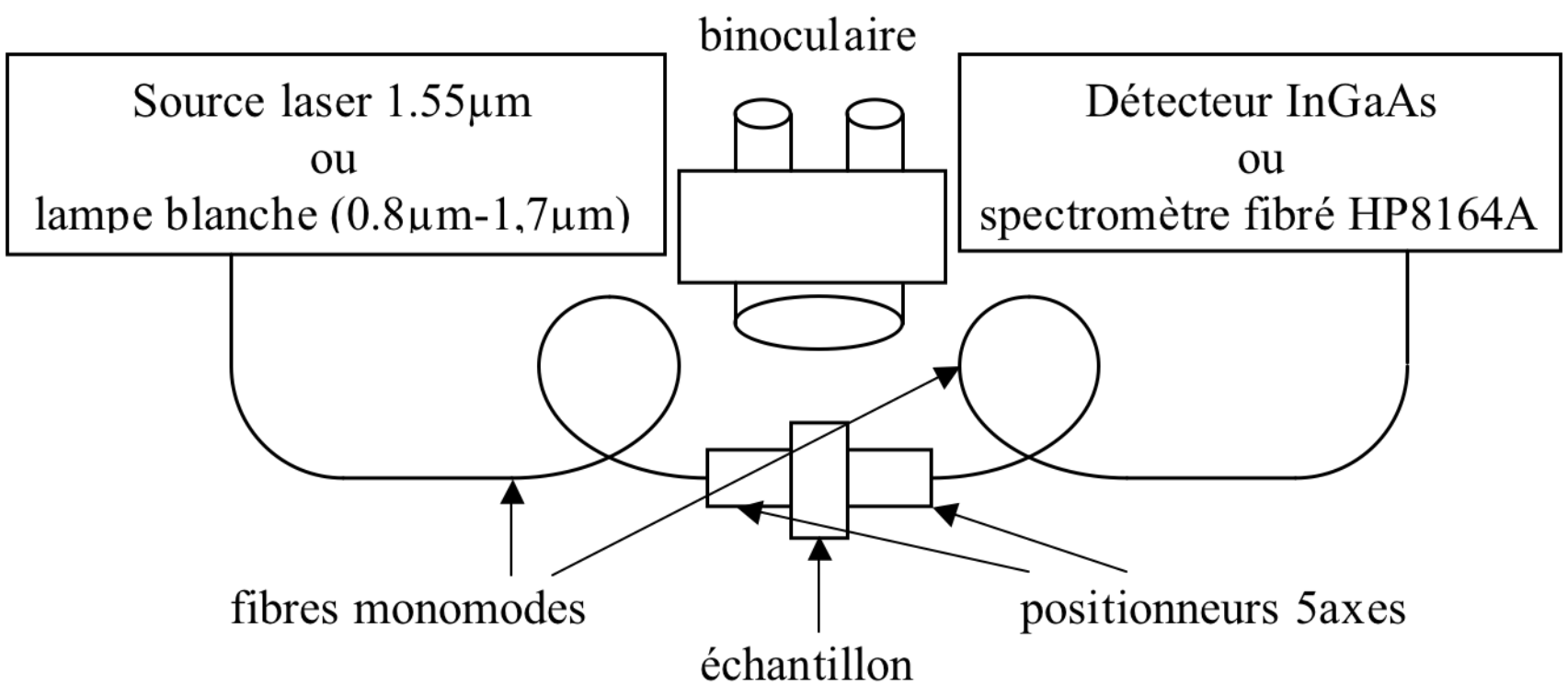

*Figure III-7 : banc de mesure photométrique*

Pour chaque mesure, nous effectuons une mesure de référence fibre à fibre sans puce, et une mesure photométrique avec puce. Dans tous les cas, nous optimisons le couplage en utilisant du gel d'indice entre les fibres ou entre les fibres et le composant.

## III – D – 2 – Pertes de propagation et transmission des puces

Les pertes de propagation ont été évaluées suivant la méthode exposée au paragraphe III–B- 4. Pour cela, on a utilisé les guides droits des « puces 4T » et des « puces 4T-ABCD » qui permettent une mesure de transmission pour des longueurs de 36mm et 80mm.

Les résultats de pertes obtenus à 1,55µm en fonction de la largeur du guide sont regroupés dans le tableau III-1.

| pertes de propagation | | | |
|---|---|---|---|
| largeur du cœur [µm] | 6 | 6,5 | 7 |
| pertes linéiques [dB/cm] | 0,085±0,012 | 0,022±0,012 | 0,015±0,012 |

*Tableau III-1: pertes de propagation mesurées sur guides droits de différentes longueurs.*

Les valeurs sont obtenues en prenant les moyennes des transmissions de tous les guides dont nous disposions (dix guides de chaque longueur et de chaque largeur de coeur). Les pertes des guides de 7µm sont en moyenne de 1,5 dB/m ± 1,2dB/m ce qui est un résultat meilleur que ceux obtenus précédemment. Cependant, on observe une remontée des pertes pour les guides de 6µm à 8,5dB/m ±1,5dB/m.

Les transmissions globales des puces 4T ont aussi été mesurées par la même méthode. Les moyennes des résultats sont reportés dans le tableau III-2.

Transmission des puces 4T et 4TABCD

| puce | 4T 6µm | 4T 6,5µm | 4T 7µm | 4TABCD 7µm |
|---|---|---|---|---|
| transmission | 68±2% | 72±2% | 70±2% | 61±2% |
| pertes d'insertion | 1,65 dB | 1,43 dB | 1,55 dB | 2,15dB |
| pertes des fonctions optiques | 0,65 dB | 0,4 dB | 0,5 dB | 0,95 dB |

*Tableau III-2 : transmission globale des dispositifs fibre à fibre à 1,55µm. Les pertes d'insertion sont les pertes globales introduites par « l'insertion » du dispositif entre les fibres.*

La transmission globale des puces 4T simples est d'environ 70%, soit des pertes d'insertion fibre à fibre de 1,55dB. La transmission relative de ces puces par rapport à celle d'un guide droit de même longueur est d'environ 90%. Les pertes en excès sont donc de l'ordre de 0,5dB pour l'ensemble des guides courbes, des croisements de guide, du coupleur asymétrique et du tricoupleur. Ces valeurs sont tout à fait similaires aux résultats obtenus sur les dispositifs réalisés précédemment en technologie silice sur silicium[*64*].

La transmission globale des 4T-ABCD est d'environ 61% soit des pertes d'insertion fibre à fibre de 2,15dB. La transmission par rapport à celle d'un guide droit de même longueur est cette fois de 80%. Les pertes en excès des fonctions optiques sont donc cette fois d'environ 1dB, le chemin optique parcouru comportant des guides courbes supplémentaires, deux croisements de guides supplémentaires et surtout une jonction Y.

En première approximation et en tenant compte de résultats déjà obtenus sur des jonctions Y précédemment réalisées, nous avons évalué les pertes des croisements de guides à environ 0,05dB et les pertes de la jonction Y à 0,3dB.

Ces valeurs auraient du être affinées par les mesures des puces test. Malheureusement, ces puces de 1cm de long présentaient de la lumière parasite autour des structures de mesure qui nous ont empêché d'obtenir des résultats exploitables. En effet, même si les guides d'onde en silice sur silicium sont bien adaptés aux fibres optiques, le couplage de l'un à l'autre comporte des pertes de transition de l'ordre de 0,2dB en théorie et d'environ 0,25dB dans la pratique suivant la qualité des flancs des puces, ceci même lorsque l'on utilise du liquide d'indice pour optimiser le couplage. Cette lumière parasite qui n'est pas couplée au mode fondamental du guide d'onde reste cependant piégée dans la structure globale silice sur silicium qui agit comme un guide d'onde planaire à faibles pertes. Le graphe de la figure III-8 montre l'exemple d'un des modes de propagation dans une couche de 25µm de silice sur silicium.

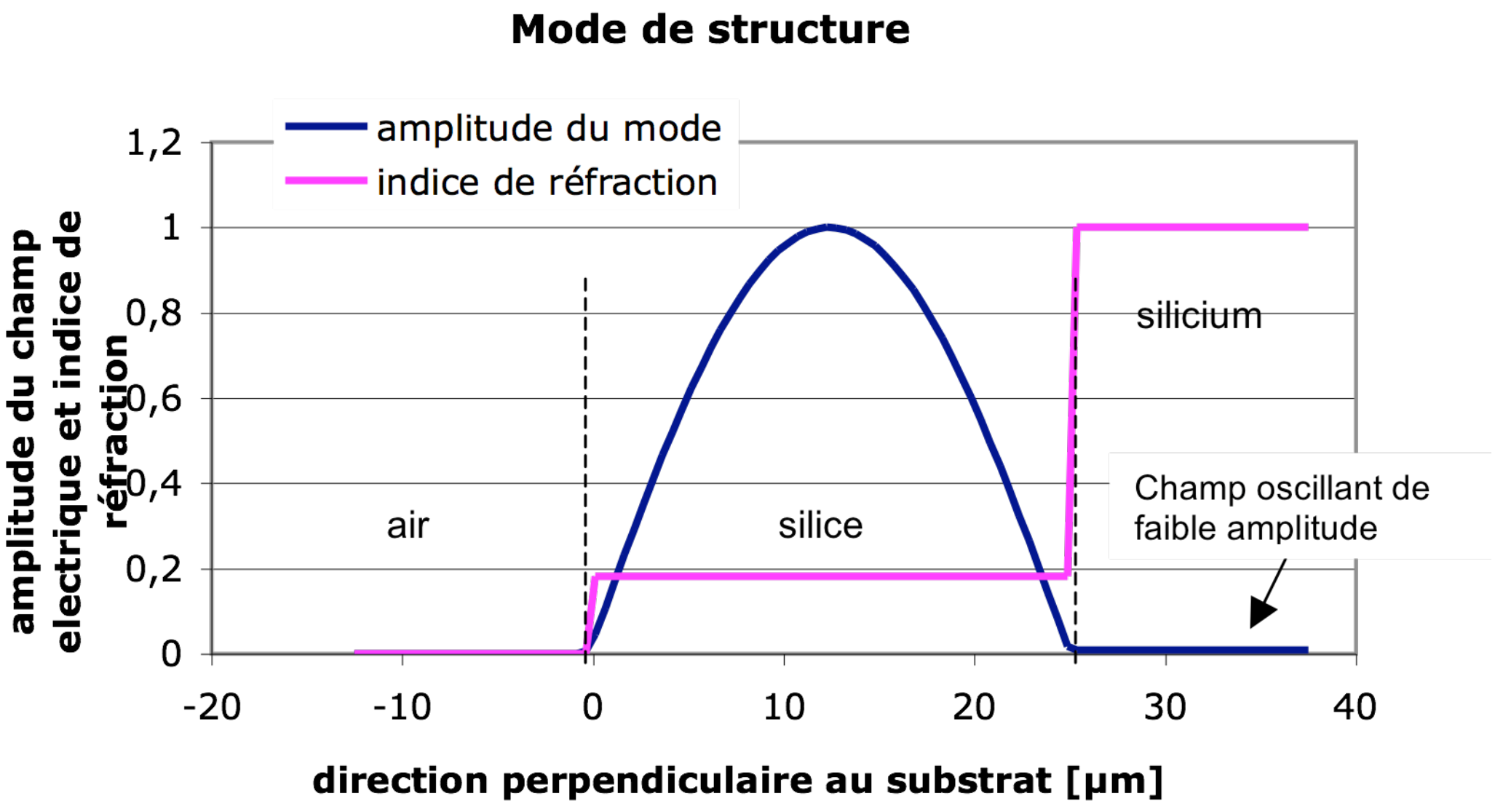

*Figure III-8 : mode de structure dans une couche de 25µm de silice sur silicium. L'indice est normalisé à 1 pour la représentation graphique.*

Les pertes théoriques d'un tel mode sont de 0,7dB/cm. Il existe dans la structure un grand nombre de ce type de mode avec des pertes de propagation croissantes. La lumière parasite générée à l'entrée de la puce (de l'ordre de quelques pourcents) est couplée à ces modes et vient perturber la mesure à la sortie de la puce. Ce phénomène est beaucoup moins important pour les puces recombineurs qui sont beaucoup plus longues.

## III – D – 3 – Validation du tricoupleur et du coupleur asymétrique

Les mesures photométriques présentées au paragraphe précédent ont permis de déterminer les transmissions des coupleurs asymétriques ainsi que celles des tricoupleurs.

Pour les coupleurs asymétriques, on procède comme suit. Si $M_1$ et $M_2$ sont les deux mesures photométriques correspondant aux deux voies de sortie d'un coupleur, les transmissions $T_1$ et $T_2$ sont simplement déterminées par les ratios :

$$T_1 = \frac{M_1}{M_1 + M_2} \qquad \text{(III-4)}$$

$$T_2 = \frac{M_2}{M_1 + M_2} \qquad \text{(III-5)}$$

Les mesures des tricoupleurs sont obtenues par une méthode similaire. Dans le cas d'une puce 4T-simple, chaque tricoupleur alimente trois coupleurs. Si $S_1$, $S_2$, et $S_3$ représentent la somme des deux sorties de chacun des trois coupleurs, les transmissions $T_1$, $T_2$, et $T_3$ des tricoupleurs seront données par :

$$T_1 = \frac{S_1}{S_1 + S_2 + S_3} \qquad \text{(III-6)}$$

$$T_2 = \frac{S_2}{S_1 + S_2 + S_3} \qquad \text{(III-7)}$$

$$T_3 = \frac{S_3}{S_1 + S_2 + S_3} \qquad \text{(III-8)}$$

Dans le cas des puces 4T-ABCD, on somme les quatre sorties de chaque cellule au lieu de sommer les deux sorties de chaque coupleur.

Les moyennes des résultats sont reportées dans le tableau III-3.

Taux de couplage des coupleurs asymétriques et des tricoupleurs

| puce | théorie | 4T 6µm | 4T 6,5µm | 4T 7µm | 4TABCD 7µm |
|---|---|---|---|---|---|
| coupleur asymétrique (%) | 52/48 | 83/17 | 69/31 | 60/40 | 58/42 |
| tricoupleur (%) | 33/33/33 | 15,5/69/15,5 | 24/52/24 | 32,5/35/32,5 | 33/34/33 |
| Précision (%) | | | 1 | | |

*Tableau III-3 : taux de couplage des coupleurs asymétriques et du tricoupleur à 1,55µm.*

Deux points apparaissent dans ces résultats. Premièrement, il semble que la perte de côte des guides est d'environ 1µm. Ainsi, les dispositifs dessinés avec des guides de 7µm sont les plus proches des valeurs théoriques. Les tricoupleurs dessinés à 7µm sont notamment très proches de la valeur théorique de 33,3/33,3/33,3 à 1,55µm. Les coupleurs asymétriques fournissent la même tendance : les dispositifs dessinés avec les guides de 7µm sont les plus proches des valeurs théoriques. Cependant, ces derniers présentent un écart non négligeable par rapport à la valeur théorique de 52/48 à 1,55µm.

Afin de confirmer ces résultats, nous avons effectué des mesures spectrales de ces mêmes puces.

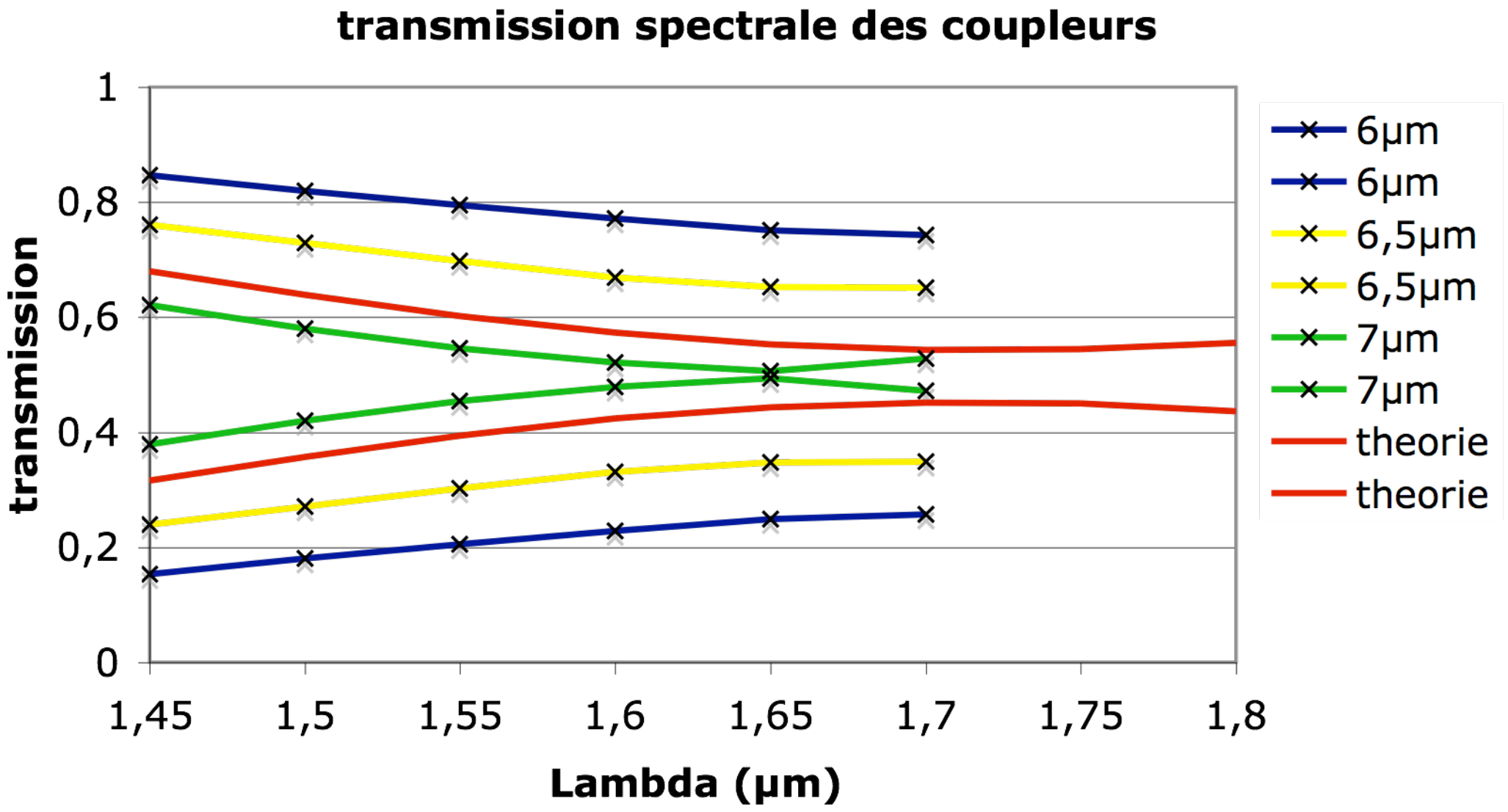

*Figure III – 9 : transmission spectrale des coupleurs asymétriques. Les guides de 7µm de large fournissent les résultats les plus proches de la théorie.*

En vérifiant le dessin du masque de photolithographie, nous avons découvert une erreur dans le dessin du coupleur asymétrique : la base de données des fonctions élémentaires n'avait pas été remise correctement à jour, et le coupleur utilisé était issu d'un ancien « design » effectué en 2D par la méthode de l'indice effectif. Nous avons simulé ce coupleur en 3D et comparé avec les résultats de mesure. Il apparaît que les coupleurs avec des guides de 7µm sont très proches des coupleurs simulés. Les courbes en rouge de la figure III-9 prennent en compte cette erreur de dessin.

Enfin, le graphe de la figure III-10 présente la transmission spectrale des tricoupleurs.

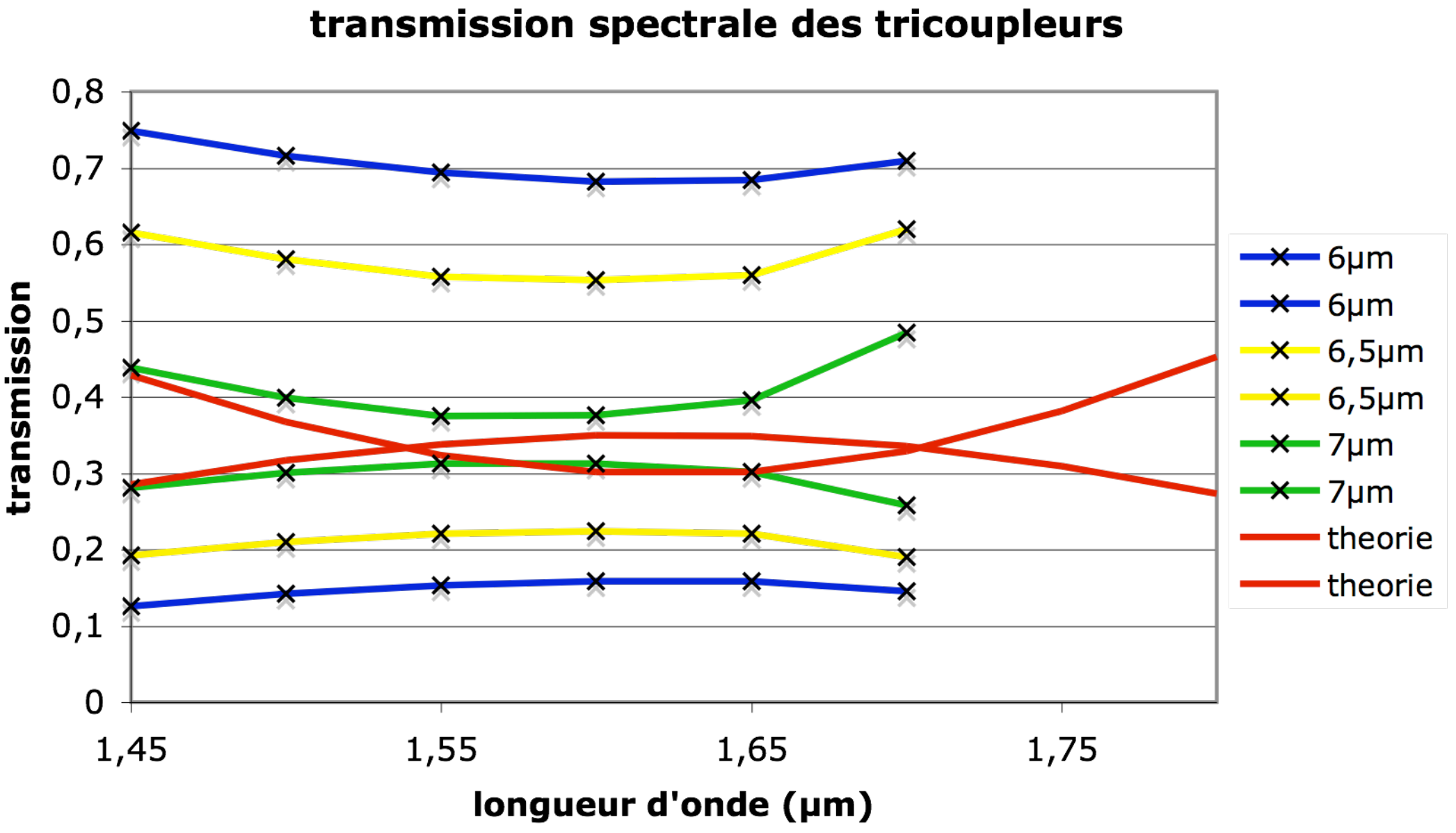

*Figure III-10 : réponse spectrale des tricoupleurs pour différentes largeurs de guides. Les guides de 7µm sont les plus proches des valeurs simulées.*

On retrouve bien que ce sont les guides d'onde ayant un cœur de 7µm de large qui fournissent des tricoupleurs les plus proches des valeurs recherchées. Les valeurs à 1700nm sont en bord de spectre de transmission de l'instrumentation et correspondent à un mauvais rapport signal sur bruit.

Enfin, nous avons cherché à vérifier la diaphonie des croisements de guide, c'est à dire le taux de lumière couplée d'un guide d'onde sur les guides d'onde le croisant. Nous n'avons pu que vérifier que le taux de lumière couplée était inférieur au minimum détectable ce qui nous donne un taux de diaphonie inférieur à -50dB.

# III – E - Mesures interférométriques

## III – E – 1 - Description du banc du LAOG

Dans le cadre de ces travaux, le LAOG (Laboratoire d'Astrophysique de Grenoble) a développé un banc interférométrique de simulation d'interféromètre stellaire [65]. Un schéma général de ce banc est représenté sur la figure III-11.

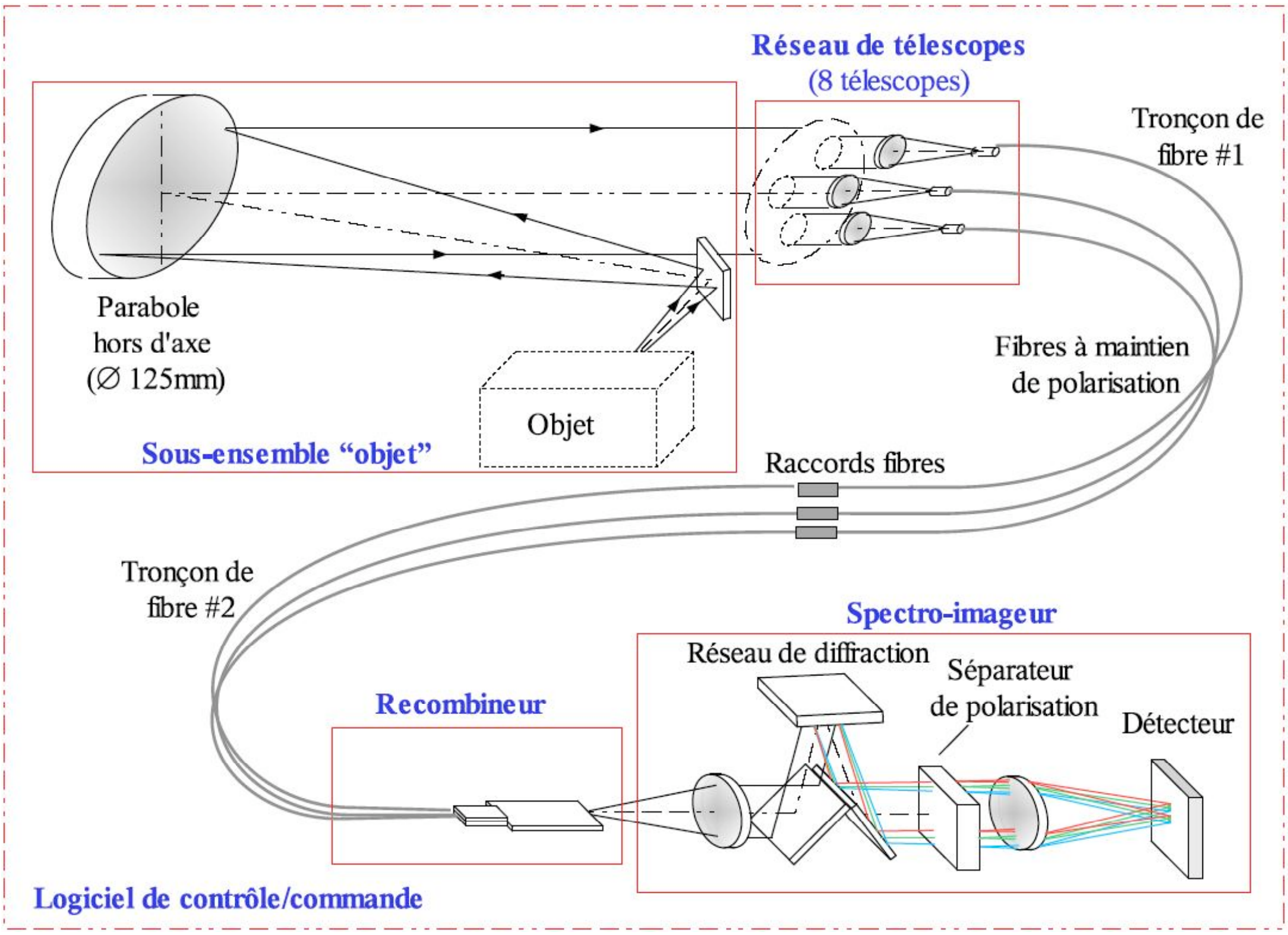

*Figure III-11 : schéma de l'interféromètre du LAOG d'après [65].*

La source stellaire est simulée par l'extrémité d'une fibre optique (ou deux dans le cas d'une simulation d'étoile binaire) placée au foyer d'un collimateur à miroir de 1m de focale. Des petits objectifs (doublets achromatiques) simulant les télescopes sont placés à la sortie dans la pupille du collimateur et sont montés sur des platines motorisées afin de pouvoir balayer la différence de marche optique. Au foyer de chacun de ces petits objectifs, la lumière est injectée dans une fibre optique à maintien de polarisation. Les entrées de ces fibres à maintien de polarisation sont réglées de manière à positionner les axes de biréfringence des fibres optiques dans le plan et perpendiculairement au plan du marbre optique. Ces fibres optiques sont ensuite rassemblées en une nappe de fibres au pas de 250µm connectée directement à l'entrée de la puce recombineur. Les axes de biréfringence des fibres sont réglés parallèlement et perpendiculairement au plan de la puce de recombinaison. En sortie de puce, la lumière est tout d'abord recollimatée, puis dispersée spectralement par un montage comportant deux miroirs et un réseau de diffraction, avant d'être séparée en polarisation à l'aide d'un prisme de Wollaston. Enfin, la lumière est refocalisée sur une caméra infrarouge reliée à un ordinateur servant à récupérer les données et à piloter les moteurs permettant de balayer le chemin optique sur chacun des bras.

La figure III-12 donne deux exemples d'images enregistrées par la caméra infrarouge. Le premier (a) donne les 24 sorties de la puce directement imagées sur la caméra. Le deuxième (b) donne ces mêmes 24 sorties dispersées en longueur d'onde puis séparées en polarisation. On observe dans ce cas, 48 petits spectres cannelés correspondant aux franges d'interférence.

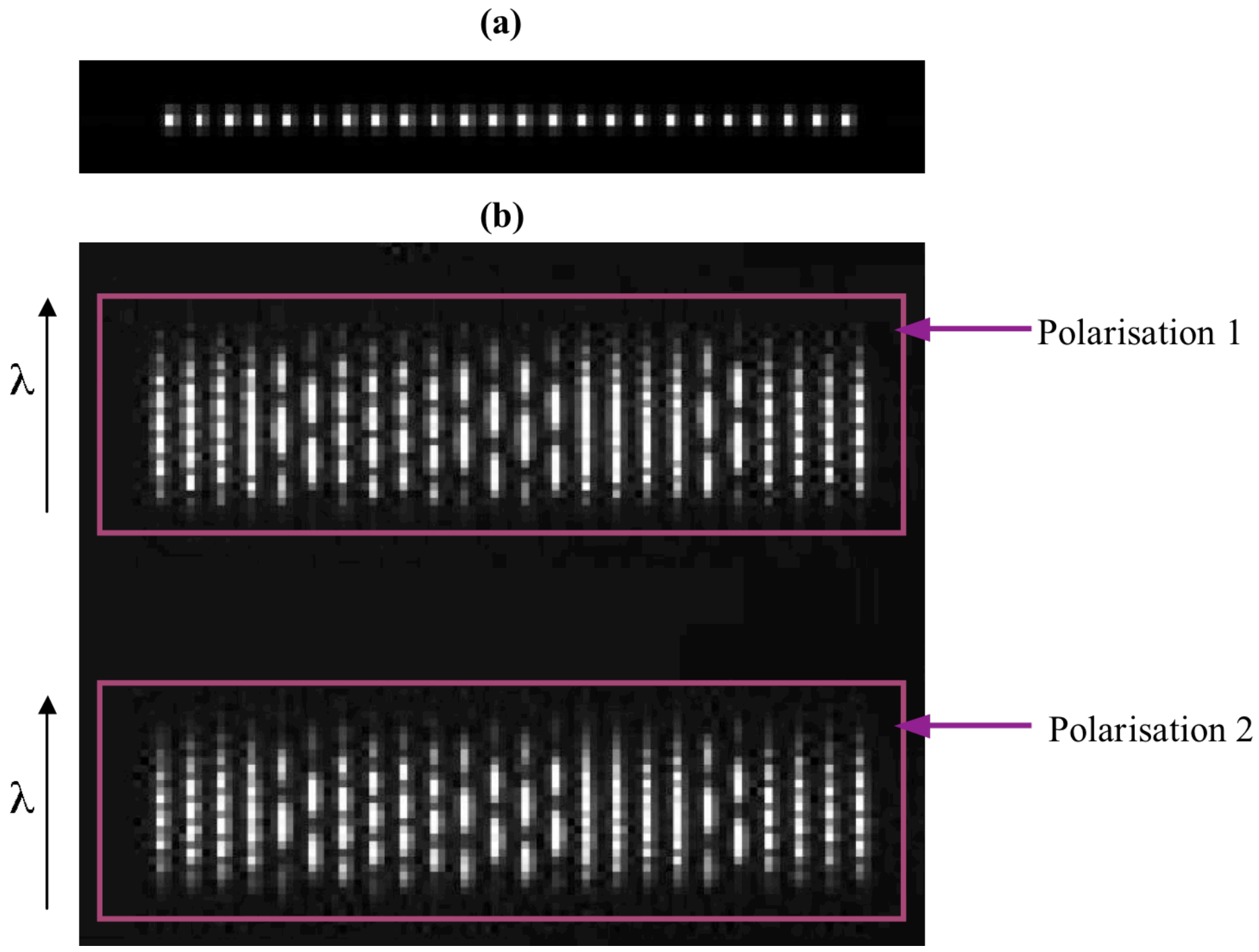

*Figure III-12 : image obtenue sur la caméra. (a) image non dispersée et non séparée en polarisation. On observe les 24 sorties de la puce. (b) image avec dispersion chromatique et séparation de polarisation. On observe 48 petits spectres cannelés.*

Une prise de mesures consiste à prendre successivement une image en masquant toutes les entrées afin d'avoir le niveau de noir, puis quatre prises photométriques en ne laissant passer qu'une voie différente à chaque fois. Ceci permet de calibrer l'ensemble du système d'un point de vue photométrique (notamment la transmission spectrale sur chaque voie). Enfin, on effectue un balayage des franges autour de la teinte plate en pilotant trois des quatre moteurs à des vitesses différentes.

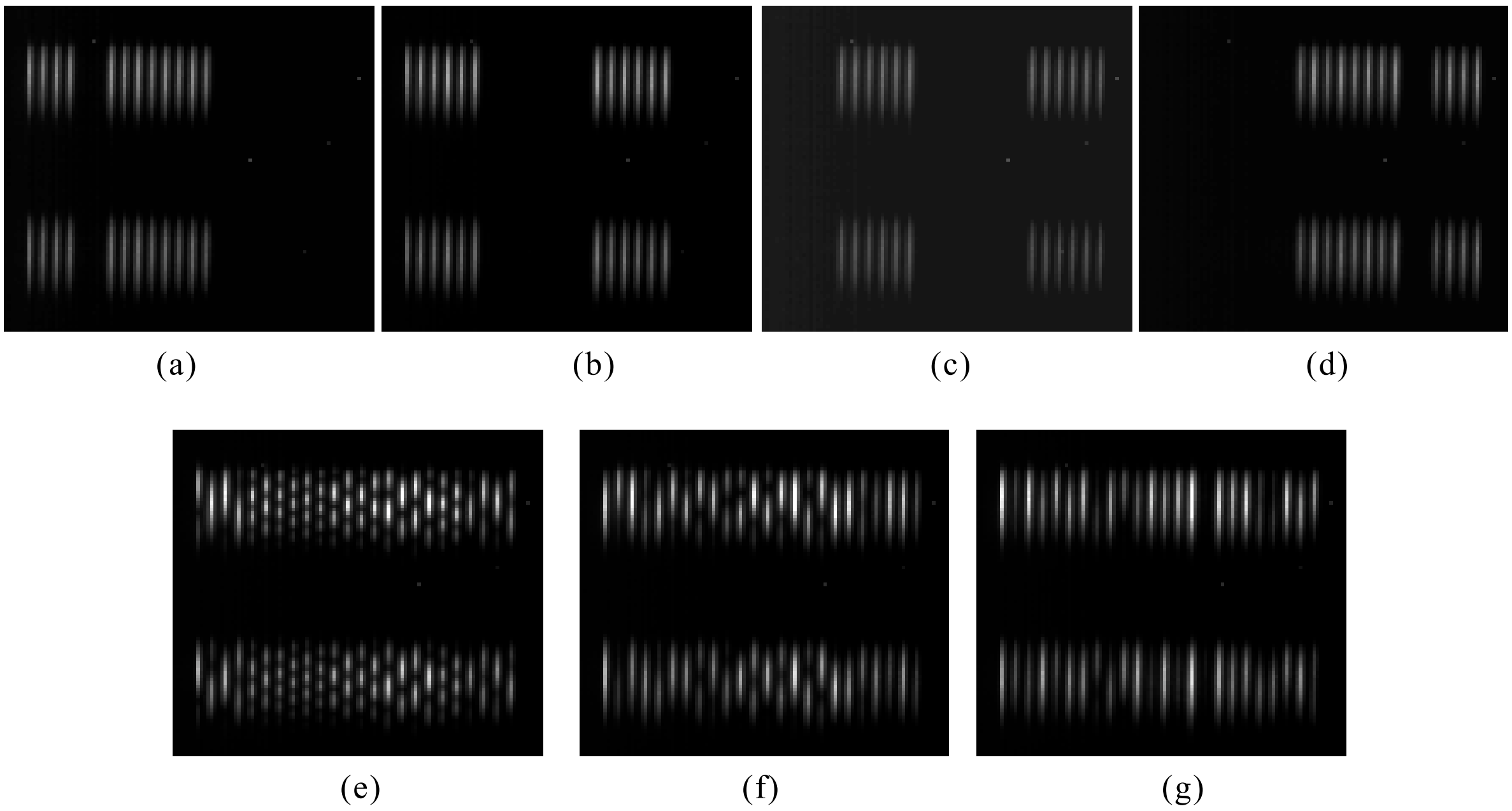

*Figure III-13 : images constituant une mesure complète de franges. (a)-(d) : signaux photométriques pour chacune des entrées. (e)-(f) : exemples d'images de franges, (e) loin de la teinte plate, (f) proche de la teinte plate, (g) quasiment la teinte plate.*

Le signal de chaque pixel est donc un interférogramme en fonction de la position des moteurs, c'est à dire de l'écart de chemin optique entre chaque paire de voies. Une mesure typique de franges consiste à faire quelques balayages de franges (un balayage fait typiquement 1024 images) pour vérifier la reproductibilité des mesures. Une semaine de mesure correspond en moyenne à plus de 30Go de données à traiter.

## III – E – 2 – Mesures photométriques spectrales

Les données issues d'une série de mesure contiennent toutes les informations nécessaires à la détermination des caractéristiques de la puce. En particulier, les signaux photométriques pour chacune des entrées permettent de remonter aux coefficients de transmission des coupleurs et tricoupleurs que nous pouvons alors comparer aux mesures que nous avons effectuées au LETI sur banc fibré. La calibration en longueur d'onde de chaque pixel de la caméra est obtenue à partir des mesures interférométriques et du pas des moteurs (nous avons observé une légère non linéarité des moteurs qui conduit à une erreur d'environ 10nm sur la détermination de la longueur d'onde). La figure III-14 montre les transmissions spectrales moyennes du coupleur asymétrique recalculées à partir des données obtenues sur ce banc.

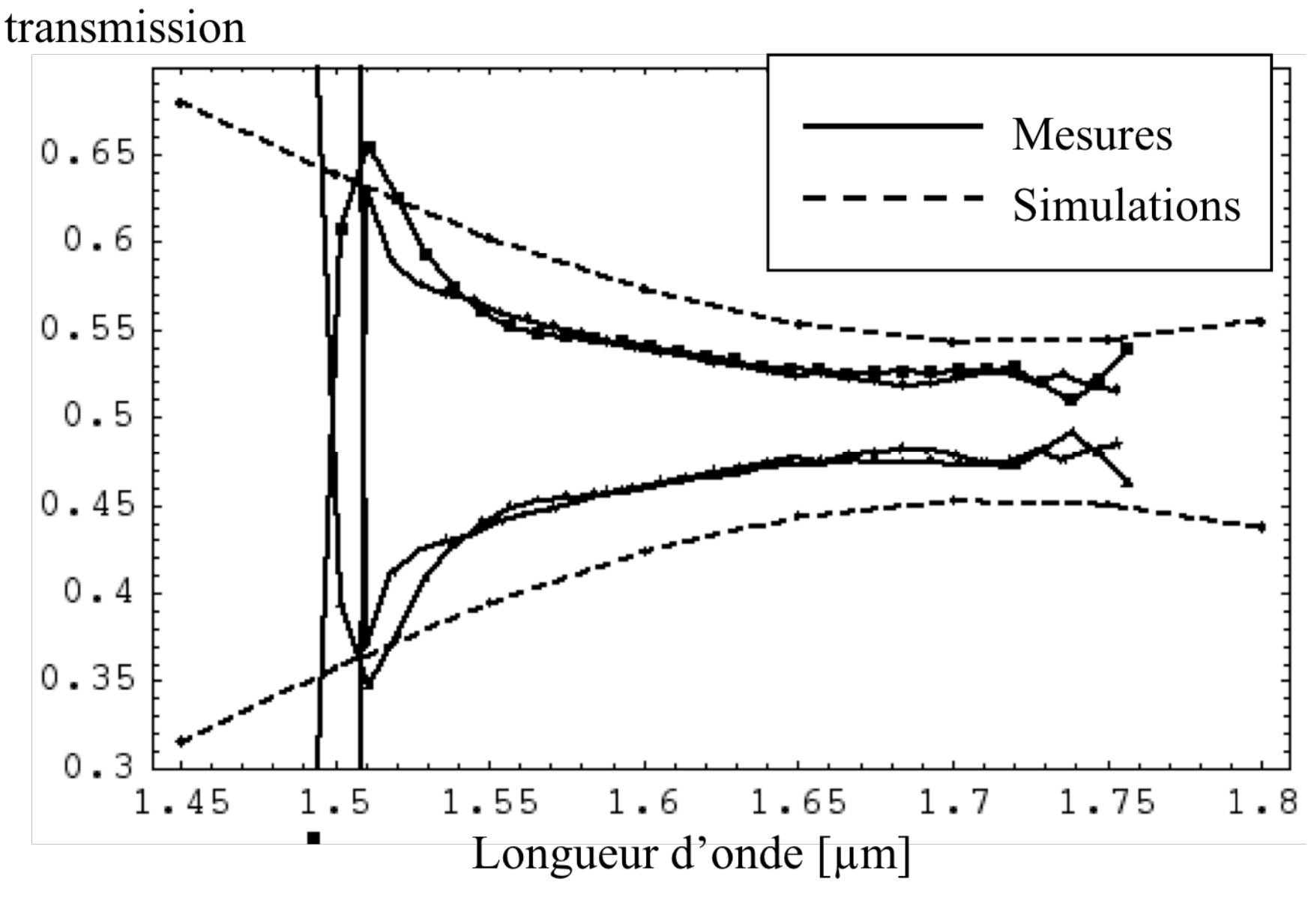

*Figure III-14 : transmission spectrale moyenne des coupleurs obtenues sur le banc du LAOG pour chaque état de polarisation.*

Les valeurs en bord de spectre correspondent à la limite d'émission de la diode superluminescente et ont un rapport signal sur bruit très faible. On observe que les transmissions sont quasiment identiques pour les deux états de polarisation et sont très proches des valeurs obtenues sur le banc fibré du LETI.

La figure III-15 montre les résultats obtenus pour le tricoupleur. Il s'agit de la moyenne des quatre tricoupleurs d'une des puces « 4T-ABCD »

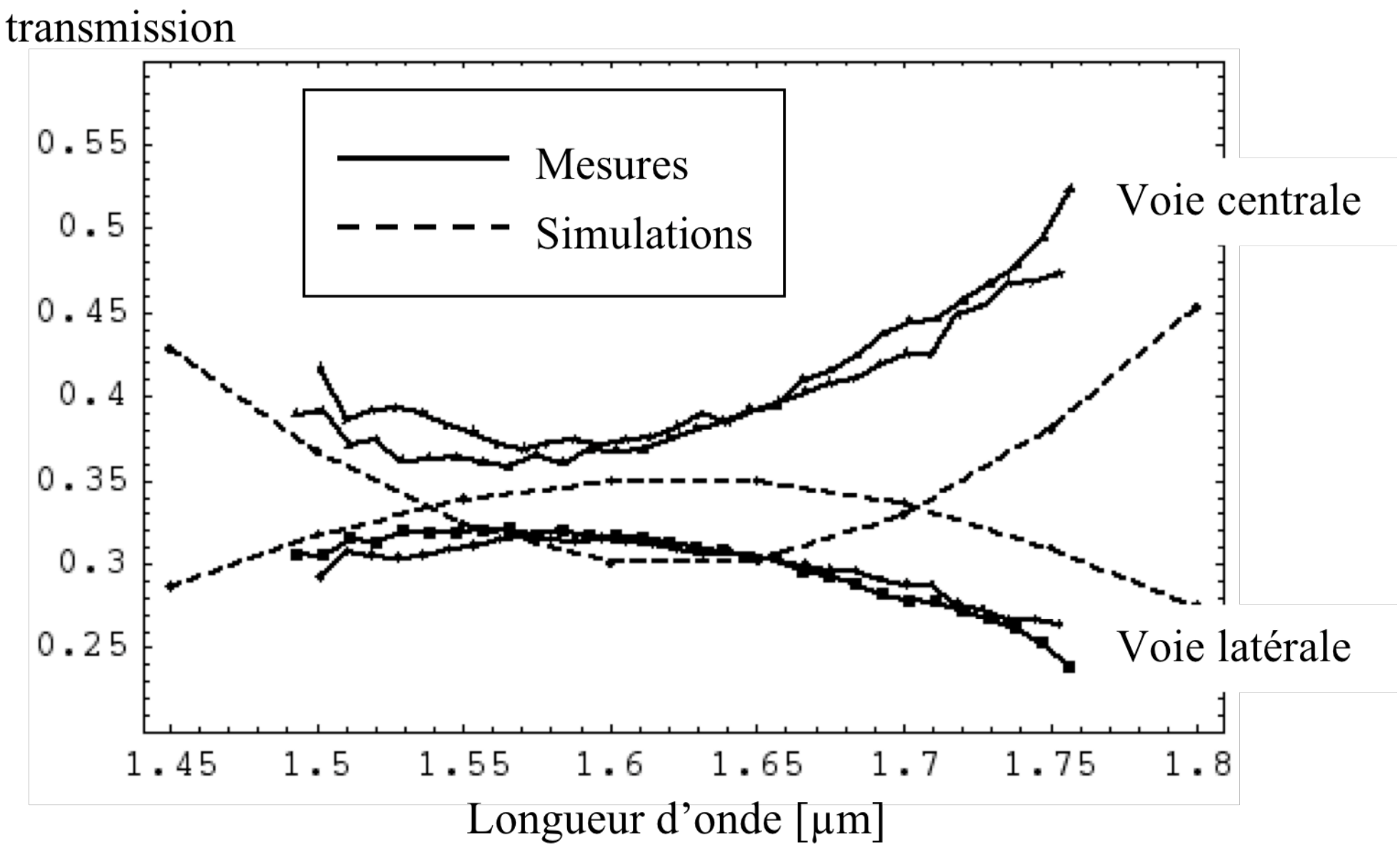

*Figure III-15 : transmission spectrale des tricoupleurs obtenues sur le banc du LAOG pour chaque état de polarisation.*

Là encore on n'observe aucun effet de polarisation. Les résultats montrent une transmission équilibrée entre les voies centrales et latérales, même si l'écart avec la théorie est cette fois légèrement plus important.

## III – E – 3 – Contraste instrumental

Des mesures de contraste instrumental ont été effectuées sur ces mêmes puces et sur le même banc par Myriam Benisty[*66*]. Le contraste instrumental $C$ est défini par :

$$C = \frac{I_{\max} - I_{\min}}{I_{\max} + I_{\min}} \qquad \text{(III-9)}$$

où $I_{min}$ et $I_{max}$ sont les valeurs minimales et maximales du signal interférométrique. C'est un paramètre qui varie entre 0 et 1 (ou entre 0 et 100%) et qui affecte directement la sensibilité globale de l'interféromètre au même titre que la transmission globale du dispositif.

A partir des données interférométriques, M. Benisty a obtenu des valeurs de contraste comprises entre 80 et 95% ±1%, ce qui se situe au meilleur niveau par rapport à l'état de l'art. La figure III-16 montre la variation du contraste obtenu en fonction de la longueur d'onde.

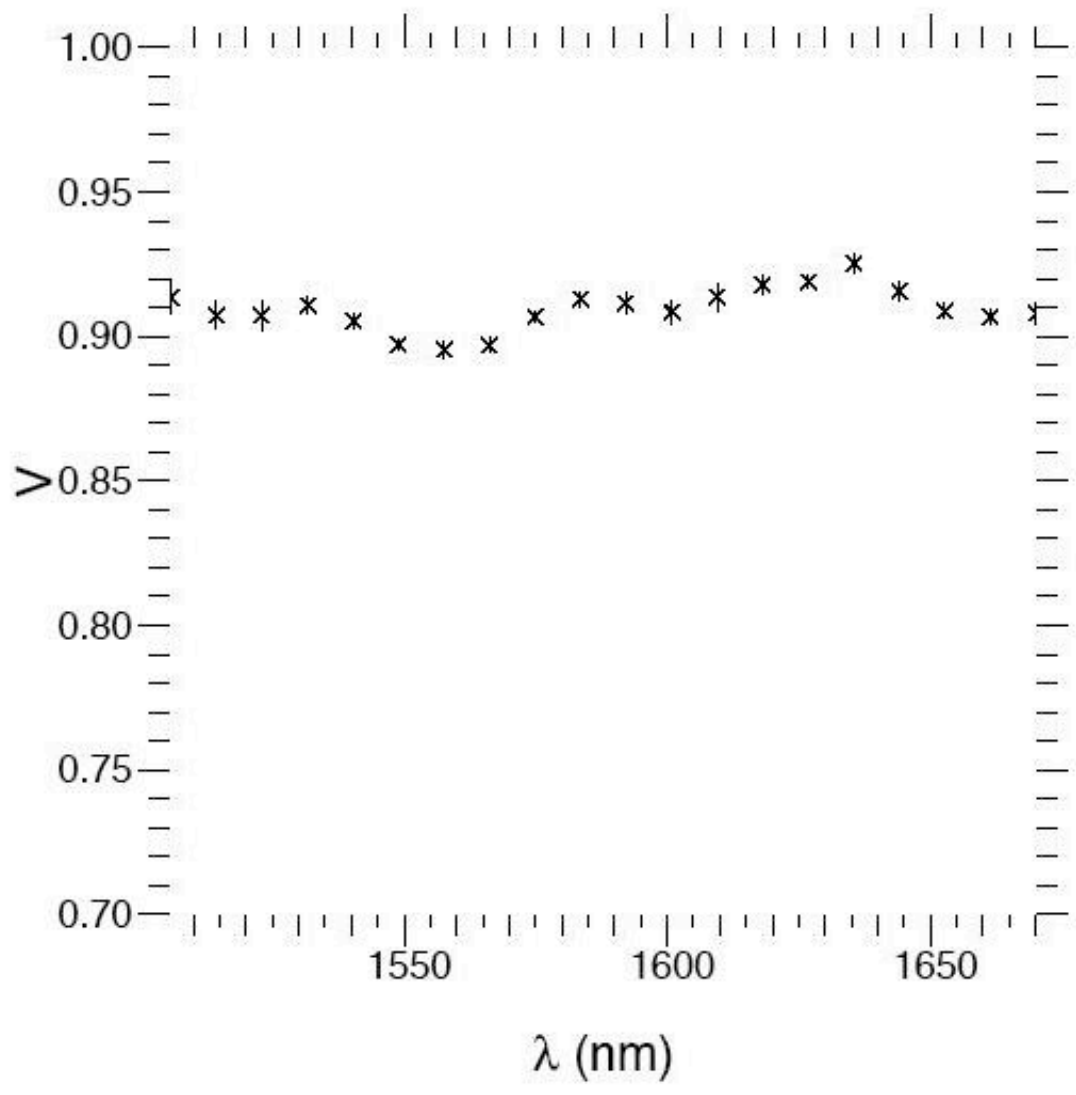

*Figure III-16 : variation du contraste (noté ici V) en fonction de la longueur d'onde. D'après [66].*

La variation du contraste en fonction de la longueur d'onde est aussi de l'ordre de 1%.

Enfin, ces mêmes mesures ont permis de montrer une stabilité de 1% à l'échelle de la journée, ce qui constitue un réel progrès par rapport aux interféromètres à base d'optique de volume où la stabilité instrumentale est plutôt mesurée sur des périodes de quelques minutes.

## III – E - 4 – Etude des déphaseurs

La campagne de mesures porte sur les déphaseurs de 6 puces 2T-ABCD comprenant chacune 6 dispositifs (3 déphaseurs simples et 3 déphaseurs achromatiques), ainsi que deux puces 4T-ABCD comprenant chacune 3 dispositifs, chaque dispositif 4T comprenant 6 déphaseurs achromatiques. Ainsi, un total de 72 déphaseurs a été mesuré.

### III – E – 4 – a – Méthode de mesure

Pour chaque cellule ABCD, nous mesurons, pour chaque polarisation et pour chaque longueur d'onde, 4 interférogrammes simultanément. On numérote les interférogrammes de chaque cellule selon le schéma de la figure III-17 :

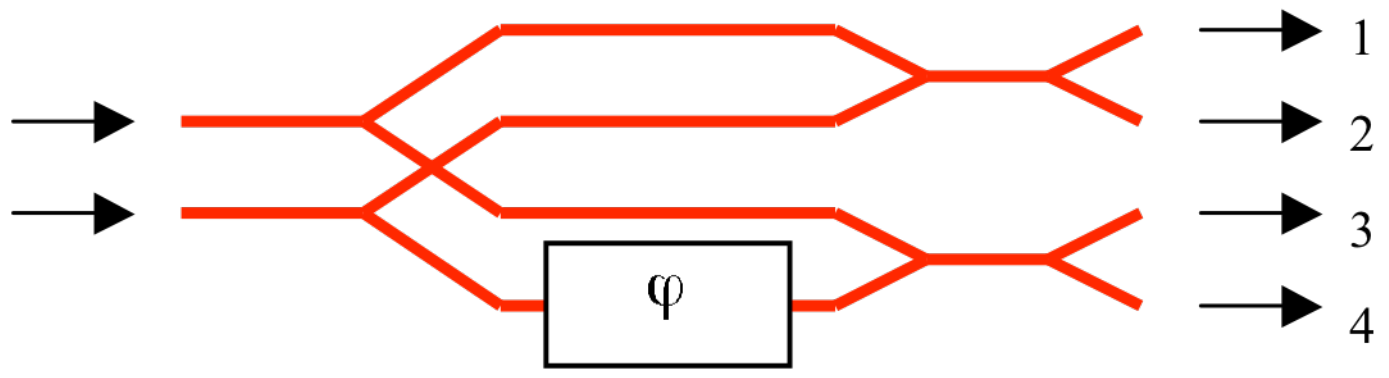

*Figure III-17 : numérotation des sorties d'une cellule ABCD.*

Par conservation d'énergie à la traversée d'un coupleur, les interférogrammes 1 et 2 sont en opposition de phase, ainsi que les interférogrammes 3 et 4 qui sont par ailleurs déphasés respectivement de $\varphi$ par rapport aux interférogrammes 1 et 2. Pour chaque série d'interférogrammes, nous avons donc le déphasage $\varphi$ par la relation :

$$\varphi = arctg\left[TF\left(S_1 - S_2\right)_{\nu\max} \cdot TF\left(S_3 - S_4\right)^*_{\nu\max}\right] \qquad \text{(III-10)}$$

où $S_i$ représente chaque interférogramme, et $\nu\ max$ est la fréquence du maximum de chaque spectre. Une attention particulière a été donnée au traitement (fenêtrage des données, filtrage, interpolation fine du spectre pour obtenir la fréquence avec la plus grande précision possible)

### III – E – 4 – b – Résultats des 2T-ABCD.

Les résultats en termes de phase sont reportés pour une polarisation sur les graphes de la figure III-18. Les résultats pour l'autre polarisation sont très similaires (voir tableau III-3). Les résultats de tous les déphaseurs sont présents afin de bien voir la dispersion des mesures pour chacune des puces (notées p1, p3, p5, p6, p8, p10).

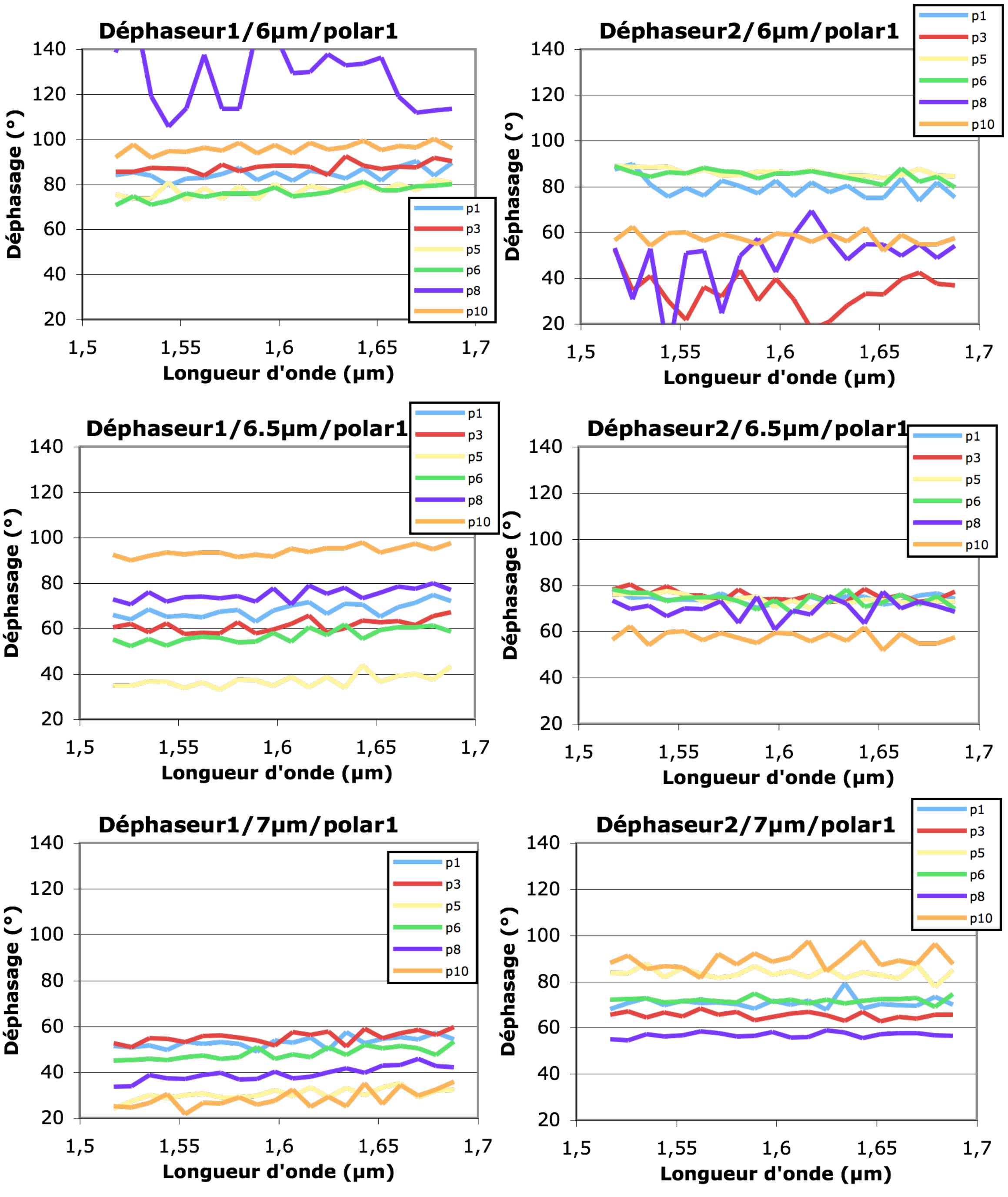

*Figure III-18 : résultats des déphaseurs des puces 2T-ABCD. La colonne de gauche présente les résultats pour le déphaseur simple. La colonne de droite présente les résultats pour le déphaseur achromatique.*

Sur 36 déphaseurs mesurés, seulement trois présentent de gros défauts tandis que les autres ont des valeurs dispersées selon leur type ou la largeur du guide. Les erreurs de mesure proviennent principalement de la vitesse de balayage des moteurs qui n'est pas parfaitement linéaire. Ainsi, la fréquence des franges mesurées n'est pas constante sur tout l'interférogramme et entraîne une imprécision sur la mesure de phase qui est de ±2°.

Ces résultats font tout d'abord apparaître que les déphaseurs achromatiques sont bien plus achromatiques que les déphaseurs simples, ceci quelle que soit la largeur du guide. Hormis quelques déphaseurs présentant manifestement un gros défaut, les valeurs sont assez regroupées dans une fourchette d'environ 20° autour d'une valeur dépendant de la largeur du guide. Enfin pour les guides d'onde avec un cœur de 7µm de large qui, nous l'avons vu précédemment, ont le meilleur comportement, les déphaseurs achromatiques sont regroupés autour d'une valeur d'environ 75° soit assez proche de la valeur théorique de 85°. Le comportement en longueur d'onde étant très plat, nous avons intégré ces valeurs en longueur d'onde et regroupé dans le tableau III-4, les valeurs moyennes et les écarts types pour chaque série de déphaseurs.

| | | dephaseur simple | | dephaseur compensé | |
|---|---|---|---|---|---|
| | | moyenne | ecart type | moyenne | ecart type |
| | | | | guides 6µm | |
| phase (°) | polar 1 | *84,4* | *7,2* | *83,6* | *2,9* |
| dispersion (°/nm) | polar 1 | *0,029* | *0,009* | *-0,029* | *0,007* |
| phase (°) | polar 2 | *91,1* | *7,3* | *85,4* | *2,2* |
| dispersion (°/nm) | polar 2 | *0,018* | *0,007* | *-0,027* | *0,002* |
| ecart en polarisation (°) | | *6,8* | *0,5* | *1,9* | *0,9* |
| | | | | guides 6,5µm | |
| phase (°) | polar 1 | *65,3* | *17,3* | *70,8* | *6,2* |
| dispersion (°/nm) | polar 1 | *0,035* | *0,004* | *-0,013* | *0,010* |
| phase (°) | polar 2 | *71,0* | *15,1* | *74,5* | *6,9* |
| dispersion (°/nm) | polar 2 | *0,039* | *0,010* | *-0,004* | *0,017* |
| ecart en polarisation (°) | | *5,6* | *2,6* | *3,7* | *1,5* |
| | | | | guides 7µm | |
| phase (°) | polar 1 | *42,4* | *10,5* | ***73,0*** | *10,9* |
| dispersion (°/nm) | polar 1 | *0,037* | *0,009* | ***0,003*** | *0,012* |
| phase (°) | polar 2 | *47,9* | *11,3* | ***76,9*** | *11,2* |
| dispersion (°/nm) | polar 2 | *0,030* | *0,010* | ***-0,006*** | *0,016* |
| ecart en polarisation (°) | | *5,5* | *0,9* | ***3,9*** | *0,8* |

*Tableau III-4 : valeur moyenne et écart type pour chacune des séries de déphaseurs. Les trois déphaseurs présentant un gros défaut ont été retirés des statistiques.*

La phase est la phase moyenne intégrée en longueur d'onde pour chaque série. La dispersion est la pente moyenne en fonction de la longueur d'onde. L'écart en polarisation est la différence des phases moyennes obtenues pour chacune des deux polarisations.

Les déphaseurs simples avec des guides de 7µm sont ceux qui s'éloignent le plus de la valeur théorique. En ce qui concerne les pentes, on retrouve bien que les déphaseurs achromatiques ont un comportement très plat en longueur d'onde, l'écart type sur la dispersion étant supérieur à la valeur moyenne. Enfin, la différence de phase entre les deux polarisations est toujours du même signe et de quelques degrés pour tous les déphaseurs, nous verrons que cela est différent pour les puces 4T-ABCD.

### *III – E – 4 – c – Résultats des 4T-ABCD.*

Nous ne disposions que de 2 puces 4T-ABCD, chacune comportant 3 dispositifs, mais chaque dispositif présente 6 recombinaisons donc 6 déphaseurs, donc nous avons aussi un total de 36 déphaseurs. Les résultats sont quasiment identiques à ceux des 2T-ABCD à l'exception de six déphaseurs tous situés au même endroit de la puce représentée sur la figure III-19.

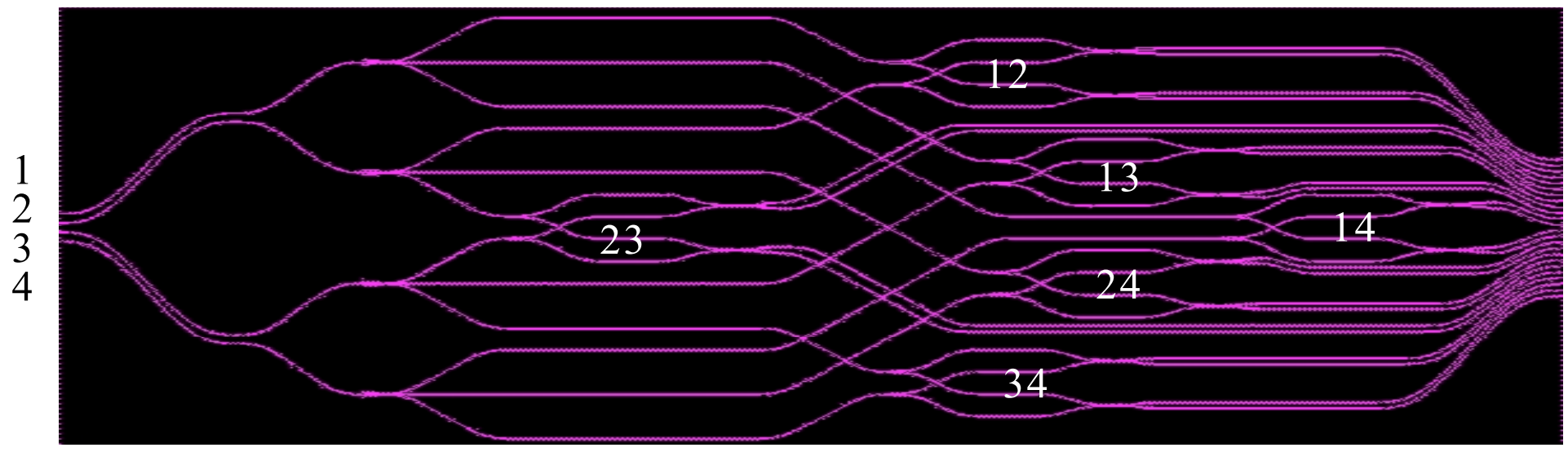

*Figure III-19 : numérotation des déphaseurs sur le 4T-ABCD*

Les résultats moyens sont reportés dans le tableau III-5.

| | | tous sauf sortie 14 | | sortie 14 | |
|---|---|---|---|---|---|
| | | moyenne | ecart type | moyenne | ecart type |
| | | guides 6µm | | | |
| phase (°) | polar 1 | *84,8* | *10,4* | *56,2* | *10,4* |
| dispersion (°/nm) | polar 1 | *0,020* | *0,006* | - | - |
| phase (°) | polar 2 | *91,7* | *11,8* | *55,8* | *10,9* |
| dispersion (°/nm) | polar 2 | *0,012* | *0,008* | - | - |
| ecart en polarisation (°) | | *6,9* | *1,7* | ***-0,5*** | *0,4* |
| | | guides 6,5µm | | | |
| phase (°) | polar 1 | *68,2* | *8,2* | *47,6* | *8,5* |
| dispersion (°/nm) | polar 1 | *0,029* | *0,007* | - | - |
| phase (°) | polar 2 | *75,0* | *8,5* | *46,3* | *8,9* |
| dispersion (°/nm) | polar 2 | *0,023* | *0,009* | - | - |
| ecart en polarisation (°) | | *6,8* | *0,7* | ***-1,3*** | *0,4* |
| | | guides 7µm | | | |
| phase (°) | polar 1 | ***69,7*** | *7,4* | ***32,6*** | *3,5* |
| dispersion (°/nm) | polar 1 | ***0,025*** | *0,004* | - | - |
| phase (°) | polar 2 | ***77,2*** | *7,9* | ***30,1*** | *3,6* |
| dispersion (°/nm) | polar 2 | ***0,019*** | *0,005* | - | - |
| ecart en polarisation (°) | | ***7,5*** | *1,4* | ***-2,5*** | *0,1* |

*Tableau III-5 : Déphasages obtenus sur les 4T-ABCD. Les déphaseurs '14' situés à l'extrémité de la puce ont un comportement différent des autres déphaseurs et ont été séparés des autres dans le tableau. Les valeurs de dispersion pour les déphaseurs '14' sont similaires à celles des autres déphaseurs et n'ont pas été reportées dans le tableau.*

Globalement, les déphaseurs ont un bon comportement notamment très achromatique, et les puces sont toutes totalement fonctionnelles, c'est à dire qu'elles permettent de mesurer le contraste et la phase des franges issues des quatre télescopes. Quelle que soit la largeur du coeur du guide, on retrouve un comportement quasiment identique aux déphaseurs mesurés sur les puces 2T-ABCD avec toutefois une dispersion très légèrement supérieure en moyenne.

Cependant, on observe sur tous ces dispositifs, et ce, quelle que soit la largeur du cœur des guides d'onde, que les déphaseurs '14' ont des valeurs plus faibles que les autres. Leur comportement est surtout différent en fonction de la polarisation. La différence de phase en polarisation obtenue sur ces déphaseurs '14' est systématiquement de signe opposé comparé à la différence de phase en polarisation de tous les autres déphaseurs (2T-ABCD compris). Deux explications sont envisagées pour expliquer ce comportement. La première explication est de supposer que le problème peut venir d'une inhomogénéité sur plaque. En effet, du fait de la grande taille de ces puces, les déphaseurs '14' sont géométriquement situés plus en bord de plaque que tous les autres déphaseurs mesurés. Les couches peuvent avoir en bord de plaque une épaisseur légèrement différente, ou encore, du fait de la courbure résiduelle de la plaque due aux contraintes, la photolithographie est peut-être légèrement dégradée en bord de plaque (défocalisation). Cependant, des observations au microscope à balayage n'ont pas permis de mettre en évidence un tel défaut, et par ailleurs, quelques simulations complémentaires où nous avons fait varier les paramètres d'épaisseur et de largeur de guide n'ont pas permis d'expliquer un tel écart.

L'autre explication est de supposer une inhomogénéité de contraintes, le comportement en polarisation de ces déphaseurs étant différent de celui de tous les autres déphaseurs. Le problème est qu'à notre connaissance, il n'existe aucun moyen pour mettre en évidence un tel phénomène. En attendant de pouvoir le faire, nous avons simplement décidé de réorganiser la disposition des puces sur le masque à la prochaine conception afin de rapprocher du centre de la plaque les puces comportant des déphaseurs.

## III – F – Bilan des caractérisations

Les résultats présentés dans ce chapitre font apparaître plusieurs points très positifs :

- En tout premier lieu, les résultats sont globalement très conformes aux résultats attendus si l'on considère que la technologie induit une perte de côte des guides de 1µm. Ceci est extrêmement encourageant et permet d'envisager la conception de circuits complexes dont on peut prédire les performances de manière fiable. Ainsi, les fonctions photométriques élémentaires (jonction Y, coupleur, tricoupleur) ont toutes été validées que ce soit photométriquement ou spectralement. Elles peuvent donc maintenant être considérées comme les éléments d'une boîte à outils utilisable pour d'autres dispositifs. De plus, les dispositifs ont un très bon comportement photométrique global, sans lumière parasite détectable, ni diaphonie détectable, ce qui valide aussi l'utilisation de croisements de guides pour réaliser des recombineurs intégrés.
- La transmission des dispositifs est globalement très bonne, même s'il est difficile d'envisager la comparaison avec l'optique de volume, aucun montage offrant de telles fonctionnalités n'ayant été réalisé à ce jour en optique de volume.

- Les déphaseurs qui étaient la principale inconnue de cette réalisation ont fourni d'excellents résultats en affichant un comportement très achromatique et des valeurs de déphasage relativement proches des valeurs calculées. Cela nous permet de valider le principe de ces déphaseurs que nous réutiliserons au chapitre suivant, mais d'une manière un peu différente.
- Enfin, ces résultats valident le transfert de la technologie optique intégrée sur silicium sur substrat huit pouces qui a de plus conduit à une réalisation présentant un nombre extrêmement réduit de défauts : toutes les fonctions photométriques sont opérationelles, et seulement trois déphaseurs parmi soixante douze présentent un défaut si l'on exclut le problème particulier des déphaseurs en bord de plaque.

La caractérisation exhaustive de tous les déphaseurs disponibles a aussi permis de mettre en évidence une grande sensibilité à leur position sur la plaquette de silicium. Même s'il est encore difficile de conclure sur ce point, on peut espérer une amélioration des performances en modifiant la répartition des puces sur la plaquette de silicium à la prochaine conception.

Enfin, à la suite de ces résultats, le LAOG envisage de collaborer avec l'université du Michigan qui collabore avec l'interféromètre CHARA, afin d'effectuer des mesures sur ciel avec un "4T-ABCD".

# III – G – Conclusion

Dans ce chapitre, nous avons présenté les caractérisations optiques effectuées sur les dispositifs réalisés. Globalement, ces résultats extrêmement encourageants nous ont incité à aller plus loin et à réflechir à l'extension des fonctionnalités de l'optique intégrée pour le recombinaison astronomique, notamment dans le cadre du projet VSI qui concerne l'étude d'un instrument de recombinaison pour le VLTi[*67*]. Plusieurs voies ont été explorées : les améliorations possibles des fonctions réalisées, l'extension à un plus grand nombre de télescopes, la possibilité d'évoluer vers d'autres bandes spectrales, et enfin, le problème du suivi de franges qui demande des fonctionnalités légèrement différentes. C'est le sujet du chapitre IV.

# Chapitre IV : Prochaine génération d'instrument pour le VLTi

## IV – A – Introduction

Ce chapitre décrit l'étude menée dans le contexte du projet VSI à la suite des résultats obtenus sur les puces réalisées et exposés au chapitre III. Le projet VSI a pour but l'étude et la réalisation d'un banc recombineur de prochaine génération pour le VLTi. Dans une première partie, nous décrirons rapidement le principe de VSI. Dans une deuxième partie, nous présentons l'étude conduite sur l'extension des fonctionnalités des recombineurs (en terme de nombre de télescopes et de bande spectrale). Enfin, dans une troisième partie, nous détaillons les possibilités offertes par l'optique intégrée pour réaliser un dispositif de suivi de franges qui demande des fonctionnalités légèrement différentes.

## IV – B – Description de l'instrument

Le projet VSI (VLTi Spectro Imager), proposé par des instituts européens pour l'ESO (European Southern Observatory) a pour but, dans sa première phase, de définir et de dimensionner un instrument recombineur pour l'interféromètre VLTi situé sur le Mont Paranal au Chili. Il regroupe les partenaires suivants :

- le Laboratoire d'Astrophysique de Grenoble (LAOG, France)
- le Cavendish Laboratory de l'université de Cambridge (Royaume Uni)
- le Max-Planck-Institüt für Radioastronomie in Bönn (MPIfR, Allemagne)
- le Centro de Astrofisica da Universidade do Porto (CAUP, Portugal)
- l'Istituto Nazionale di Astrofisica (INAG, Italie)
- l'Institut d'Astrophysique et de Géophysique de Liège (IAGL, Belgique)
- L'Institut für Astronome, Universität Wien (IfA, Autriche)
- L'Astrophysikalisches Institüt und Universitäts-stenwarte (AIU Jena, Allemagne)

Afin de pouvoir couvrir un maximum de domaines d'observation scientifiques et d'utiliser pleinement les possibilités offertes par l'interféromètre du VLT, l'instrument recombineur doit avoir les fonctionnalités suivantes :

- Recombinaison de 4 télescopes (version minimale) et 6 télescopes (but)
- Résolution temporelle de l'ordre du jour
- 2 ou 3 résolutions spectrales allant de 100 à 10000 canaux spectraux
- Suiveur de franges interne
- Dynamique de l'image allant de 100 à 1000
- Champ de vue correspondant à quelques centaines de millisecondes d'angle
- Etendue spectrale allant de 1 à 2,5µm

A la suite des résultats obtenus en laboratoire ainsi que des résultats obtenus par IONIC, le LAOG a proposé dès le début du projet l'utilisation de l'optique intégrée pour effectuer la recombinaison optique des faisceaux issus des télescopes. J'ai donc étudié le dimensionnement des puces envisagées pour cet instrument afin d'en prévoir les performances.

Le concept de l'instrument recombineur basé sur l'utilisation de nos composants optiques intégrés est schématisé sur la figure IV-1 :

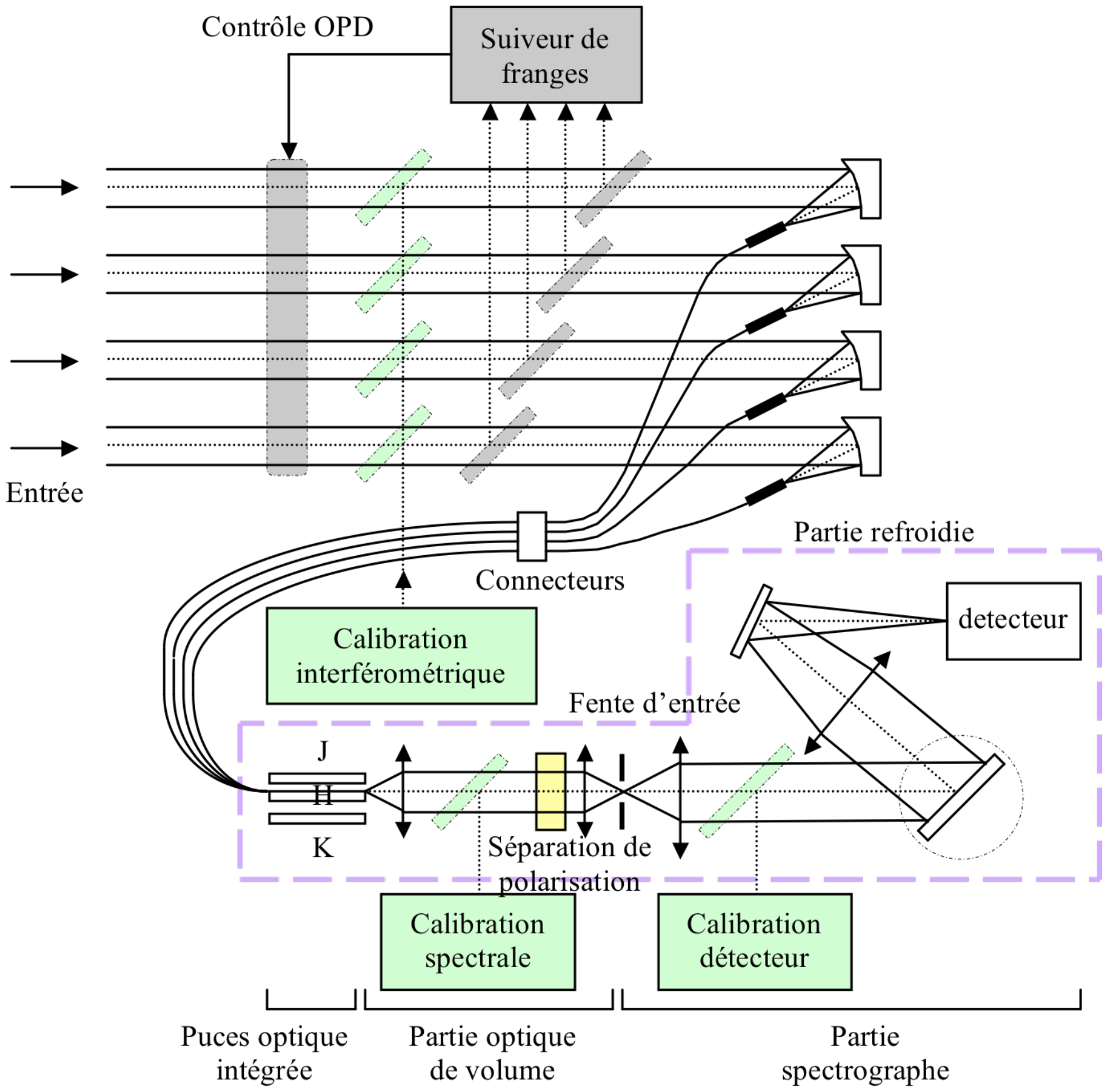

*Figure IV-1 : concept du recombineur VSI. On utilise plusieurs puces interchangeables optimisées à la configuration du VLTi (4T, 6T, bande spectrale,…).*

La partie suiveur de franges est séparée de la partie scientifique. Les faisceaux issus des télescopes sont injectées dans des fibres monomodes à maintien de polarisation connectées à l'une des puces recombineur choisie en fonction de la bande spectrale d'observation et la configuration de l'interféromètre (4 ou 6 télescopes). Les signaux de sortie de la puce sont ensuite collimatés, séparés en polarisation à l'aide d'un prisme de Wollaston, puis imagés sur la fente d'entrée d'un spectrographe. Le réseau du spectrographe est interchangeable afin d'adapter la résolution spectrale voulue. L'ensemble optique allant de la puce au détecteur est sous enceinte refroidie afin de minimiser le bruit, notamment en bande K. Enfin, il est possible d'intercaler des sources de lumière à différents endroits afin d'aligner et de calibrer l'instrument.

L'intérêt majeur de ce concept d'instrument basé sur l'utilisation de recombineurs optiques intégrés réside dans l'exceptionnelle souplesse de l'instrument. En effet, par une simple déconnection/connexion de fibres optiques, on utilise un schéma de recombinaison optimal quelle que soit la configuration de l'interféromètre. Une telle fonctionnalité en optique de volume aurait un coût prohibitif, des phases d'alignement et de calibration beaucoup plus longues, et un encombrement incompatible avec la place disponible dans la salle de recombinaison du VLTi.

Nous allons maintenant décrire les puces que nous envisageons pour cet instrument.

## IV – C - Vers les bandes J et K

Les bandes J (1,11µm – 1,33µm) et K (2,0µm – 2,4µm) sont situées de part et d'autre de la bande H (1,47µm – 1,78µm) et représentent les bandes de transparence de l'atmosphère dans le proche infrarouge.

Nous avons d'abord envisagé de réaliser un recombineur recouvrant les deux bandes J et H, puis nous avons rejeté cette option. La première raison concerne les fonctions qui doivent être le plus achromatique possible. Il est bien plus difficile de concevoir des coupleurs, tricoupleurs ou déphaseurs achromatiques de 1,1µm à 1,8µm plutôt que par bande spectrale séparée. De plus, on peut s'attendre à des performances plus élevées si on optimise des fonctions pour chaque bande séparément plutôt que sur l'ensemble du domaine de longueur d'onde. L'autre raison concerne la plage de monomodicité des guides. Un guide monomode pour les deux bandes spectrales sera (à écart d'indice constant) plus petit qu'un guide optimisé pour la bande H seulement. Il en résulte que les fonctions de routage devront utiliser un rayon de courbure plus grand pour les différents chemins parcourus, et conduiront à des puces nettement plus grosses, donc avec des pertes de propagation plus importantes. Concevoir un dispositif par bande spectrale permet donc d'obtenir de bien meilleures performances.

## IV – C – 1 – Bande J

Même si nous n'avons encore jamais réalisé de recombineurs pour l'interférométrie astronomique en bande J, les dispositifs réalisés précédemment au LETI pour les télécommunications ont été vérifiés dans un domaine de longueur d'onde proche, où nous n'avons jamais observé de phénomène particulier gênant.

En utilisant la même méthode déjà décrite au chapitre II, et en conservant un écart d'indice de réfraction entre le cœur et la gaine du guide à 0,01, on ajuste la dimension des guides pour être monomodes sur la bande J. On trouve alors des guides carrés de 4,3µm de côté.

Les guides ayant des dimensions inférieures à ceux conçus pour la bande H, la réalisation technologique des dispositifs se trouvera facilitée car elle nécessitera des épaisseurs de couches plus faibles, donc des problèmes de déformation de plaquette moindres. En contrepartie, les pertes par diffusion étant inversement proportionnelles à la longueur d'onde [*68*], la rugosité des guides risque d'avoir une influence plus grande sur les pertes globales du dispositif.

## IV – C – 2 – Bande K

En bande K, la limite de monomodicité étant plutôt située à une longueur d'onde de 2µm, il est plus judicieux d'utiliser des guides de dimensions plus importantes. Cependant, afin de ne pas compliquer les étapes de photolithographie et gravure du cœur du guide, nous avons conservé l'épaisseur standard de 4,85µm±0,15µm. On trouve alors que les guides les mieux adaptés à la bande K ont une largeur de 9µm (avec toujours le même écart d'indice de 0,01).

Au tout début du travail présenté dans ce document, nous avons étudié et réalisé un recombineur à deux télescopes pour la bande K en utilisant de tels guides. Les composants furent connectés à des fibres optiques en verre fluoré au LAOG puis testés en laboratoire avant d'être installés sur l'instrument VINCI du VLTi. En considérant la numérotation des sorties de la figure IV-2, les performances du dispositif sont résumées dans le tableau IV-1.

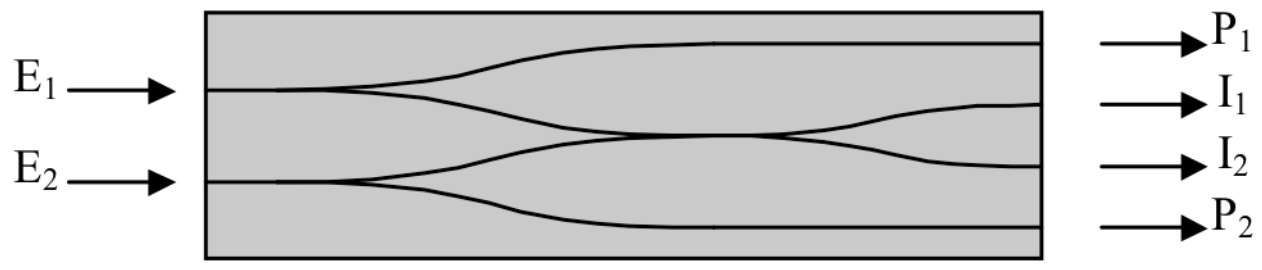

*Figure IV-2 : schéma du composant 2T-K et numérotation des sorties*

**Caractéristiques du 2TK**

| Entrée | E1 | E2 |
|---|---|---|
| Sortie P1 (%) | *37,2* | - |
| sortie I1 (%) | *17,2* | *17,3* |
| Sortie I2 (%) | *17* | *17,5* |
| Sortie P2 (%) | - | *37,2* |
| Transmission totale (%) | *71,4 ± 0,6* | *72,0 ± 0,6* |
| Contraste (%) | *91,6 ± 0,2* | |
| Contraste en lumière polarisée (%) | *> 95* | |

*Tableau IV-1 : résultats mesurés en bande K par le LAOG en laboratoire après collage à la nappe de fibres optiques.*

Les transmissions sont très légèrement inférieures à celles que l'on obtient en bande H, les contrastes restant très bons. Ce composant a été installé sur l'instrument VINCI du VLTi où il a remplacé le recombineur à fibres FLUOR. Un exemple de franges d'interférence obtenues sur l'étoile λ-scorpi est représenté en figure IV-3.

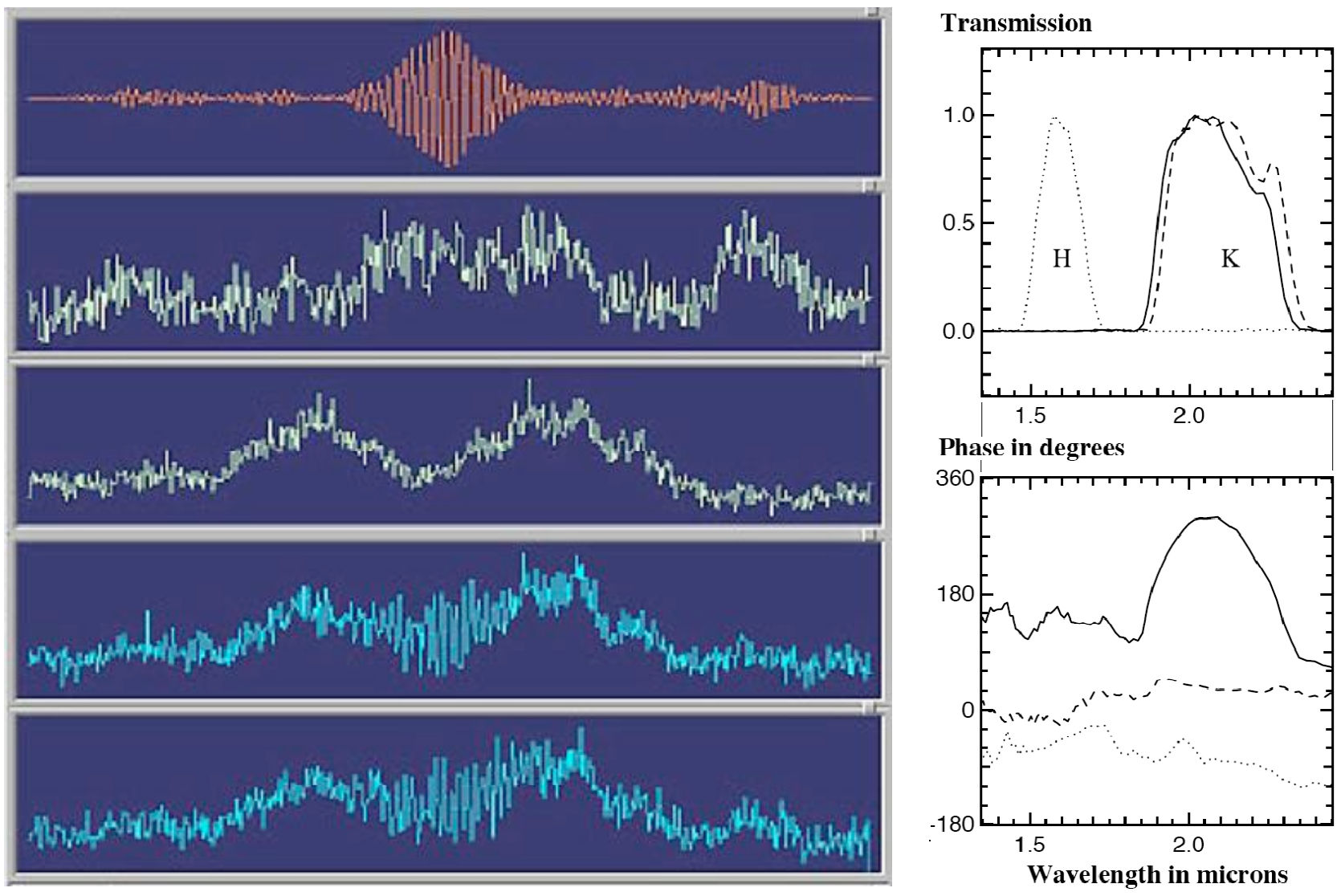

*Figure IV-3 : image de gauche : signaux obtenus sur l'étoile λ-scorpi au VLTi. A droite, de haut en bas : signal interférométrique corrigé des voies photométriques, les deux signaux photométriques, les deux signaux interférométriques bruts. Graphes de droite : en trait plein, transmission et phase des franges du 2TK, en traits pointillés, transmission et phase du recombineur à fibre FLUOR. En pointillés, transmission et phase du 2TH à jonction Y IMEP.*

Le 2TK présente une bonne transmission et de bons contrastes. Cependant, il présente un problème de phase chromatique des franges. Lebouquin [39] a montré qu'un traitement du signal approprié permettait de corriger les mesures et d'obtenir les bonnes valeurs de contraste. A l'heure actuelle, nous n'avons toujours pas réussi à discerner si le problème provient de la puce elle même, ou des fibres en verre fluoré connectées à l'entrée et qui présentent des contraintes aux abords du collage. Enfin, l'équipe scientifique du VLTi a salué l'excellente stabilité de l'instrument qui lui permet notamment de l'utiliser comme outil de calibration pour AMBER.

## IV – C – 3 – Extension des résultats aux bandes J et K

De par l'expérience que nous avons en bande J, et les résultats obtenus en bande K, on peut donc raisonnablement envisager d'utiliser l'optique intégrée dans ces deux bandes spectrales. En prenant en compte les impératifs technologiques, la méthode décrite au chapitre II conduit aux résultats regroupés dans le tableau IV-2.

| **bande spectrale** | J | | | H | | | K | | |
|---|---|---|---|---|---|---|---|---|---|
| **constrate d'indice** | *0,01* | | | | | | | | |
| **largeur guide [µm]** | *4,3* | | | *6* | | | *9* | | |
| **hauteur guide [µm]** | *4,3* | | | *4,85* | | | | | |
| **rayon de courbure limite [µm]** | *5300* | | | *8000* | | | *14100* | | |
| **biréfringence** | *0,001* | | | | | | | | |
| **pertes [dB/cm]** | *0,05* | | | | | | *?* | | |
| **couplage fibre [dB]** | *0,25* | | | | | | *?* | | |
| **longueur d'onde [µm]** | *1,1* | *1,22* | *1,35* | *1,45* | *1,6* | *1,75* | *2* | *2,2* | *2,4* |
| **largeur mode [µm]** | *5,24* | *5,61* | *6,04* | *7,16* | *7,64* | *8,16* | *10,6* | *11,3* | *12,1* |
| **hauteur mode [µm]** | *5,24* | *5,61* | *6,04* | *6,46* | *6,94* | *7,46* | *8,14* | *8,92* | *9,82* |
| **O.N. horizontale** | *0,13* | *0,14* | *0,14* | *0,13* | *0,13* | *0,14* | *0,12* | *0,12* | *0,13* |
| **O.N. verticale** | *0,13* | *0,14* | *0,14* | *0,14* | *0,14* | *0,15* | *0,16* | *0,16* | *0,15* |

*Tableau IV-2 : récapitulatif des caractéristiques des guides silice sur silicium calculés pour la recombinaison. Les largeurs, hauteurs, et ouvertures numériques des modes sont calculées sur la meilleure gaussienne approximant la propagation guidée (voir chapitreII) et servent à la conception des systèmes optiques après la puce.*

A partir de ces valeurs, nous avons conçu les différents éléments du 4T-ABCD pour chaque bande spectrale. Les résultats sont regroupés dans le tableau IV-3.

| **Bande spectrale** | **J** | **H** | **K** |
|---|---|---|---|
| **angle jonctions X [°]** | *30* | | |
| **pertes jonctions X simulées [dB]** | *0,09* | *0,05* | *0,06* |
| **pertes jonctions X mesurées [dB]** | | *~0,05* | |
| **diaphonie jonctions X [dB]** | *>60* | *>60* | *52* |
| **pertes jonctions Y simulées [dB]** | *0,19* | *0,18* | *0,16* |
| **pertes jonctions Y mesurées [dB]** | | *~0,25* | |
| **pertes coupleur simulées [dB]** | *0,04* | *0,04* | *0,04* |
| **pertes tricoupleur simulées [dB]** | *0,04* | *0,04* | *0,04* |
| **Longueur puce 4T-ABCD [mm]** | *55* | *80* | *140* |
| **pertes totales simulées [dB]** | *1,45* | *1,46* | *2,45* |
| **transmission totale simulée** | *0,72* | *0,71* | *0,57* |
| **transmission totale mesurée** | | *0,65* | |

*Tableau IV-3 : Performances des 4T-ABCD à tricoupleurs simulées par BPM pour les trois bandes spectrales de VSI. Les valeurs mesurées disponibles sont rajoutées lorsqu'elles sont disponibles.*

Il apparaît qu'un 4T-ABCD en bande K comporte des dimensions importantes et sera difficilement réalisable. La transmission en bande K étant légèrement moins bonne, la transmission globale de ces puces sera aussi moins bonne. Afin de réduire la taille de ces puces, nous avons envisagé un routage différent basé sur une séparation par coupleur 66/33 suivi d'une jonction Y comme expliqué au chapitre II. Nous avons gardé l'égalité des chemins optiques après séparation des faisceaux, c'est à dire que chaque paire recombinée a parcouru exactement la même longueur de guide droit ainsi que les mêmes portions courbes et le même nombre de croisements, mais nous n'avons pas conservé la symétrie globale de la puce afin de raccourcir globalement le dispositif. Le dessin détaillé en bande H d'un tel dispositif est représenté sur la figure IV-4.

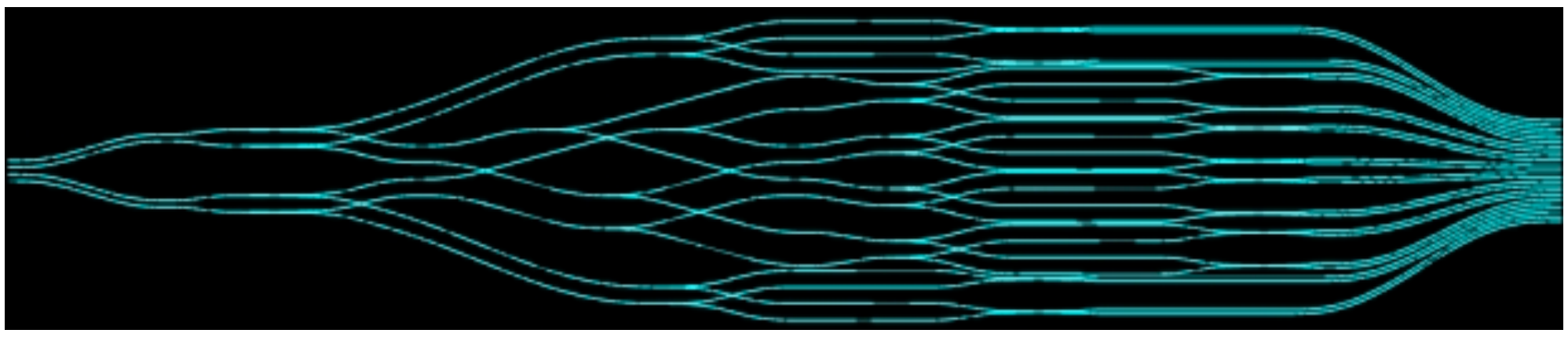

*Figure IV-4 : autre routage du 4T-ABCD. L'égalité des chemins optiques est gardée, mais la symétrie de la puce n'est pas conservée.*

La puce fait environ 58mm de long au lieu de 80mm en utilisant le routage précédent à tricoupleur, et la largeur est sensiblement la même (12mm). Les performances que l'on peut attendre d'un tel dispositif sont regroupées et comparées à celles de l'autre routage dans le tableau IV-4.

| **Recombineur** | **4T ABCD (1)** | | | **4T ABCD (2)** | | |
|---|---|---|---|---|---|---|
| **Bande spectrale** | J | H | K | J | H | K |
| **Rayon de courbure limite (mm)** | *5,3* | *8* | *14,1* | *5,3* | *8* | *14,1* |
| **Longueur du dispositif (mm)** | *50* | *80* | *140* | *40* | *60* | *106* |
| **Nombre de croisements** | *5* | | | *4* | | |
| **Nombre de jonctions Y** | *1* | | | *2* | | |
| **Nombre de coupleurs** | *1* | | | *2* | | |
| **Nombre de tricoupleurs** | *1* | | | *0* | | |
| **Longueur des parties courbes (mm)** | *20* | *30* | *50* | *20* | *30* | *50* |
| **Pertes simulées (dB)** | *1,46* | *1,46* | *2,45* | *1,49* | *1,42* | *2,20* |
| **Transmission** | *0,72* | *0,71* | *0,57* | *0,71* | *0,72* | *0,60* |
| **Transmission mesurée** | - | *0,65* | - | - | - | - |

*Tableau IV-4 : comparaison des performances attendues des 2 types de 4T-ABCD. (1) correspond au routage utilisant des tricoupleurs, (2) correspond au routage utilisant un coupleur 66/33 suivi d'une jonction Y sur la 'route 66'.*

Les résultats font apparaître que le gain en terme de pertes n'est pas élevé, les pertes de propagation plus petites dues à une longueur plus courte de la puce étant rattrapées par les pertes introduites par la jonction Y supplémentaire dans le deuxième routage. Par contre le gain en encombrement permet une longueur de puce plus réaliste en bande K où par ailleurs, nous n'avons qu'une idée approximative des pertes de propagation. Si l'on désire à tout prix diminuer encore les pertes, il pourrait être avantageux de remplacer les jonctions Y par des coupleurs asymétriques 50/50, mais ceci conduira à une séparation des faisceaux plus chromatique.

En conclusion, nous avons montré dans ce paragraphe que les recombineurs 4T-ABCD pouvaient être envisagés dans les bandes spectrales J et K situées autour de la bande H et surtout situées dans la bande de transparence de la silice dont nos guides sont constitués. En ce qui concerne les performances, on peut s'attendre à obtenir des caractéristiques équivalentes à celles obtenues en bande H avec des transmissions légèrement inférieures en bande K du fait de la taille plus importante des puces. Nous allons maintenant aborder l'extension de ces concepts à plus de télescopes.

# *IV – D - Vers plus de télescopes*

Dans une première partie, nous présenterons l'extension des possibilités des recombinaisons par paires à six télescopes voire huit télescopes. Mais si Lebouquin et al. ont montré dans [*59*] que le codage par paires matriciel ABCD était le plus adapté pour quatre télescopes, ceci n'est plus forcément vrai à 6 ou plus de télescopes où le signal est étalé sur plus de canaux. C'est pourquoi nous exposerons dans une deuxième partie les possibilités en recombinaison multiaxiale. Enfin nous comparerons les deux types de recombineurs.

## IV – D – 1 - Recombinaison par paires

### *IV – D – 1 – a – Recombinaison à six télescopes*

Une recombinaison par paires à six entrées conduit à une séparation des faisceaux d'entrée en cinq canaux et à quinze recombinaisons en tout. Nous avons envisagé la séparation en cinq voies de la figure IV-5.

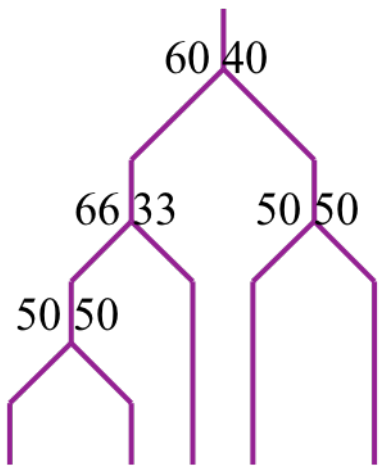

*Figure IV-5 : exemple de séparation d'un faisceau en cinq en optique intégrée.*

Le schéma est constitué d'un coupleur asymétrique 60/40 suivi d'un coupleur asymétrique 66/33 et de deux jonctions Y et permet d'obtenir 20% sur chaque voie.

Il existe un grand nombre de schémas de recombinaison possibles à six télescopes, et nous n'avons pas effectué une analyse exhaustive de toutes les possibilités. Parmi les quelques configurations étudiées, nous avons retenu le schéma de la figure IV-6 qui présentait un nombre minimal de croisements de guides. Ce schéma conserve la symétrie donc l'égalité des chemins optiques après séparation des faisceaux. On peut noter qu'il utilise deux recombineurs à trois télescopes de chaque coté et une zone centrale plus complexe. On obtient pour ce schéma huit croisements de guides au maximum sur certaines voies.

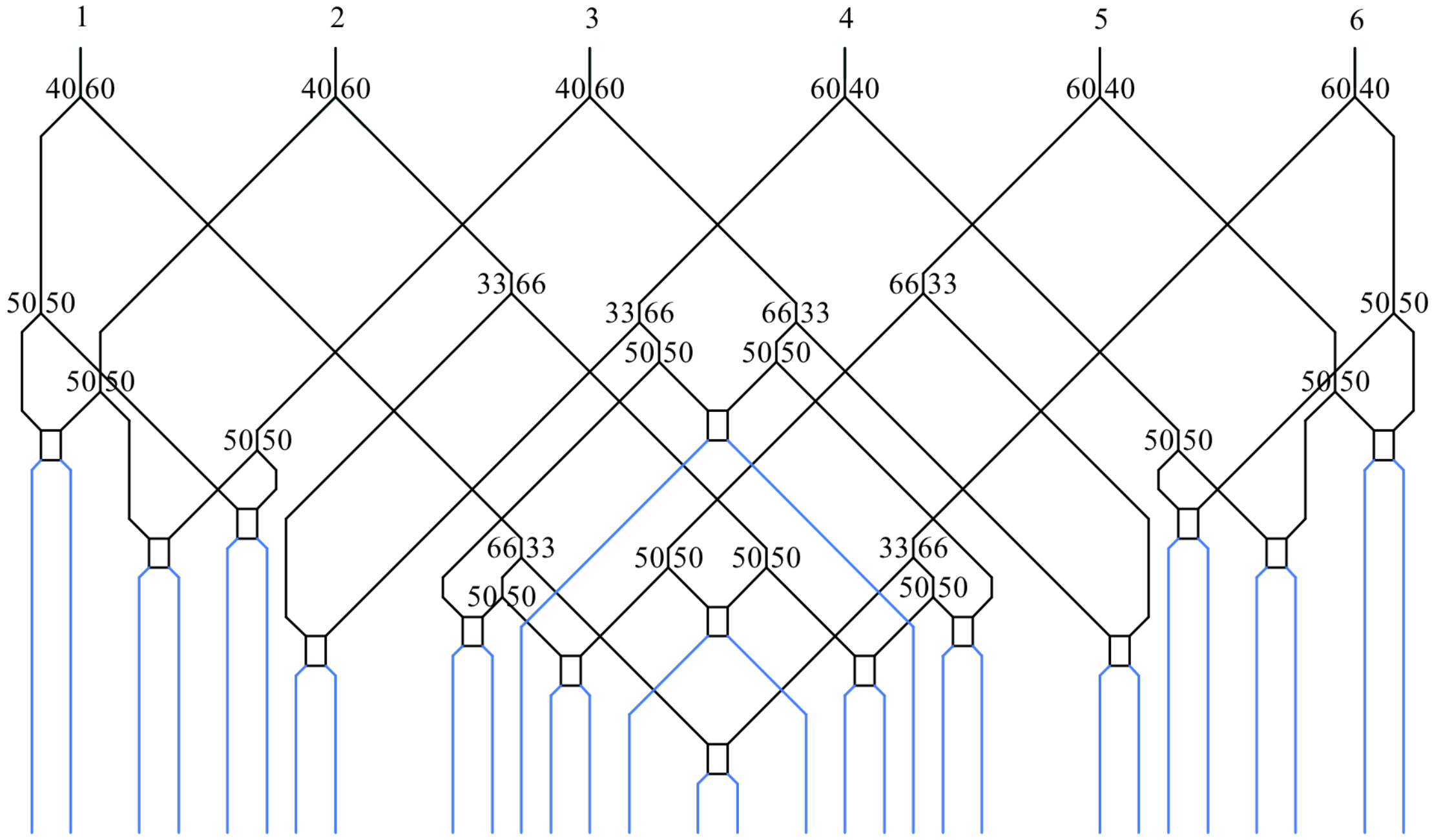

*Figure IV-6 : schéma de recombinaison à 6 télescopes. Les faisceaux sont en noir avant recombinaison et en bleu après recombinaison.*

A partir de ce schéma, si l'on conserve l'égalité des chemins optiques par paire, mais que l'on ne conserve pas la symétrie globale en déportant les deux recombinaisons centrales (recombinaisons 34 et 25), on peut alors (comme pour les recombineurs à quatre télescopes) gagner en encombrement et en nombre de croisements. On obtient finalement le schéma plus réaliste de la figure IV-7.

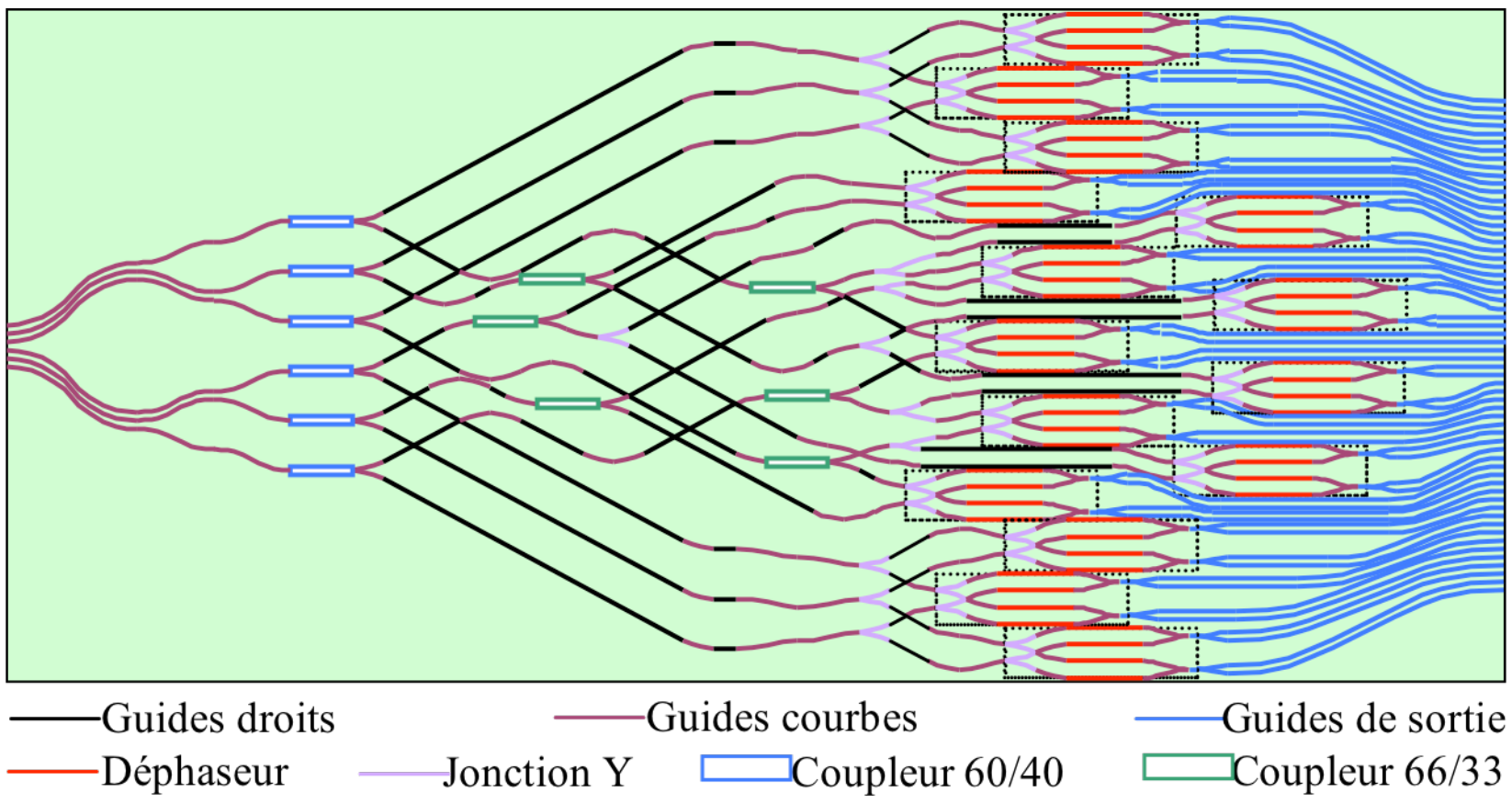

*Figure IV-7 : schéma du recombineur 6T-ABCD. Le schéma est anamorphosé pour plus de clarté (échelle ~1 en horizontal, échelle ~ 2 en vertical si on considère la bande H). On reconnaît en haut et en bas deux recombineurs 3 télescopes et une zone centrale pour les autres paires.*

La puce comporte 6 entrées et 60 sorties (15 recombinaisons à 4 sorties). Chaque paire recombinée a parcouru la même longueur de guide droit ainsi que les mêmes parties courbes. Le nombre maximum de croisements est de 8 suivant les voies, en incluant le croisement de la cellule ABCD. Un tel schéma permet d'évaluer la taille de la puce et les performances que l'on peut espérer obtenir pour les 6T-ABCD. Les résultats sont reportés dans le tableau IV-5.

| Recombineur | 6T ABCD | | |
|---|---|---|---|
| Bande spectrale | J | H | K |
| Rayon de courbure limite (mm) | 5,3 | 8 | 14,1 |
| Longueur du dispositif (mm) | 66 | 100 | 175 |
| Nombre de croisements | 8 | | |
| Nombre de jonctions Y | 2 | | |
| Nombre de coupleurs | 3 | | |
| Nombre de tricoupleurs | 0 | | |
| Longueur des parties courbes (mm) | 30 | 50 | 85 |
| Pertes simulées (dB) | 2,03 | 1,88 | 3,21 |
| Transmission | 0,63 | 0,65 | 0,48 |

*Tableau IV-5 : performances simulées des 6T-ABCD.*

S'il est envisageable de réaliser des 6T-ABCD en bande J ou H où les tailles de puces sont encore raisonnables et les transmissions supérieures à 60%, la taille du dispositif en bande K est peu réaliste (même si des substrats huit pouces permettent une telle réalisation), et conduit à des pertes plus importantes.

### IV – D – 1 – b – Recombinaison à huit télescopes

Le cahier des charges du futur instrument VSI ne comporte pas de recombineur à huit télescopes. En effet, même si le VLTi possède bien huit télescopes en tout (4 télescopes de 8,2m et 4 télescopes auxiliaires configurables de 1,8m dédiés à l'interférométrie), il n'y a pas suffisamment de lignes à retard opérationnelles ni prévues pour envisager un fonctionnement à huit télescopes.

A titre exploratoire, nous avons cependant regardé les possibilités de recombinaison par paires en optique intégrée. Il s'agit cette fois de diviser chaque faisceau incident en sept canaux et d'effectuer 28 recombinaisons. Un exemple de séparation en sept faisceaux est donné sur la figure IV-8.

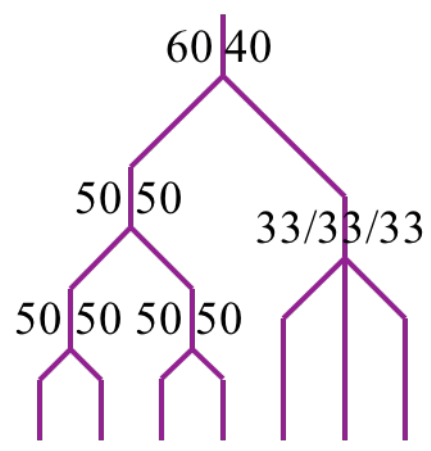

*Figure IV-8 : exemple de séparation en sept faisceaux en optique intégrée en utilisant un coupleur asymétrique, un tricoupleur et des jonctions Y.*

Là encore, nous n'avons pas exploré les possibilités de manière exhaustive, mais simplement retenu une configuration que nous avons trouvée afin d'évaluer si elle était réalisable ou non. Nous avons trouvé le schéma de recombinaison symétrique de la figure IV-9.

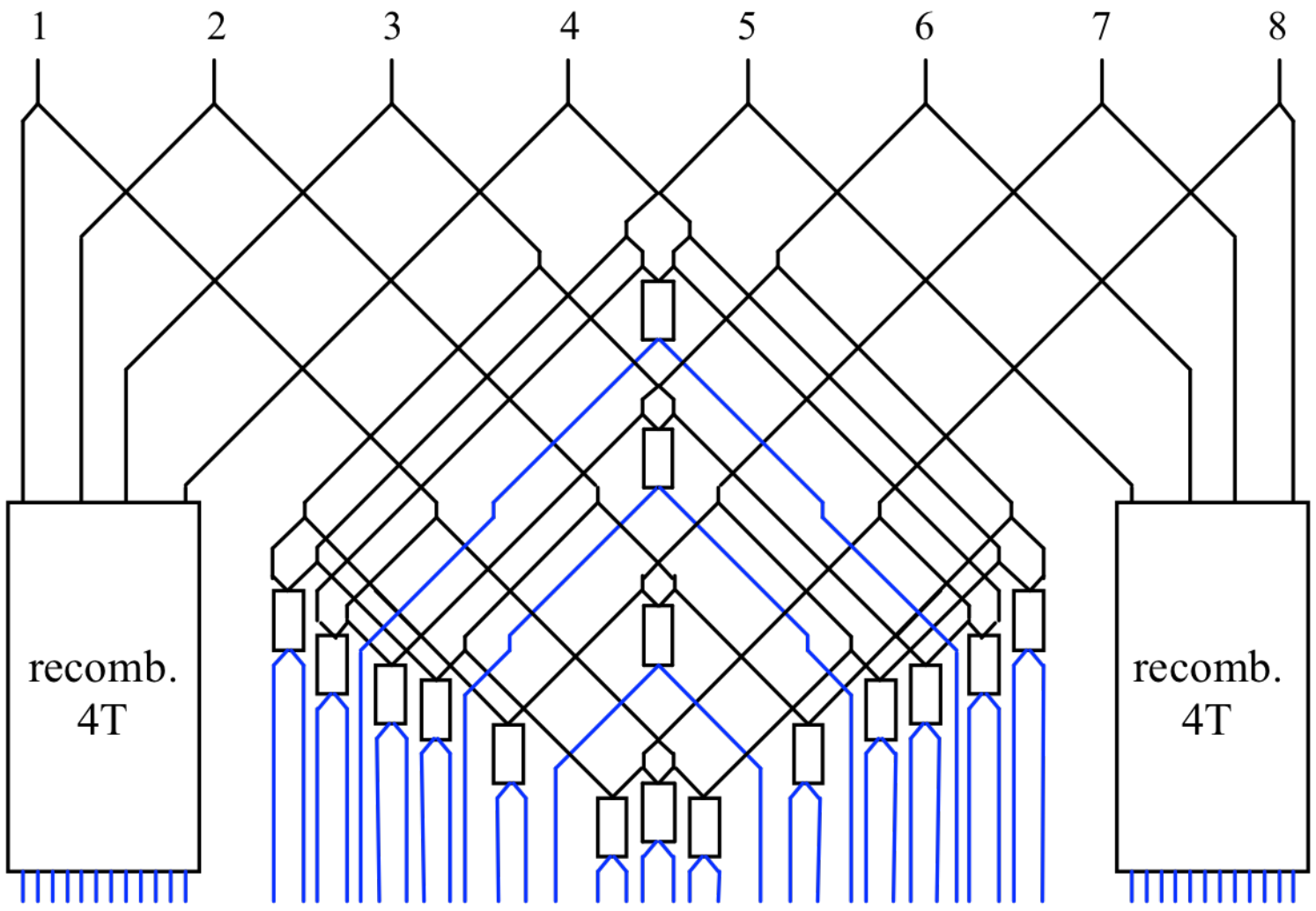

*Figure IV-9 : exemple de schéma de recombinaison symétrique à huit télescopes. Le schéma utilise deux recombineurs à quatre télescopes et une partie centrale plus complexe.*

En conservant l'égalité des chemins optiques par paires et la symétrie du schéma, nous avons réussi à faire le schéma plus réaliste suivant, mais en considérant une recombinaison par paire simple, sans cellule ABCD.

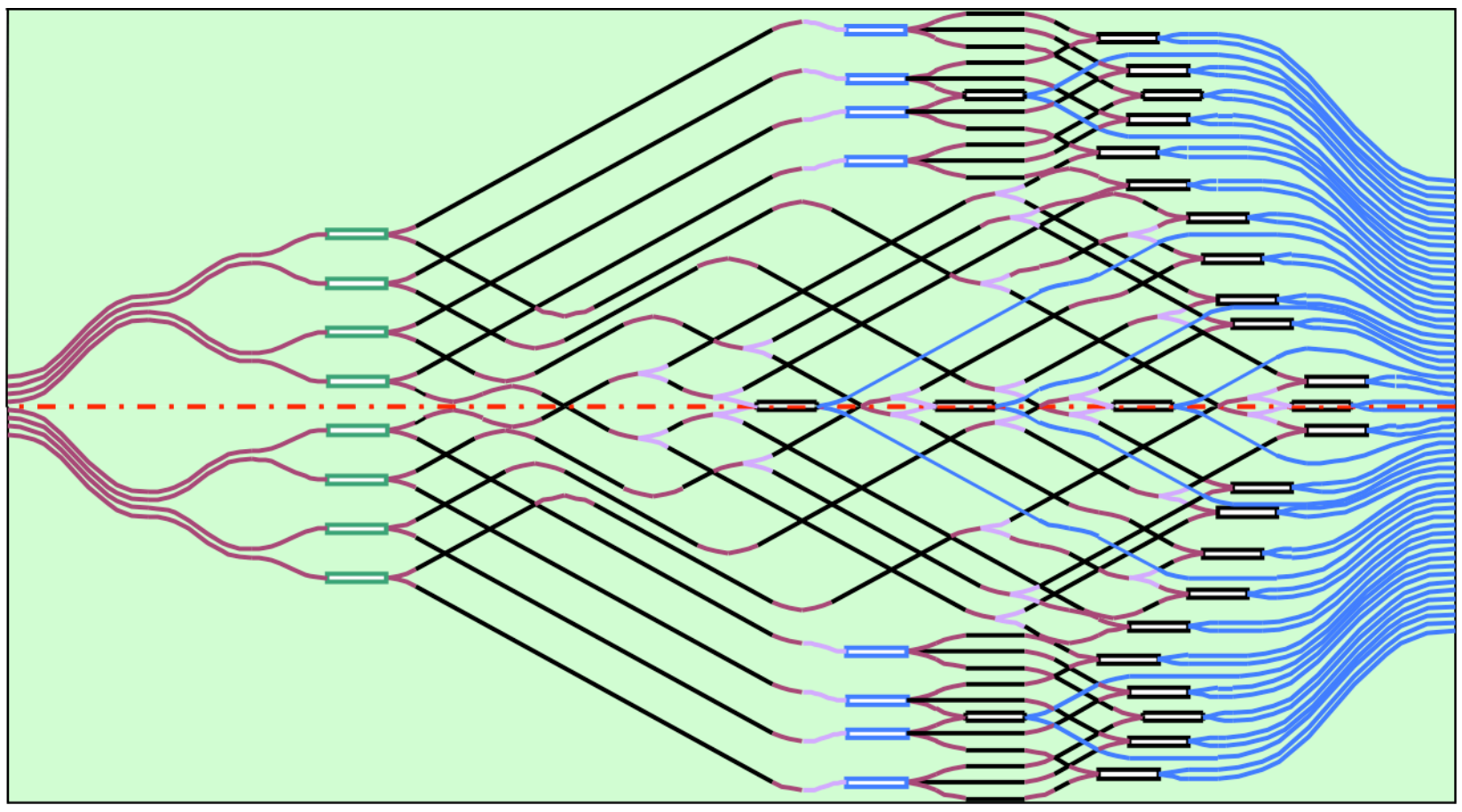

*Figure IV-10 : recombineur à 8 télescopes par paires. Là encore, le schéma est anamorphosé pour plus de clarté (échelle ~1 en horizontal, échelle ~ 2 en vertical si on considère la bande H). On reconnaît en haut et en bas deux recombineurs 4 télescopes à tricoupleurs et une zone centrale pour les autres paires.*

Un tel dispositif comporte donc 8 entrées pour 56 sorties (28 paires recombinées à 2 sorties). En bande H, un tel dispositif mesure 110mm de long par 30mm de large, et il peut y avoir jusqu'à 12 croisements de guide sur certaines voies. Une évaluation des performances est résumée dans le tableau IV-6.

| **Recombineur** | **8T par paires** | | |
|---|---|---|---|
| **Bande spectrale** | J | H | K |
| **Rayon de courbure limite (mm)** | *5,3* | *8* | *14,1* |
| **Longueur du dispositif (mm)** | *80* | *110* | *185* |
| **Nombre de croisements** | *12* | | |
| **Nombre de jonctions Y** | *2* | | |
| **Nombre de coupleurs** | *2* | | |
| **Nombre de tricoupleurs** | *1* | | |
| **Longueur des parties courbes (mm)** | *30* | *40* | *60* |
| **Pertes simulées (dB)** | *2,42* | *2,22* | *3,48* |
| **Transmission** | *0,57* | *0,59* | *0,45* |

*Tableau IV-6 : Evaluation des performances des recombineurs à huit télescopes par paires.*

Là encore la taille de puce en bande K est très grande et conduit à des pertes plus importantes que dans les bandes J et H. Il n'en reste pas moins que ces dispositifs sont réalisables sur un substrat huit pouces en technologie silice sur silicium.

Nous n'avons pas poussé plus loin l'étude des recombineurs par paires. Nous avons en effet prouvé qu'il était bien possible de réaliser de tels composants en technologie silice sur silicium et nous avons même été en mesure de fournir une évaluation des performances que l'on pouvait espérer, mais il est clair que la complexité (et donc la taille) des circuits croît rapidement avec le nombre de télescopes.

Concernant les 4T-ABCD, nous avons montré que l'on pouvait diminuer la taille des dispositifs, mais sans pour autant gagner de manière significative sur le coefficient de transmission global, la diminution des pertes de propagation étant contrebalancée par les pertes de la jonction Y supplémentaire. Par ailleurs, nous avons montré qu'il était possible de réaliser des 6T-ABCD par paires, mais avec une transmission plus faible en bande K à cause des tailles de puces très importantes. En bande J, nous avons montré que les dispositifs peuvent avoir des performances comparables à celles des dispositifs bande H et avec des encombrements plus réduits. Enfin, nous avons réussi à trouver un premier schéma de recombineur à huit télescopes par paires simples. Sans optimisation particulière, ce composant est réalisable dans les trois bandes J, H, K.

Il nous semble difficile de généraliser ce principe de recombinaison à un encore plus grand nombre de télescopes. C'est pourquoi nous avons envisagé une autre méthode de recombinaison plus adaptée à un grand nombre de télescopes et déjà mentionnée au chapitre II : la recombinaison multi axiale « tout-en-un » que nous allons maintenant exposer.

## IV – D – 2 - Recombinaison tout en un

### *IV – D – 2 – a – Description*

La recombinaison multiaxiale, déjà présentée au chapitre II, utilise une zone de recombinaison à guide planaire, dans laquelle on injecte des faisceaux avec des angles différents. La figure IV-11 en montre le principe de base ainsi que les notations utilisées par la suite.

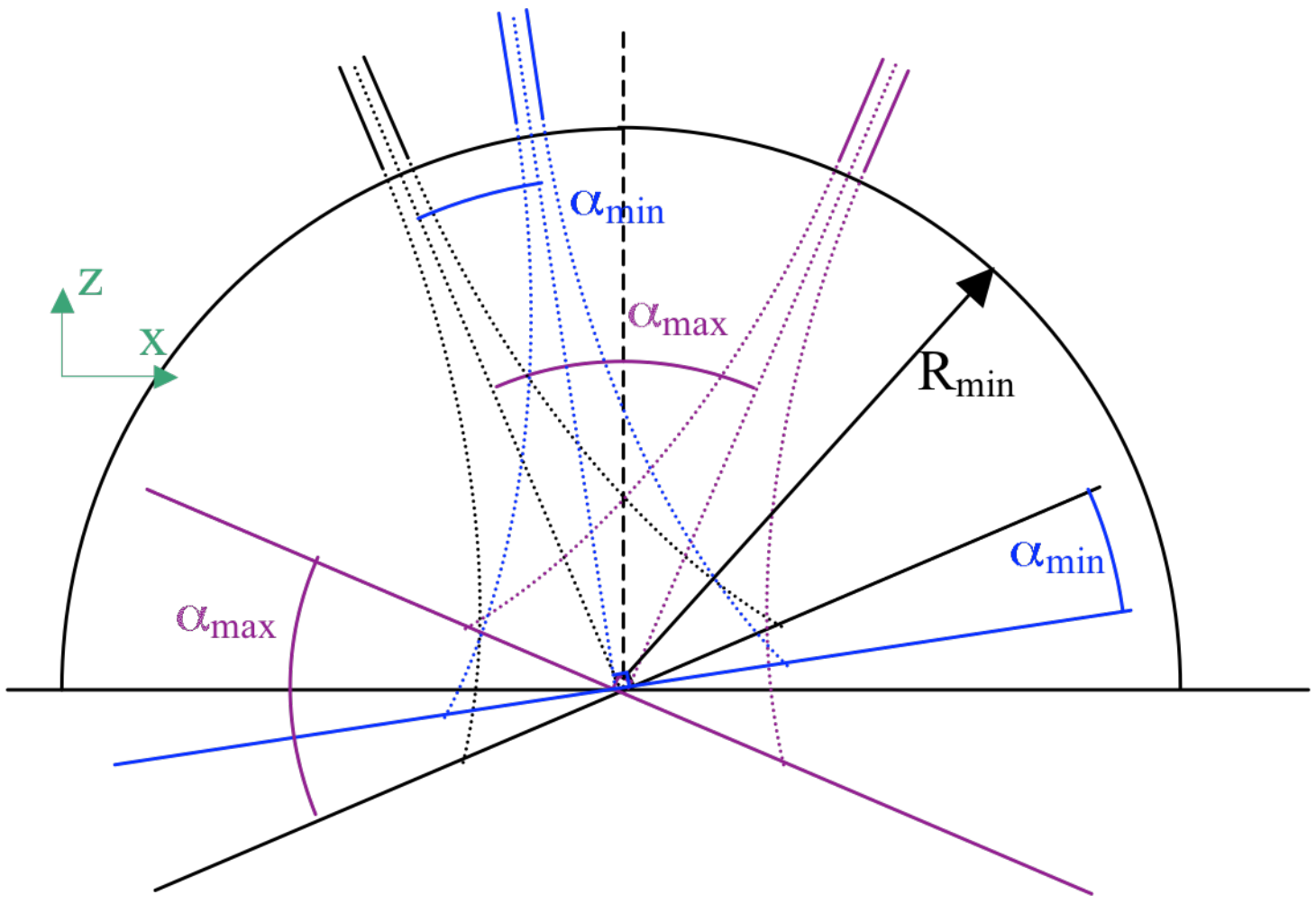

*Figure IV-11 : recombinaison planaire tout-en-un. Schéma de principe et notations utilisées.*

Les faisceaux sont injectés à différents angles. Les interférences *I(x)* pour une paire recombinée s'écrivent :

$$I(x) = I_e(x)\left(\frac{1+\cos(\phi(x))}{2}\right) \qquad \text{(IV-1)}$$

$$\phi(x) = \frac{4\pi}{\lambda} n_{eff}\, x \sin(\alpha) \qquad \text{(IV-2)}$$

où $\alpha$ représente $\alpha_{min}$ ou $\alpha_{max}$ sur la figure IV-11. Ainsi, si chaque paire recombinée s'effectue à un angle différent, il est possible de séparer les signaux des interférences en effectuant une analyse spectrale des fréquences spatiales du signal en sortie de puce.

### *IV – D – 2 – b – Distribution angulaire*

La méthode la plus simple pour optimiser la répartition angulaire des guides injectant les signaux dans la zone planaire consiste à considérer que chaque angle entre deux guides d'entrée est égal à un nombre entier fois un angle $\alpha_{min}$ qui est l'angle minimum que nous pouvons accepter entre deux guides d'entrée (nous verrons juste après comment dimensionner cet angle minimum).

$$\alpha_{paire} = p\alpha_{\min} \text{ avec p entier.} \qquad \text{(IV-3)}$$

Ainsi, chaque paire recombinée sera codée sur une fréquence spatiale égale à un nombre entier fois une fréquence minimale $f_{min}$ dépendant de $\alpha_{min}$. Un programme de recherche informatique nous a permis d'explorer toutes les répartitions angulaires pour six et huit télescopes. Les résultats font apparaître les points suivants :

- A six télescopes, la fréquence maximale $f_{max}$ minimale que l'on peut utiliser pour obtenir une répartition angulaire sans recouvrement de spectre dans le domaine des fréquences spatiales est de dix-sept fois la fréquence minimale. Sachant qu'il y a quinze paires recombinées, l'espace spectral occupé est quasiment occupé à 100%. Pour cette fréquence optimale, les répartitions angulaires possibles des angles des faisceaux d'entrée sont au nombre de quatre : $\alpha$=[0,1,4,10,12,17].$\alpha_{\min}$, $\alpha$=[0,1,4,10,15,17].$\alpha_{\min}$, $\alpha$=[0,1,8,11,13,17].$\alpha_{\min}$, $\alpha$=[0,1,8,12,14,17].$\alpha_{\min}$. On retrouve ici un résultat déjà trouvé par Lebouquin et al. dans [*69*].
- Nous verrons juste après que le dimensionnement de l'angle $\alpha_{min}$ influe beaucoup sur l'encombrement total de la puce. Si l'on désire minimiser cet encombrement, il peut être intéressant d'envisager une répartition angulaire où l'angle minimal entre deux entrées est de $2\alpha_{min}$ et non *pas* $\alpha_{min}$. Une telle répartition réduit en effet la taille de la zone planaire par deux ce qui diminue la taille de la puce et la taille du champ à imager en sortie. On trouve que la fréquence maximale $f_{max}$ minimale pour arriver à une telle configuration est de vingt fois la fréquence minimale. Pour cette fréquence optimale, il existe une seule configuration possible (parmi 35) répondant à ce critère : $\alpha$=[0,2,8,13,17,20].$\alpha_{\min}$. Une telle configuration occupe certes un espace spectral plus important mais nécessite aussi moins de pixels pour échantillonner le signal à la sortie du dispositif. Le choix d'une telle configuration est donc extrêmement dependant des autres paramètres du système complet.
- A huit telescopes, la fréquence maximale $f_{max}$ minimale que l'on peut utiliser pour obtenir une répartition angulaire sans recouvrement de spectre dans le domaine des fréquences spatiales est de trente-quatre fois la fréquence minimale. On peut noter ici un léger gain par rapport à ce qui est connu dans la littérature [59]. Pour cette fréquence optimale, l'unique répartition possible est : $\alpha$=[0,1,4,9,15,22,32,34].$\alpha_{min}$.

Nous allons maintenant voir comment dimensionner un dispositif à partir de ces répartitions angulaires.

### *IV – D – 2 – c – Dimensionnement du dispositif*

La sortie du dispositif à recombinaison planaire correspond à une ligne imagée sur la fente d'entrée d'un spectrographe à l'aide d'un système optique standard, puis envoyée après dispersion spectrale sur une caméra infrarouge. Cette caméra effectue l'échantillonnage du signal dans la direction 'x' correspondant aux franges d'interférence et dans la direction 'y' perpendiculaire au plan du dispositif et correspondant au domaine de longueurs d'onde. Dans la direction 'y', l'épaisseur de la ligne correspond à la taille $\omega_y$ du mode fondamental de la zone planaire. La taille de pixel $l_{pix}$ impose alors le grandissement du système optique et donc aussi l'échantillonnage dans la direction 'x' des franges d'interférence si l'on considère un système optique sans anamorphose.

$$l_{pix} = \omega_y \qquad \text{(IV-4)}$$

Ensuite, la taille du pixel étant fixée, l'interfrange minimum $i_{min}$ correspondant aux faisceaux d'entrée faisant un angle maximal entre eux doit être échantillonné sur quatre pixels si l'on veut être en mesure de déterminer l'amplitude et la phase des franges d'interférence. Ainsi, la répartition angulaire est entièrement déterminée par le critère d'échantillonnage des franges. On trouve :

$$\alpha_{\min} = \frac{\alpha_{\max}}{17} = \frac{2}{17}\sin^{-1}\left(\frac{\lambda}{8nl_{pix}}\right) \qquad \text{(IV-5)}$$

où $n$ est l'indice effectif du mode planaire et $\lambda$ est la longueur d'onde.

Le dernier dimensionnement concerne la longueur de la zone planaire (le paramètre $R_{min}$ de la figure IV-11). Puisque l'angle entre les guides d'entrée est déterminé, ce paramètre va fixer la distance $d_{min}$ entre les guides d'entrée qui doit être suffisamment grande pour que les guides d'entrée n'échangent pas d'énergie entre eux avant la zone planaire.

$$R_{\min} = \frac{d_{\min}}{2\sin\left(\frac{\alpha_{\min}}{2}\right)} \approx \frac{d_{\min}}{\alpha_{\min}} \qquad \text{(IV-6)}$$

Le dimensionnement du dispositif est entièrement contraint par ces trois critères. On trouve alors en considérant une propagation de faisceau gaussien dans la zone planaire la taille du champ total $l_{champ}$ à $1/e^2$ en intensité qui permet de dimensionner le système optique de sortie :

$$l_{champ} = \sqrt{2}\omega_x \sqrt{1 + \left( \frac{17\lambda d_{\min}}{2n\pi\omega_x^2 \sin^{-1}\left( \lambda / 8nl_{pix} \right)} \right)^2} \qquad \text{(IV-7)}$$

où $\omega_x$ représente la taille du mode fondamental du guide d'entrée dans la direction du substrat.

Le tableau ci-dessous regroupe les valeurs pour des recombineurs multiaxiaux à six télescopes pour nos guides dans les bandes H et K.

| recombineur | 6T multiaxial | | | |
|---|---|---|---|---|
| bande spectrale | *H* | | *K* | |
| echantillonnage fréquence maximum | *4 pixels* | *6 pixels* | *4 pixels* | *6 pixels* |
| largeur pixel dans le plan objet [µm] | *3,5* | | | |
| distance minimum entre guides [µm] | *40* | | *60* | |
| angle minimum [°] | *0,27* | *0,18* | *0,37* | *0,25* |
| angle maximum [°] | *4,55* | *3,04* | *6,35* | *4,17* |
| longueur zone planaire [mm] | *8,55* | *12,8* | *9,2* | *14* |
| largeur du champ en sortie [mm] | *1,11* | *1,67* | *1,43* | *2,16* |
| largeur zone planaire en sortie [mm] | *3* | *4,5* | *4* | *5* |
| nombre de franges à fréquence min. | *3,9* | | *5* | |
| nombre de franges à fréquence max. | *66,7* | | *85,6* | |

*Tableau IV-7 : dimensionnement des multiaxiaux dans les bandes H et K. L'échantillonnage est identique en bande H et K car le système optique sera conçu pour les deux bandes spectrales. Le nombre de franges varie très peu en fonction de l'échantillonnage.*

La taille de la zone planaire reste raisonnable en bande H comme en bande K. Si on désire échantillonner le signal sur six pixels par franges à la fréquence de franges la plus élevée, on trouve un champ évidemment plus grand puisque le grandissement du système optique reste inchangé.

### *IV – D – 2 – d – Vérification numérique*

Nous avons simulé à la BPM un tel composant en bande H afin de confirmer ces calculs. La figure IV-12 montre des exemples de propagation dans un tel dispositif.

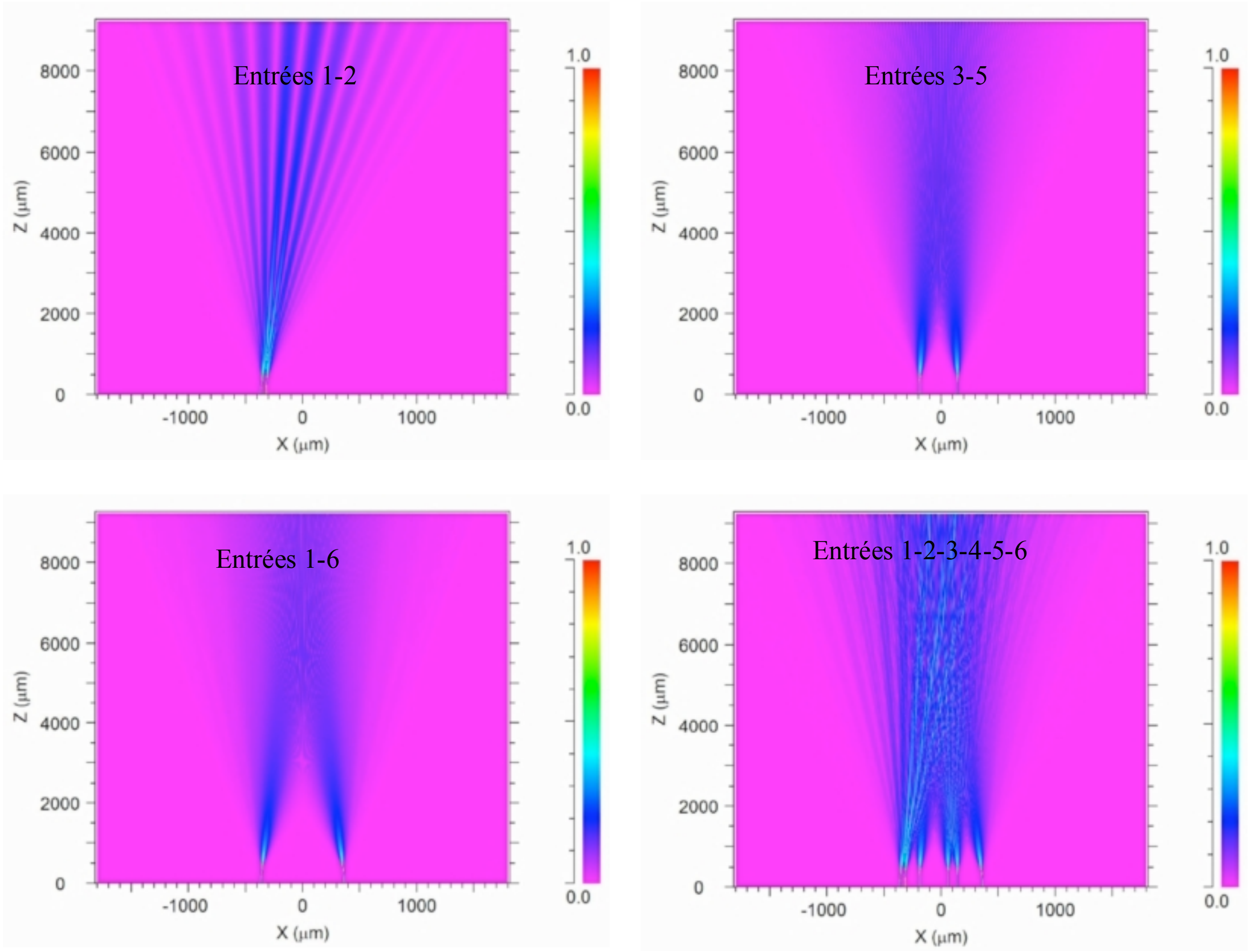

*Figure IV-12 : exemples de propagation BPM d'un recombineur multiaxial 6T en bande H. La propagation utilise un maillage à 0.1µm dans la direction transverse à la propagation. La longueur d'onde est de 1,6µm. L'image en haut à gauche correspond aux entrées les plus proches donc à la fréquence de franges la plus basse. La résolution de l'image en bas à gauche ne permet pas de résoudre les franges à la fréquence la plus haute.*

Le cas le plus intéressant correspond à la propagation lorsque l'on excite simultanément les six entrées. Dans ce cas, nous avons récupéré la répartition du champ en sortie que nous avons traitée de manière à simuler une pixellisation par une caméra. Le graphe de gauche de la figure IV-13 montre la répartition d'intensité obtenue sur la caméra en supposant des pixels de 3,5µm. Le graphe de droite montre sa transformée de Fourier (limitée à 1024 pixels).

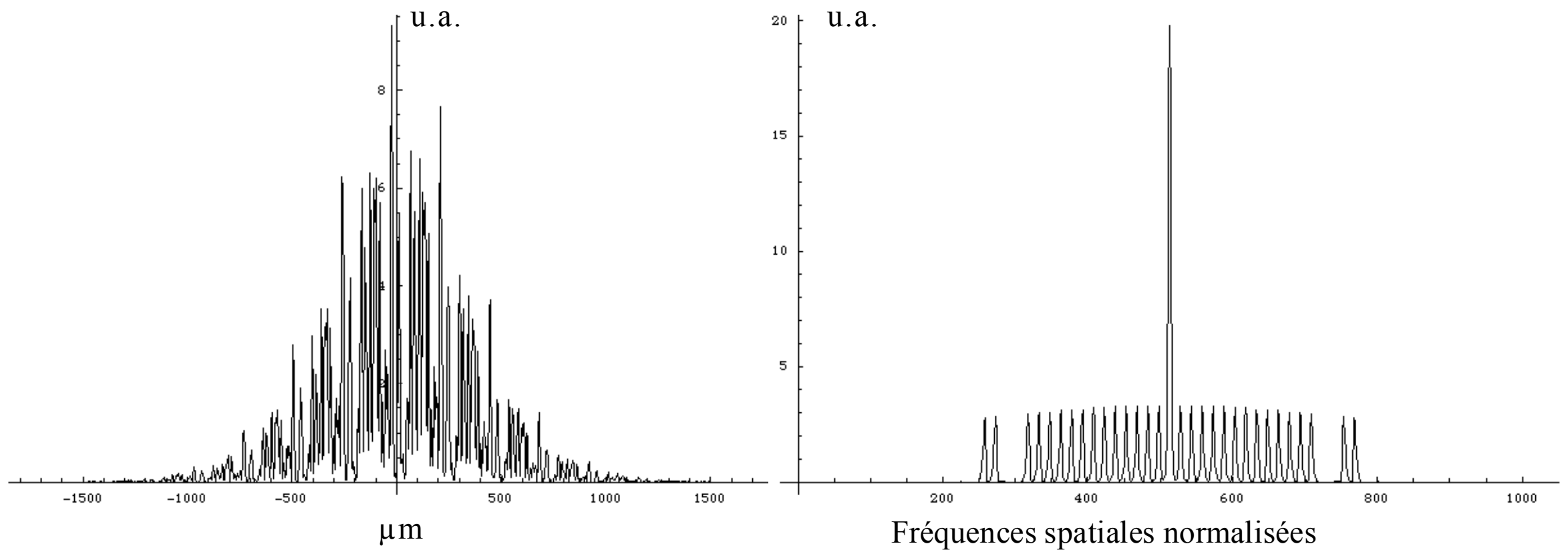

*Figure IV-13 : intensité en sortie de zone planaire et sa transformée de Fourier. Les pics fréquentiels pour chaque paire recombinée se détachent clairement et ne se recouvrent pas.*

Les pics de fréquences spatiales de chaque paire recombinée apparaissent clairement, et ne comportent aucun recouvrement. On note une légère baisse de ces pics fréquentiels aux hautes fréquences spatiales dues à l'échantillonnage du signal.

Si l'on simule le composant conçu pour un échantillonnage à 6 pixels par franges, on obtient les graphes de la figure IV-14.

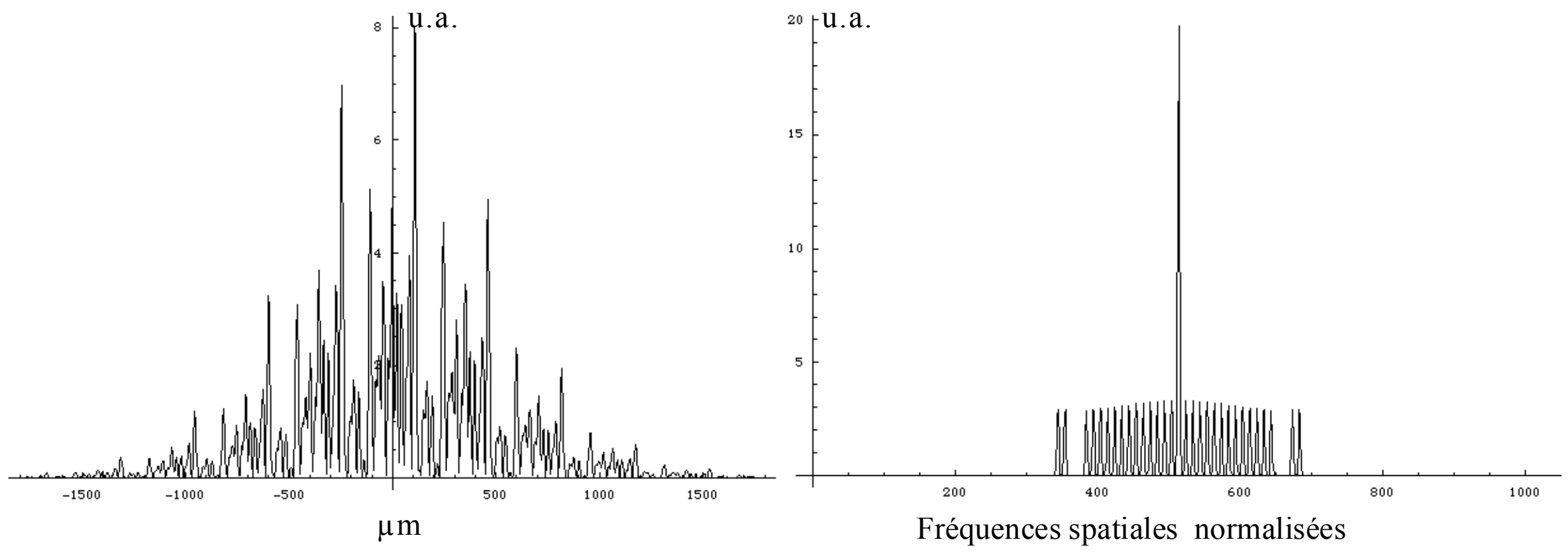

*Figure IV-14 : résultat pour un 6T dimensionné pour un échantillonnage à 6 pixels par franges.*

On note que les pics fréquentiels sont plus resserrés, mais ne se recouvrent toujours pas.

### *IV – D – 2 – e – Influence de la longueur d'onde*

Les dispositifs sont conçus en prenant des valeurs à la longueur d'onde moyenne de la bande considérée H ou K. Nous avons simulé leur comportement toujours à la BPM à des valeurs de longueur d'onde en bordure de spectre afin de vérifier que nous n'avions pas de recouvrement de spectres. Les graphes de la figure IV-15 montrent les spectres obtenus pour différentes longueurs d'onde toujours pour un dispositif en bande H.

Il apparaît que la fréquence spatiale des franges diminue lorsque la longueur d'onde augmente mais qu'aucun recouvrement de spectre n'a lieu même aux longueurs d'onde les plus élevées, ceci même dans le cas du dispositif dimensionné pour un échantillonnage à six pixels par frange.

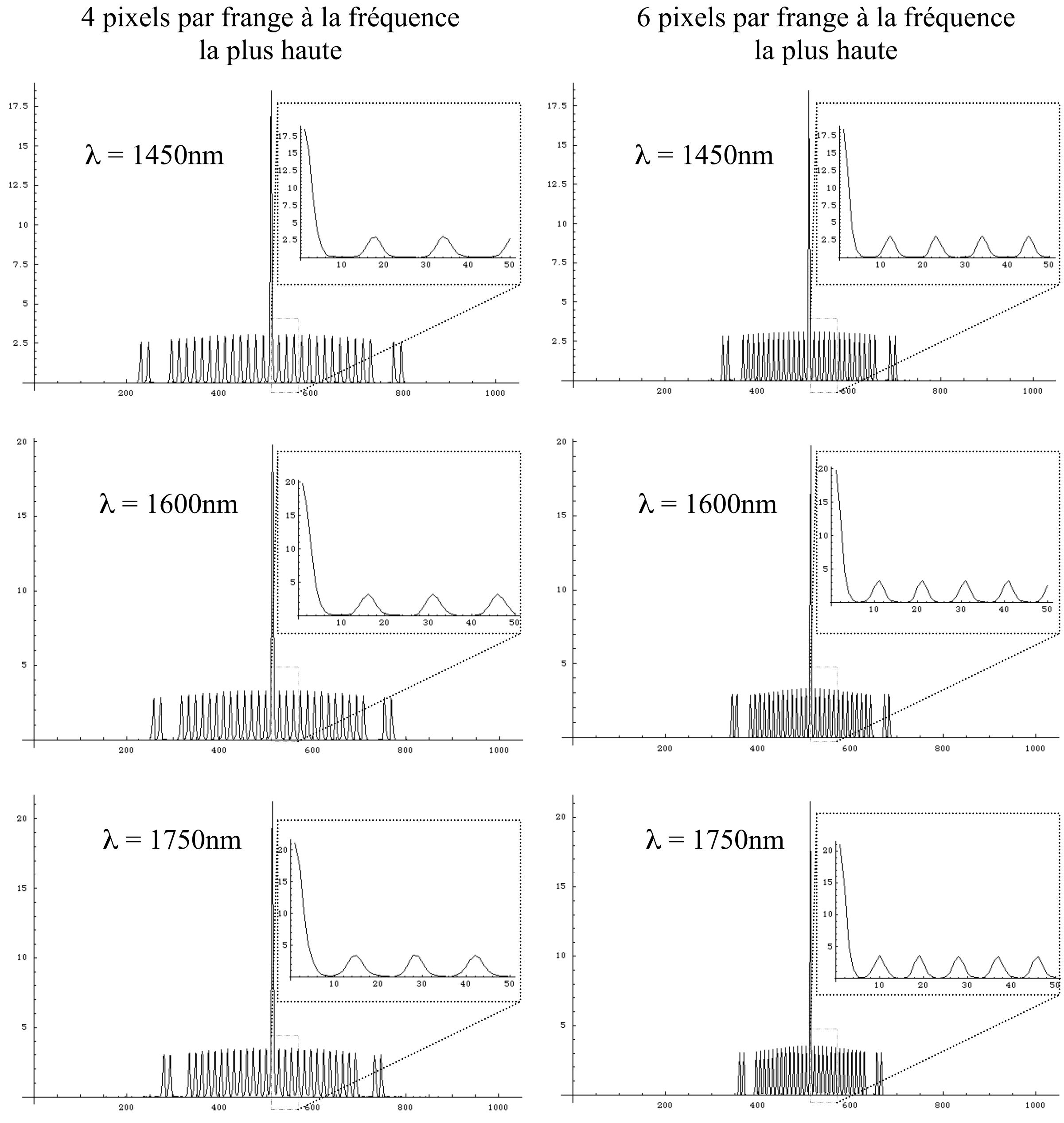

*Figure IV-15 : évolution des spectres de franges en fonction de la longueur d'onde. La fréquence spatiale diminue lorsque la longueur d'onde augmente mais les pics restent bien séparés même aux longueurs d'onde les plus grandes.*

### *IV – D – 2 – f – Calcul de la bande spectrale*

Si pour une longueur d'onde donnée, il n'apparaît aucun recouvrement de pic de fréquence spatiale, la position de ces pics dépend de la longueur d'onde. Les pics se resserrent lorsque la longueur d'onde augmente. Ainsi, pour des longueurs d'onde plus grandes, le pic à la fréquence spatiale la plus élevée va se rapprocher du pic de fréquence spatiale adjacent pour une longueur d'onde plus courte. On peut ainsi définir une bande spectrale d'utilisation en fixant une valeur limite à ce recouvrement (figure IV- 16)

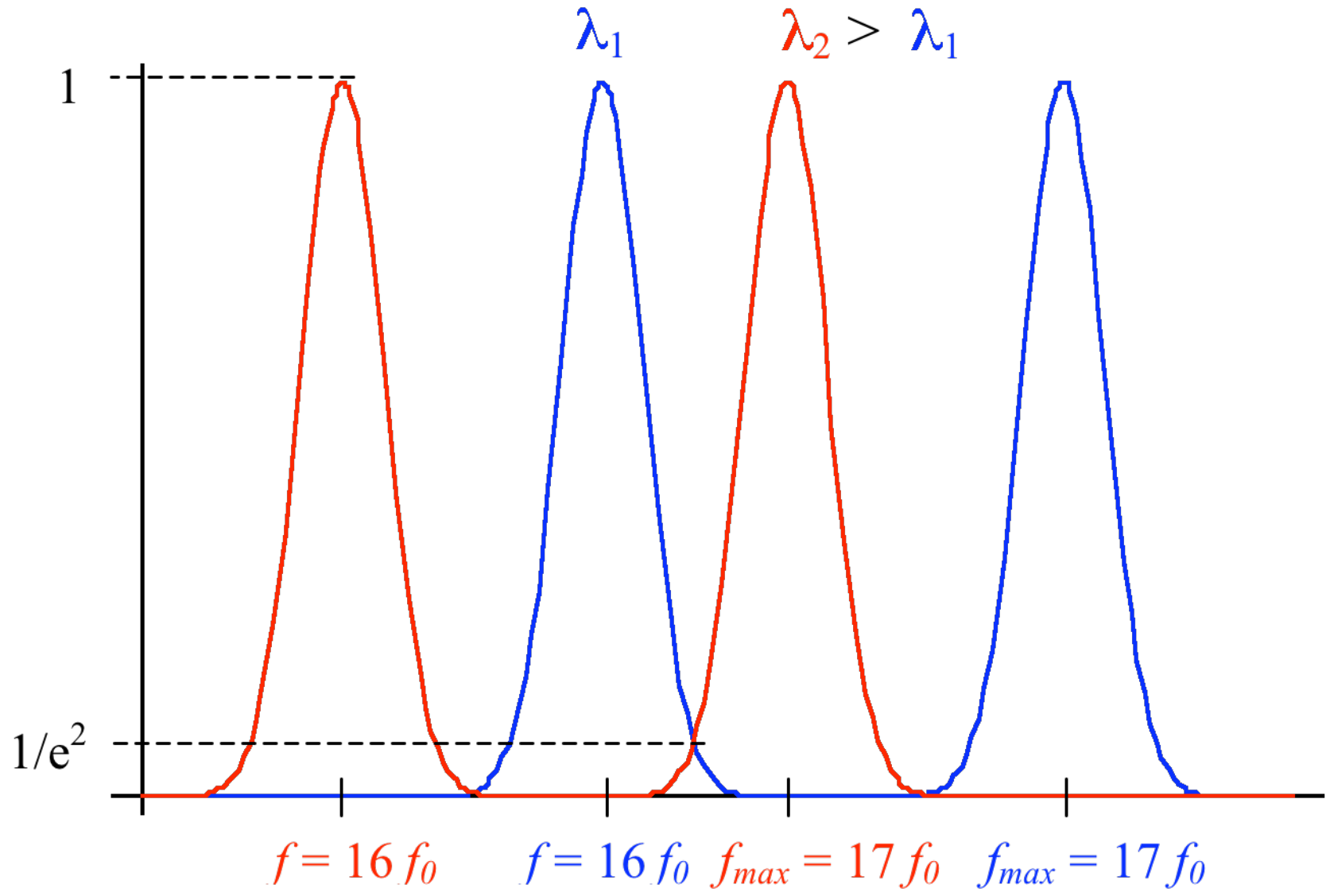

*Figure IV- 16 : recouvrement de spectre pour deux longueurs d'onde. Pour la longueur d'onde la plus grande, les pics de fréquence spatiale des franges se déplacent vers des fréquences plus basses et viennent recouvrir le pic de fréquence adjacent.*

Nous avons vu que l'intensité en sortie de zone planaire pouvait s'apparenter à une gaussienne modulée par les franges d'interférence dont les caractéristiques sont regroupées dans le tableau IV- 7. Les pics de fréquences spatiales sont donc aussi apparentés à des gaussiennes de largeur $\omega_f$ à 1/e séparés d'une fréquence constante $f_0$ données par :

$$\omega_f = \frac{2\sqrt{2}}{\pi l_{champ}} \qquad \text{(IV-8)}$$

$$f_0 = \frac{2n}{\lambda}\sin\left(\frac{\alpha_{\min}}{2}\right) \approx \frac{n\alpha_{\min}}{\lambda} \qquad \text{(IV-9)}$$

où $\alpha_{min}$ et $l_{champ}$ sont données par (IV-5) et (IV-7). En fixant que les pics doivent être au minimum éloignés de la largeur à $1/e^2$ d'un pic fréquentiel (ce qui correspond approximativement à un recouvrement de densité spectrale de puissance de 1%), on obtient une largeur de bande spectrale $\delta\lambda$ :

$$\delta\lambda = \frac{\lambda}{16}\left(1 - \frac{8\lambda}{\pi l_{champ} n\alpha_{\min}}\right) \qquad \text{(IV-10)}$$

Les résultats sont reportés dans le tableau IV-8.

| **recombineur** | **6T multiaxial** | | | |
|---|---|---|---|---|
| **bande spectrale** | *H* | | *K* | |
| **échantillonage fréquence maximum** | *4 pixels* | *6 pixels* | *4 pixels* | *6 pixels* |
| **largeur de bande (nm)** | *46* | *46* | *79* | *81* |

*Tableau IV-8 : largeur de bande spectrale d'utilisation des recombineurs multiaxiaux.*

Les valeurs sont relativement larges mais nécessitent tout de même une dispersion spectrale en sortie de puce pour une observation sur l'ensemble d'une fenêtre de transparence H ou K.

### *IV – D – 2 – g – Dispositif complet*

Contrairement aux dispositifs par paires où les signaux photométriques des entrées peuvent être obtenus à partir des sorties interférométriques, il est ici nécessaire de prélever une partie des faisceaux incidents pour obtenir la photométrie des signaux.

Le signal interférométrique étant étalé en sortie sur un grand nombre de pixels, chaque pixel recevra seulement une petite quantité de photons. Si le calcul du ratio de prélèvement photométrique optimal n'est pas encore possible, car dépendant de paramètres du système qui ne sont pas encore déterminés, on peut cependant prévoir un pourcentage sur les voies photométriques plutôt bas. Pour une première réalisation, nous avons fixé le ratio de prélèvement de 10%. Là encore, nous pouvons obtenir une telle fonction en utilisant un coupleur asymétrique optimisé. La figure IV-17 donne les résultats de simulation d'un tel coupleur en bande H.

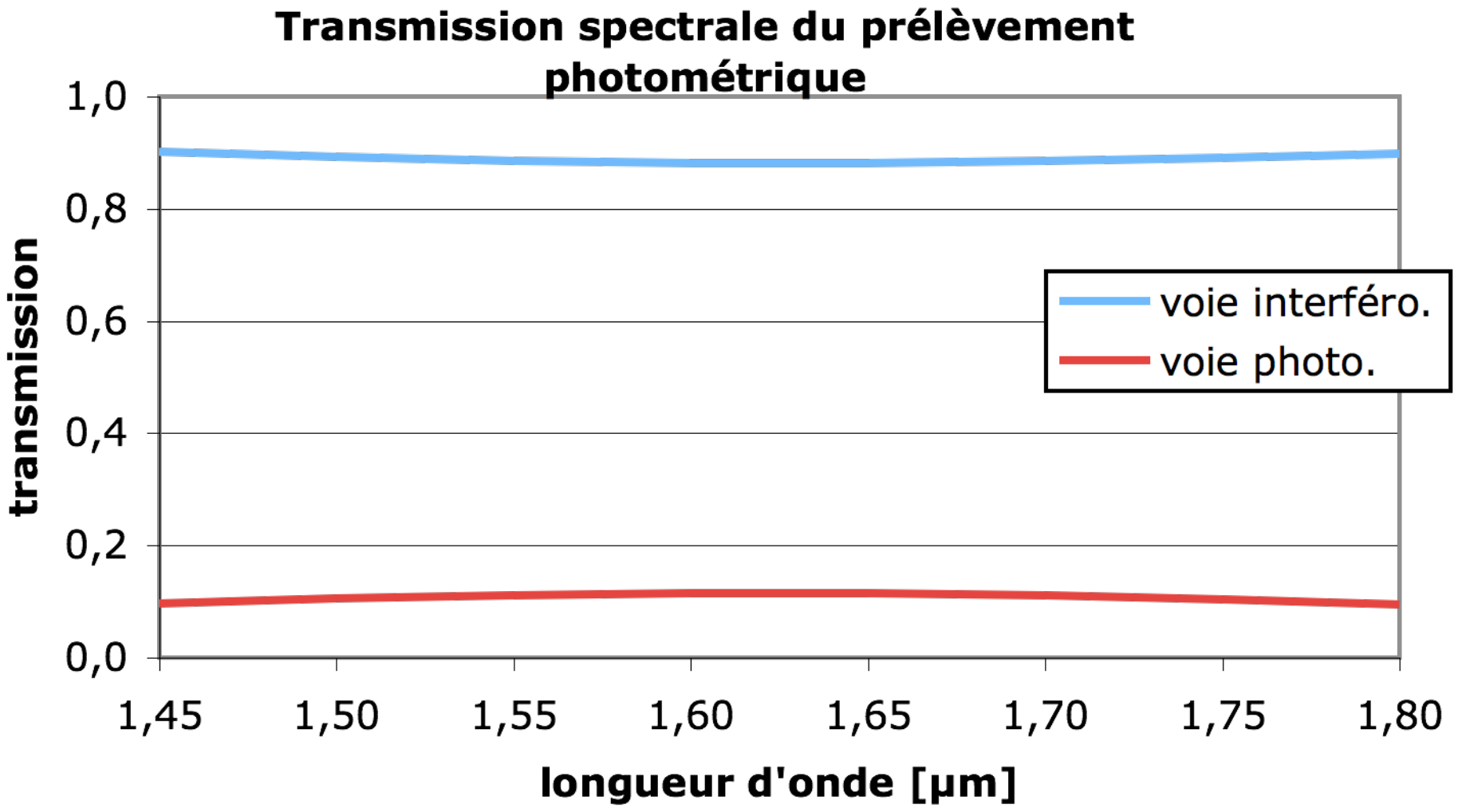

*Figure IV-17 : ratio de prélèvement des voies photométriques en fonction de la longueur d'onde.*

Enfin la figure IV-18 montre une conception détaillée du dispositif complet en bande H. La puce fait 28mm de long sur 4mm de large (le dispositif complet en bande K sur le même principe fait 45mm de long sur 7mm de large). Les voies photométriques sont toutes reportées du même coté afin de limiter le champ sur la caméra et sont espacées de 60µm les unes des autres. Les croisements de guide s'effectuent à 30°, et pour cette première réalisation, les chemins optiques ne seront pas égalisés à l'intérieur de la puce.

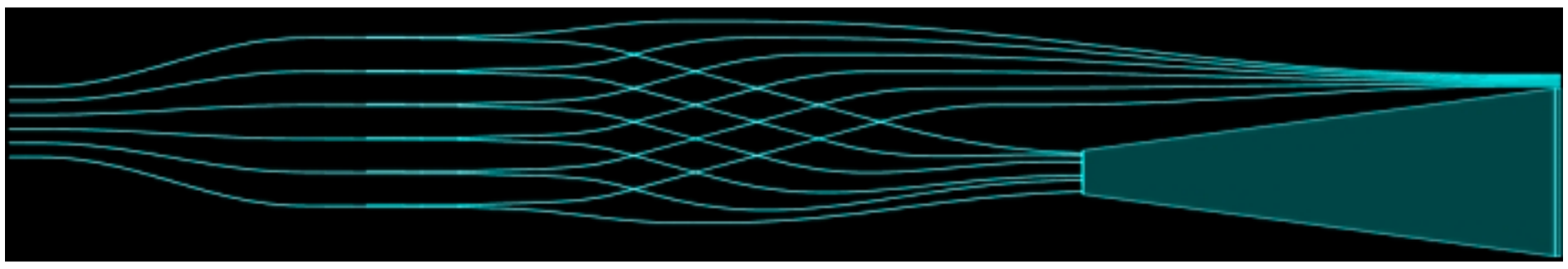

*Figure IV-18 : schéma complet du recombineur 6T multi axial en bande H.*

En conclusion, les recombineurs multiaxiaux à six télescopes sont parfaitement réalisables en optique intégrée sur silicium quelle que soit la bande spectrale J, H, ou K. Un premier dimensionnement optimisant la sortie de la puce à l'entrée d'un spectrographe conduit à une zone planaire de dimension environ égale à un centimètre de long et fait intervenir des angles faibles. En conséquence, on peut déjà prédire que de tels composants seront de dimensions bien plus réduites que des recombineurs par paire de type ABCD, et devraient donc présenter moins de pertes de propagation.

## IV – D – 3 - Choix de la configuration

Le choix définitif du mode de recombinaison du futur instrument VSI n'est bien sûr pas encore arrêté à l'heure actuelle.

D'une part, l'instrument VSI est consitué d'un bon nombre de sous-systèmes interdépendants dont les caractéristiques ne sont pas toutes encore déterminées et de ce point de vue, l'étude des dispositifs exposés dans ce chapitre permet de participer à la définition de ces sous-systèmes.

En effet, le choix d'un mode de recombinaison par paires conduit à une puce complexe mais fournissant en sortie des signaux interférométriques parfaitement séparés dans différents canaux ordonnés. Si ce mode est optimal à quatre télescopes, il est encore réalisable à six télescopes voire huit télescopes, mais conduit à des tailles de puces extrêmes et un nombre de croisements de guides élevé limitant la transmission globale. A l'inverse, le choix d'un mode de recombinaison tout en un multiaxial conduit à une puce beaucoup plus simple, de taille plus réduite et a priori à des transmissions globales plus élevées. Mais ce mode de recombinaison impose un système optique de sortie ayant un plus grand champ et travaillant à plus grande ouverture numérique, un spectrographe capable de travailler aussi à plus grande ouverture, et enfin une caméra infrarouge comportant un plus grand nombre de pixels.

D'une part, cette étude est pour l'instant théorique et doit être complétée par une étude expérimentale comprenant des caractérisations sur banc optique de dispositifs réalisés. C'est pourquoi nous projetons dans les mois qui viennent, de réaliser des 4T-ABCD modifiés en bande H, ainsi que des dispositifs dédiés aux bandes J et K, et des dispositifs multiaxiaux.

# IV – E – Perspectives d'améliorations possibles

## IV – E – 1 - Amélioration de la cellule ABCD

On peut considérer que les résultats de mesure sur les déphaseurs ont validé le principe de notre composant. Le fait de concaténer plusieurs segments de guides de largeurs différentes conduit en effet à aplatir la réponse en longueur d'onde de ces déphaseurs. Cependant, nous n'avons pas pris en compte un autre effet chromatique susceptible de dégrader le contraste instrumental de la cellule ABCD qui est le chromatisme du coupleur asymétrique lui-même. La figure IV-19 montre une simulation des franges d'interférences obtenues en sortie de coupleur lorsque l'on injecte un signal sur chaque voie du coupleur à différentes valeurs de déphasage, ceci pour différentes longueurs d'onde.

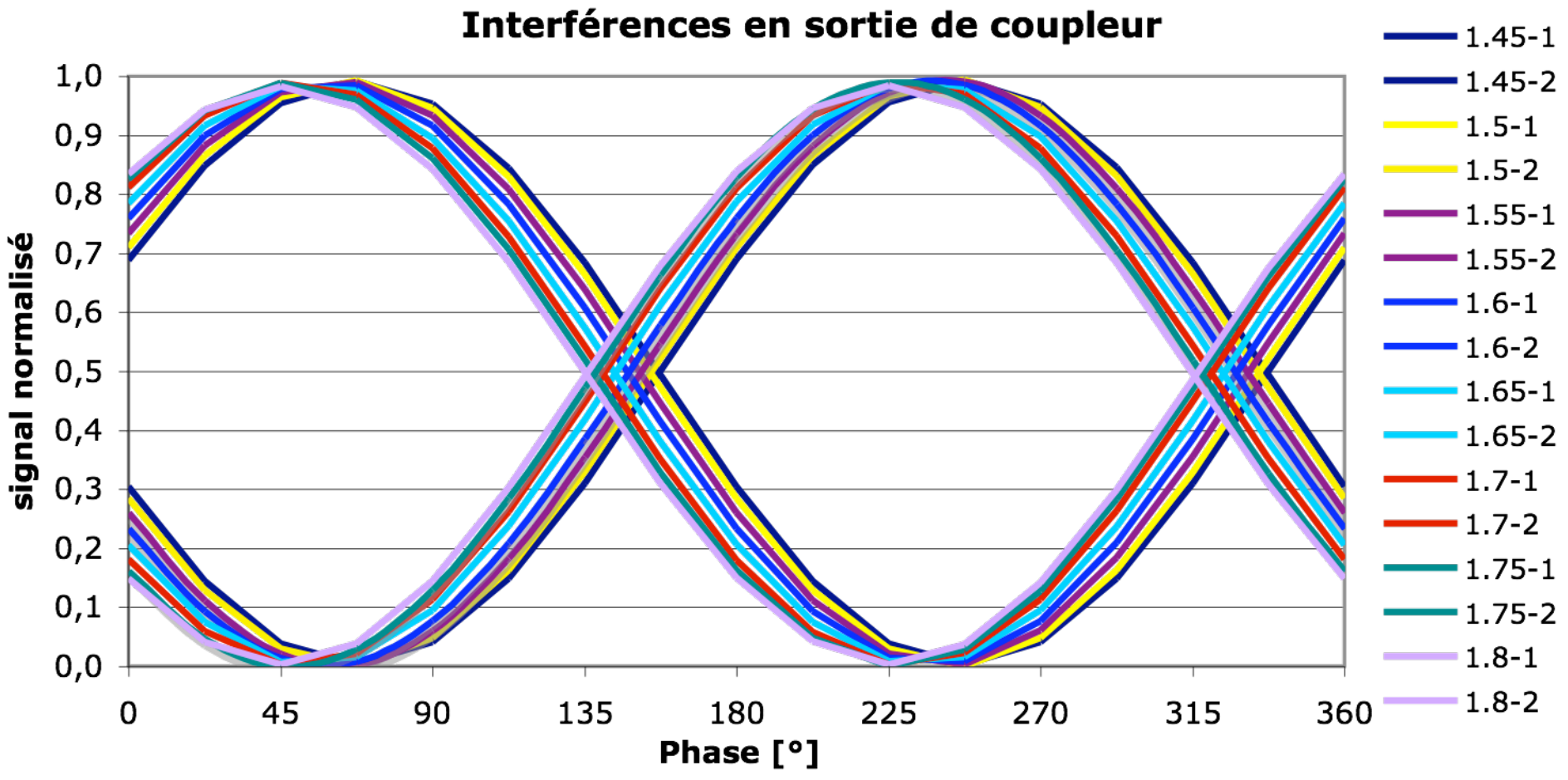

*Figure IV-19 : franges d'interférence en sortie du coupleur asymétrique en fonction de la longueur d'onde.*

Si le coupleur a bien été optimisé en transmission de manière à être le plus près possible d'un ratio de séparation 50/50, il introduit un déphasage chromatique non négligeable sur les franges d'interférences. Ce déphasage introduit par le coupleur est reporté sur le graphe de la figure IV-20.

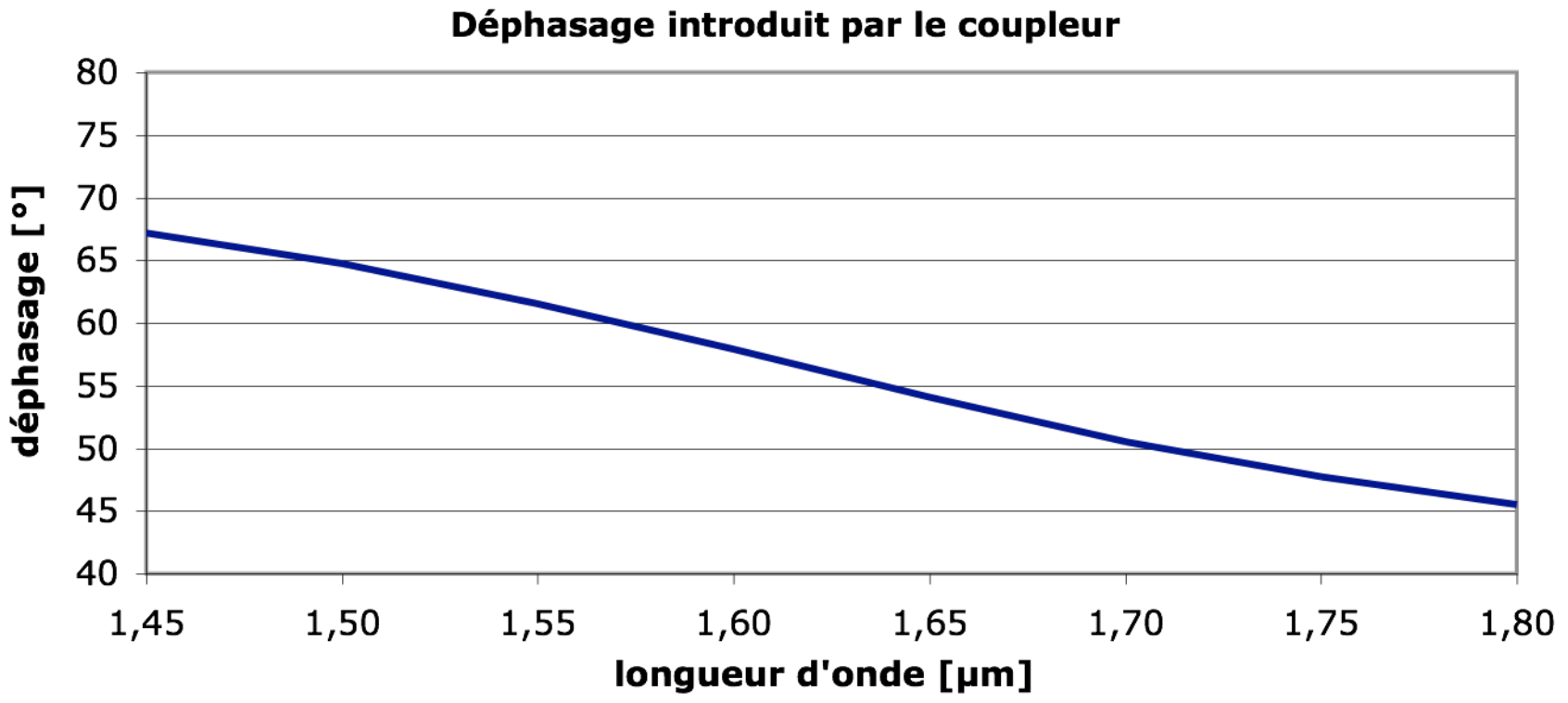

*Figure IV-20 : déphasage du coupleur en fonction de la longueur d'onde.*

Nous avons vu que la méthode de conception des déphaseurs exposée au chapitre II permettait de réaliser des fonctions de déphasage plus complexes qu'un déphasage achromatique. Elle permet entre autres de concevoir un déphaseur capable de compenser la chromaticité du coupleur. Nous avons donc conçu le déphaseur suivant.

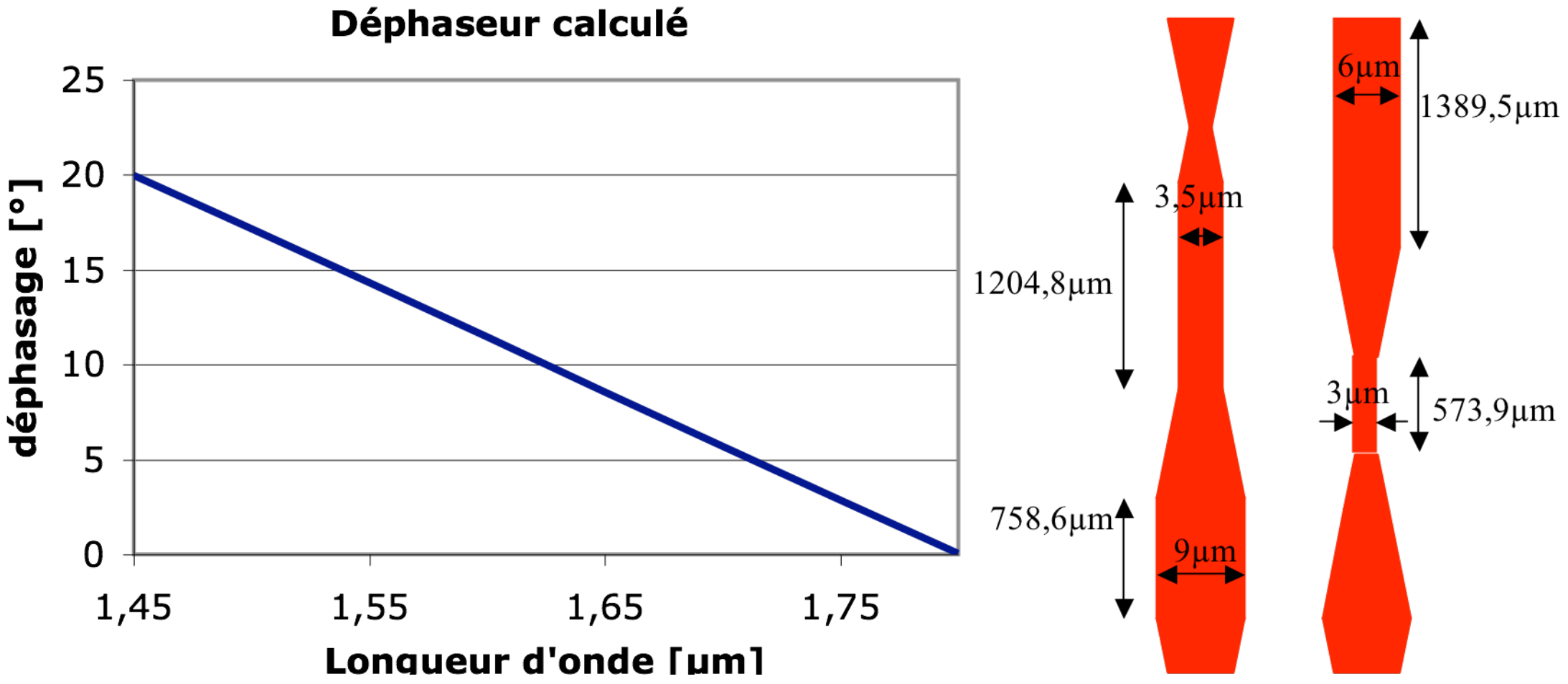

*Figure IV-21 : Déphaseur compensant le chromatisme du coupleur.*

Une simulation globale BPM de franges dans le même coupleur que précédemment incluant un tel déphaseur en entrée donne alors le résultat suivant.

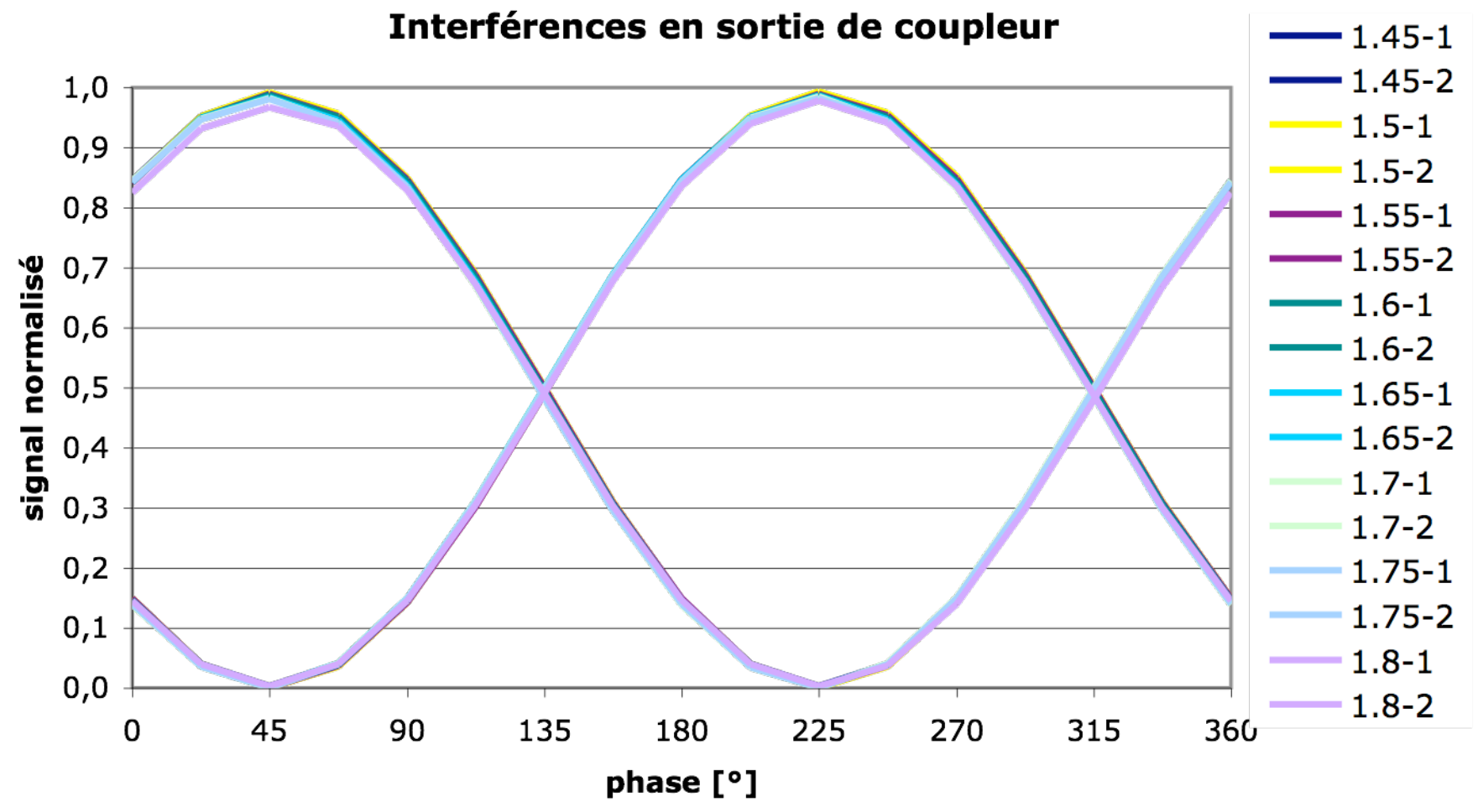

*Figure IV-22 : coupleur simulé avec son déphaseur en entrée. Les franges sont beaucoup mieux superposées en fonction de la longueur d'onde.*

On conserve la bonne optimisation du coupleur pour un taux de couplage 50/50 avec en plus une très bonne dépendance en longueur d'onde. Si l'on regarde le déphasage de ces franges en fonction de la longueur d'onde, on obtient le résultat de la figure IV-23.

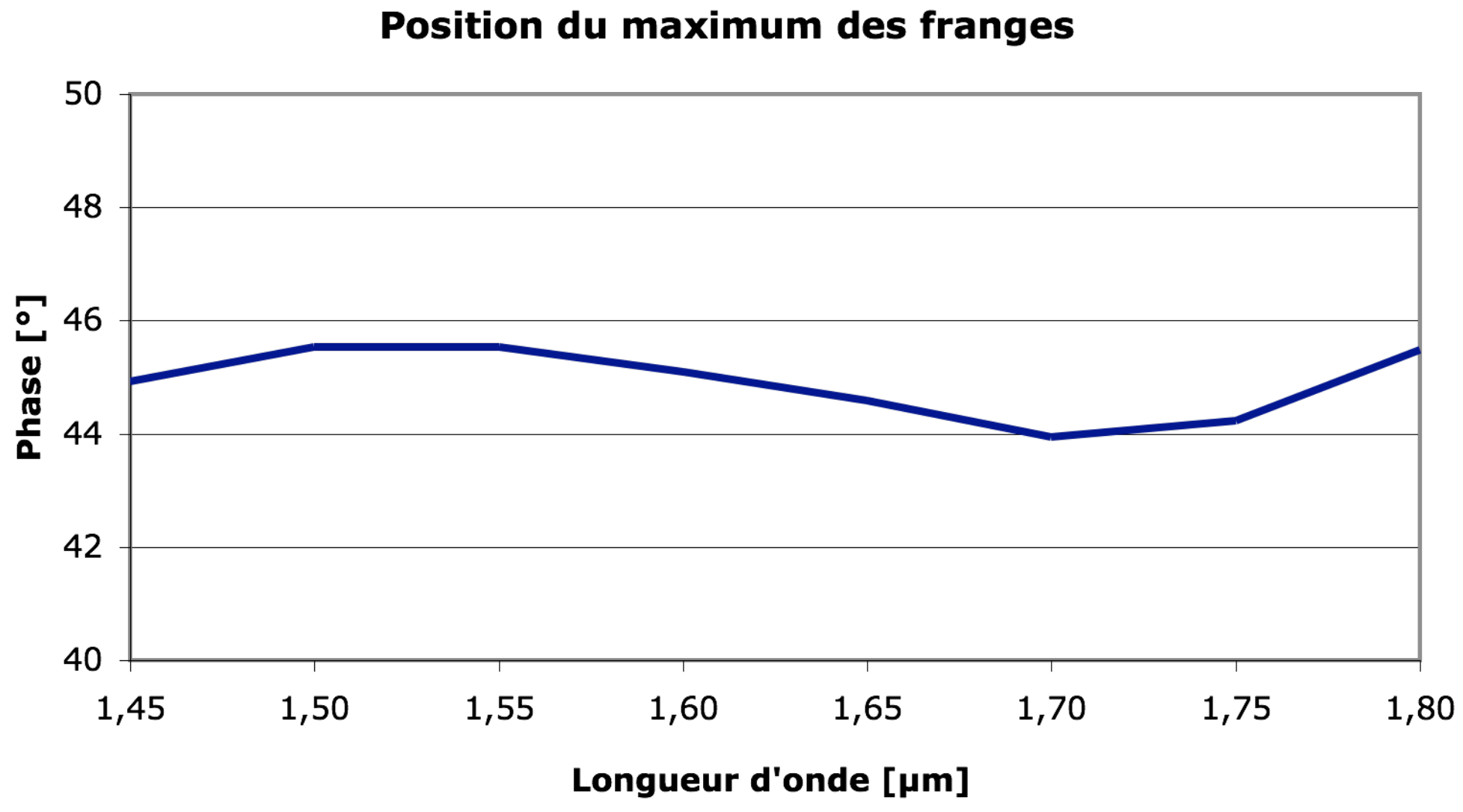

*Figure IV-23 : position du maximum des franges en fonction de la longueur d'onde. La variation résiduelle est inférieure au degré.*

La position du maximum des franges d'interférences est positionnée à 45±1° sur toute la bande H.

Enfin, puisque le déphasage moyen est de 45°, il n'est plus nécessaire d'introduire un déphaseur achromatique 90° dans la cellule ABCD, et il suffit de réaliser une cellule ABCD symétrique par rapport à son axe central (figure IV-24).

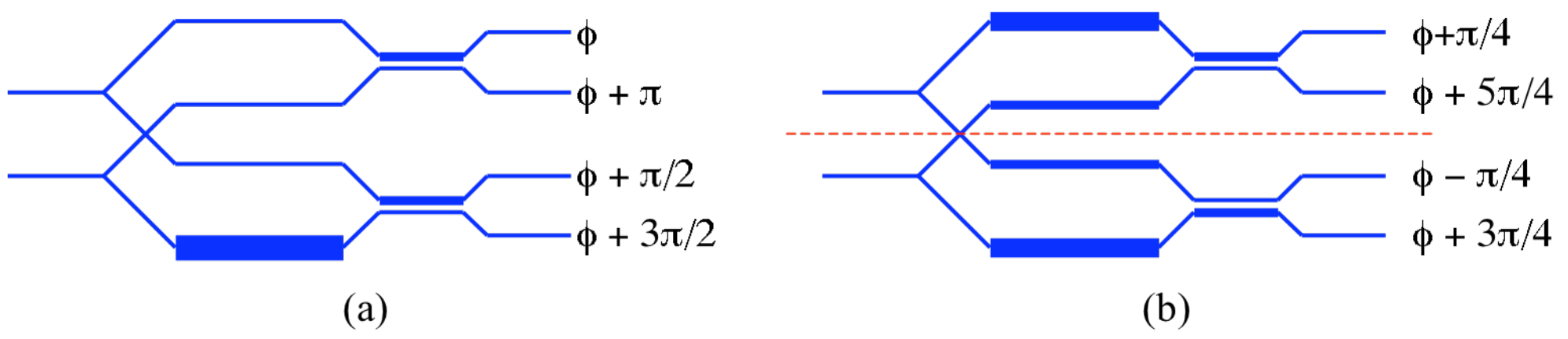

*Figure IV-24 : Cellule ABCD selon l'ancien modèle et cellule ABCD améliorée.*

Nous avons prochainement prévu de réaliser de telles cellules ABCD modifiées. En effet, si les simulations laissent entrevoir des résultats prometteurs et meilleurs que la cellule déjà réalisée et caractérisée, les incertitudes liées à la réalisation technologique risquent de modifier légèrement le comportement du coupleur. Si nous avons vu que nous obtenions des taux de couplage proches des valeurs visées, nous n'avons pour l'instant aucune mesure sur le comportement en phase des coupleurs

## IV – E – 2 – Cellule « ABC »

Nous avons vu que l'une des difficultés rencontrées pour étendre les possibilités des recombineurs à base de cellule ABCD à plus de quatre télescopes était la taille importante des puces qui conduit à des pertes de propagation de plus en plus importantes. Cet encombrement provient entre autres de l'encombrement de la cellule ABCD elle-même qui comporte un croisement de guide donc une largeur assez importante.

En théorie, quatre signaux interférométriques sont redondants pour mesurer la visibilité complexe des franges d'interférences. En effet, trois voies « ABC » devraient suffire, à condition de fournir des signaux interférométriques déphasés en sortie. Nous avons donc regardé ce que pouvait donner un tricoupleur utilisé en recombineur « ABC ».

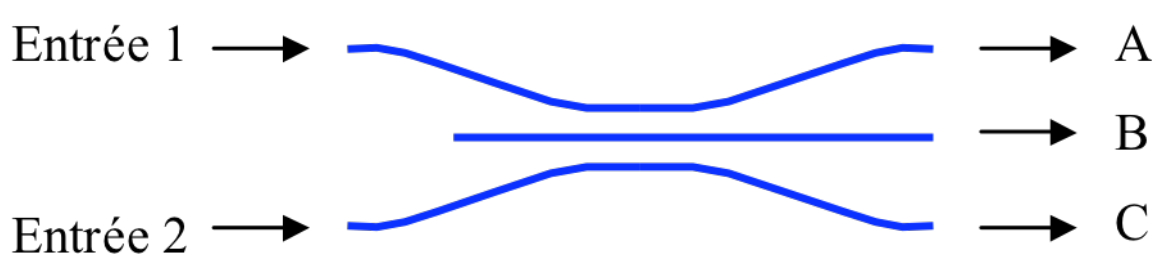

*Figure IV-25 : tricoupleur utilisé en recombineur « ABC ».*

Le graphe de la figure IV-26 montre les signaux interférométriques en sortie du tricoupleur pour différentes longueurs d'onde lorsqu'on injecte des signaux lumineux sur ses deux voies latérales.

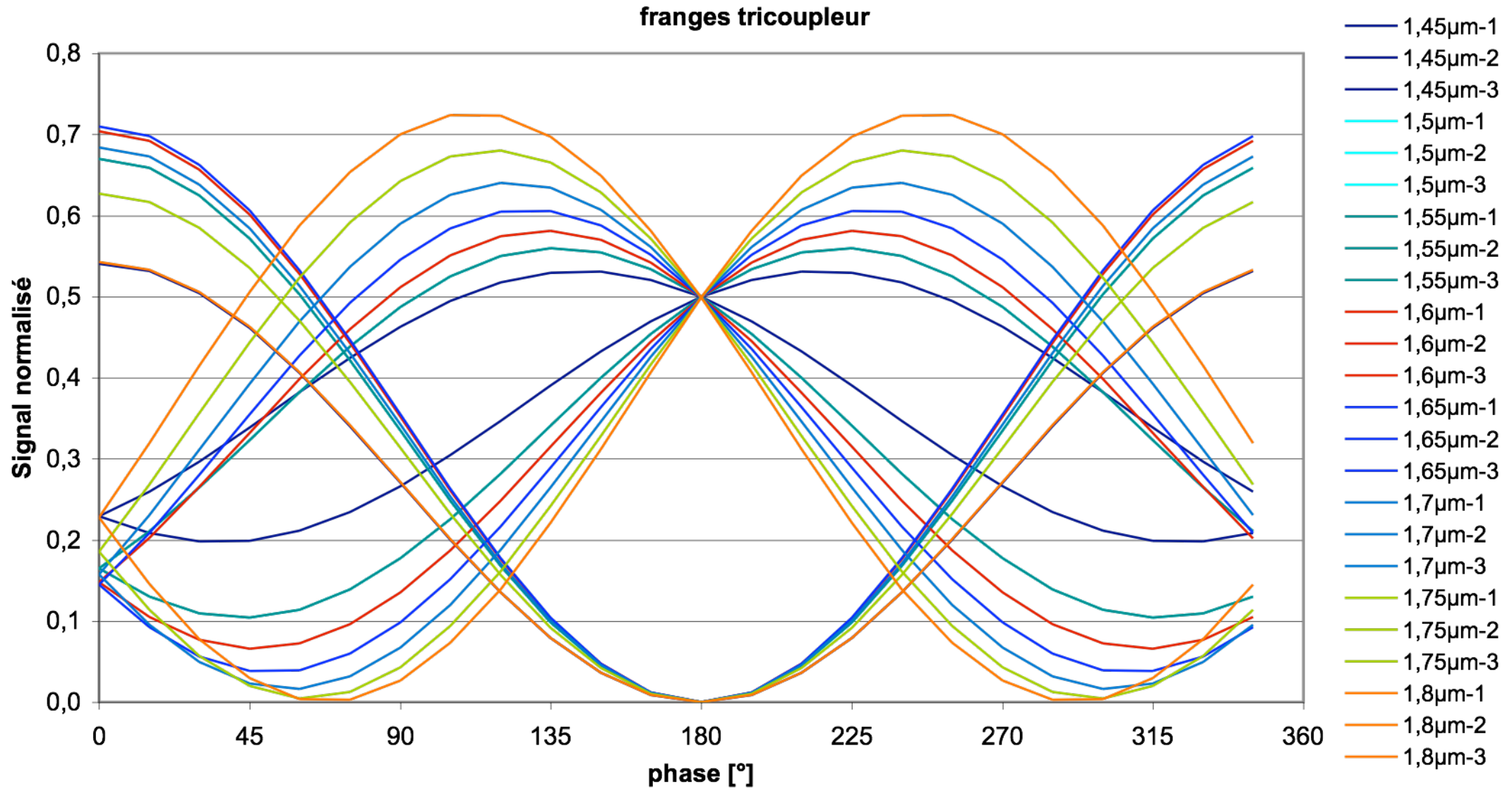

*Figure IV-26 : réponse interférométrie d'un tricoupleur excité sur ses voies latérales pour différentes longueurs d'onde.*

Le tricoupleur fournit bien trois sorties interférométriques complémentaires déphasées. Il est donc normalement possible à partir de ces trois signaux d'obtenir la phase des franges d'interférences.

Par ailleurs, le fonctionnement est similaire à celui d'un coupleur, c'est à dire que la somme des signaux des trois voies donne par conservation d'énergie la somme des intensités entrant dans le tricoupleur. On peut donc comme dans le cas des coupleurs asymétriques remonter à la photométrie des signaux incidents.

Enfin, du fait de la symétrie du dispositif, la frange noire sur la voie centrale du tricoupleur est parfaitement achromatique et conduit donc à un contraste instrumental unitaire optimal pour la mesure de visibilité. Par contre, le comportement des voies latérales est relativement chromatique mais devrait pouvoir être amélioré par optimisation du « design ». Le graphe de la figure IV-27 montre les sorties du tricoupleur en fonction de la phase entre les signaux d'entrée lorsque l'on intègre le signal en longueur d'onde.

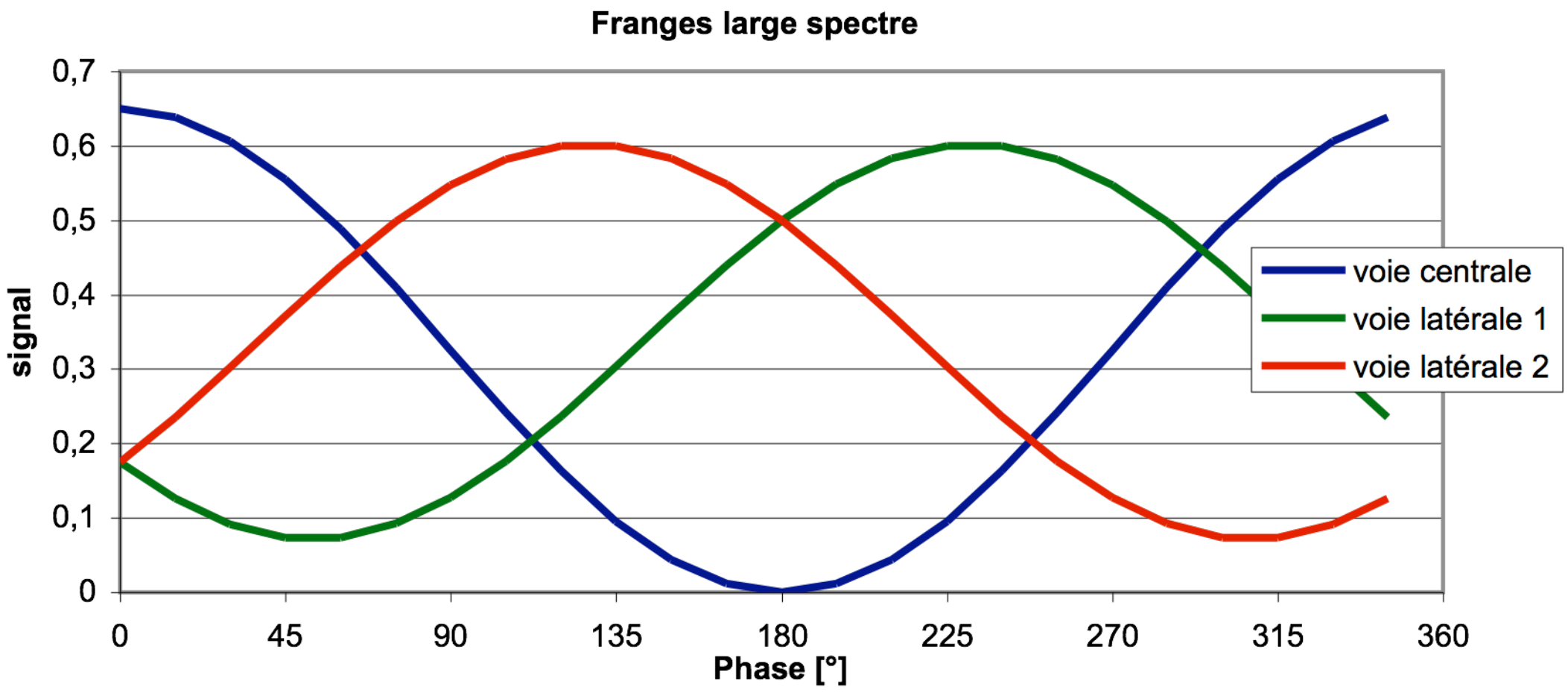

*Figure IV-27 : franges obtenues en sortie de tricoupleur en fonction de la phase entre les deux entrées lorsque l'on intègre le signal en longueur d'onde.*

On note que le contraste est bien unitaire sur la sortie centrale du tricoupleur. Malgré la chromaticité sur les voies latérales, le contraste en bande large est de 0,79 sur les voies latérales. Ce composant présente donc à notre avis un potentiel intéressant. Non seulement il présente les fonctionnalités requises pour la mesure de phase des franges mais de plus, son encombrement est beaucoup plus réduit que celui d'une cellule ABCD. Il permet donc d'envisager avec plus de sérénité l'extension des recombineurs par paires à six télescopes, voire huit télescopes. Du point de vue purement photométrique, la transmission d'un tricoupleur est forcément meilleure que celle d'une cellule ABCD puisqu'il ne comporte ni jonction Y, ni croisement de guides, et la transmission globale des dispositifs sera aussi améliorée du fait de la taille moins grande des puces.

Enfin, le nombre de sorties pour chaque paire recombinée est minimal, on étale globalement les signaux sur un nombre minimal de canaux et on atteint ainsi, au moins théoriquement, le meilleur rapport signal sur bruit photométrique qu'il soit possible d'obtenir quel que soit le mode de recombinaison envisagé.

D'un point de vue purement optique, cette solution semble meilleure. Il reste avant d'envisager son utilisation de vérifier son intérêt d'un point de vue système. En effet, du point de vue du traitement du signal, une cellule ABCD fournit des signaux idéalement répartis pour obtenir une mesure de phase robuste, même en présence de signaux fortement bruités. L'analyse doit donc être approfondie dans ce domaine.

## IV – F - Recombineur optique intégrée pour suiveur de franges

Un suiveur de franges est un système permettant de compenser les fluctuations du chemin optique dues aux turbulences atmosphériques ou à la variation des conditions environnementales de l'interféromètre. Il est composé d'un système de mesure en temps réel de la position des franges d'interférences sur une partie des photons prélevés sur les faisceaux issus des télescopes (souvent dans une autre bande spectrale que la bande d'observation) et d'un système permettant de faire varier le chemin optique en temps réel, ceci afin de maintenir ces franges stables.

Un tel système présente l'avantage considérable de pouvoir moyenner le signal sur de longues périodes, améliorant sensiblement le rapport signal sur bruit des mesures sur des faibles flux de photons et permet ainsi d'observer des sources astronomiques moins brillantes. Les performances d'un suiveur de franges conditionnent donc de manière très importante les performances globales de l'instrument de recombinaison, voire de l'interféromètre complet.

Afin d'avoir la sensibilité nécessaire pour effectuer une mesure instantanée, un suiveur de franges utilise une bande spectrale large pour mesurer les franges et utilise moins de recombinaisons que l'instrument d'observation scientifique. Pour un système à N télescopes, il n'est en effet pas nécessaire pour un suiveur de franges de mesurer les franges de toutes les paires de bases possibles, mais seulement les franges de N bases (N-1 sont théoriquement suffisantes, mais on prend généralement 1 base redondante en plus qui s'avère utile en cas de décrochage du système dû à une fluctuation de phase trop importante) [*70*].

Un recombineur pour suiveur de franges est donc un dispositif simplifié par rapport à un recombineur scientifique, mais qui doit être capable d'extraire instantanément la phase des mesures de franges effectuées sur une bande spectrale large. L'utilisation des résultats obtenus sur les recombineurs par paire de type ABCD vient donc immédiatement à l'esprit. La cellule ABCD permet de mesurer instantanément la phase des franges d'interférences et les améliorations envisagées dans ce chapitre permettent d'envisager son utilisation en bande spectrale large.

Le schéma le plus simple que l'on peut utiliser pour un suiveur de franges est représenté sur la figure IV-28 pour le cas de quatre télescopes (il est trivialement extensible à plus de télescopes)

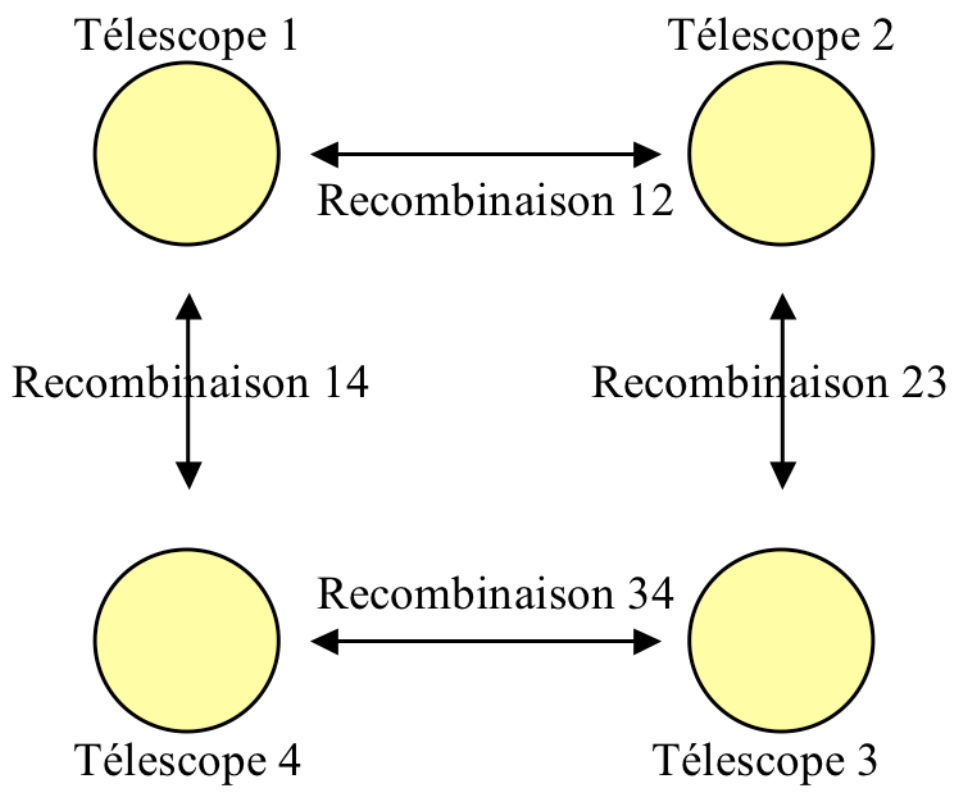

*Figure IV-28 : schéma de recombinaison pour un suiveur de franges.*

Chaque voie doit être séparée en deux avant d'aller se recombiner avec ses deux télescopes voisins. Le routage d'un tel dispositif en optique intégrée est représenté sur la figure IV-29 pour le cas de quatre télescopes.

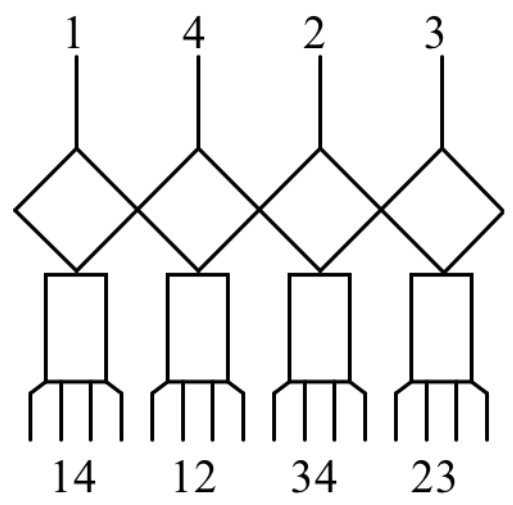

*Figure IV-29 : recombineur optique intégré pour suiveur de franges à quatre télescopes.*

Il suffit de choisir judicieusement l'ordre des entrées dans le dispositif pour minimiser le nombre de croisements dans la puce et optimiser la transmission.

Le principe est extensible à plus de télescopes sans compliquer de manière importante le routage. Une conception détaillée en bande K conduit au dispositif de la figure IV-30.

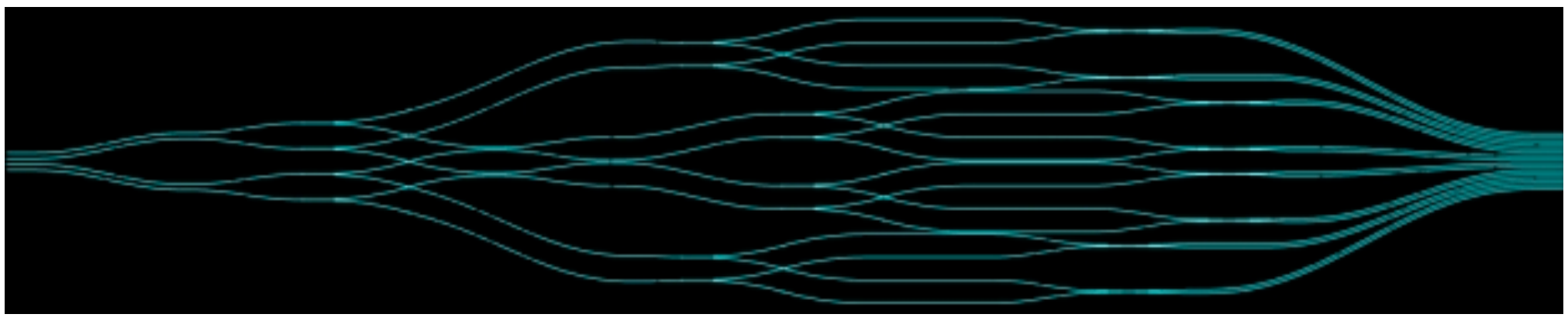

*Figure IV-30 : recombineur suiveur de franges en bande K pour quatre télescopes.*

Un tel dispositif est entièrement symétrique, utilise des jonctions Y parfaitement achromatiques pour la séparation des faisceaux, et conserve une égalité rigoureuse des chemins optiques pour chaque paire recombinée dans la puce. La taille totale de la puce est de 70mm par 13mm en bande K. Etendre le principe à 6 ou 8 télescopes ne présente pas de difficultés car il n'est pas nécessaire de recombiner toutes les paires de base comme c'est le cas pour les recombineurs scientifiques.

## IV – G - Conclusion

Dans ce chapitre, nous avons cherché à explorer vers quelles voies nous pouvions étendre les possibilités de l'optique intégrée pour la recombinaison astronomique.

Dans un premier temps, les résultats expérimentaux obtenus au chapitre III nous ayant permis de valider nos méthodes de conception, nous avons envisagé l'extension des dispositifs 4T-ABCD aux bandes J et K. Si cette extension ne pose aucune difficulté particulière en bande J, les résultats ont fait apparaître que les dimensions des dispositifs tendaient à être importantes en bande K. Nous avons donc cherché à améliorer ce point en retravaillant le routage des dispositifs pour améliorer l'intégration.

Cette amélioration sur la taille des dispositifs nous a permis d'envisager dans un deuxième temps l'extension de ce principe de recombinaison par paires matricielle à six télescopes voire, huit télescopes. Si nous avons montré que des dispositifs 6T-ABCD sont réalisables dans les trois bandes spectrales J, H, K, il apparaît là encore que la taille des dispositifs est le facteur limitant vers plus de télescopes.

Nous nous sommes alors orientés vers la méthode de recombinaison multiaxiale. Nous avons effectué le dimensionnement complet de ces dispositifs et avons montré que ce principe conduisait à des dispositifs à six télescopes beaucoup plus simples et parfaitement réalisables, la contrepartie étant que la complexité est en quelque sorte reportée à la sortie de la puce où il sera nécessaire de concevoir un système optique d'imagerie possédant un grand champ et travaillant à une ouverture numérique importante. En conséquence, le choix de la configuration finale de recombinaison pour VSI reste à déterminer en fonction des caractéristiques des autres sous-systèmes de l'instrument de recombinaison.

Dans la dernière partie, nous avons mentionné quelques voies possibles d'amélioration des performances qu'il serait intéressant d'explorer pour compléter ce travail. Nous avons proposé une voie d'amélioration des cellules « ABCD » afin de les rendre encore plus achromatiques en utilisant un déphaseur compensant la chromaticité du coupleur. Les performances obtenues sont telles qu'il devient alors envisageable d'utiliser ces cellules « ABCD » pour un suiveur de franges en optique intégrée dont nous avons décrit la géométrie. Enfin, comme alternative à la cellule « ABCD » au centre de ce travail, nous avons envisagé l'utilisation d'un tricoupleur à ondes évanescentes comme cellule de recombinaison simplifiée qui permet la mesure instantanée de la phase des franges d'interférences tout en diminuant considérablement l'encombrement global des dispositifs de recombinaison par paires.

# Conclusion

Notre travail est situé dans le cadre du développement de composants optiques intégrés pour la recombinaison de faisceaux en interférométrie astronomique. Basé sur les résultats obtenus par IONIC et son recombineur intégré à trois télescopes, le but était dans un premier temps de concevoir, réaliser, puis caractériser un recombineur pour un réseau de quatre télescopes. Une première partie de ce travail a donc été de comparer les différents modes de recombinaison afin de choisir le plus adapté aux possibilités de l'optique intégrée.

Cette première étude nous a permis de sélectionner un mode de recombinaison matricielle par paires pour lequel nous avons optimisé sur une bande spectrale large, les différents composants de base dont nous avions besoin. En particulier, nous avons mis en place une méthode générale de conception permettant de générer dans un chemin optique un déphasage contrôlé en fonction de la longueur d'onde. A l'aide de cette méthode, nous avons conçu un déphaseur achromatique intégré qui nous a permis de réaliser une recombinaison matricielle achromatique. Nous avons alors appliqué cette recombinaison matricielle à la conception d'un recombineur par paires utilisant un coupleur à trois guides dont nous avons aussi optimisé la réponse sur une bande spectrale large.

La réalisation de ces dispositifs a bénéficié des dernières améliorations que nous venions d'accomplir sur notre technologie de réalisation silice sur silicium, notamment de l'utilisation de substrats de silicium de huit pouces et nous a permis d'obtenir des composants avec un nombre extrêmement réduit de défauts. En effet, bien que quelques composants aient montré certaines performances dégradées, tous les composants de recombinaison à deux ou quatre télescopes réalisés par ce « lot » technologique se sont avérés opérationnels.

La caractérisation de ces dispositifs a été effectuée en deux temps. Nous avons d'abord effectué des caractérisations photométriques fibre à fibre sur un banc au LETI et avons obtenu des résultats très prometteurs avec de très bonnes transmissions globales (60% fibre à fibre pour un 4T-ABCD) et des ratios de coupleur, ou de coupleur à trois guides conformes aux prédictions théoriques à quelques pourcents près, validant ainsi notre méthode de conception et la technologie de réalisation. La caractérisation interférométrique de ces dispositifs a nécessité le développement d'un banc de caractérisation au LAOG, banc qui simule un interféromètre stellaire capable de recombiner jusqu'à huit faisceaux prélevés dans la pupille d'un gros collimateur. Ce banc a permis la caractérisation exhaustive de plus de soixante-dix déphaseurs. Les résultats ont montré des valeurs de déphasage peu dispersées autour de 75° au lieu des 90° visés et un excellent comportement achromatique. Ces résultats, qui nous ont permis de valider la méthode de conception de ces déphaseurs, ont aussi permis de détecter que les déphaseurs situés en bord de plaque avaient un comportement sensiblement différent des autres, en terme de polarisation, révélant un possible phénomène de gradient de contrainte dans nos couches jusque là indétectable. Malgré ce défaut, les puces « 4T-ABCD » réalisées se sont toutes avérées pleinement fonctionnelles, c'est à dire capables de mesurer instantanément l'amplitude et la phase des franges d'interférence des six paires de bases recombinées.

Ces résultats nous ont permis d'aller plus loin et d'effectuer une étude théorique complète des performances des recombineurs en bande J, H et K pour quatre puis six télescopes. Nous avons même proposé un schéma de recombinaison pour huit télescopes. S'il apparaît que des recombineurs à six télescopes sont tout à fait envisageables dans ces trois bandes spectrales, les résultats font apparaître que les tailles des puces en bande K sont très importantes même en optimisant l'intégration de nos dispositifs. Nous nous sommes alors intéressés au mode de recombinaison multiaxial pour lequel nous avons montré que nous pouvions envisager des puces plus simples aux dimensions plus réduites au prix d'un système optique de sortie plus complexe.

Enfin, nous avons proposé d'autres voies d'améliorations possibles comme l'amélioration du comportement de la cellule de recombinaison de type ABCD en introduisant un ensemble « déphaseur coupleur », où encore une simplification de cette cellule ABCD en une cellule à trois sorties déphasées permettant aussi de mesurer instantanément la phase des franges. De par son encombrement beaucoup plus faible, une telle cellule à trois sorties résoudrait les problèmes de taille des recombineurs par paires. Par ailleurs, puisque ce travail a permis la mise au point d'une mesure de phase de franges d'interférences performante, il nous semble intéressant d'appliquer ces résultats aux systèmes de suivi de franges pour lesquels nous pourrions réaliser des recombineurs simplifiés avec des transmissions globales optimisées.

Le travail effectué au cours de cette thèse confirme l'intérêt de la technologie de composants d'optique intégrée en silice sur substrat de silicium pour réaliser des circuits de recombinaison interférométrique pour la synthèse d'ouverture en astronomie. La faisabilité de la recombinaison d'un réseau possédant jusqu'à huit télescopes est en très bonne voie. Cette étude a permis de fournir tous les éléments nécessaires aux études de phase A des instruments imageurs VSI[*67*] et GRAVITY[*71*] pressentis pour le VLTI. Une comparaison détaillée des avantages et inconvénients des solutions classiques en optique de volume et solutions en optique intégrée basée sur ce travail de thèse a conclu à la supériorité de l'optique intégrée. En particulier, le concept de circuit d'optique intégrée 4T-ABCD a été reconnu par les deux consortiums pour être au cœur des deux instruments. Ces études ont de plus révélé que les concepts d'instruments « suiveurs de franges » capables d'assurer le cophasage, à la fraction de longueur d'onde près, d'un réseau de télescopes gagneraient fortement à utiliser des circuits de recombinaison en optique intégrée. Des pré-études de tels systèmes sont envisagées à l'horizon 2009. Enfin, « l'European Interferometry Initiative » a entâmé un travail de réflexion prospective sur un « Kilometric Optical Interferometer » constitué d'une vingtaine de télescopes sur des bases kilométriques. En se fondant sur les résultats présentés ici, elle a retenu les technologies optique intégrée comme un élément clé de la future instrumentation focale d'un tel interféromètre.

# Liste des figures

# Liste des tableaux

# Références

### *Autres publications associées à ce travail*